\newif\ifconfver
\confverfalse      
\confvertrue        

\newif\ifplainver  
\plainvertrue

\ifplainver
    \confverfalse   
\fi

\ifconfver
     \documentclass[10pt,twocolumn,twoside]{IEEEtran}
\else
    \ifplainver
        \documentclass[11pt]{article}
        \usepackage{fullpage}
    \else
        \documentclass[12pt,draftcls,onecolumn]{IEEEtran}
    \fi
\fi

\usepackage{calc,amsfonts,amssymb,amsmath,bm,url,color,theorem,graphicx,cite}
\usepackage{enumerate}
\usepackage{psfrag,float}
\usepackage{algorithm}
\usepackage{array}
\usepackage{algorithmic}
\usepackage{subcaption,epstopdf}
\usepackage{xr}
\usepackage{soul}

\definecolor{orange}{RGB}{255,107,0}

\newtheorem{Fact}{Fact}
\newtheorem{Lemma}{Lemma}
\newtheorem{Prop}{Proposition}

\theorembodyfont{\rmfamily}

\newtheorem{Remark}{Remark}

\newcommand\bq{\ensuremath{{\bm q}}}
\newcommand\bx{\ensuremath{{\bm x}}}
\newcommand\by{\ensuremath{{\bm y}}}

\newcommand\bh{\ensuremath{{\bm h}}}

\newcommand\be{\ensuremath{{\bm e}}}

\newcommand\bR{\ensuremath{{\bm R}}}

\newcommand\bX{\ensuremath{{\bm X}}}

\newcommand\bC{\ensuremath{{\bm C}}}
\newcommand\bc{\ensuremath{{\bm c}}}
\newcommand\ba{\ensuremath{{\bm a}}}

\newcommand\bA{\ensuremath{{\bm A}}}

\newcommand\bb{\ensuremath{{\bm b}}}

\newcommand\bB{\ensuremath{{\bm B}}}
\newcommand\blam{\ensuremath{{\bm \lambda}}}

\newcommand\bD{\ensuremath{{\bm D}}}
\newcommand\bu{\ensuremath{{\bm u}}}

\newcommand\bY{\ensuremath{{\bm Y}}}

\newcommand\bs{\ensuremath{{\bm s}}}

\newcommand{\Zbb}{\mathbb{Z}}

\newcommand{\Cbb}{\mathbb{C}}

\newcommand{\setS}{\mathcal{S}}

\newcommand{\Exp}{\mathbb{E}}
\newcommand{\jj}{\mathfrak{j}}
\newcommand{\dec}{\mathrm{dec}}
\newcommand{\Diag}{\mathrm{Diag}}

\newcommand\br{\ensuremath{{\bm r}}}

\newcommand{\bzero}{{\bm 0}}
\newcommand{\bone}{{\bm 1}}
\newcommand{\bI}{{\bm I}}

\newcommand\sgn{\ensuremath{{\rm sgn}}}

\newcolumntype{M}[1]{>{\centering\arraybackslash}m{#1}}

\makeatletter
\def\bstctlcite{\@ifnextchar[{\@bstctlcite}{\@bstctlcite[@auxout]}}
\def\@bstctlcite[#1]#2{\@bsphack
  \@for\@citeb:=#2\do{%
    \edef\@citeb{\expandafter\@firstofone\@citeb}%
    \if@filesw\immediate\write\csname #1\endcsname{\string\citation{\@citeb}}\fi}%
  \@esphack}
\makeatother
\begin{document}

\bibliographystyle{IEEEtran}

\newcommand{\papertitle}{
One-Bit Sigma-Delta MIMO Precoding}

\newcommand{\paperabstract}{
Coarsely quantized MIMO signalling methods have gained popularity in the recent developments of massive MIMO as they open up opportunities for massive MIMO implementation using cheap and power-efficient radio-frequency front-ends.
This paper presents a new one-bit MIMO precoding approach using spatial Sigma-Delta ($\Sigma\Delta$) modulation.
In previous one-bit MIMO precoding research,
one mainly focuses on using optimization to tackle the difficult binary signal optimization problem that arises from the precoding design.
Our approach attempts a different route.
Assuming angular MIMO channels, we apply $\Sigma\Delta$ modulation---a classical concept in  analog-to-digital conversion of temporal signals---in space.
The resulting $\Sigma\Delta$ precoding approach has two main advantages:
First, we no longer need to deal with binary optimization in $\Sigma\Delta$ precoding design.
Particularly, the binary signal restriction is replaced by peak signal amplitude constraints.
Second, the impact of the quantization error can be well controlled via modulator design and under appropriate operating conditions.
Through symbol error probability analysis, we reveal that the very large number of antennas in massive MIMO provides favorable operating conditions for $\Sigma\Delta$ precoding.
In addition, we develop a new $\Sigma\Delta$ modulation architecture that is capable of adapting the channel to achieve nearly zero quantization error for a targeted user.
Furthermore, we consider multi-user $\Sigma\Delta$ precoding using the zero-forcing and symbol-level precoding schemes.
These two $\Sigma\Delta$ precoding schemes perform considerably better than their direct one-bit quantized counterparts, as simulation results show.
}


\ifplainver


    \title{\papertitle}

    \author{
    Mingjie Shao$^\dag$, Wing-Kin Ma$^\dag$, Qiang Li$^\ddag$, and Lee Swindlehurst$^\star$ \\ ~ \\
    $^\dag$Department of Electronic Engineering, The Chinese University of Hong Kong, \\
    Hong Kong SAR of China \\ ~ \\
    $^\ddag$School of Information and Communications Engineering, \\
    University of Electronic Science and Technology of China, China \\ ~ \\
    $^\star$Department of Electrical Engineering and Computer Science, \\
    University of California, Irvine, USA \\ ~ \\
    E-mails: mjshao@ee.cuhk.edu.hk, wkma@ee.cuhk.edu.hk, lq@uestc.edu.cn,  \\
    swindle@uci.edu
    }

    \maketitle

    \begin{abstract}
    \paperabstract
    \end{abstract}

\else
    \title{\papertitle}

    \ifconfver \else {\linespread{1.1} \rm \fi

    \author{Mingjie Shao,  Wing-Kin Ma, Qiang Li, and Lee Swindlehurst}

    \maketitle

    \ifconfver \else
        \begin{center} \vspace*{-2\baselineskip}
        \end{center}
    \fi

    \begin{abstract}
    \paperabstract
    \\
    \end{abstract}

    \begin{IEEEkeywords}\vspace{-0.0cm}
        massive MIMO, one-bit MIMO, Sigma-Delta modulation, MIMO precoder design
    \end{IEEEkeywords}

    \ifconfver \else \IEEEpeerreviewmaketitle} \fi

 \fi

%

%

\section{Introduction}

Recently there has been growing interest in coarsely quantized multi-input multi-output (MIMO) transceiver
implementations for massive MIMO communications systems
that employ very large antenna arrays. These studies are strongly motivated by the
need to reduce the hardware cost and power consumption of radio-frequency (RF)
front-ends---which grow rapidly under massive MIMO---and the idea is to use
low-resolution
analog-to-digital converters (ADCs)/digital-to-analog converters (DACs)
and energy-efficient low-dynamic-range power amplifiers.
A number of researchers have investigated MIMO channel estimation and
MIMO detection using one-bit or low-resolution ADCs
\cite{FanJWZ5,juncil2015near,MollenCLH17,LiTSMSL17,JacobssonDCGS17,ShaoLL18,JeonLHH18},
and it has been found that the very large number of antennas in massive MIMO
indeed helps recover information lost due to the coarsely quantized signals.

MIMO precoding using one-bit DACs is another emerging topic in this area.
A natural direction is to simply quantize the output of a conventional linear precoder,
such as zero forcing (ZF), and the question is how the coarse quantization effects impact
system performance \cite{Nossek2009Transmit,SaxenaFS17,LiTSML17} using, for example,
the Bussgang decomposition as an analysis tool. More recently, there has been emphasis on
directly designing a one-bit precoder, rather than following the aforementioned
precode-then-quantize direction. The direct one-bit precoding designs use criteria such
as minimum mean-square error and minimum symbol error probability
\cite{CastanedaJDCGS17,Jacobsson2017,Swindlehurst2017,LandauL17,JeddaMSN18,li2018massive,Sohrabi2018,shao2018framework},
and numerically these designs were found to yield significantly improved performance.
The challenge with direct one-bit precoding design is mainly centered on the optimization,
which requires finding a good non-convex algorithm to handle a large-scale binary
optimization problem. Promising numerical results have been reported with the direct
one-bit precoding designs, but there is still much to be understood, {\em e.g.}, are the
good numerical results an indication that most of the local minima have good quality,
and if yes when can we guarantee this to happen? We refer the reader to
\cite{Sohrabi2018,shao2018framework} for further descriptions of the various design approaches.

Since we have mentioned one-bit ADCs/DACs for MIMO,
we should also mention the classical one-bit approach for analog-to-digital conversion---Sigma-Delta ($\Sigma\Delta$) modulation.
The $\Sigma\Delta$ modulation approach
exploits the use of oversampled, or low-frequency, signals
in order to reduce the impact of the quantization
noise.
The  $\Sigma\Delta$ principle is to
employ a feedback loop
to quantize the
accumulated error between the input and the one-bit quantized output.
The net effect is to shape
the quantization noise to the high end of the frequency spectrum, where it can be
separated from the signal of interest using a simple low-pass filter and decimator. For background
on the $\Sigma\Delta$ approach and its various generalizations, the reader is referred
to the tutorial article~\cite{aziz1996overview}.

Alternatively, or in addition to quantization noise shaping in temporally oversampled
systems, one can employ the $\Sigma\Delta$ effect using signals oversampled in space
using an array of antennas. In such {\em spatial} $\Sigma\Delta$ architectures, the
feedback signal is derived from the delayed and quantized outputs of adjacent antennas
rather than or in addition to those of the given antenna. Oversampling in this context
means that the elements of a uniform linear array would be spaced closer than
one-half wavelength apart. As a result, the quantization
error can be pushed to higher spatial frequencies, mitigating the distortion
for signals of interest that might arrive from lower spatial frequencies, i.e., those
near the broadside of the array. This idea has been exploited recently by a number
of researchers \cite{Corey_Sig,baracspatial,nikoofard,madan2017}.
Venkateswaran and van der Veen \cite{Venk2011} use the concept in a different way, by beamforming the
one-bit ADC outputs and using this as the feedback signal to each antenna, with the goal
of removing interfering sources. The spatial $\Sigma\Delta$ approach should not be
confused with the multi-antenna architecture of \cite{palguna2016}, in which each
antenna output is modulated by a different Hadamard sequence prior to $\Sigma\Delta$
quantization in time. This is a variation of the approach originally proposed in \cite{GaltonJ95},
that uses a parallel bank of $\Sigma\Delta$ ADCs in order to obviate the need
for temporal oversampling.

The $\Sigma\Delta$ idea has also been used for transmit signal processing. Scholnik {\em et~al.}
\cite{scholnik2004spatio} use space-time $\Sigma\Delta$ DACs to generate one-bit outputs
that directly drive each of the antennas, focusing the resulting quantization noise
to directions and frequencies that do not impact the signal at the desired receiver.
Krieger {\em et al.} \cite{kriegerDense} considered designs of analog beamforming weights
for phased arrays when low-resolution phase shifters are employed. The goal there
is to reduce the error that results from quantization
of the weights of a transmit beamformer, and the weights are generated via $\Sigma\Delta$
quantization assuming a ``dense'' (oversampled) linear array.

Curiously, to the best of the authors' knowledge, the current developments of one-bit massive
MIMO precoding do not seem to have touched upon the possibility of spatial $\Sigma\Delta$ modulation.
It is therefore interesting to explore and understand what opportunities spatial $\Sigma\Delta$ modulation
can bring to one-bit massive MIMO precoding---this is the main objective of this paper.
We summarize our contributions, and compare them with existing literature, below.

\begin{enumerate}[1.]
	\item Our study reveals that one-bit massive MIMO precoding using spatial $\Sigma\Delta$ modulation,
	or simply $\Sigma\Delta$ precoding for short, allows us to effectively mitigate the quantization noise
	effects. More precisely, we consider uniform linear arrays with user angles being within a certain
	``tolerable'' range, say, $[-30^\circ,30^\circ]$. We show that the quantization noise can be substantially
	suppressed when the number of antennas is large. This conclusion resembles that for analog beamforming by Krieger {\em et. al}
	\cite{kriegerDense}, although the context of this work is completely different from that of \cite{kriegerDense}.
	
	\item We generalize the concept of spatial $\Sigma\Delta$ modulation.
	The spatial $\Sigma\Delta$ modulation concept used in the aforementioned literature usually considers direct application of the basic $\Sigma\Delta$ modulation for low-pass temporal signals.
	In this application, the best noise shaping result, in terms of nearly zero quantization noise, is possible only when the signal of interest comes from the broadside.
	We question whether the broadside angle can be altered.
	We develop a new $\Sigma\Delta$ modulation architecture whose angle for nearly zero quantization noise can be changed to any angle,
	and in the single-user case this new modulator allows us to adapt the user angle for achieving nearly zero quantization noise.
	Furthermore, we generalize this angle-steering concept to any type of channel, rather than just the angular channel.
	
	\item The $\Sigma\Delta$ precoding approach allows us to revisit the easier precode-then-quantize approach,
	this time with much better controlled quantization noise. We show that the ``precode'' part of the
	precode-then-quantize operation is to design precoders under peak amplitude constraints.
	Leveraging this advantage, we develop  multi-user $\Sigma\Delta$ precoding schemes using ZF and symbol-level precoding
	(SLP) for both the PSK and QAM cases. Efficient optimization algorithms for SLP, with the design emphasis of
	operating under the assumption of a large number of antennas, are also derived.
	
\end{enumerate}

The organization of this paper is as follows.
Section \ref{sec:problem} describes the massive MIMO one-bit precoding problem.
Section \ref{sec:basics_sigma_delta} reviews the basics of $\Sigma\Delta$ modulation.
Sections \ref{sec:sigma_delta_SU} and \ref{sec:sigma_delta_MU} describe our $\Sigma\Delta$ precoding developments for the single-user and multi-user cases, respectively.
Section \ref{sec:sim} provides simulation results.
Section \ref{sec:conclusion} concludes this work.

\section{Problem Settings}
\label{sec:problem}

The scenario we consider is the multiuser MISO downlink over a quasi-static frequent-flat channel and under one-bit transmitted signal constraints.
The model is given by
\begin{equation} \label{eq:base_model}
y_{i,t}  = \sqrt{\frac{P}{2N}} \bh_i^T \bx_t + v_{i,t}, \quad t=1,\ldots,T,
\end{equation}
and for $i=1,\ldots,K$,
where
$y_{i,t} \in \Cbb$ represents the complex baseband received signal of the $i$th user at symbol time $t$;
$K$ denotes the number of users;
$T$ is the transmission block length;
$P$ is the total transmission power;
$N$ is the number of antennas of the BS;
$\bh_i \in \Cbb^N$ is the channel from the BS to the $i$th user;
$\sqrt{P/(2N)} \bx_t$, with $\bx_t \in \{ \pm 1 \pm \jj \}^N$, represents the complex baseband one-bit transmitted signal;
$v_{i,t}$ is noise and is assumed to be i.i.d. circular complex Gaussian with mean zero and variance $\sigma_v^2$.

The BS aims to blast parallel data symbols to the users.
To put into context,
let $s_{i,t} \in \setS$ denote the symbol to be transmitted to the $i$th user at symbol time $t$, where $\setS$ denotes the symbol constellation set.
For convenience with our development later, we will assume that
\[ \max_{s \in \setS} |s| = 1;
\]
or, the symbols are normalized such that the above equation holds.
The challenge is to find $\bx_t \in \{ \pm 1 \pm \jj \}^N$, for $t=1,\ldots,T$, such that
\begin{equation} \label{eq:sym_shape}
\bh_i^T \bx_t \approx c_{i,t} s_{i,t}, \quad \text{for all $i,t$,}
\end{equation}
where $c_{i,t} > 0$ denotes a scaling factor;
or, in words, we aim to shape the symbols at the user side under the one-bit transmitted signal constraints.
As a more technical note, we should mention that i) if the decision of the symbols at the user side depends on the signal amplitude, e.g., $M$-ary QAM, we should also make $c_{i,1} = \cdots = c_{i,T}$ for every $i$; see
\cite{Jacobsson2016Nonlinear,Sohrabi2018,shao2018framework} (also \cite{JeddaMSN18} for a further discussion); and that
ii) if the decision involves only signal phase, e.g., $M$-ary PSK, the $c_{i,t}$'s are allowed to be different.
In the currently available literature, this one-bit precoding challenge is formulated as a binary optimization problem---which is hard to solve by nature.
For details, read the recently growing body of literature \cite{Jacobsson2017,Swindlehurst2017,Jedda2017,Swindlehurst2018,Sohrabi2018,shao2018framework,li2018massive}.

We are interested in the single-path angular array channel.
The settings that lead to such channels are that the antennas at the BS are arranged as a uniform linear array, and that there is only one propagation path from the BS to each user;
the extension to
other channels
will be given later.
For the single-path angular channel,  each $\bh_i$ is characterized as
\begin{equation}
\bh_i = \alpha_i \ba(\theta_i),
\end{equation}
where $\alpha_i \in \Cbb$ is the complex channel gain;
$\theta_i \in [-\pi/2,\pi/2]$ denotes the angle of departure from the BS to the $i$th user;
\begin{equation}
\ba(\theta) = [~ 1, e^{-\jj \frac{2\pi d}{\lambda} \sin(\theta)}, \ldots,  e^{-\jj (N-1)\frac{2\pi d}{\lambda} \sin(\theta)}  ~]^T
\end{equation}
denotes the array response vector at $\theta$,
 in which $\lambda$ is the carrier wavelength and  $d \leq \lambda/2$  is the inter-antenna spacing.

\section{Basics of $\Sigma\Delta$ Modulation}
\label{sec:basics_sigma_delta}

\begin{figure}[htb]
	\centering
	\includegraphics[width=0.7\linewidth]{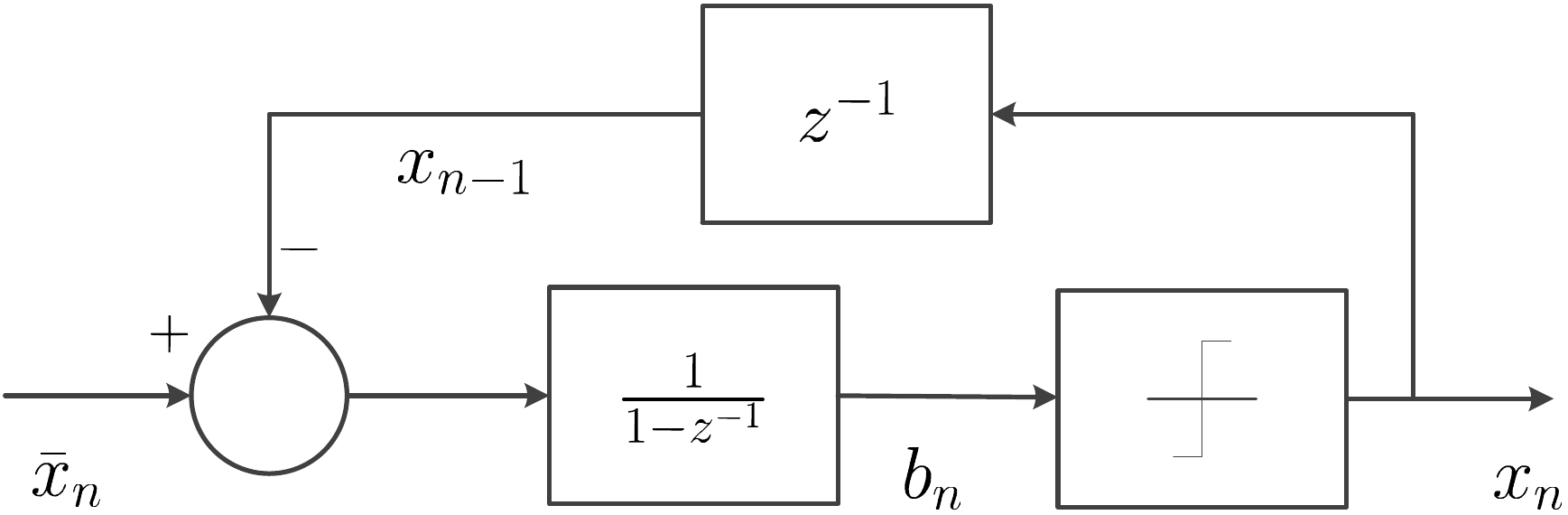}
	\caption{The first-order $\Sigma\Delta$ modulator.}
	\label{fig:sigmadelta}
\end{figure}

In this section we review the basic concepts of $\Sigma\Delta$ modulation \cite{aziz1996overview}.
We will focus on the notion of noise shaping, and will pay less attention to aspects that have little relevance to the one-bit precoding context.
Consider the system in Fig.~\ref{fig:sigmadelta}, which is called the first-order $\Sigma\Delta$ modulator.
We have a discrete-time real-valued signal sequence $\{ \bar{x}_n \}_{n \in \Zbb_+}$ as the modulator input.
In the application of temporal DACs,
$\bar{x}_n$ is a significantly oversampled version of some signal.
Here, it is sufficient to know that $\bar{x}_n$ is a low-pass signal.
The problem is to one-bit quantize $\{ \bar{x}_n \}_{n}$ in a way that the resulting quantization noise is high-pass.
Doing so satisfactorily will result in negligible quantization noise effects on the low-pass frequency region of the desired signal $\bar{x}_n$.
The $\Sigma\Delta$ modulator output sequence, denoted by $\{ x_n \}_{n \in \Zbb_+}$,
is generated as
\begin{subequations} \label{eq:sig_del_basic}
\begin{align}
x_n & = \sgn(b_n), \label{eq:sig_del_basic_a} \\
b_n & = b_{n-1} + (\bar{x}_n - x_{n-1}),  \label{eq:sig_del_basic_b}
\end{align}
\end{subequations}
for $n=0,1,\ldots$, and with $b_{-1} = x_{-1}= 0$.
Let $q_n = x_n - b_n$, $n \in \Zbb_+$, denote the quantization noise,
 and let $q_{-1} = 0$ for convenience.
From \eqref{eq:sig_del_basic} one can show that
\[
x_n = \bar{x}_n + q_n - q_{n-1}, \quad n \in \Zbb_+,
\]
and subsequently
\[
X(z) = \bar{X}(z) + (1-z^{-1}) Q(z),
\]
where $X(z) = \sum_{n=0}^\infty x_n z^{-n}$ denotes the $z$-transform.
Since $1-z^{-1}$ is a high-pass response, the quantization noise is suppressed at low frequency.

A key issue in $\Sigma\Delta$ modulation is the effect of overloading.
Overloading refers to the situation when the quantizer input $b_n$ has amplitude greater than $2$.
The consequence is that the corresponding quantization noise $q_n$ goes beyond the range $[-1,1]$.
As an example of showing what problem overloading can bring,
consider
\[
\bar{x}_n  = 1 + \epsilon, \quad \text{for all $n \in \Zbb_+$},
\]
where $\epsilon > 0$. This is an instance in which the signal amplitude is greater than one.
One can verify from \eqref{eq:sig_del_basic} that $b_n = 1 + (n+1) \epsilon$ and $q_n= -(n+1)\epsilon$.
We see that the quantization noise is unbounded as $n \rightarrow \infty$.
A sufficient condition under which overloading can be safely avoided is to limit the input signal range as
\begin{equation} \label{eq:amp_1}
-1 \leq \bar{x}_n \leq 1, \quad \text{for all $n \in \Zbb_+$}.
\end{equation}
Under the above condition it is guaranteed that
$|b_n| \leq 2$
for all $n \in \Zbb_+$, and consequently,
\[
-1 \leq q_n \leq 1, \quad \text{for all $n \in \Zbb_+$}.
\]
To see this, suppose $|b_{n-1}| \leq 2$.
Then, we see from \eqref{eq:sig_del_basic_b} that
\[
|b_n| \leq |\bar{x}_n| + | b_{n-1} - x_{n-1} | \leq 2,
\]
where we have used $|b_{n-1} - x_{n-1} | \leq 1$, implied by \eqref{eq:sig_del_basic_a}.


Under the no-overload condition \eqref{eq:amp_1}, it is very common to assume that the quantization noise $q_n$ is i.i.d., uniformly distributed on $[-1,1]$, and independent of $\{ \bar{x}_n \}$.
This assumption is widely adopted for signal-to-quantization-noise ratio (SQNR) prediction in the $\Sigma\Delta$-DAC/ADC literature.
We should, however, emphasize that the uniform  i.i.d. assumption is only a convenient approximation for the sake of tractable analysis.
Quantization noise analysis in $\Sigma\Delta$ modulation is a complicated topic,
and we refer the reader to the Introduction of \cite{gray1990quantization} which provides an excellent discussion.
Simply speaking, from a theoretical viewpoint, $\Sigma\Delta$ quantization noise analysis is very difficult owing to the feedback and coarse quantization nature of the $\Sigma\Delta$ modulator.
Some analysis results are available, e.g., in \cite{gray1990quantization} and the references therein, but they are very complicated for practical use.
From a practical viewpoint, it has been found by experiments and simulations that the uniform  i.i.d. assumption yields reasonable approximations in many applications, but it can also be a poor approximation for some specific signals.
For the latter the remedial solution is to apply dithering, which will be discussed later.
In this paper we will apply the uniform  i.i.d. assumption,
and the reader should bear in mind that the uniform  i.i.d. assumption can fail sometimes.

There are three further aspects we would like to discuss.
First, while the no-overload condition \eqref{eq:amp_1} is widely adopted for ensuring bounded quantization noise,
 overloading does not necessarily imply unbounded quantization noise.
An example is $\bar{x}_n = (-1)^n (1+\epsilon)$ for some $0 < \epsilon < 1$.
It can be verified that $q_n = - \epsilon$ for even $n$, and $q_n = 0$ for odd $n$.
In fact, one can argue that a moderate amount of overloading could be acceptable in practice, since not all kinds of overloaded input signals trigger the occurrence of unbounded quantization noise.
For example, the second-order $\Sigma\Delta$ modulator \cite{aziz1996overview} cannot avoid overloading for any input signal range (unless $\bar{x}_n = 0$ for all $n$) \cite{gray1990quantization}, and yet it is still used in practice.
That being said, there seems to be little theoretical work on understanding the quantization noise bound
under overloading.

Second, we previously mentioned that the uniform i.i.d. assumption is far from true for some specific signals.
Among them,
DC and pure sinusoidal signals are most well-known \cite{norsworthy1992effective,gray1990quantization}.
A popular way to handle
 the non-i.i.d. issue
is to apply dithering.
For example, as described in \cite{norsworthy1992effective}, consider modifying \eqref{eq:sig_del_basic_a} as
\begin{equation} \label{eq:dit}
x_n = \sgn(b_n + u_n),
\end{equation}
where $u_n$, called a dither signal, is uniform i.i.d. generated on $[-\delta,\delta]$ for some constant $\delta > 0$.
Intuitively, the idea is to use artificial noise to make the overall quantization noise $q_n = x_n - b_n$ more random, thereby attempting to destroy correlated patterns that $q_n$ may exhibit in the no-dithering case.
Empirically, it has been found that dithering works to a certain extent  \cite{norsworthy1992effective}.
However, dithering also increases the quantization noise level.
It can be verified from \eqref{eq:sig_del_basic_b},
\eqref{eq:amp_1}
and \eqref{eq:dit} that $-1-\delta \leq q_n \leq 1 + \delta$.

Third, better noise-shaping, in terms of further suppressing the low-pass region of the quantization noise, can be achieved by employing more advanced $\Sigma\Delta$ modulators, e.g., the higher-order and multi-stage versions of the first-order $\Sigma\Delta$ modulator.
The issue arising would be with overloading, however,
and in some cases multi-bit quantization is used to avoid overloading.

Readers are referred to the literature \cite{aziz1996overview,gray1990quantization} for further details of the above three aspects.
To keep our forthcoming development simple, we will consider only the first-order $\Sigma\Delta$ modulator without overloading and without dithering, unless otherwise specified.

\section{$\Sigma\Delta$ Precoding: Single-User Case}
\label{sec:sigma_delta_SU}

This section and the subsequent sections describe how we apply $\Sigma\Delta$ modulation to perform one-bit precoding.
In this section we consider the single-user case.

\subsection{Spatial $\Sigma\Delta$ Modulation}
\label{sec:sig_del_mod_basic}

Consider the basic model \eqref{eq:base_model} for the single-user case.
For the sake of notational simplicity, we remove the time index $t$ and user index $i$ from  \eqref{eq:base_model} and write
\begin{equation} \label{eq:model_su}
y = \sqrt{\frac{P}{2N}} \bh^T \bx + v,
\end{equation}
with $\bh = \alpha \ba(\theta)$; $\theta$ is the user's angle.
Let $\bar{\bx} = [~\bar{x}_1,\ldots,\bar{x}_N ~]^T$, with $-1 \leq \Re(\bar{x}_n) \leq 1$ and $-1 \leq \Im(\bar{x}_n) \leq 1$ for all $n$, be the signal we wish to $\Sigma\Delta$-modulate.
We apply first-order $\Sigma\Delta$ modulation (as described in the preceding section) to $\{ \bar{x}_n \}_{n=1}^N$ to obtain $\{ x_n \}_{n=1}^N$.
The resulting $\bx = [~ x_1, \ldots, x_N ~]^T$ then serves as the one-bit transmitted signal.
More precisely, we use two first-order $\Sigma\Delta$ modulators, one for the real part and another for the imaginary part, to get $\bx$.
By doing so, we perform $\Sigma\Delta$ modulation in space.
The advantages of doing so will become clear as we analyze the subsequent quantization noise effects below.

Following the preceding section, we can write
\begin{equation} \label{eq:model_x_su}
\bx = \bar{\bx} + \bq - \bq^-
\end{equation}
where $\bq = [~ q_1, q_2 \ldots, q_N ~]^T$; $\bq^- = [~ 0, q_1, \ldots, q_{N-1} ~]^T$; each $q_i$ is complex quantization noise with $-1 \leq \Re(q_n) \leq 1$ and $-1 \leq \Im(q_n) \leq 1$ (the aforesaid noise range is guaranteed when $-1 \leq \Re(\bar{x}_n) \leq 1$ and $-1 \leq \Im(\bar{x}_n) \leq 1$).
For the sake of analysis, we model the $q_n$'s as i.i.d. uniform noise on the unit box interval $\{ q = a + \jj b \mid a, b \in [-1,1] \}$.
Putting \eqref{eq:model_x_su} into \eqref{eq:model_su} gives
\begin{subequations} \label{eq:model_su2}
\begin{align}
y & = \sqrt{\frac{P}{2N}} \bh^T \bar{\bx} +  w, \\
w & = \sqrt{\frac{P}{2N}} \bh^T(\bq - \bq^-) + v,
\end{align}
\end{subequations}
where $w$ denotes a noise term that combines quantization noise and background noise.
We are interested in knowing how the noise power scales with the system parameters.
Let $z = e^{\jj \frac{2\pi d}{\lambda} \sin(\theta)}$ for convenience.
We see that
\begin{align*}
\ba^T(\bq - \bq^-) & = (1-z^{-1}) \sum_{n=0}^{N-2} z^{-n} q_{n+1}  +z^{-(N-1)} q_N,
\end{align*}
and consequently, $\Exp[ \ba^T(\bq - \bq^-) ] = 0$ and
\begin{align*}
\Exp[ | \ba^T(\bq - \bq^-) |^2 ] & = |1 - z^{-1}|^2 (N-1) \sigma_q^2 + \sigma_q^2,
\end{align*}
where $\sigma_q^2 = \Exp[ |q_n|^2 ] = 2/3$ due to the assumption of uniform i.i.d. quantization noise.
It follows that $\Exp[ w ] = 0$ and
\begin{align*}
\sigma_w^2 & = \Exp[ |w|^2 ]  = \frac{|\alpha|^2 P}{3 N} ( |1-z^{-1}|^2 (N-1) +  1) + \sigma_v^2.
\end{align*}
By assuming large $N$, the above quantization noise variance formula can be simplified to
\begin{subequations} \label{eq:w_power}
\begin{align}
\sigma_w^2
&
\approx \frac{|\alpha|^2 P}{3} |1-z^{-1}|^2  + \sigma_v^2  \label{eq:w_power_b} \\
& = \frac{4|\alpha|^2 P}{3} \left| \sin\left( \frac{\pi d}{\lambda} \sin(\theta)  \right) \right|^2 + \sigma_v^2. \label{eq:w_power_c}
\end{align}
\end{subequations}


Eq.~\eqref{eq:w_power_c} reveals interesting behaviors with the quantization noise effects at the user side.
\begin{enumerate}[1.]
\item First, the quantization noise power at the user side is independent of the number of antennas $N$.
This will give us substantial advantages in using massive MIMO to suppress the quantization noise, as we will further show in the next subsection.

\item
Second, the quantization noise power increases as the absolute value of the angle $|\theta|$ increases;
broadside ($\theta= 0$) is the best, while endfire ($\theta= \pi/2$ or $\theta= -\pi/2$)  is the worst.
This suggests that spatial $\Sigma\Delta$ modulation serves users with smaller $|\theta|$ better.
This also suggests that if we work on sectored antenna arrays, where we only need to deal with a restricted angular range, say, from $-30^\circ$ to $30^\circ$,
spatial $\Sigma\Delta$ modulation has an advantage.

\item Third, the quantization noise power decreases as we decrease the inter-antenna spacing $d$.
This means that
we may want to employ more densely spaced antennas.
In practice, however,
it is infeasible to have very small inter-antenna spacing as that will introduce mutual coupling effects.
Also, the physical dimensions of the antennas prevent small spacing. We will have to rely on large $N$ and smaller operating angular ranges to reduce the quantization noise.

\end{enumerate}
A further comment is as follows.

\begin{Remark}
We should also draw connections between conventional $\Sigma\Delta$ modulation for discrete-time signals and the spatial $\Sigma\Delta$ modulation proposed above.
Simply speaking, frequency in the temporal case becomes angle in the spatial case.
$\Sigma\Delta$ modulation in time and space serve low frequency and low angle signals better, respectively.
Also, applying small $d$ in the spatial case is essentially the same as oversampling in the temporal case.
In fact, the latter typically considers a very large oversampling factor, such as $128$, such that quantization noise becomes almost negligible \cite{aziz1996overview}.
Such extreme oversampling is however inapplicable to the spatial case; as mentioned above, mutual coupling and the physical dimension constraint prevent us from doing so.
\end{Remark}

\subsection{$\Sigma\Delta$ Maximum Ratio Transmission}
\label{sec:mrt_su}

In the preceding subsection we have presented a different paradigm to deal with one-bit precoding:
Using spatial $\Sigma\Delta$ modulation, we can convert the one-bit precoding problem to a precoding problem for an amplitude-limited signal $\bar{\bx}$, specifically, $-\bone \leq \Re(\bar{\bx}) \leq \bone$ and $-\bone \leq \Im(\bar{\bx}) \leq \bone$.
Let us consider a simple precoding scheme, namely, the maximum ratio transmission (MRT) approach
\begin{equation} \label{eq:mrt}
\bar{\bx} = \frac{\alpha^* s}{|\alpha|}   \ba^*(\theta),
\end{equation}
where $s \in \setS$ is a symbol.
Note that $\bar{\bx}$ satisfies the aforementioned amplitude-limit constraints since $| [ \ba(\theta) ]_n |=1$ for all $n$ and $|s| \leq 1$.
We are interested in performing a symbol-error probability (SEP) analysis of this $\Sigma\Delta$ MRT scheme.
Plugging \eqref{eq:mrt} into the model \eqref{eq:model_su2}, we get
\[
y = c\cdot  s + w, \quad c= |\alpha| \sqrt{\frac{P N}{2}}.
\]
Let us make an approximation, namely, that $w$ is circular Gaussian distributed with mean zero and variance given by \eqref{eq:w_power_c}.
Let $\hat{s} = {\rm dec}(y)$ be the decision of $s$, where ${\rm dec}$ denotes the decision function associated with $\setS$.
The SEP can be characterized as
\begin{equation} \label{eq:theo_psk_sep}
{\rm P}( \hat{s} \neq s ) \leq \beta Q \left( \chi_{_{\large M}}  \sqrt{ {\sf SNR}_{{\sf eff}}  } \right),
\end{equation}
where
$(\beta,\chi_{_{\large M}}) = (2, \sqrt{2} \sin(\pi/M))$ if $\setS$ is the $M$-ary PSK constellation set, and
$(\beta,\chi_{_{\large M}}) = (4, 1/(\sqrt{M}-1))$ if  $\setS$ is the $M$-ary QAM constellation set and $M$ is a power of $4$;
$Q(t) = \int_{t}^\infty ( e^{-z^2/2} /\sqrt{2\pi} ) dz $;
\begin{align}
{\sf SNR}_{{\sf eff}}
& = \frac{  c^2 }{ \sigma_w^2 }
\nonumber
\end{align}
denotes the effective SNR \cite{proakis2001digital}.
The effective SNR plays the main role in determining the SEP performance.
From the above derivations, we see that
\begin{align}
{\sf SNR}_{{\sf eff}}
& = \frac{ |\alpha|^2 P N }{ \frac{8|\alpha|^2 P}{3} \left| \sin\left( \frac{\pi d}{\lambda} \sin(\theta)  \right) \right|^2 + 2\sigma_v^2   }.
\label{eq:SNR_eff}
\end{align}

Let us extract some insights from the effective SNR derivation \eqref{eq:SNR_eff}.
\begin{enumerate}[1.]
\item First, increasing the power $P$ is not helpful in reducing quantization noise power.
In fact, we have $\lim_{P \rightarrow \infty} {\sf SNR}_{{\sf eff}}  = 3N/(8\left| \sin\left( \frac{\pi d}{\lambda} \sin(\theta)  \right) \right|^2) )$.

\item Second, the effective SNR increases linearly with the number of antennas $N$.
In particular we observe that under a fixed power $P$, increasing $N$---which also means less power per antenna---is effective in improving the effective SNR.
This suggests that $\Sigma\Delta$ precoding is particularly suitable for massive MIMO.
\end{enumerate}

In Appendix~\ref{supp:IQ_plots}, we provide additional numerical results to give readers some intuitive feeling on the noise shaping performance of $\Sigma\Delta$ MRT.
One will see that, in general, the symbol shaping error of $\Sigma\Delta$ MRT reduces with $N$ and increases with $|\theta|$.

\subsection{Quantization Noise Zeroing by $\Sigma\Delta$ Angle Steering}
\label{sec:angles}

We have seen that the quantization noise tends to increase as the angle $\theta$ is further away from $0$.
It is natural to question whether we can reduce the quantization noise by re-designing the $\Sigma\Delta$ modulator.
The answer turns out to be yes.

\begin{figure}[htb]
	\centering
	\ifconfver
		\includegraphics[width=\linewidth]{fig/anglesteering}
	\else
		\includegraphics[width=.75\linewidth]{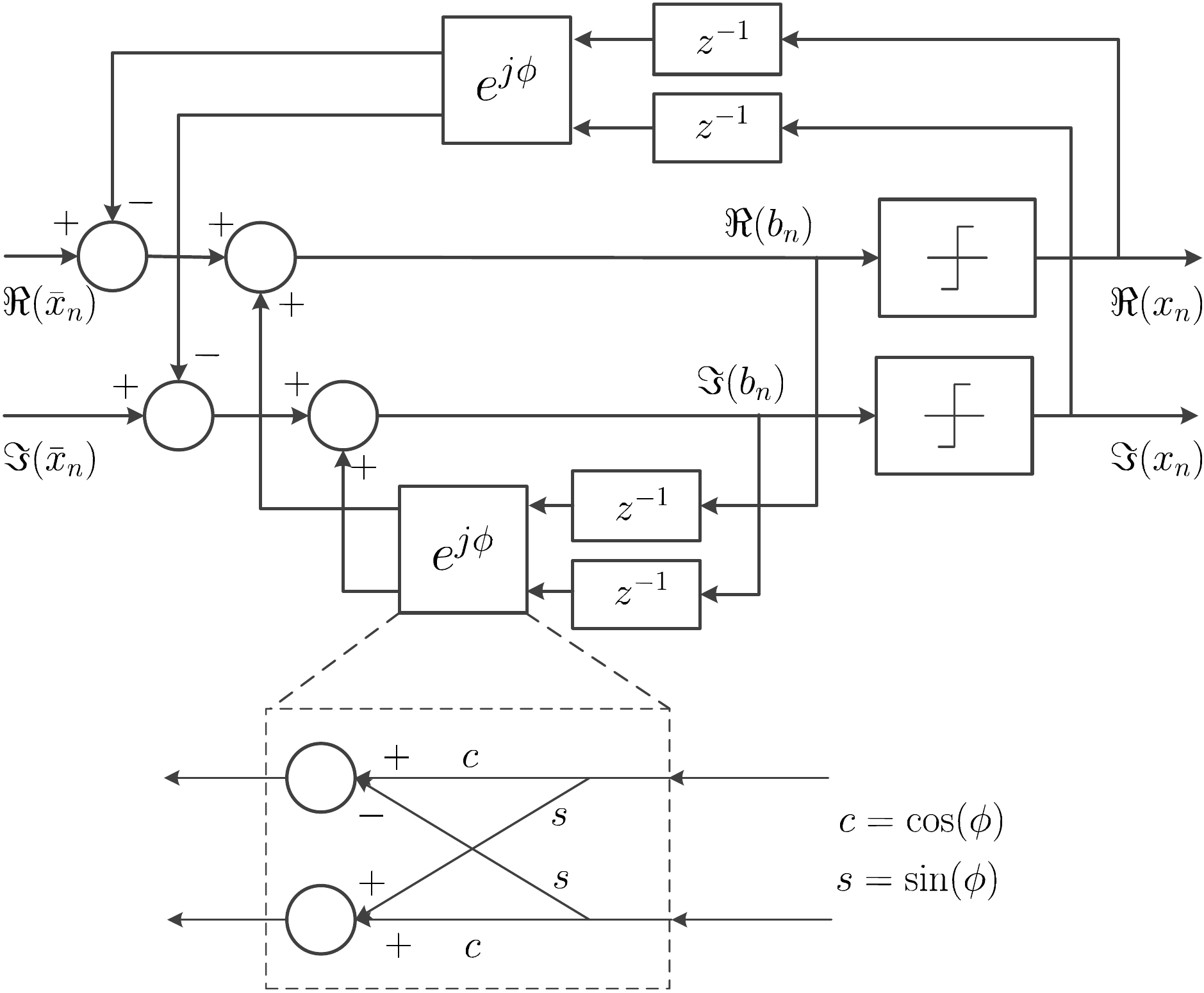}
	\fi
	\caption{The angle-steered first-order $\Sigma\Delta$ modulator.}
	\label{fig:anglesteer}
\end{figure}

Our idea borrows insight from bandpass $\Sigma\Delta$ modulation \cite{aziz1996overview}, although our task is still different from that of the latter.
Consider the modified first-order $\Sigma\Delta$ modulator in Fig.~\ref{fig:anglesteer}, which we refer to as an {\em angle-steered} $\Sigma\Delta$ modulator.
In this system, $\bar{x}_n$, $b_n$ and $x_n$ are all complex-valued,
and $\phi \in [-\pi,\pi]$ is given.
The modulation process is
described by
\begin{subequations} \label{eq:sig_del_ang}
\begin{align}
x_n & = \sgn(\Re(b_n)) + \jj \cdot \sgn(\Im(b_n)),
\label{eq:sig_del_ang_a}
\\
b_n & = e^{\jj \phi} b_{n-1} + (\bar{x}_n - e^{\jj \phi} x_{n-1}),
\label{eq:sig_del_ang_b}
\end{align}
\end{subequations}
Let $q_0 = 0$, and let $q_n = x_n - b_n$ be the quantization noise.
From \eqref{eq:sig_del_ang} one can show that
\begin{equation} \label{eq:sig_del_ang2}
x_n = \bar{x}_n + q_n - e^{\jj \phi} q_{n-1},
\end{equation}
where the difference compared with the previous first-order $\Sigma\Delta$ modulator is the inclusion of the phase shift term $e^{\jj \phi}$.
We are concerned with the range of $\bar{x}_n$ under which no overloading will occur.
\begin{Fact} \label{fact:A_angles}
	Consider the angle-steered $\Sigma\Delta$ modulator in Fig.~\ref{fig:anglesteer} or in \eqref{eq:sig_del_ang}.
	Let
	\begin{equation}
	A = 2 - | \cos(\phi) | - | \sin(\phi) |.
	\end{equation}
	If $| \Re(\bar{x}_n) | \leq A$ and  $| \Im(\bar{x}_n) | \leq A$ for all $n$,
	then $b_n$ is not overloaded, and the quantization noise $q_n$ is bounded with
	$| \Re(q_n) | \leq 1$ and  $| \Im(q_n) | \leq 1$.
\end{Fact}
{\em Proof:} \
We prove Fact \ref{fact:A_angles} by induction.
Assume  $| \Re(\bar{x}_n) | \leq A$ and  $| \Im(\bar{x}_n) | \leq A$ for all $n$.
It is easy to see that $| \Re(q_{1}) | \leq 1$ and  $| \Im(q_{1}) | \leq 1$.
Now, suppose that $| \Re(q_{n-1}) | \leq 1$ and  $| \Im(q_{n-1}) | \leq 1$ are true.
Using $b_n = \bar{x}_n - e^{\jj \phi} q_{n-1}$, which can be shown from \eqref{eq:sig_del_ang}, we have
\begin{align*}
| \Re(b_n) | & \leq | \Re(\bar{x}_n) |  + | \cos(\phi) \Re(q_{n-1})| + | \sin(\phi) \Im(q_{n-1})| \\
& \leq A + | \cos(\phi) | + | \sin(\phi) | = 2,
\end{align*}
and similarly, $| \Im(b_n) | \leq 2$.
Consequently,
we
must have $| \Re(q_n) | \leq 1$ and  $| \Im(q_n) | \leq 1$.
The proof is complete.
\hfill $\blacksquare$

\medskip

We should mention that the largest value of $A$ is $A=1$, which happens when
$\phi \in \{ 0, \pm \pi/2, \pm \pi \}$.
The smallest value of $A$ is $A= 0.59$, which happens when
$\phi \in \{ \pm \pi/4, \pm 3\pi/4 \}$.
This means that there is a mild compromise with the signal range if no overloading is desired.

However, the aforementioned compromise brings a significant advantage, namely, quantization noise zeroing.
Following the same noise analysis in Section \ref{sec:sig_del_mod_basic},
we can show that
\begin{align*}
	\sigma_w^2 & \approx \frac{|\alpha|^2 P}{3} |1- e^{\jj \phi} z^{-1} |^2  + \sigma_v^2   \\
	& = \frac{4|\alpha|^2 P}{3} \left|  \sin\left( \frac{ \phi-  \frac{2\pi d}{\lambda} \sin(\theta)}{2}  \right) \right|^2 + \sigma_v^2.
\end{align*}
Hence, by selecting $\phi= 2\pi d \sin(\theta) / \lambda$,
we can eliminate the quantization noise effects.
To get more insight,
let us consider MRT under such angle-steered $\Sigma\Delta$ modulation.
The corresponding MRT scheme is $\bar{\bx} = \frac{A \alpha^* s}{|\alpha|} \ba(\theta)$.
The effective SNR under angle steering is
\begin{align}
{\sf SNR}_{{\sf eff}}
& = \frac{ A^2 |\alpha|^2 P N }{  2\sigma_v^2   },
\label{eq:SNR_eff_angles}
\end{align}
with $A= 2 - | \cos( 2\pi d \sin(\theta) / \lambda )| - | \sin(2\pi d \sin(\theta) / \lambda) |$.
We see that the sole factor of performance reduction is $A$, which is reduced to $0.59$ (equivalently, $-4.64$dB SNR loss relative to $A=1$) in the worst case.
Thus, we see that the angle corresponding to the minimum quantization noise in the previous $\Sigma\Delta$ modulator, that is, the broadside angle $\theta= 0$, can be steered to any desired angle using the angle-steered $\Sigma\Delta$ modulation approach.

Again, to give readers some intuition, Appendix~\ref{supp:IQ_plots} provides an additional numerical result that shows that the angle-steered $\Sigma\Delta$ modulation approach leads to almost zero symbol shaping error.

\begin{Remark} \label{rem:angles}
It is worthwhile to note that the angle-steered $\Sigma\Delta$ MRT scheme described above does not require the uniform i.i.d. assumption with the quantization noise.
From \eqref{eq:model_su2}, \eqref{eq:sig_del_ang2}, and with $\phi= 2\pi d \sin(\theta) / \lambda$, one can show that the overall noise term $w$ is actually given by
\[
w= \sqrt{\frac{P}{2N}} \alpha z^{N-1} q_N + v;
\]
we will show the details and insight of the above expression under a more general setting in the subsequent subsection.
Note that the same phenomenon also happens with the basic $\Sigma\Delta$ MRT scheme when the user angle is $\theta= 0$.
As such, there is no need to assume that the $q_n$'s are i.i.d., and the remaining factor lies only in the surviving quantization noise term $q_N$ in the above equation.
That surviving term is small compared with the signal term for large $N$, and thus may be ignored.
\end{Remark}

\begin{Remark}
The angle-steered $\Sigma\Delta$ modulation architecture can be used to change the angular range the system serves.
Previously, we mentioned that the basic spatial $\Sigma\Delta$ modulation is more appropriate for serving users under a smaller angular range, say, from $-30^\circ$ to $30^\circ$.
Now, with angle steering, we can easily alter the center of the angular range, say, to $60^\circ$, thereby serving users from $30^\circ$ to $90^\circ$.
\end{Remark}

\subsection{Angle-Steered $\Sigma\Delta$ Modulation for Any Channels}
\label{sec:gen_angles}
It is intriguing to further question whether the angle steering idea in the last subsection can be generalized to any arbitrary channel $\bh$, rather than just the one-path angular channel under uniform linear arrays.
The answer turns to be also yes.

Without loss of generality, assume $h_n \neq 0$ for all $n$.
Also, assume the elements of the antenna array to be indexed such that $0 < |h_1| \leq |h_2| \leq \cdots \leq |h_N|$.
Consider modifying the  angle-steered $\Sigma\Delta$ modulator \eqref{eq:sig_del_ang} as follows:
\begin{subequations} \label{eq:sig_del_gen_ang}
	\begin{align}
	x_n & = \sgn(\Re(b_n)) + \jj \cdot \sgn(\Im(b_n)),
	\label{eq:sig_del_gen_ang_a}
	\\
	b_n & = \tfrac{h_{n-1}}{h_n} b_{n-1} + \left( \bar{x}_n - \tfrac{h_{n-1}}{h_n} x_{n-1} \right),
	\label{eq:sig_del_gen_ang_b}
	\end{align}
\end{subequations}
for $n=1,\ldots,N$ and
with $h_0 = 0$.
From the above equations,
one can readily show that
\begin{equation} \label{eq:x_gen_angles}
x_n = \bar{x}_n + q_n - \tfrac{h_{n-1}}{h_n} q_{n-1},
\end{equation}
where $q_0 = 0$; $q_n = x_n - b_n$ for $n=1,\ldots,N$.
By observing
\begin{align*}
\bh^T \bx & = \sum_{n=1}^N h_n \bar{x}_n +  \sum_{n=1}^N h_n \left( q_n - \tfrac{h_{n-1}}{h_n} q_{n-1} \right) \\
& = \bh^T \bar{\bx} + h_N q_N,
\end{align*}
where the quantization noise terms $q_1,\ldots,q_{N-1}$ are successively canceled,
the signal model reduces to
\begin{subequations}
	\begin{align}
	y & = \sqrt{\frac{P}{2N}} \bh^T \bar{\bx} +  w, \\
	w & = \sqrt{\frac{P}{2N}} h_N q_N + v. \label{eq:w_gen_angles}
	\end{align}
\end{subequations}
Suppose that the $\Sigma\Delta$ modulator is not overloaded such that $|q_N| \leq 1$.
Then, for most massive MIMO cases of interest in which $|h_N| \ll \sum_{n=1}^{N-1} |h_n|$,
the quantization noise term in $w$ can be neglected.
We call this modulator a {\em generalized angle-steered} $\Sigma\Delta$ modulator.
The sufficient condition for no overloading is as follows.
\begin{Fact} \label{fact:A_gen_angles}
	Consider the generalized angle-steered $\Sigma\Delta$ modulator in \eqref{eq:sig_del_ang_a} and \eqref{eq:sig_del_gen_ang}.
	Let, for $n=1,\ldots,N$,
	\begin{equation} \label{eq:A_gen_sig_del}
	A_n = 2 - \tfrac{|h_{n-1}|}{|h_n|} ( |\cos(\phi_n)| + |\sin(\phi_n)| ),
	\end{equation}
	where $\phi_n$ denotes the phase of $h_{n-1}/h_n$.
	If $| \Re(\bar{x}_n) | \leq A_n$ and  $| \Im(\bar{x}_n) | \leq A_n$ for all $n$,
	then $b_n$ is not overloaded, and the quantization noise $q_n$ is bounded with
	$| \Re(q_n) | \leq 1$ and  $| \Im(q_n) | \leq 1$.
\end{Fact}
The proof of Fact \ref{fact:A_gen_angles} is essentially the same as that of Fact \ref{fact:A_angles}, and we shall thus omit it.
Note that $0.59 \leq A_n < 2$.
Also, since the signal range \eqref{eq:A_gen_sig_del} varies with $n$, it makes sense to modify the MRT scheme accordingly:
\begin{equation} \label{eq:MRT_gen_angles}
\bar{\bx}_n = \br s,
\end{equation}
where $r_n = A_n h_n^*/\max\{ |\Re(h_n) |, |\Im(h_n) | \}$ for all $n$.

\section{$\Sigma\Delta$ Precoding: Multi-User Case}
\label{sec:sigma_delta_MU}

The study in the preceding section provides us with vital insights into how the performance of $\Sigma\Delta$ precoding scales with the system parameters, assuming a single user.
Now we turn to the multi-user case.

The development follows exactly the same spirit as the preceding section.
We simplify the notation of the basic signal model \eqref{eq:base_model} by removing the index $t$, i.e.,
\begin{equation*}
y_{i}  = \sqrt{\frac{P}{2N}} \bh_i^T \bx + v_{i}, \quad i=1,\ldots,K.
\end{equation*}
For simplicity, we apply $\Sigma\Delta$ modulation without angle steering.
Adaptation to the angle-steered case is straightforward.
The corresponding model is
\begin{equation} \label{eq:model_mu}
y_{i}  = \sqrt{\frac{P}{2N}} \bh_i^T  \bar{\bx} + w_{i}, \quad i=1,\ldots,K,
\end{equation}
where $\bar{\bx} \in \Cbb^N$ is an amplitude-limited desired signal, with $-\bone \leq \Re(\bar{\bx}) \leq \bone$ and $-\bone \leq \Im(\bar{\bx}) \leq \bone$;
$w_i$ is a term combining quantization noise and background noise.
The noise term $w_i$ is modeled as mean-zero circular complex Gaussian. The variance of $w_i$, denoted by $\sigma_{w,i}^2$, is evaluated as
\begin{equation} \label{eq:sig_sq_w_i}
\sigma_{w,i}^2 = \frac{4|\alpha_i|^2 P}{3} \left| \sin\left( \frac{\pi d}{\lambda} \sin(\theta_i)  \right) \right|^2 + \sigma_v^2,
\end{equation}
where large $N$ has again been assumed;
note that \eqref{eq:sig_sq_w_i} directly follows from the noise variance formula \eqref{eq:w_power}.

In the first two subsections below,
we will describe two design schemes for $\bar{\bx}$ under the assumption of $M$-ary PSK constellations.
Then, the third subsection will consider the adaptation of the two schemes to the $M$-ary QAM constellation case.
The final subsection will discuss the extension to the multi-path angular channel case.

\subsection{$\Sigma\Delta$ Zero-Forcing}
\label{sec:sig_del_zf}

The first scheme we consider is ZF.
For notational convenience, define
\[
\| \bx \|_{IQ-\infty} = \max\{ |\Re(x_1)|, |\Im(x_1)|, \ldots, |\Re(x_N)|, |\Im(x_N)|   \};
\]
that is, the infinity norm applied on the in-phase and quadrature-phase components of a vector.
Also, assume $M$-ary PSK constellations.
The ZF precoding scheme implements
\begin{equation} \label{eq:zf}
\bar{\bx} = \gamma \bA^\dag \bD  \bs,
\end{equation}
where $\bs \in \setS^K$ is the symbol vector, with $s_i$ representing the symbol for the $i$th user;
\begin{align*}
\bD & = \Diag(\sigma_{w,1} \alpha_1^* / | \alpha_1 |^2, \ldots, \sigma_{w,K} \alpha_K^* / | \alpha_K |^2), \\
\bA & = [~ \ba_1, \ldots, \ba_K ~]^T, \quad \ba_i = \ba(\theta_i);
\end{align*}
and $\gamma$ is a normalization constant such that $\| \bar{\bx} \|_{IQ-\infty} = 1$.
It is easy to see that
\begin{equation} \label{eq:gamma}
\gamma = \frac{1}{\| \bA^\dag \bD  \bs \|_{IQ-\infty}  }.
\end{equation}
This ZF precoding scheme is designed such that every user has the same effective SNR, and consequently, uniform SEP performance.
To see this, consider putting \eqref{eq:zf} into \eqref{eq:model_mu}. It can be shown that
\[
y_i = c_i \cdot s_i + w_i, \quad c_i = \sqrt{\frac{P}{2N}} \gamma \sigma_{w,i}.
\]
Following the effective SNR concept used in the preceding section, the effective SNR of the $i$th user is
\begin{equation} \label{eq:SNR_eff_zf}
{\sf SNR}_{{\sf eff},i} = \frac{c_i^2}{\sigma_{w,i}^2} = \frac{P}{2N} \gamma^2.
\end{equation}
Clearly, the effective SNRs of all the users are identical.

In the simulation results section we will show the performance of this $\Sigma\Delta$ ZF precoding scheme.
Here, we are interested in analyzing how the effective SNRs scale with the system parameters.
The result is as follows.
\begin{Prop} \label{prop:eff_snr_zf}
Consider the $\Sigma\Delta$ ZF precoding scheme described above.
	Let $k = \arg \max_{i=1,\ldots,K} \sigma_{w,i}/|\alpha_i|$.
	The users' effective SNRs are bounded by
	\begin{align}
	{\sf SNR}_{{\sf eff},i} & \geq \frac{ PN |\alpha_k|^2 \lambda_{\rm min}^2(\bR) }{2 K^3 \sigma_{w,k}^2} \nonumber  \\
	& = \frac{ PN |\alpha_k|^2 \lambda_{\rm min}^2(\bR) }{2 K^3 \left( \frac{4|\alpha_k|^2 P}{3} \left| \sin\left( \frac{\pi d}{\lambda} \sin(\theta_k)  \right) \right|^2 + \sigma_v^2 \right)}, \label{eq:snr_eff_zf_ana}
	\end{align}
	for all $i$,
	where $\bR = \bA \bA^H / N$; $\lambda_{\rm min}(\bR)$ denotes the smallest eigenvalue of $\bR$.
	Also, it holds that
	\begin{equation} \label{eq:lam_R}
		1 \geq \lambda_{\rm min}(\bR)  \geq 1 - (K-1) \rho,
	\end{equation}
	where
	\[
	\rho = \max_{i \neq j } \left| D_N\left( \frac{\pi d}{\lambda} ( \sin(\theta_i) - \sin(\theta_j) ) \right) \right|,
	\]
	and  $D_N(\phi) = \sin(N \phi)/(N \sin(\phi))$ is the digital sinc function.	
\end{Prop}

{\em Proof:} \
From \eqref{eq:gamma}--\eqref{eq:SNR_eff_zf},  we see that the problem is to analyze $\| \bA^\dag \bD  \bs \|_{IQ-\infty}$.
Let $\| \cdot \|_p$ denote either the $p$-norm for vectors or the induced $p$-norm for matrices.
We have
\begin{align*}
\| \bA^\dag \bD  \bs \|_{IQ-\infty} & \leq \| \bA^\dag \bD  \bs \|_{\infty}
\leq \| \bA^\dag \|_\infty \| \bD  \bs \|_{\infty}  \\
& = \| \bA^\dag \|_\infty \max_i \sigma_{w,i}/|\alpha_i|,
\end{align*}
where we have used $\| \bx \|_{IQ-\infty} \leq \| \bx \|_\infty$, $\| \bA x \|_\infty \leq \| \bA \|_\infty \| \bx \|_\infty$, and $|s_i| \leq 1$ for all $i$.
By using
\[
\bA^\dag = \bA^H (\bA \bA^H)^{-1} = \frac{1}{N} \bA^H \bR^{-1},
\]
we further get
\begin{align*}
\| \bA^\dag \|_\infty & \leq  \frac{1}{N} \| \bA^H \|_\infty \| \bR^{-1} \|_\infty  \\
& \leq  \frac{1}{N} K ( \sqrt{K} \| \bR^{-1} \|_2 ) \\
& = \frac{K^{3/2}}{N} \lambda_{\rm min}^{-1}(\bR),
\end{align*}
where the first inequality is due to $\| \bX \bY \|_\infty \leq \| \bX \|_\infty \| \bY \|_\infty$;
the second inequality is due to $\| \bA^H \|_\infty = \max_{i} \sum_{j=1}^K |a_{ji}|= K$ and due to $\| \bX \|_\infty \leq \sqrt{n} \| \bX \|_2$ for $\bX \in \Cbb^{m \times n}$;
the third inequality is due to the fact that for a positive definite $\bX$, it is true that $\| \bX \|_2 = \lambda_{\rm max}(\bX)$ and $\lambda_{\rm max}(\bX^{-1})  = 1/ \lambda_{\rm min}(\bX)$.
By putting the above inequalities into  \eqref{eq:gamma} and \eqref{eq:SNR_eff_zf},
the desired inequality \eqref{eq:snr_eff_zf_ana} is obtained.
The proof of \eqref{eq:lam_R}
is relegated to Appendix~\ref{appen:1}.
\hfill $\blacksquare$

\medskip

Let us discuss the implications of the theoretical result in \eqref{eq:snr_eff_zf_ana}--\eqref{eq:lam_R}.
First, the quantization noise effects are the same as what we see in the single-user case; a larger absolute value of the angle means a larger quantization noise power.
Second, the lower bound of the effective SNRs increases linearly with the number of antennas $N$.
Again, this suggests that $\Sigma\Delta$ precoding is favorable for massive MIMO.
Third, $\lambda_{\rm min}(\bR)$, which appears in the signal power part of the effective SNR, is large if the user angles are well separated, but small if some of the angles are close.
This factor is relative to $N$. Fixing the angles, larger $N$ brings $\lambda_{\rm min}(\bR)$ closer to its largest value, $1$.
 Fourth, we are interested in how $N$ should scale with the number of users $K$.
Very intuitively, by reading \eqref{eq:snr_eff_zf_ana}, there is an indication that $N$ should increase cubically with $K$; doing so keeps $N/K^3$ constant in the effective SNR bound.
However, note that this is a prediction from a performance bound that is safe, but also pessimistic, by its nature.
For instances where $\ba_1,\ldots,\ba_K$ are orthogonal---which one can expect it to be approximately true when $N$ is very large, one can redo the proof of Proposition~\ref{prop:eff_snr_zf} to obtain a better bound
\[
{\sf SNR}_{{\sf eff},i} \geq \frac{ PN |\alpha_k|^2  }{2 K \left( \frac{4|\alpha_k|^2 P}{3} \left| \sin\left( \frac{\pi d}{\lambda} \sin(\theta_k)  \right) \right|^2 + \sigma_v^2 \right)},
\]
which is merely the single-user effective SNR \eqref{eq:SNR_eff} downscaled by $K$.
In such instances it suffices to scale $N$ linearly with $K$.

\subsection{$\Sigma\Delta$ Symbol-Level Precoding}
\label{sec:sig_del_slp}

The second scheme we consider is SLP.
The idea is to design, on a per-symbol-time basis, an amplitude-limited $\bar{\bx}$ such that the SEP performance of the users is improved.
It is interesting to first draw a connection between SLP and ZF.
As shown in \cite{liu2018symbol},
any $\bar{\bx} \in \Cbb^N$ can be expressed as
\begin{equation} \label{eq:slp_int}
\bar{\bx} = \bA^\dag ( \bD \bs + \bu ) + {\bm \eta},
\end{equation}
where ${\bm \eta}$ lies in the nullspace of $\bA$, $\bD = \Diag(\beta_1,\ldots,\beta_K)$, with $\beta_i > 0$ for all $i$, and $\bu \in \Cbb^K$.
Putting \eqref{eq:slp_int} into the model \eqref{eq:model_mu} gives
\begin{equation} \label{eq:slp_int2}
y_i = \sqrt{\frac{P}{2N}} \alpha_i (\beta_i s_i + u_i) + w_i, \quad i=1,\ldots,K,
\end{equation}
where the nullspace term $\bm \eta$ has no impact on the received signals, the $u_i$'s appear as symbol perturbation terms, and the $\beta_i$'s appear as symbol gains.
There are two main ideas.
First, conditioned on $s_i$, we can use $u_i$ to purposely push the shaped symbol away from the decision boundaries.
SEP performance can thereby be improved.
Second, while the nullspace term $\bm \eta$ seems useless at first glance, it plays a hidden role in improving energy efficiency.
Intuitively, from \eqref{eq:slp_int}, we may hope that some particular $\bm \eta$ can cancel some of the signal components of $\bA^\dag ( \bD \bs + \bu )$, possibly reducing the subsequent IQ amplitude limit $\| \bar{\bx} \|_{IQ-\infty}$.
In the related context of per-antenna power constrained linear precoding,
it has been alluded to that using the nullspace term can be beneficial \cite{wiesel2008zero}.

Having shed light on the intuition, we turn to the design.
We formulate the design as a minimax SEP problem.
Assume $M$-ary PSK constellations.
Let ${\sf SEP}_i = {\rm P}(\hat{s}_i \neq s_i )$, with $\hat{s}_i = {\rm dec}(y_i)$, be the SEP of the $i$th user.
The problem is
\begin{equation} \label{eq:minimax_sep}
\begin{aligned}
\min_{ \| \bar{\bx} \|_{IQ-\infty} \leq 1 } \ \max_{i=1,\ldots,K} {\sf SEP}_i.
\end{aligned}
\end{equation}
Our first challenge is to find a tractable characterization of ${\sf SEP}_i$.
Consider the following result.
\begin{Lemma}[\cite{shao2018globsip}]\footnote{As a technically subtle note, the SEP result for the case of $z = c \cdot s$, where $c > 0$, is available in the classical communications literature.
However, the same result for arbitrary $z$ does not seem to be as readily available.}
	\label{Lem:PSK_SEP}
	Let $\setS$ be the $M$-ary PSK constellation set.
	Let $y= z + w$, where $w$ is circular complex Gaussian with mean zero and variance $\sigma_w^2$, and $z \in \Cbb$ is arbitrary.
	Let $s \in \setS$, and let $\hat{s} = \dec(y)$.
	It holds that
	\[
	{\rm P}(\hat{s}\neq s) \leq 2 Q \left( \chi_{_{\large M}} \frac{\psi}{\sigma_w}  \right),
	\]
	where $\chi_{_{\large M}} = \sqrt{2} \sin(\pi/M)$, and
	\[
	\psi = \Re(z s^*) - | \Im(z s^*) | \cot(\pi/M).
	\]	
\end{Lemma}
Applying Lemma~\ref{Lem:PSK_SEP} to the signal model \eqref{eq:model_mu}, we characterize the users' SEPs as
\[
{\sf SEP}_i \leq 2 Q (\chi_{_{\large M}} \sqrt{ {\sf SNR}_{{\sf eff},i} } ),
\]
where
\[
{\sf SNR}_{{\sf eff},i}  = \frac{P ( \Re( \bh_i^T \bar{\bx} s_i^*) - | \Im(\bh_i^T \bar{\bx} s_i^* ) | \cot(\pi/M)   )}{2 N \sigma_{w,i}^2}.
\]
Since the above bound on ${\sf SEP}_i$ decreases as ${\sf SNR}_{{\sf eff},i}$ increases, and since this relationship is monotone, it makes sense to consider
\begin{equation} \label{eq:maximin_snr}
\begin{aligned}
\max_{ \| \bar{\bx} \|_{IQ-\infty} \leq 1 } \ \min_{i=1,\ldots,K} {\sf SNR}_{{\sf eff},i}
\end{aligned}
\end{equation}
as a convenient and reasonable approximation of the minimax SEP problem \eqref{eq:minimax_sep}.
As a slight abuse of notation, redefine the variable $\bx$ as
\[
\bx = [~ \Re(\bar{\bx})^T, \Im(\bar{\bx})^T ~]^T.
\]
Through some efforts, problem \eqref{eq:maximin_snr} can be rewritten as
\begin{equation} \label{eq:minimax_key}
\begin{aligned}
\min_{ \bx \in [-1,1]^{2N} } f(\bx) \triangleq
\max\{ \bc_1^T \bx, \cdots, \bc_{2K}^T \bx \},
\end{aligned}
\end{equation}
where
\begin{align*}
\bc_i & = \left\{  \begin{array}{ll} -\bb_i + \br_i & i=1,\ldots,K \\
-\bb_i - \br_i & i=K+1,\ldots,2K
\end{array} \right. \\
\bb_i & = \sigma_{w,i}^{-1}  [~ \Re(s_i^* \bh_i^T), -\Im(s_i^* \bh_i^T) ~]^T, \\
\br_i & = \sigma_{w,i}^{-1} \cot(\pi/M)  [~ \Im(s_i^* \bh_i^T), \Re(s_i^* \bh_i^T) ~]^T.
\end{align*}
It is worthwhile to note that problem \eqref{eq:minimax_key} is convex.

Our second challenge is to find a suitable algorithm for computing the optimal solution to problem \eqref{eq:minimax_key}; note that we consider large $N$.
Since problem \eqref{eq:minimax_key} can be formulated as a linear program, one could use general-purpose conic optimization software to complete the task.
However, we argue that this is not preferred for large $N$.
Here we give two solutions; both exploit the problem structure.
One is to apply the smoothed accelerated projected gradient (APG) method, previously developed for the non-convex one-bit precoding problem \cite{shao2018framework}.
Concisely, the method works as follows.
We first approximate the non-differentiable $f$ by
\[
\hat{f}(\bx) = \mu \log \left( \sum_{i=1}^{2K} e^{\bc_i^T \bx /\mu} \right),
\]
where $\mu > 0$;
note that $\hat{f}$ is smooth and it is a tight approximation of $f$ when $\mu \rightarrow 0$.
Then, we apply the APG method \cite{nesterov1983method,beck2009fast,beck2017first} on the smoothed problem.
This gives rise to the following algorithm
\begin{equation} \label{eq:primal_apg}
\bx^{k+1} = \left[ \bx_{\sf ex}^k - \beta^k  \nabla \hat{f}(\bx_{\sf ex}^k  )  \right]_{-\bone}^\bone, \quad k=0,1,2,\ldots
\end{equation}
where $\beta^k > 0$ is the step size; $\nabla \hat{f}(\bx)$ is the gradient of $\hat{f}$ at $\bx$; $\bx_{\sf ex}^k$, called an extrapolated point, is given by
\[
\bx_{\sf ex}^k = \bx^k + \gamma_k ( \bx^k - \bx^{k-1}),
\]
with $\gamma_k = (\xi_{k-1} - 1)/\xi_k$, $\textstyle \xi_i = (1 + \sqrt{1+4\xi_{k-1}^2})/2$ and $\xi_{-1}= 0$;
the notation $\left[ \cdot  \right]_{-\bone}^\bone$ denotes the projection onto $[-1,1]^n$.
Note that $\left[ \cdot  \right]_{-\bone}^\bone$ is merely an element-wise clipping function; i.e., if $\by = \left[ \bx  \right]_{-\bone}^\bone$ then $y_i = \max\{ -1, \min\{ x_i, 1 \}   \}$  for all $i$.
We choose $\beta^k$ as the reciprocal of the Lipschitz constant of $\nabla \hat{f}$, a choice that guarantees convergence to an optimal solution.
The Lipschitz constant of $\nabla \hat{f}$ can be shown to be $\| \bC \|_2^2/\mu$ \cite{beck2017first},
where  $\bC = [~ \bc_1, \ldots, \bc_{2K} ~]$, and $\| \cdot \|_2$ denotes the spectral matrix norm.
We will call the algorithm in \eqref{eq:primal_apg} the primal APG method.

The second solution considers a dual form of problem \eqref{eq:minimax_key}.
The primal APG method has $2N$ decision variables, which is large,
and the motivation of the dual method is to see if we can use a smaller number of variables to solve problem \eqref{eq:minimax_key}.
More accurately, consider a regularized form of problem \eqref{eq:minimax_key}
\begin{equation} \label{eq:minimax_key_reg}
\begin{aligned}
\min_{ -\bone \leq \bx \leq \bone } f(\bx) + \frac{\tau}{2}  \| \bx \|_2^2
\end{aligned}
\end{equation}
for some small $\tau > 0$.
As a key observation, we note that
\[
f(\bx) = \max_{\blam \geq \bzero, \blam^T \bone = 1} \blam^T \bC^T \bx.
\]
The above alternative expression of $f$ leads us to
\begin{subequations} \label{eq:dual_eq}
\begin{align}
\text{\eqref{eq:minimax_key_reg}}
 & =
	\min_{ -\bone \leq \bx \leq \bone } \ \max_{\blam \geq \bzero, \blam^T \bone = 1}
	\ \blam^T \bC^T \bx + \frac{\tau}{2} \| \bx \|_2^2
	\\
	& = \max_{\blam \geq \bzero, \blam^T \bone = 1} \ \min_{ -\bone \leq \bx \leq \bone }
	\ \blam^T \bC^T \bx + \frac{\tau}{2} \| \bx \|_2^2
	 \label{eq:dual_eq_b} \\
	& = \max_{\blam \geq \bzero, \blam^T \bone = 1} \ g(\blam) \triangleq \sum_{i=1}^{2N} - \varphi_\tau(\bar{\bc}_i^T \blam ), \label{eq:dual_eq_c}
\end{align}
\end{subequations}
where $\bar{\bc}_i$ denotes the $i$th row of $\bC$; $\varphi_\tau$ is the Huber function and is given by
\[
\varphi_\tau(y) = \left\{ \begin{array}{ll}  y^2/(2\tau) & |y| \leq \tau \\ |y|- \tau/2 & \text{otherwise} \end{array} \right.
\]
Note that
\eqref{eq:dual_eq_b} is due to Sion's minimax theorem \cite{sion1958general},
and \eqref{eq:dual_eq_c} is due to $\min_{-1 \leq x \leq 1 } \ y x + \tau x^2 /2 = -\varphi_\tau(y)$ which one can easily show.
Consider the dual problem in \eqref{eq:dual_eq_c}, which has $2K$ decision variables.
In the same vein as the previously introduced APG method, we use APG to solve problem \eqref{eq:dual_eq_c}
\begin{equation} \label{eq:dual_apg}
\blam^{k+1} = \Pi_{\{ \blam \geq \bzero \mid \blam^T \bone =1 \}} \left( \blam_{\sf ex}^k + \beta^k  \nabla g(\blam_{\sf ex}^k  )  \right),
\end{equation}
for $k=0,1,2,\ldots$,
where $\Pi_{\{ \blam \geq \bzero \mid \blam^T \bone =1 \}}$ denotes the projection onto the unit simplex;
$\blam_{\sf ex}^k$ is defined in the same way as $\bx_{\sf ex}^k$; $\beta^k$ is the step size.
Note that there exist very efficient algorithms for computing the unit simplex projection \cite{condat2016fast}.
Also, the Lipschitz constant of $\nabla g$ is shown to be $\| \bC \|_2^2/\tau$.

Once we compute the optimal solution $\blam^\star$ to problem \eqref{eq:dual_eq_c}, the question that remains is how we can use $\blam^\star$ to recover the optimal solution $\bx^\star$ to problem \eqref{eq:minimax_key_reg}.
From the study of minimax theory \cite{bertsekas2003convex}, it is understood that $\bx^\star$ must be an optimal solution to
\begin{equation} \label{eq:inn_sol}
\min_{ -\bone \leq \bx \leq \bone }  (\blam^\star)^T  \bC^T \bx + \frac{\tau}{2} \| \bx \|_2^2.
\end{equation}
Since problem \eqref{eq:inn_sol} has one optimal solution only, owing to the strong convexity of its objective function,
the optimal solution to problem \eqref{eq:inn_sol} must be $\bx^\star$ itself.
The optimal solution to \eqref{eq:inn_sol} is simply
\begin{equation} \label{eq:dual_recover}
\bx^\star = \left[ - \tfrac{1}{\tau} \bC \blam^\star \right]_{-\bone}^\bone.
\end{equation}
We will call the method in \eqref{eq:dual_apg} and \eqref{eq:dual_recover} the dual APG method.

From our numerical experience, the primal and dual APG methods are both competitive.
This will be shown in the simulation results section.

\subsection{The QAM Case}
\label{sec:qam}

Having studied the $\Sigma\Delta$ ZF and SLP schemes for the $M$-ary PSK constellation case in the preceding subsections,
we now consider the $M$-ary QAM constellation case.
There is an aspect we need to discuss first.
Previously we ignore the time index $t$ in the basic signal model \eqref{eq:base_model}.
This is without loss of generality since PSK symbols do not require signal amplitude information for detection, and it is unnecessary to coordinate the scalings of the received signals over time.
But this is no longer true in the QAM case.
More technically, in \eqref{eq:sym_shape}, we need to make the received signal scaling coefficients $c_{i,t}$'s to be consistent for every symbol, namely, by having $c_{i,1} = \cdots = c_{i,T} \triangleq c_i$ for every $i$
\cite{Jacobsson2016Nonlinear,Sohrabi2018,shao2018framework,JeddaMSN18}.

Let us consider the ZF scheme under the above design consideration.
We modify the ZF scheme in Section~\ref{sec:sig_del_zf} as
\begin{equation} \label{eq:zf_sig_del_qam}
\bar{\bx}_t = \frac{\bA^\dag \bD  \bs_t}{\max_{t=1,\ldots,T}  \| \bA^\dag \bD  \bs_t \|_{IQ-\infty}}, \quad t=1,\ldots,T.
\end{equation}
Following the same development as before,
one can show that the corresponding received signals are given by
\begin{equation} \label{eq:zf_sig_del_qam_rx}
y_{i,t} = c_i \cdot s_{i,t} + w_{i,t}, \quad c_i = \sqrt{\frac{P}{2N}} \gamma \sigma_{w,i},
\end{equation}
where $\gamma = 1/ \max_{t=1,\ldots,T}  \| \bA^\dag \bD  \bs_t \|_{IQ-\infty}$.
It can be shown that the same result in Proposition \ref{prop:eff_snr_zf} applies  here.

For the SLP scheme, essentially the same problem was considered in \cite{shao2018framework}.
The latter considers the minimax SEP design in non-convex one-bit precoding,
and does so by joint optimization of all $\bar{\bx}_1,\ldots,\bar{\bx}_T$ and all scalings $c_1,\ldots,c_K$.
This amounts to a large-scale problem, but enhanced SEP was also observed.
The algorithm proposed there is similar to the primal APG method in Section~\ref{sec:sig_del_slp}, with a non-convex penalty term for forcing a binary solution.
By removing that penalty term, the algorithm will be applicable to our $\Sigma\Delta$ SLP design.
We omit the details due to space limitation.

We propose one more scheme that strikes a balance between ZF and SLP.
Consider the following nullspace-assisted ZF scheme
\ifconfver
\begin{equation} \label{eq:null_zf_sig_del_qam}
\bar{\bx}_t \!= \! \frac{\bA^\dag \bD  \bs_t + \bm \eta_t}{\max_{t=1,\ldots,T}  \| \bA^\dag \bD  \bs_t + \bm \eta_t \|_{IQ-\infty}}, ~ t=1,\ldots,T,
\end{equation}
\else
\begin{equation} \label{eq:null_zf_sig_del_qam}
\bar{\bx}_t = \frac{\bA^\dag \bD  \bs_t + \bm \eta_t}{\max_{t=1,\ldots,T}  \| \bA^\dag \bD  \bs_t + \bm \eta_t \|_{IQ-\infty}}, \quad t=1,\ldots,T,
\end{equation}
\fi
where every $\bm \eta_t$ lies in the nullspace of $\bA$.
The scheme \eqref{eq:null_zf_sig_del_qam} is a more general version of the ZF scheme \eqref{eq:zf_sig_del_qam},
taking advantage of the design simplicity of the latter.
It is also a special case of the SLP scheme.
From the alternative SLP interpretation \eqref{eq:slp_int}--\eqref{eq:slp_int2},
one can see that \eqref{eq:null_zf_sig_del_qam} is an SLP scheme that drops the symbol perturbation terms $\bu_t$'s, adopts the simple way to decide the received signal gains $\beta_i$'s in ZF, but keeps the nullspace term $\bm \eta_t$.
The received signals of the scheme \eqref{eq:null_zf_sig_del_qam} are the same as \eqref{eq:zf_sig_del_qam_rx}, with $\gamma$ replaced by $\gamma= 1/\max_{t=1,\ldots,T}  \| \bA^\dag \bD  \bs_t + \bm \eta_t \|_{IQ-\infty}$.
Now, the problem is to find $\bm \eta_1,\ldots,\bm \eta_T$ such that $\gamma$ is maximized.
It is readily seen that we can achieve this by solving, in a time decoupled manner,
\begin{equation} \label{eq:prob_inf_norm}
 \min_{\bm \xi_t \in \Cbb^{N-K}} \| \br_t + \bB \bm \xi_t \|_{IQ-\infty},
 \quad t=1,\ldots,T,
\end{equation}
where we apply change of variable ${\bm \eta}_t = \bB \bm \xi_t$;
$\bB \in \Cbb^{N \times (N-K)}$ is an orthogonal basis of the nullspace of $\bA$;
$\br_t = \bA^\dag \bD  \bs_t$.
We will show by simulation results that this nullspace-assisted ZF scheme provides order-of-magnitude SEP improvement over the ZF scheme.

We finish by mentioning how we solve the problems in \eqref{eq:prob_inf_norm}.
We first reformulate each problem in \eqref{eq:prob_inf_norm} in a form similar to problem \eqref{eq:minimax_key}, but without the constraint $\bx \in [-1,1]^{2N}$.
Then we apply the smoothed APG method in \eqref{eq:primal_apg} (without projection) to find the solution.
We omit the details for the sake of brevity.

\subsection{The Multi-Path Case}

Our preceding developments can also be extended to the case of multi-path angular channels.
Consider the multi-path channel model
\begin{equation}
\bh_i = \sum_{l=1}^{L_i} \alpha_{il} \ba(\theta_{il}),
\label{eq:ch_mpath}
\end{equation}
where $\alpha_{il}$ and $\theta_{il}$ correspond to the complex channel gain and angle of the $l$th path to the $i$th user, respectively; $L_i$ is the number of paths associated with the $i$th user.
Following the same development as in the preceding sections, it can be shown that the basic signal model takes the same form as \eqref{eq:model_mu}, i.e.,
\[
y_{i}  = \sqrt{\frac{P}{2N}} \bh_i^T  \bar{\bx} + w_{i}, \quad i=1,\ldots,K.
\]
The difference is that the expression of the noise variance $\sigma_{w,i}^2$ is replaced by
\begin{align} \label{eq:sigma_w_mp}
\sigma_{w,i}^2 & \approx \frac{ P}{3N}
\left( \sum_{n=0}^{N-1} \left| \sum_{l=1}^{L_i} \alpha_{il} (z_{il}^{-n} - z_{il}^{-n-1}) \right|^2 \right)
+ \sigma_v^2,
\end{align}
where $z_{il} =  e^{\jj \frac{2\pi d}{\lambda} \sin(\theta_{il})}$.
As a result, the ZF and SLP schemes developed above can be applied to the multi-path case (with minor modifications).
The detailed derivations of \eqref{eq:sigma_w_mp} are relegated to Appendix \ref{app:noise_va_mpath}.


We
should mention that $\sigma_{w,i}^2$ does not increase with $N$.
Using $| x_1 + \cdots + x_L |^2 \leq L (| x_1 |^2 + \cdots + |x_L|^2 )$, we show from \eqref{eq:sigma_w_mp} that
\begin{align*}
\sigma_{w,i}^2 & \leq \frac{4  P L_i}{3} \sum_{l=1}^{L_i}|\alpha_{il}|^2 \left| \sin\left( \frac{\pi d}{\lambda} \sin(\theta_{il})  \right) \right|^2 + \sigma_v^2.
\end{align*}
As can be seen, the above bound does not depend on $N$.

\section{Simulation Results}
\label{sec:sim}

This section shows our simulation results for $\Sigma\Delta$ precoding.

\subsection{Single-User Case with Basic $\Sigma\Delta$ Modulation}

We start with the single-user case, specifically, the basic $\Sigma\Delta$ MRT scheme in Section~\ref{sec:mrt_su}.
The simulation settings are as follows.
The number of antennas and the inter-antenna spacing are $N=256$ and $d= \lambda/8$, respectively.
The complex channel gain $\alpha$ has unit amplitude and phase uniformly drawn from $[-\pi,\pi]$ in each simulation trial.
The symbol constellation is $8$-ary PSK.
For benchmarking we also evaluate the theoretical SEP bound of the basic $\Sigma\Delta$ MRT scheme, i.e., \eqref{eq:theo_psk_sep}--\eqref{eq:SNR_eff}, and the simulated symbol error rate (SER) performance of the unquantized MRT scheme.
Here, unquantized MRT  (or any precoding) refers to the case where one applies MRT (or any precoding) without the one-bit signal restriction.

\ifconfver
	\begin{figure}[hbt!]
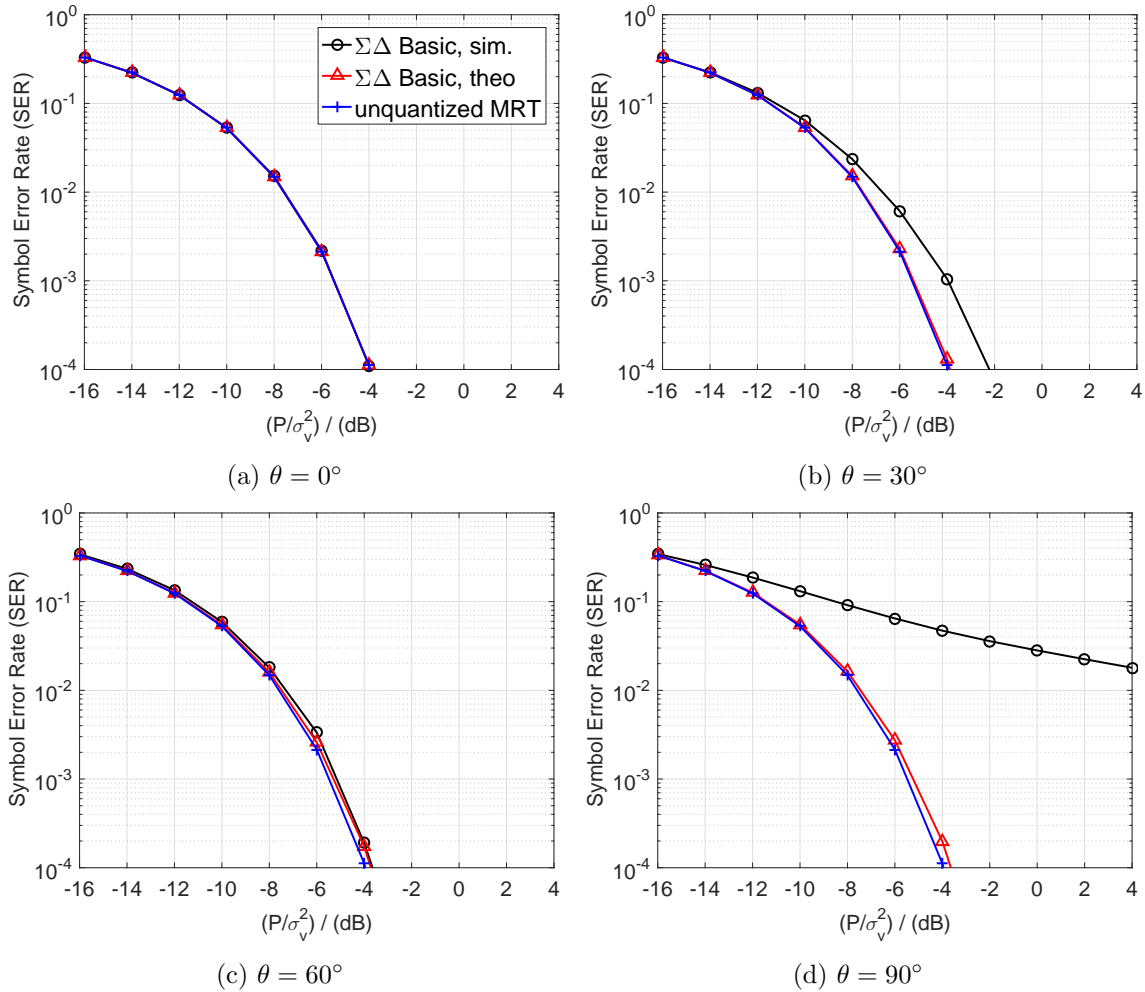

		\centering	
	\begin{subfigure}[b]{0.49\linewidth}
			\includegraphics[width=\textwidth]{fig_temp/theta0.eps}
			\caption{$\theta=0^\circ$}
		\end{subfigure}
	\begin{subfigure}[b]{0.49\linewidth}
			\includegraphics[width=\textwidth]{fig_temp/thetapi6.eps}
			\caption{$\theta=30^\circ$}
		\end{subfigure}
	\begin{subfigure}[b]{0.49\linewidth}
			\includegraphics[width=\textwidth]{fig_temp/thetapi3.eps}
			\caption{$\theta=60^\circ$}
		\end{subfigure}
	\begin{subfigure}[b]{0.49\linewidth}
			\includegraphics[width=\textwidth]{fig_temp/thetapi2.eps}
			\caption{$\theta=90^\circ$}
		\end{subfigure}
		\caption{SERs for different $\theta$.}\label{fig:phi0theta}
	\end{figure}
\else
	\begin{figure}[H]
		\centering	
		\begin{subfigure}[b]{0.45\linewidth}
			\includegraphics[width=\textwidth]{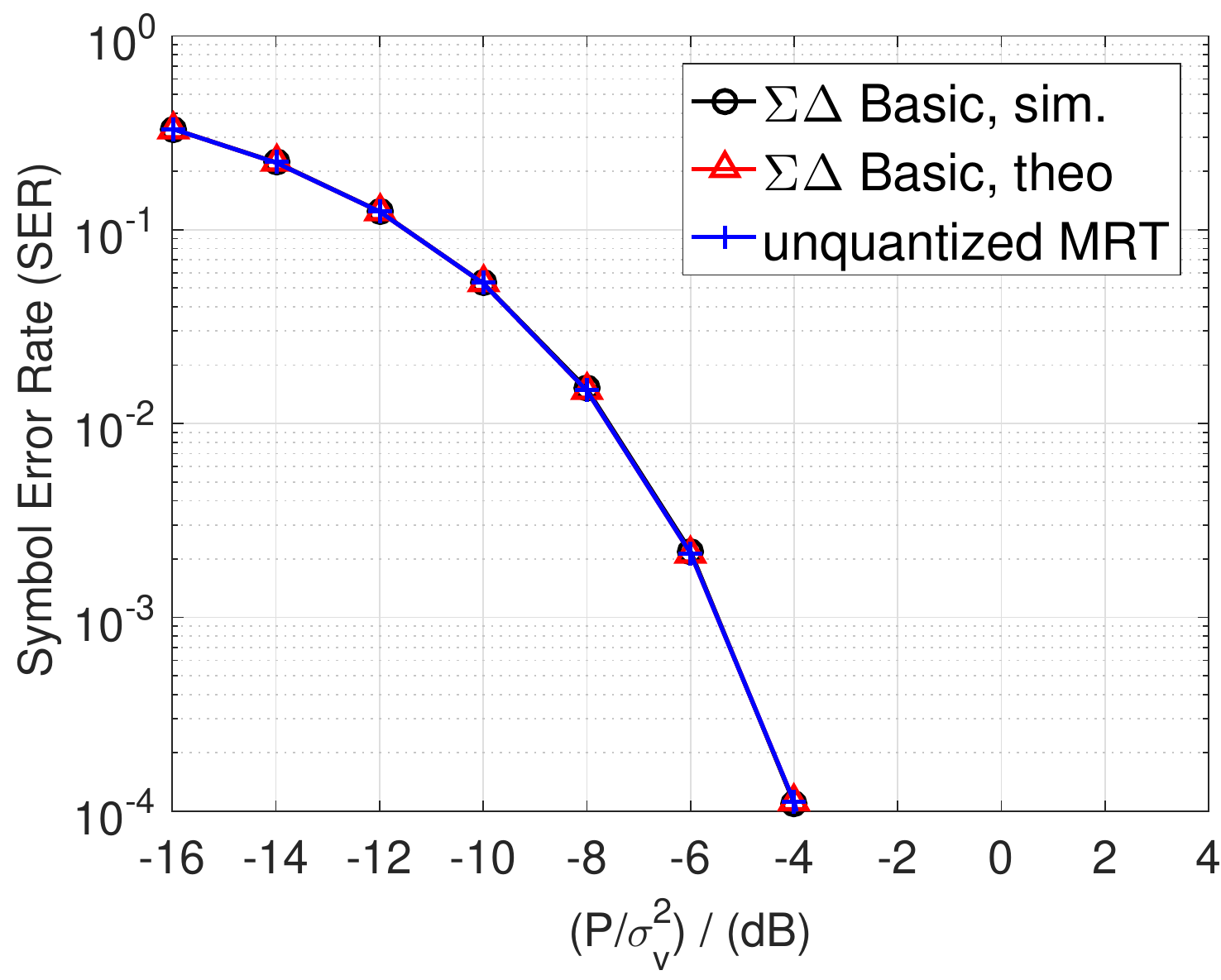}
			\caption{$\theta=0^\circ$}
		\end{subfigure}
		~
		\begin{subfigure}[b]{0.45\linewidth}
			\includegraphics[width=\textwidth]{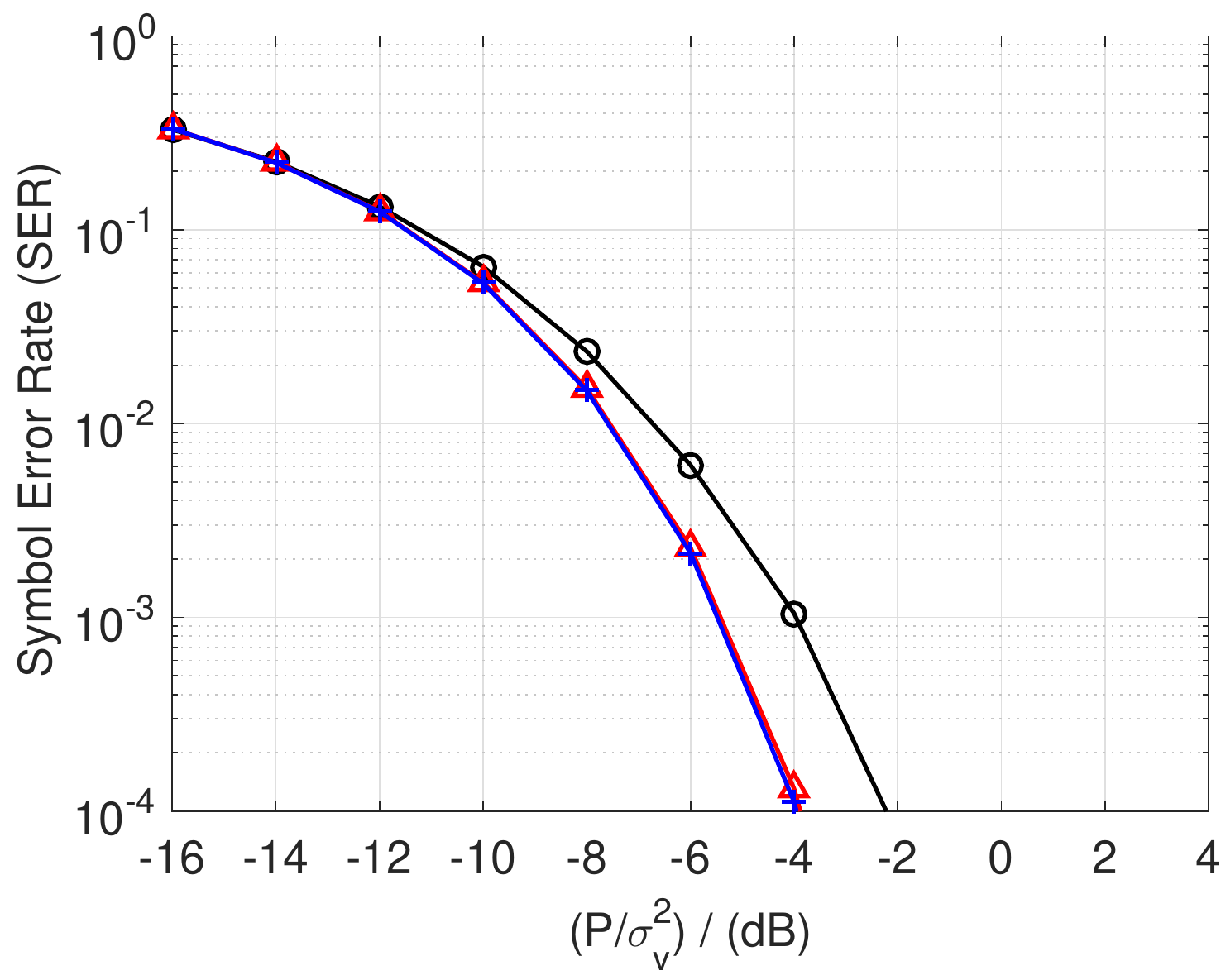}
			\caption{$\theta=30^\circ$}
		\end{subfigure}
		~
		\begin{subfigure}[b]{0.45\linewidth}
			\includegraphics[width=\textwidth]{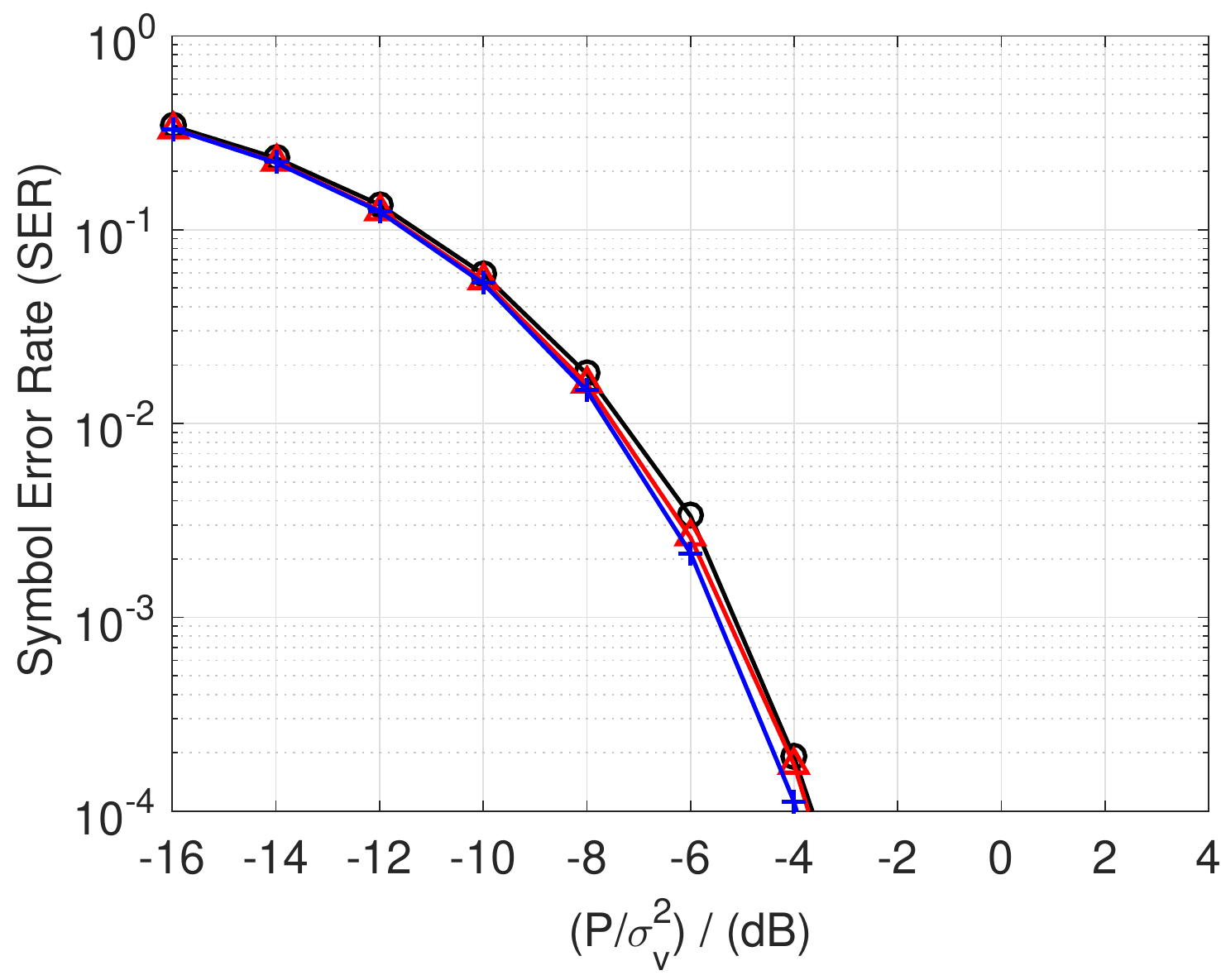}
			\caption{$\theta=60^\circ$}
		\end{subfigure}
		~
		\begin{subfigure}[b]{0.45\linewidth}
			\includegraphics[width=\textwidth]{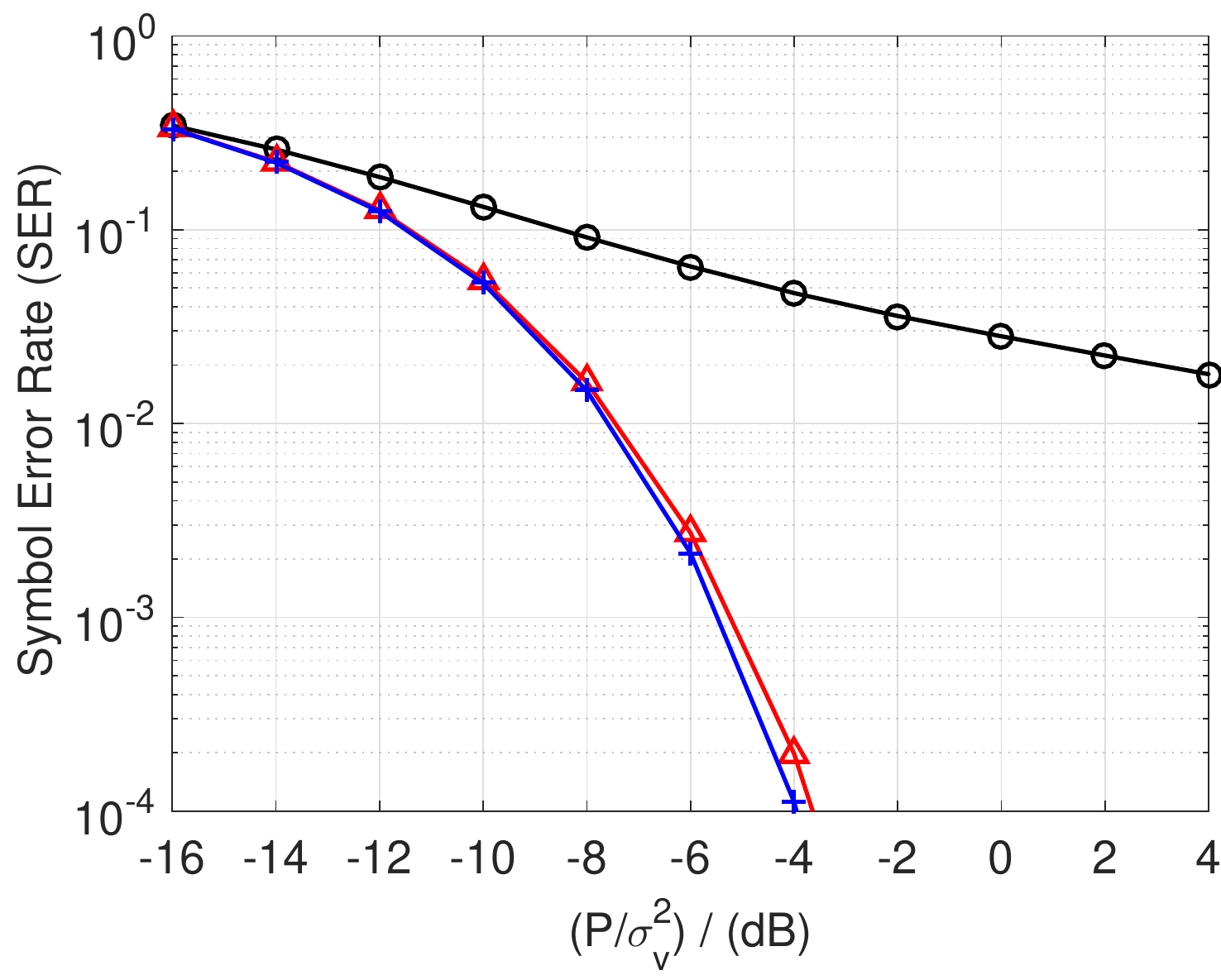}
			\caption{$\theta=90^\circ$}
		\end{subfigure}
		~
		\caption{SERs for different $\theta$.
		}\label{fig:phi0theta}
	\end{figure}
\fi

Fig. \ref{fig:phi0theta} shows the SER performance under several different values of the user angle $\theta$.
We see that, for the cases of $\theta=0^\circ$ and $\theta=60^\circ$, the simulated SER performance of the basic $\Sigma\Delta$ MRT scheme is almost the same as the theoretical.
For the case of $\theta=30^\circ$, we observe a small gap between the simulated and theoretical SER performance of the basic $\Sigma\Delta$ MRT scheme.
The reason, as we found out, is that the quantization noise could have its behavior deviating from the i.i.d. assumption in some specific cases, and $\theta=30^\circ$ happens to fall into one such case.
We may mitigate the non-i.i.d. effect by dithering, although it may not be worthwhile to try dithering in this case since the performance gap is small and dithering increases the quantization noise level.
Moreover, for the case of $\theta=90^\circ$, we notice that the simulated and theoretical SER performance has a significant gap.
Again, this is because the quantization noise is not i.i.d., and the non-i.i.d. effect is severe in this case.

Next, we show the radiation patterns of the basic $\Sigma\Delta$ MRT scheme.
The simulation settings are the same as the previous.
Fig.~\ref{fig:angular_spectrum} plots the angular power spectrum $P(\vartheta)= \Exp[ | \ba(\vartheta)^T \bar{\bx} |^2 ]$ for several values of the desired angle $\theta$.
We see that the actual angular power spectrum does not always look like what theory ideally suggests, i.e., superposition of the highpass quantization noise spectrum and the MRT signal spectrum, the latter of which appears as a spike at $\theta$.
We see a highpass response with the actual angular power spectrum for the instances of $\theta = 30^\circ, 60^\circ, 90^\circ$,
but this is not seen for the instance of $\theta = 0^\circ$.
We expect a single peak at the desired angle $\theta$, but we also see some smaller peaks at other angles for the instances of $\theta = 0^\circ, 30^\circ, 60^\circ$.
Also, we do not see a peak for the instance of $\theta = 90^\circ$.
The non-ideal phenomena we see are due to the non-i.i.d. effects, and they are identical to those in the temporal $\Sigma\Delta$ modulation of DC and sinusoidal input signals \cite{norsworthy1992effective,gray1990quantization}.
Nonetheless we can also argue that the actual angular spectrum roughly follows the theoretical, say, for $\theta = 30^\circ, 60^\circ$.
As an aside, while our interest is to apply precoding to a target user, the quantization noise of the $\Sigma\Delta$ scheme also causes interference to other angles---an issue one needs to be careful when operating under multi-cell interfering channel environments.
\ifconfver
	\begin{figure}[hbt!]
		\centering	
	\begin{subfigure}[b]{0.49\linewidth}
			\includegraphics[width=\textwidth]{fig_temp/angle_spectrum0.eps}
			\caption{$\theta=0^\circ$}
		\end{subfigure}
	\begin{subfigure}[b]{0.49\linewidth}
			\includegraphics[width=\textwidth]{fig_temp/angle_spectrum30.eps}
			\caption{$\theta=30^\circ$}
		\end{subfigure}
	\begin{subfigure}[b]{0.49\linewidth}
			\includegraphics[width=\textwidth]{fig_temp/angle_spectrum60.eps}
			\caption{$\theta=60^\circ$}
		\end{subfigure}
	\begin{subfigure}[b]{0.49\linewidth}
			\includegraphics[width=\textwidth]{fig_temp/angle_spectrum90.eps}
			\caption{$\theta=90^\circ$}
		\end{subfigure}
		\caption{The angular power spectrum for different $\theta$.}\label{fig:angular_spectrum}
	\end{figure}
\else
	\begin{figure}[hbt!]
		\centering	
		\begin{subfigure}[b]{0.45\linewidth}
			\includegraphics[width=\textwidth]{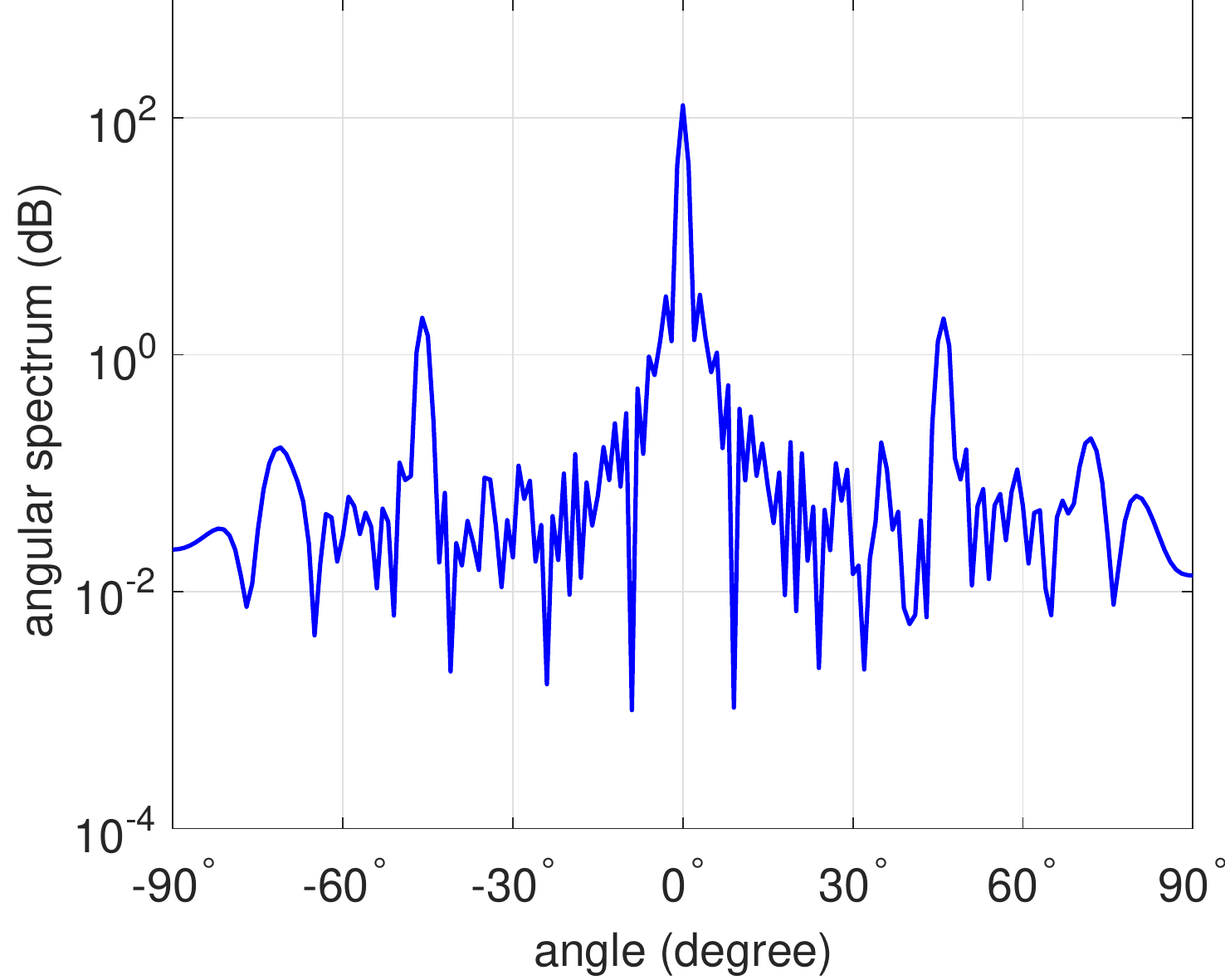}
			\caption{$\theta=0^\circ$}
		\end{subfigure}
		~
		\begin{subfigure}[b]{0.45\linewidth}
			\includegraphics[width=\textwidth]{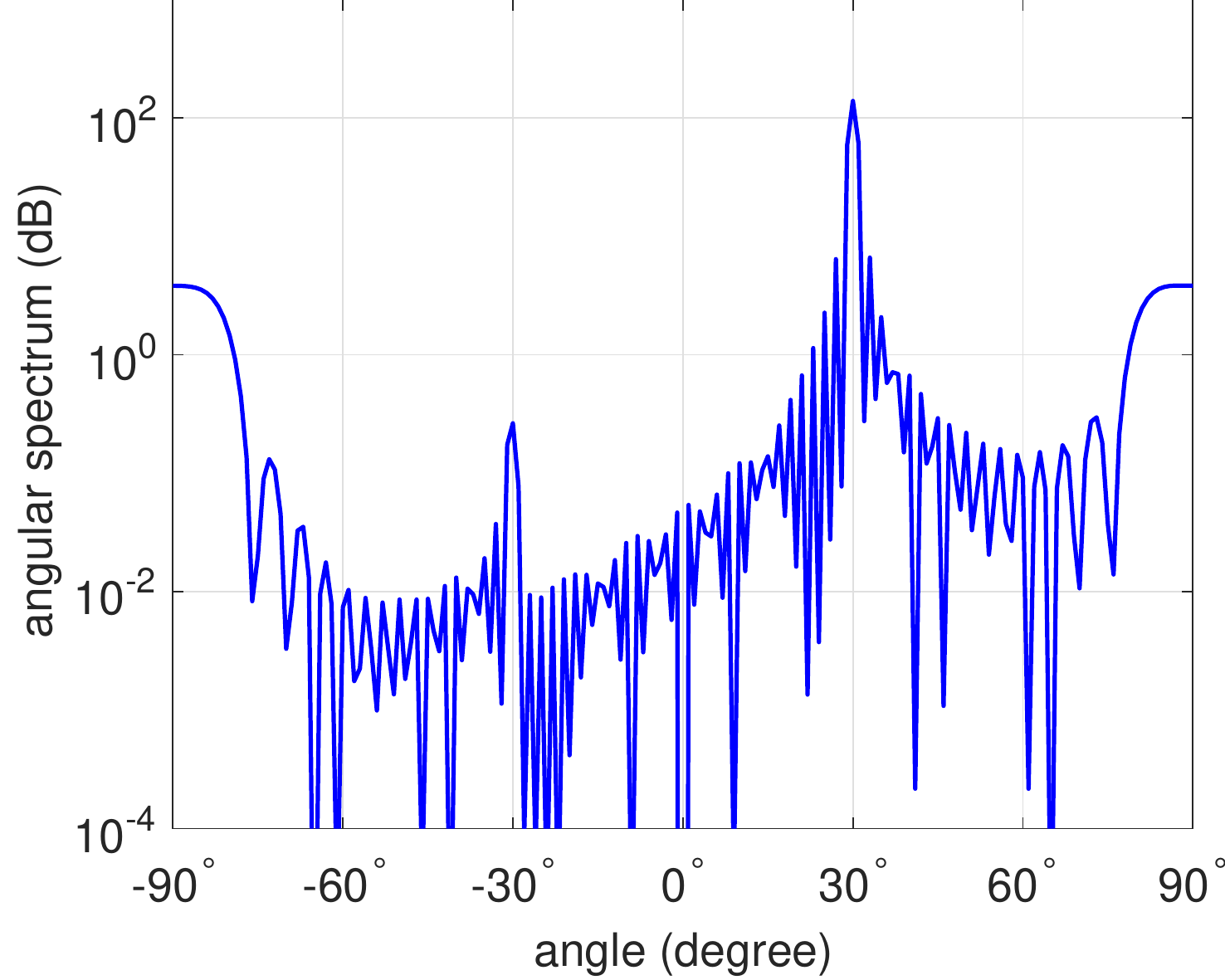}
			\caption{$\theta=30^\circ$}
		\end{subfigure}
		~
		\begin{subfigure}[b]{0.45\linewidth}
			\includegraphics[width=\textwidth]{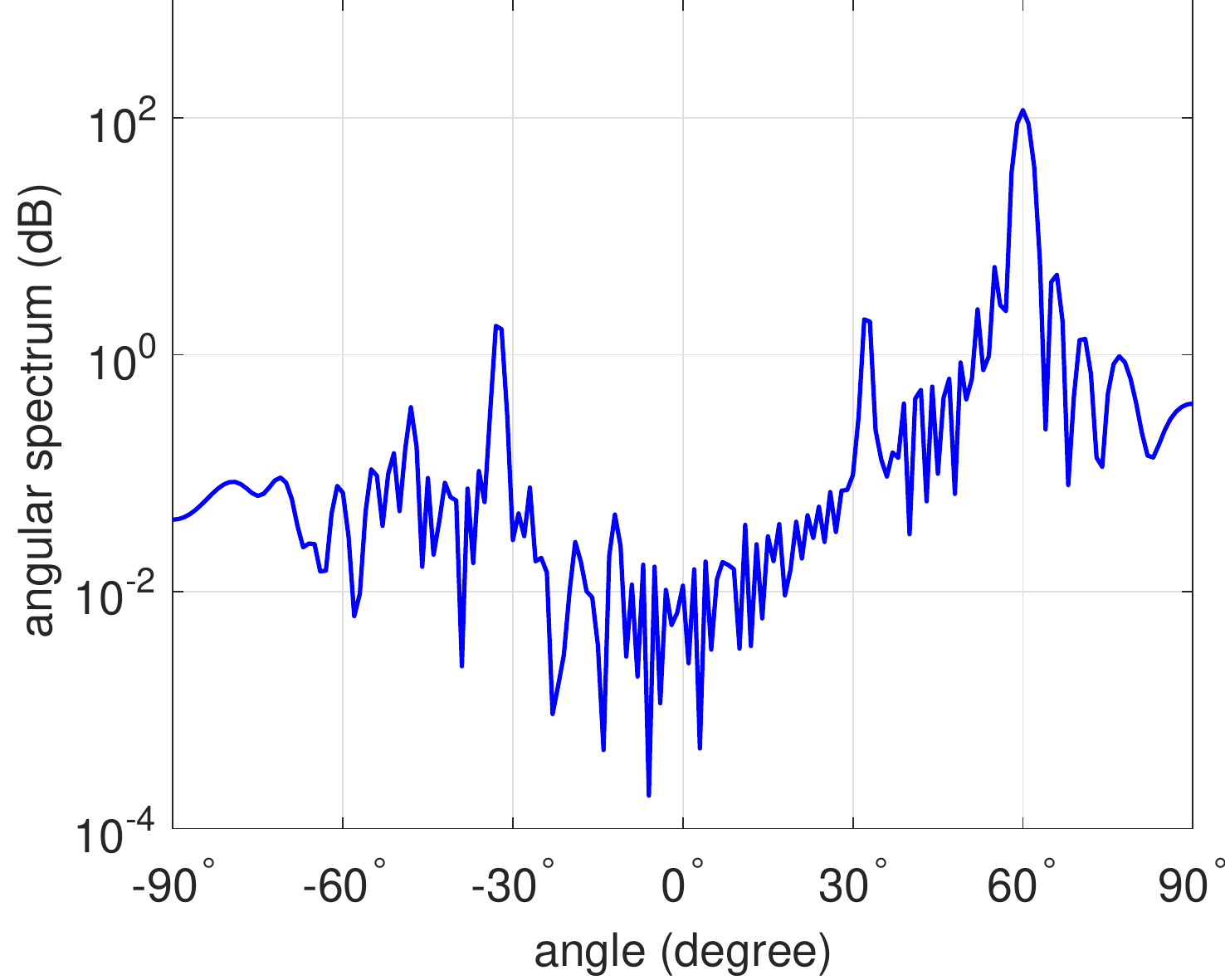}
			\caption{$\theta=60^\circ$}
		\end{subfigure}
		~
		\begin{subfigure}[b]{0.45\linewidth}
			\includegraphics[width=\textwidth]{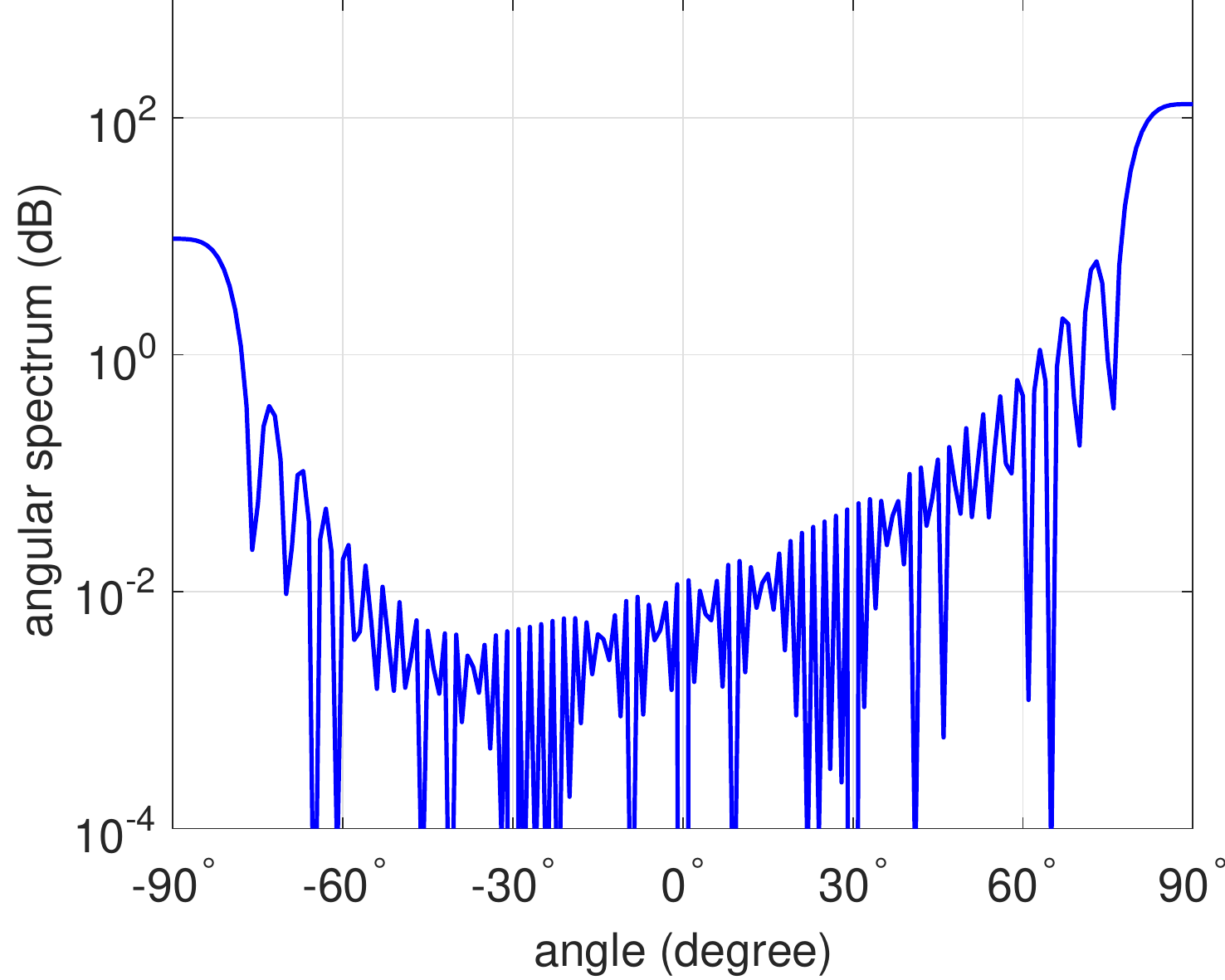}
			\caption{$\theta=90^\circ$}
		\end{subfigure}
		~
		\caption{Angular spectrum for different $\theta$.}\label{fig:angular_spectrum}
	\end{figure}
\fi

Finally,
we examine the SER performance under different numbers of antennas $N$.
The user angle is fixed at $\theta= 60^\circ$.
The result in Fig. \ref{fig:N} illustrates that, under the same SNR level $P/\sigma_v^2$, increasing $N$ reduces the SERs substantially.
This numerical observation is in agreement with the SEP analysis result in Section~\ref{sec:mrt_su}.

\ifconfver
\begin{figure}[htb]
	\centering
	\includegraphics[width=0.85\linewidth]{fig_temp/N.eps}
	\caption{SERs for different $N$.}\label{fig:N}
\end{figure}
\else
\begin{figure}[htb]
	\centering
	\includegraphics[width=0.55\linewidth]{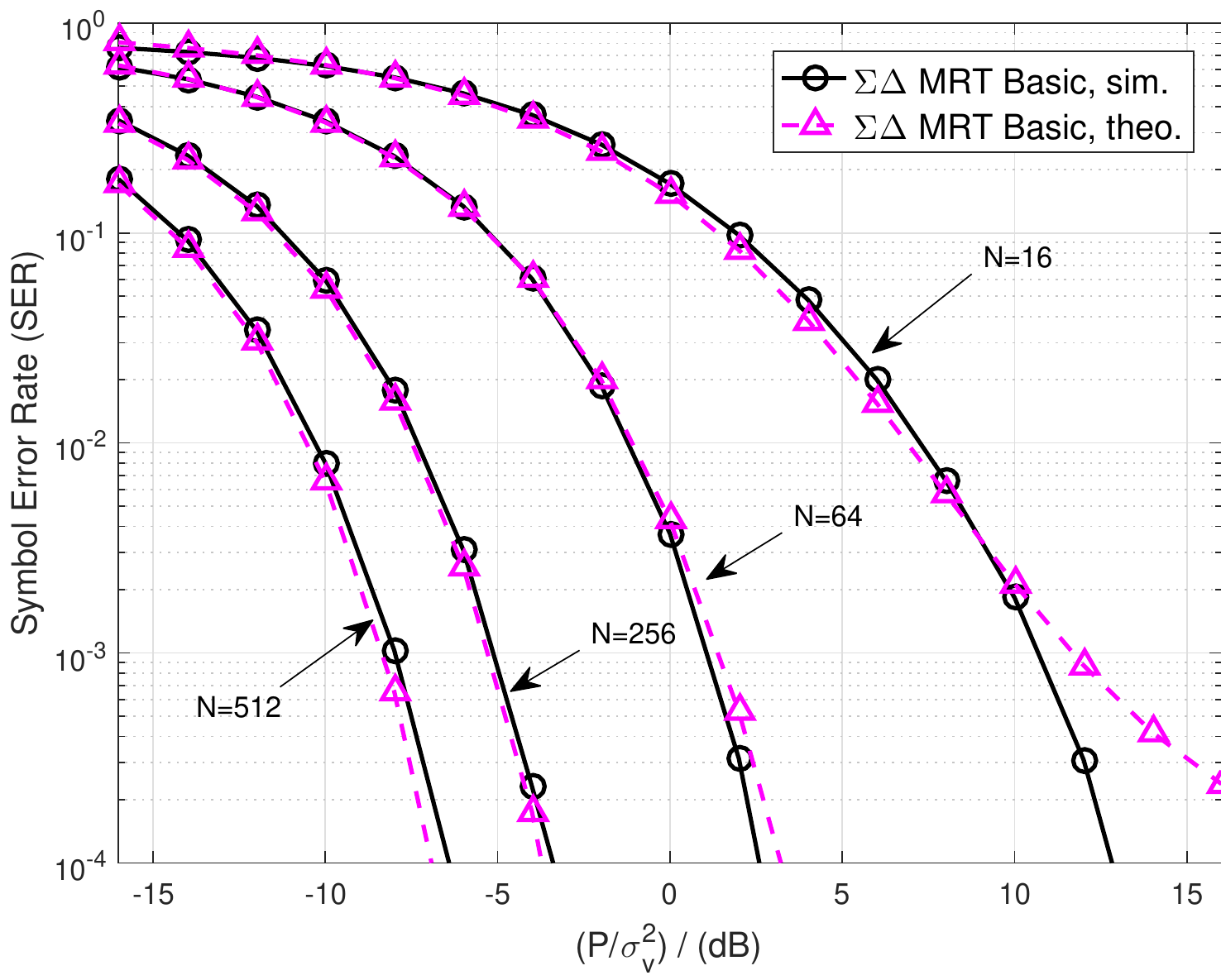}
	\caption{SERs for different $N$.}\label{fig:N}
\end{figure}
\fi

\subsection{Single-User Case with Angle-Steered $\Sigma\Delta$ Modulation}

We turn our attention to the angle-steered $\Sigma\Delta$ MRT scheme in Section~\ref{sec:angles}.
The simulation settings are essentially identical to those in the preceding subsection, and the difference is that
we reduce the  number of antennas to $N=128$, increase the inter-antenna spacing to $d=\lambda/2$, and try a large angle of $\theta = 90^\circ$.
The basic $\Sigma\Delta$ MRT scheme is expected to work poorly, as suggested by the analysis in Section~\ref{sec:mrt_su}.
Also, as we have seen in the simulation results in the preceding subsection, the non-i.i.d. quantization noise effect may become significant.
To mitigate the non-i.i.d. effect, we try the dithered $\Sigma\Delta$ MRT scheme, specifically, by applying the dithering procedure \eqref{eq:dit} to the basic spatial $\Sigma\Delta$ modulator.
The dithering level $\delta$ in \eqref{eq:dit} is set to $\delta=0.8$.

\ifconfver
	\begin{figure}[hbt]
		\centering	
			\includegraphics[width=0.85\linewidth]{fig_temp/AS_thetapi2.eps}
		\caption{SERs under angle-steering and dithering; $\theta=90^\circ$.}\label{fig:extreme}
	\end{figure}
\else
	\begin{figure}[hbt]
		\centering	
			\includegraphics[width=0.55\linewidth]{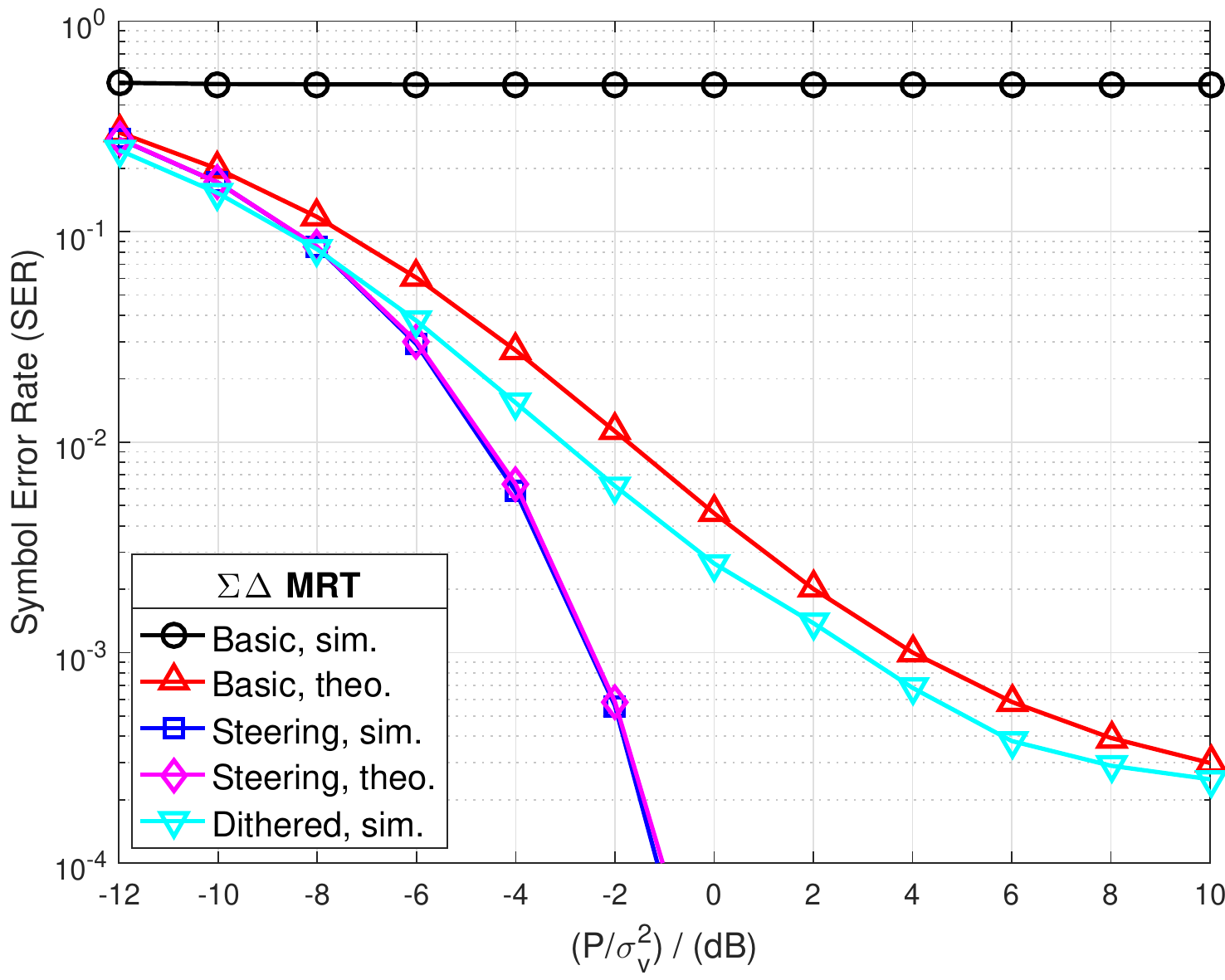}
		\caption{SERs under angle-steering and dithering; $\theta=90^\circ$.}\label{fig:extreme}
	\end{figure}
\fi

Fig. \ref{fig:extreme} shows the SER performance of the basic, dithered and angle-steered $\Sigma\Delta$ MRT schemes.
As seen previously in Fig. \ref{fig:phi0theta}(d), the basic $\Sigma\Delta$ MRT scheme suffers from the non-i.i.d. quantization noise effect when $\theta=90^\circ$.
The situation now is even worse.
We see from Fig. \ref{fig:extreme} that the basic $\Sigma\Delta$ MRT scheme completely fails,
and does not perform as the theoretical SER performance says.
The dithered $\Sigma\Delta$ MRT scheme yields significantly improved performance,
and this indicates that dithering can reduce the non-i.i.d. effect.
However, it is the angle-steered $\Sigma\Delta$ MRT scheme that gives the best performance.
Also, the theoretical SER performance of the angle-steered $\Sigma\Delta$ MRT scheme accurately predicts the simulated SER performance.

Next, we consider the generalized angle-steered $\Sigma\Delta$ MRT scheme in Section \ref{sec:gen_angles} under i.i.d. Gaussian channels.
Specifically, in each simulation trial, the channel $\bh$ is i.i.d. complex circular Gaussian generated with mean zero and unit variance.
The number of antennas is $N=256$, and the symbol constellation is $16$-ary QAM.
The benchmark scheme is the unquantized MRT.
The unquantized MRT scheme we consider is the one under the peak IQ amplitude constraint $\| \bx \|_{IQ-\infty} \leq 1$ and without one-bit quantization; precisely, it is implemented by \eqref{eq:MRT_gen_angles} with $A_n =1$ and with $\sigma_w^2 = \sigma_v^2$.
Also, we try the direct one-bit quantization of the unquantized MRT scheme, which we call it the quantized MRT scheme; we will use the same convention to name other direct one-bit quantized algorithms in the sequel.
In addition to the generalized angle-steered $\Sigma\Delta$ MRT scheme, we try a heuristic  where we overload the generalized angle-steered $\Sigma\Delta$ MRT scheme by setting $A_n =1$ for all $n$.
Careful readers will see from Section \ref{sec:gen_angles} that the issue will be that the surviving quantization noise term $q_N$ in \eqref{eq:w_gen_angles} may become large.
But if it does not in general, then the overloaded heuristic will
have the advantage of enhanced SNR.

\ifconfver
\begin{figure}[htb!]
	\centering
	\includegraphics[width=0.85\linewidth]{fig_temp/iidsteering16QAM.eps}
	\caption{SERs for the i.i.d. Gaussian channel.}\label{fig:iidQAM}
\end{figure}
\else
\begin{figure}[htb!]
	\centering
	\includegraphics[width=0.55\linewidth]{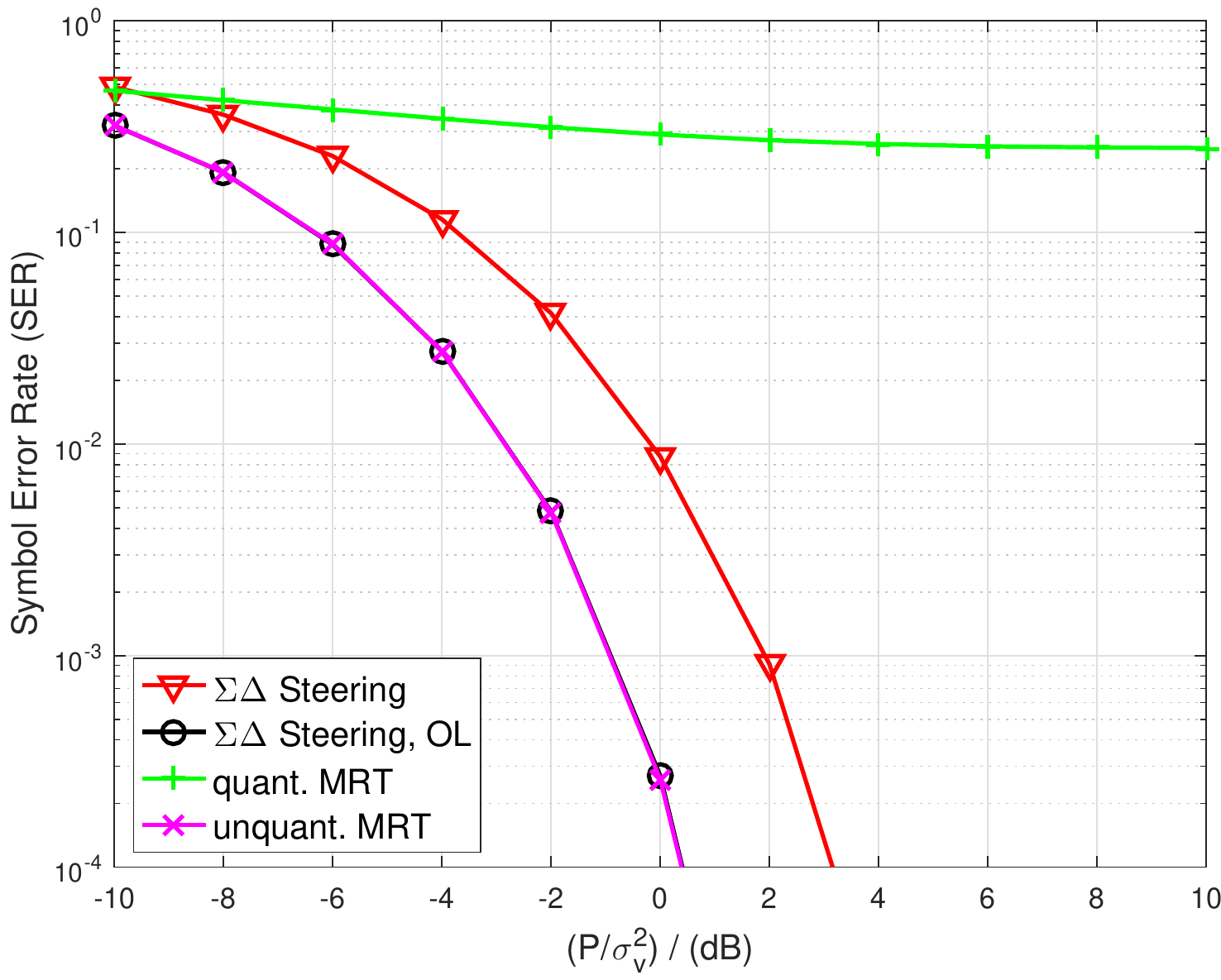}
	\caption{SERs for the i.i.d. Gaussian channel.}\label{fig:iidQAM}
\end{figure}
\fi

Fig. \ref{fig:iidQAM} shows the results.
We have the following observations.
First, the quantized MRT scheme fails to work.
Second, the generalized angle-steered $\Sigma\Delta$ MRT scheme (``$\Sigma\Delta$ Steering'' in the figure) yields SER performance that is about $3$dB away from that of the unquantized MRT scheme.
This agrees with our analysis, which suggests $4.64$dB as the worst case.
Third, the overloaded generalized angle-steered $\Sigma\Delta$ MRT scheme (``$\Sigma\Delta$ Steering, OL'') yields SER performance almost the same as that by the unquantized MRT.
While our present work only considers the no-overload case, this simulation result suggests that overloading can be beneficial.
We will leave overloading as a subject of future investigation.

\subsection{The Multi-User Case}

Now we consider the multi-user case.
The simulation settings are as follows:
The number of antennas is $N= 512$;
the inter-antenna spacing is $d= \lambda/8$;
the number of users is $K= 24$, and the users are within an angular range $[-30^\circ, 30^\circ]$;
the angles $\theta_i$'s are randomly picked from $[-30^\circ, 30^\circ]$ with inter-angle difference no  less than $1^\circ$;
the complex channel gains $\alpha_i$'s have phases uniformly drawn from $[-\pi,\pi]$, and their amplitudes are generated as $|\alpha_i| = r_0/r_i$ where $r_0 = 30$ and $r_i$ are uniformly drawn from $[20,100]$ (this is a standard free-space path-loss model, with $r_i$ being the distance from the BS to the $i$th user and $r_0$ being a reference value);
the symbol constellation is $8$-ary PSK.

The settings of the $\Sigma\Delta$ SLP scheme should also be mentioned.
For the primal APG, the smoothing parameter is $\mu =0.05$, and the algorithm stops when  $\| \bx^{k+1}-\bx^{k} \|_2\leq 10^{-5}$ or when a maximum iteration number of $2000$ is reached.
For the dual APG, the regularization parameter is $\tau=0.005$, and the algorithm stops when $\| \blam^{k+1}-\blam^{k} \|_2\leq 10^{-7}$ or when a maximum iteration number of $3000$ is reached.
	\begin{figure}[htb!]
		\centering
		\includegraphics[width=0.55\linewidth]{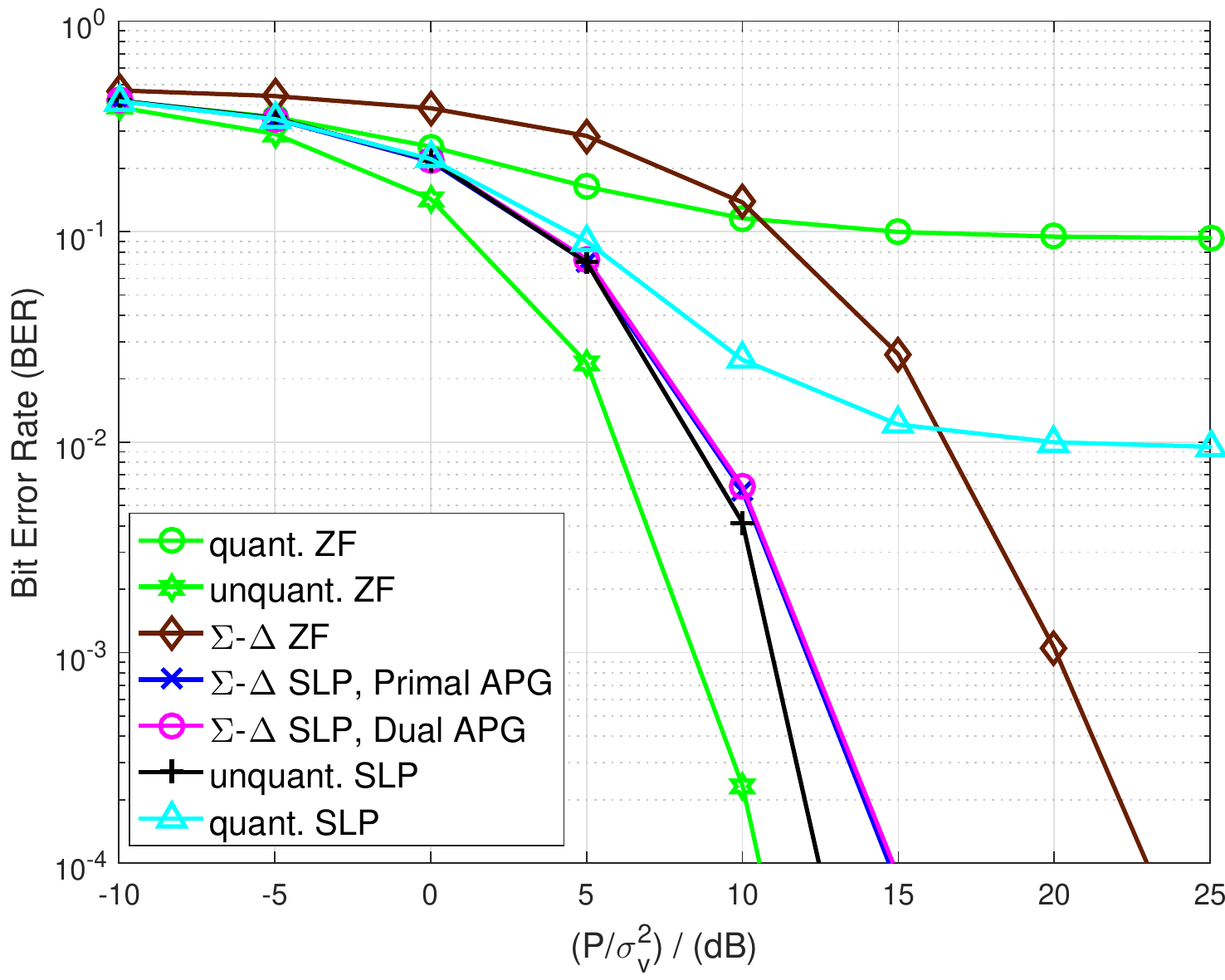}
		\caption{BERs for the multi-user case.}\label{fig:mu_1}
	\end{figure}

Fig. \ref{fig:mu_1} shows the results.
In the legend,
``unquant. ZF'' is the unquantized ZF scheme under the average power constraint;
``quant. ZF'' is the direct one-bit quantization of the unquantized ZF scheme;
``$\Sigma\Delta$ ZF'' is the $\Sigma\Delta$ ZF scheme in Section \ref{sec:sig_del_zf};
``$\Sigma\Delta$ Primal APG'' and ``$\Sigma\Delta$ Dual APG'' are the primal and dual $\Sigma\Delta$ SLP schemes in  Section \ref{sec:sig_del_slp}, respectively;
``unquant. SLP'' is the unquantized version of the SLP scheme;
``quant. SLP'' is the direct one-bit quantization of the unquantized SLP scheme.
We see that the proposed  $\Sigma\Delta$ ZF and SLP schemes work well.
The quantized ZF and SLP schemes do not, however.

Next, we perform benchmarking with some existing one-bit precoding designs.
The simulation settings are the same as the previous, except that we reduce the number of antennas to $N=256$ and the angular range to $[-22.5^{\circ}, 22.5^{\circ}]$.
The compared algorithms are the SQUID algorithm \cite{Jacobsson2017} and the maximum safety margin (MSM) algorithm \cite{Jedda2017}.
The results in Fig. \ref{fig:mu_2} show that the $\Sigma\Delta$ SLP scheme outperforms SQUID and MSM, and the $\Sigma\Delta$ ZF scheme performs better than the latter when the SNR is greater than $25$dB.
In addition to SERs, we also compare the algorithm runtimes.
Table \ref{tb:time} shows the runtime results;
the results were obtained on MATLAB, and a desktop computer with Intel i7-4770 processor and $16$GB memory was used to perform the runtime test.
We can see that the proposed $\Sigma\Delta$ SLP designs yield competitive runtime performance compared to SQUID and MSM.

\ifconfver
\begin{figure}[htb!]
	\centering
	\includegraphics[width=0.85\linewidth]{fig_temp/MUPSK_com}
	\caption{BER comparison of the multi-user $\Sigma\Delta$ precoding schemes and some existing one-bit precoding schemes.}\label{fig:mu_2}
\end{figure}
\else
\begin{figure}[hbt!]
	\centering
	\includegraphics[width=0.55\linewidth]{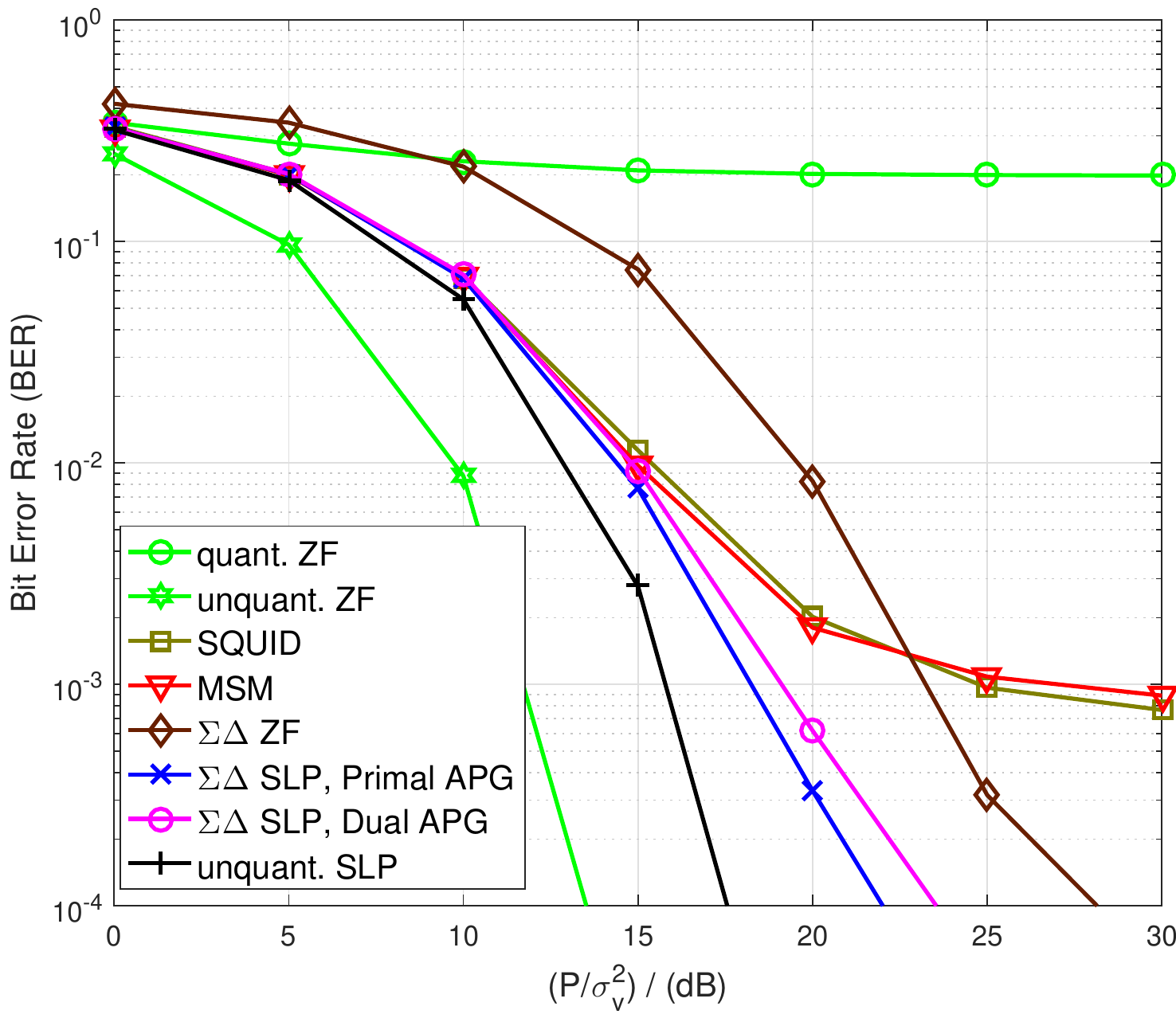}
	\caption{BERs for the multi-user case.}\label{fig:mu_2}
\end{figure}
\fi


\ifconfver
 \begin{table}[htb!]
	\centering
	\captionsetup{justification=centering}
	\caption{Average runtime (in Sec.) of different algorithms; $(N,K)=(256,24)$, $8$-ary PSK. \\[-0.4em]}\label{tb:time}
	\renewcommand{\arraystretch}{1.2}
	\resizebox{\linewidth}{!}{%
		\begin{tabular}{M{11mm}|M{10mm} M{14mm} M{13mm} M{7mm} M{7mm} }
			\hline
			Algorithm & $\Sigma$-$\Delta$ ZF & $\Sigma$-$\Delta$ Primal APG & $\Sigma$-$\Delta$ Dual APG & SQUID& MSM \\ \hline\hline
			runtime  & {\bf 0.0021} & {\bf 0.0574} & {\bf 0.0496} & 1.0324 &0.9197\\ \hline
		\end{tabular}
	}
\end{table}
\else
 \begin{table}[H]
	\centering
	\captionsetup{justification=centering}
		\caption{Average runtime (in Sec.) for different algorithms; $(N,K)=(256,24)$, $8$PSK. \\[-0.4em]}\label{tb:time}
	\renewcommand{\arraystretch}{1.2}
	\resizebox{0.9\linewidth}{!}{%
		\begin{tabular}{M{20mm}|M{20mm} M{30mm} M{30mm} M{20mm} M{20mm} }
			\hline
		Algorithm & $\Sigma$-$\Delta$ ZF & $\Sigma$-$\Delta$ Primal APG & $\Sigma$-$\Delta$ Dual APG & SQUID& MSM \\ \hline\hline
			runtime  & {\bf 0.0021} & {\bf 0.0574} & {\bf 0.0496} & 1.0324 &0.9197\\ \hline
		\end{tabular}
	}
\end{table}
\fi

Finally, we consider the QAM case.
The simulation settings are: $16$-ary QAM, $N= 256$, $K= 16$, and the transmission block length $T= 100$.
Also, the angular range is $[-30^\circ, 30^\circ],$ and the $\alpha_i$'s are generated by the same way as before.
Fig. \ref{fig:mu_QAM} shows the results.
In the plot,
``$\Sigma\Delta$ ZF'' is the $\Sigma\Delta$ ZF scheme in \eqref{eq:zf_sig_del_qam};
``$\Sigma\Delta$ null. ZF'' is the nullspace-assisted $\Sigma\Delta$ ZF scheme in \eqref{eq:null_zf_sig_del_qam}--\eqref{eq:prob_inf_norm},
and ``GEMM'' is the direct one-bit precoding design in \cite{shao2018framework}.
We see that the $\Sigma\Delta$ ZF schemes, with and without nullspace assistance, work.
Also we should pay particular attention to the nullspace-assisted $\Sigma\Delta$ ZF scheme.
It has a $5$dB gain compared to the $\Sigma\Delta$ ZF scheme,
and it is only $3$dB away from GEMM.
We should mention that GEMM handles a more complicated design problem than the nullspace-assisted $\Sigma\Delta$ ZF scheme.

\ifconfver
\begin{figure}[htb!]
	\centering
	\includegraphics[width=0.85\linewidth]{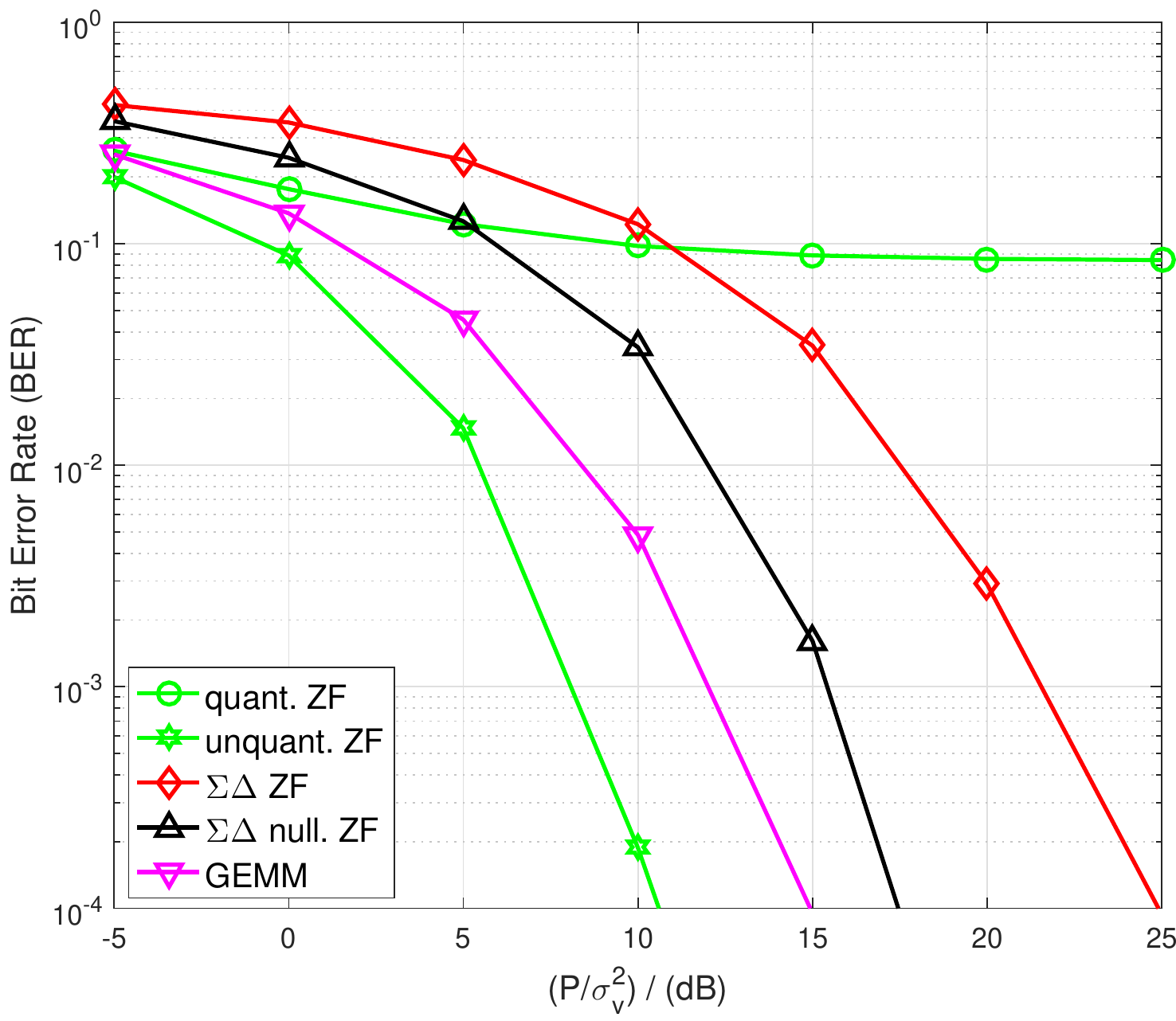}
	\caption{BERs of the multi-user $\Sigma\Delta$ precoding schemes in the $16$-ary QAM case.}\label{fig:mu_QAM}
\end{figure}
\else
\begin{figure}[htb!]
	\centering
	\includegraphics[width=0.55\linewidth]{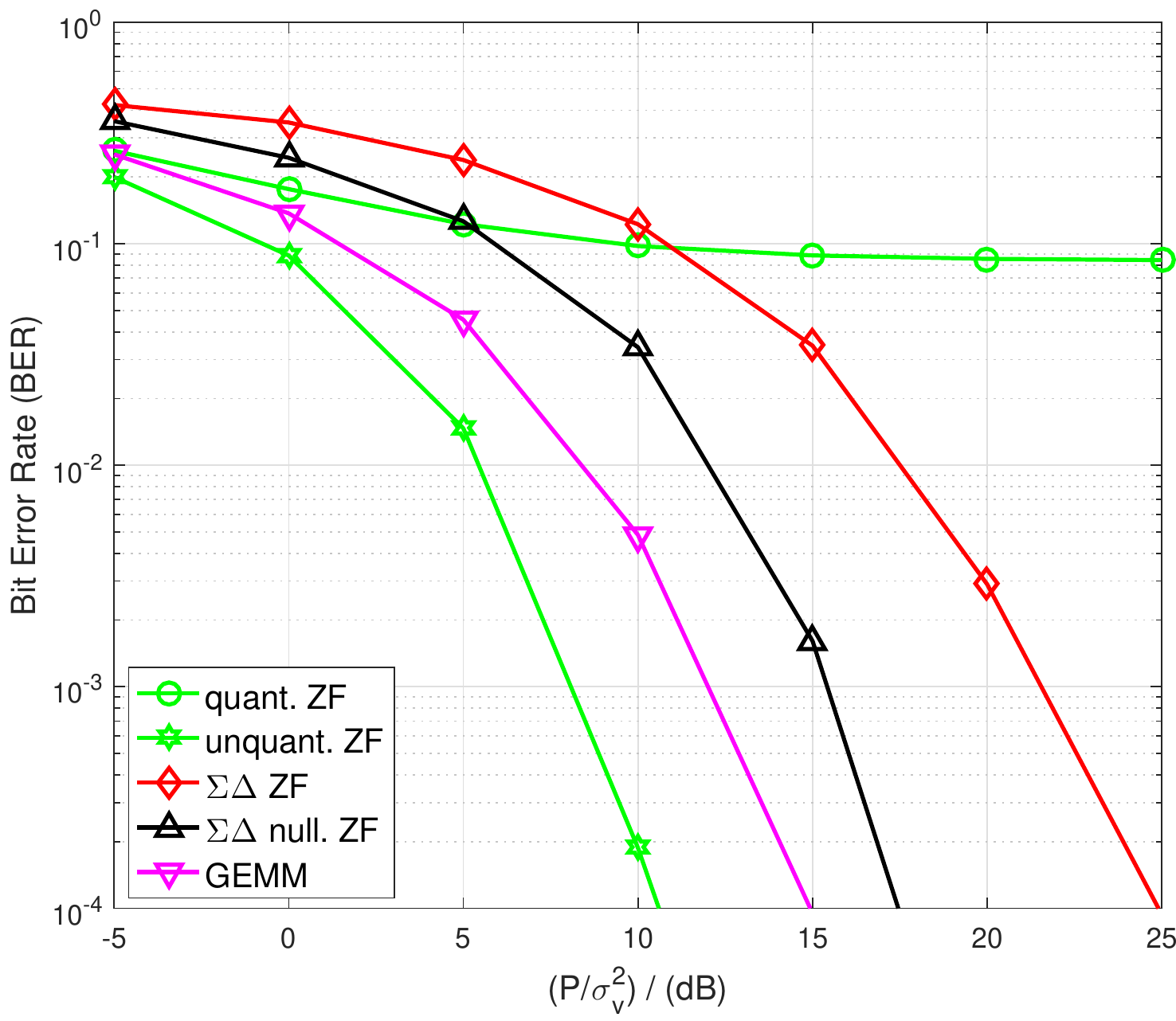}
	\caption{BERs for the multi-user case; $16$-QAM.}\label{fig:mu_QAM}
\end{figure}
\fi

\section{Conclusion}
\label{sec:conclusion}

In this paper we studied the potential of spatial $\Sigma\Delta$ modulation for one-bit MIMO precoding.
We showed that $\Sigma\Delta$ precoding is an excellent candidate when the system is equipped with a massive antenna array and when the users lie within a certain angular sector, which is a typical assumption in many cellular systems.
The major advantage of $\Sigma\Delta$ precoding is that
it can achieve good performance for relatively simple designs such as quantized linear precoding, whereas direct one-bit design requires complicated non-convex methods with binary signal constraints (or relaxed versions thereof) in order to obtain low error rates.
While our initial $\Sigma\Delta$ precoder assumed a simple angular channel, we showed how to generalize the idea to any type of channel
in the single-user case.
%

\appendix

\ifplainver
	\section*{Appendix}
	\renewcommand{\thesubsection}{\Alph{subsection}}
\else
	\section*{Appendix}
\fi

\section{Additional Numerical Results for the $\Sigma\Delta$ MRT Schemes} \label{supp:IQ_plots}

This section provides additional numerical results for the $\Sigma\Delta$ MRT schemes developed in Section~\ref{sec:sigma_delta_SU}.
In particular, we intend to give the reader some intuitive feeling by showing the in-phase quadrature-phase (IQ) scatter plots under various parameter settings.

The simulation settings are as follows.
The inter-antenna spacing is $d= \lambda/8$.
There is no background noise $v$.
In each realization, the complex channel gain $\alpha$ is randomly generated, with its phase being uniform on $[-\pi, \pi]$ and with its amplitude being $1$.
The basic $\Sigma\Delta$ MRT scheme in Section \ref{sec:mrt_su}  is employed.

Fig.~\ref{fig:PSK_con1} shows an extensive collection of the IQ scatter plots of the received signal $y$ when the symbol constellation is $8$-ary PSK.
In each IQ scatter plot, we overlay points obtained from $1,000$ channel realizations.
The red stars represent the true symbols, and the blue circles the shaped symbols at the user side (after proper normalization).
Some observations are as follows.
First, the symbol shaping errors generally reduces when the number of antennas $N$ is increased or when the angle $\theta$ is nearer to the broadside.
Second, the symbol shaping errors are seen to show structure when $\theta$ is large, e.g., $\theta \geq 80^\circ$.
It appears that the i.i.d. assumption no longer works for large $\theta$.
Third, there are a few specific angles, specifically, $\theta = 30^\circ$ and $\theta = 35^\circ$, where the symbol shaping errors are worse than expected.
Again, this appears to be due to the non-i.i.d. effect, though it is not as serious as the case of $\theta \geq 80^\circ$.
Fig. \ref{fig:QAM_con1} shows the IQ scatter plots where we change the symbol constellation to $16$-ary QAM.
We observe similar results.

\begin{figure}[H]
		\centering	
		\begin{subfigure}[b]{0.35\linewidth}
			\includegraphics[width=\textwidth]{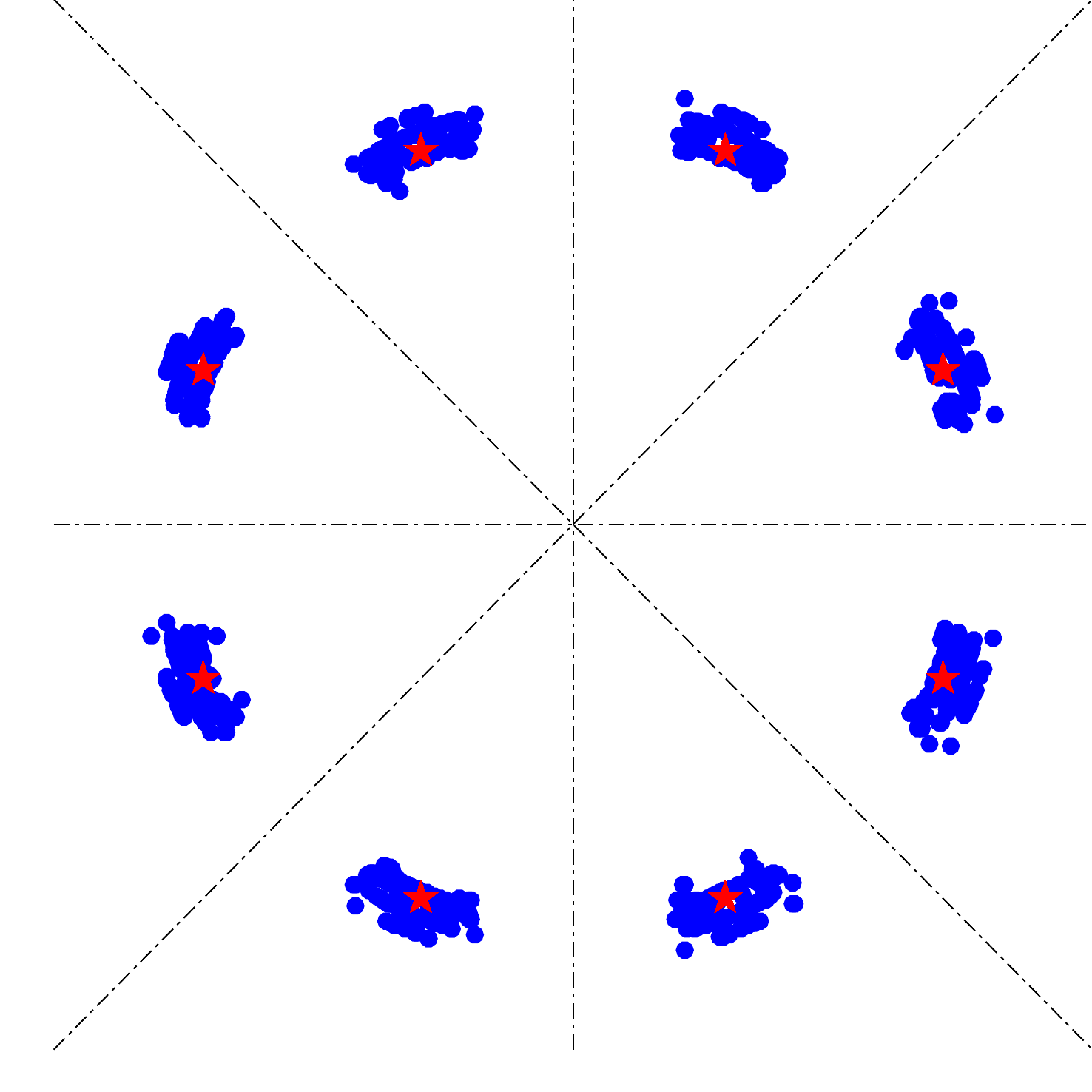}
			\caption{$\theta=30^\circ$, $N=512$, basic}
		\end{subfigure}
		\quad\quad
		\begin{subfigure}[b]{0.35\linewidth}
			\includegraphics[width=\textwidth]{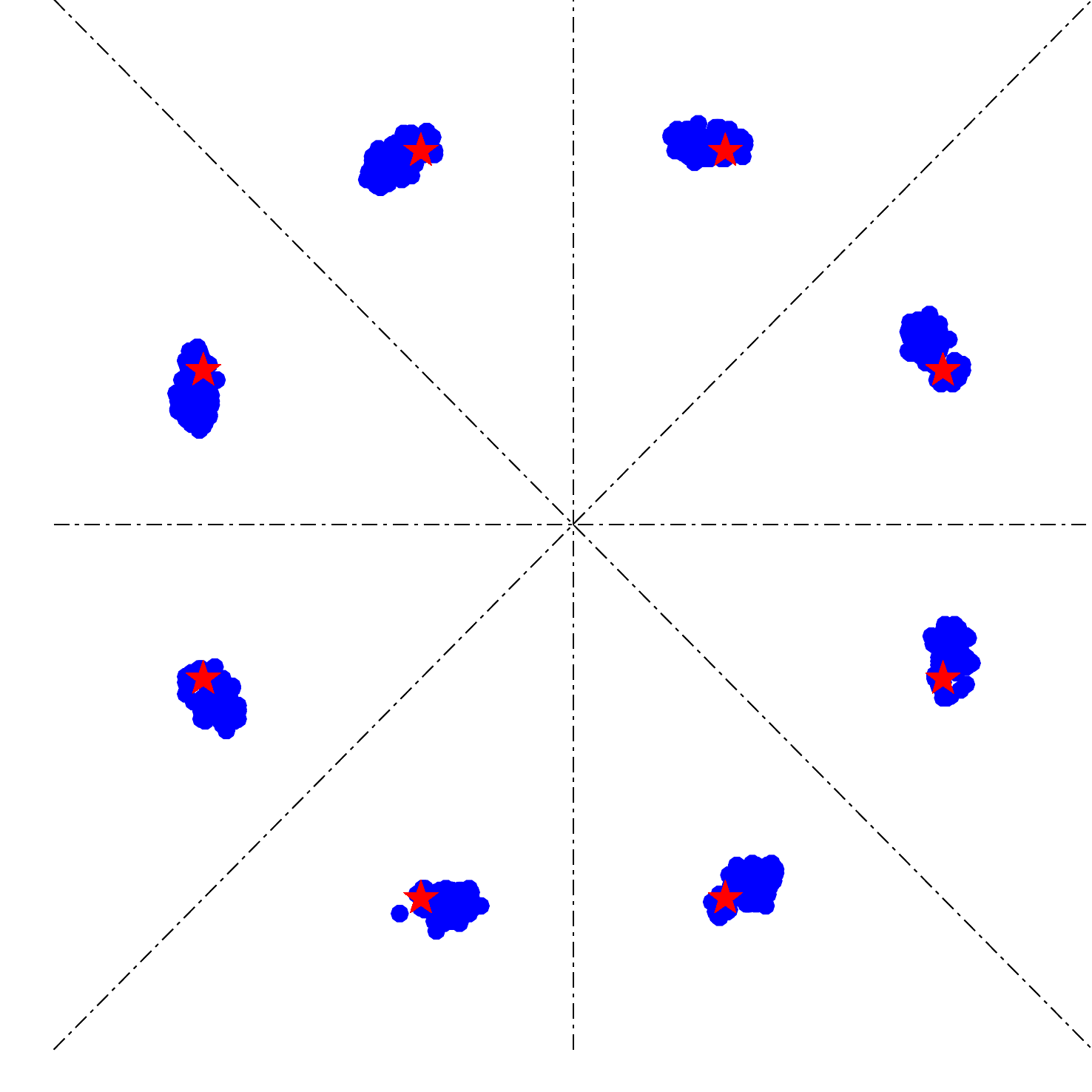}
			\caption{$\theta=30^\circ$, $N=512$, dithered}
		\end{subfigure}
		\quad\quad
		\begin{subfigure}[b]{0.35\linewidth}
			\includegraphics[width=\textwidth]{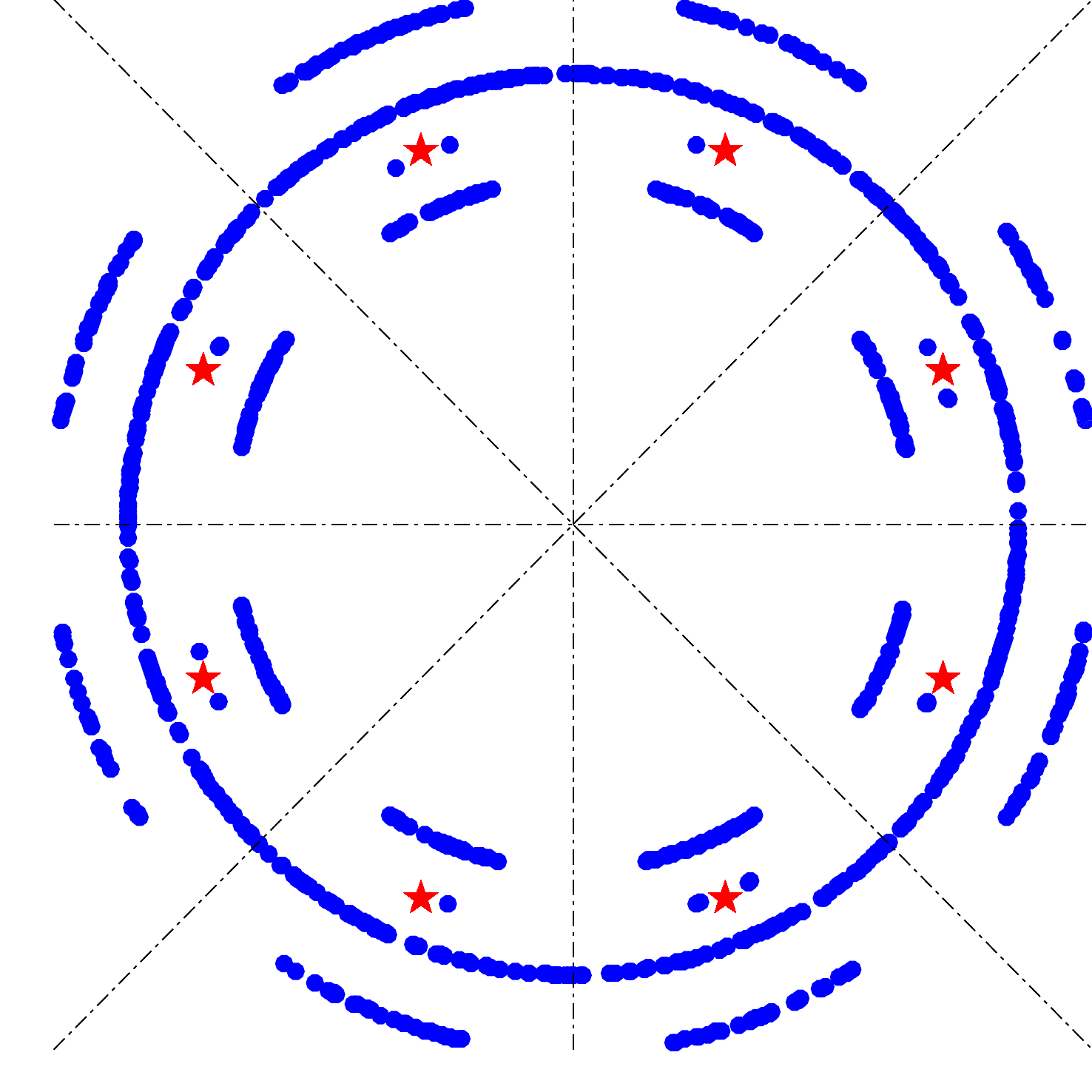}
			\caption{$\theta=90^\circ$, $N=512$, basic}
		\end{subfigure}
	\quad\quad
		\begin{subfigure}[b]{0.35\linewidth}
			\includegraphics[width=\textwidth]{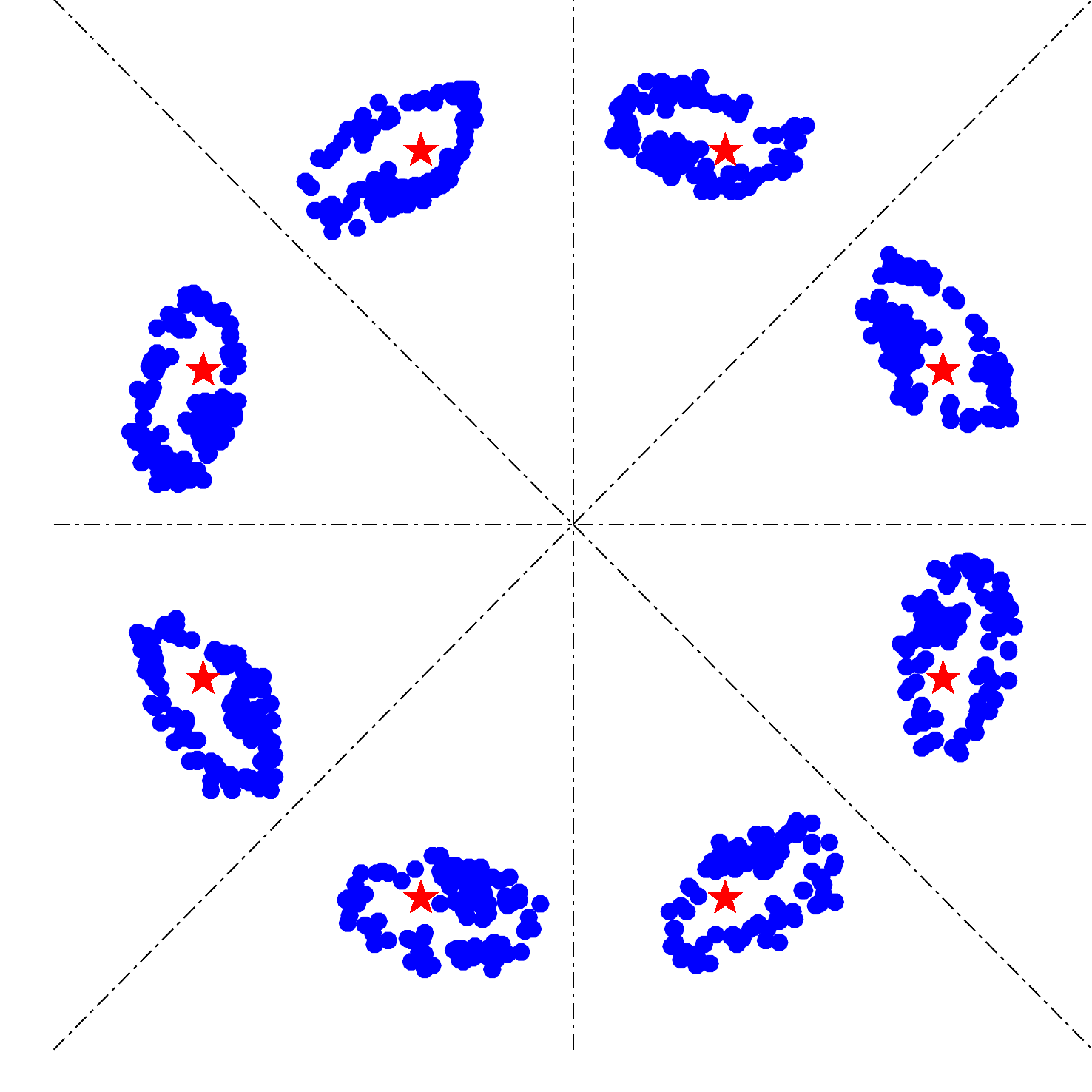}
			\caption{$\theta=90^\circ$, $N=512$, dithered}
		\end{subfigure}
		\caption{IQ scatter plots of the basic and dithered $\Sigma\Delta$ MRT schemes. }\label{fig:con_dither}
	\end{figure}

One can apply dithering to mitigate the non-i.i.d. quantization effect.
Fig. \ref{fig:con_dither} shows some results.
We apply the dithering procedure described in Section \ref{sec:basics_sigma_delta}, with dithering level $\delta= 0.6$.
We observe that dithering can in fact make the symbol shaping errors more random or less structured.
However, we should also mention that the dithering level is usually chosen by trial and error, and increasing the dithering level also increases the quantization noise range.

\begin{figure}[H]
	\centering	
	\begin{subfigure}[b]{0.3\linewidth}
		\includegraphics[width=\textwidth]{N512theta90-eps-converted-to.pdf}
		\caption{$\theta=90^\circ$,  basic}
	\end{subfigure}
	\quad
	\begin{subfigure}[b]{0.3\linewidth}
		\includegraphics[width=\textwidth]{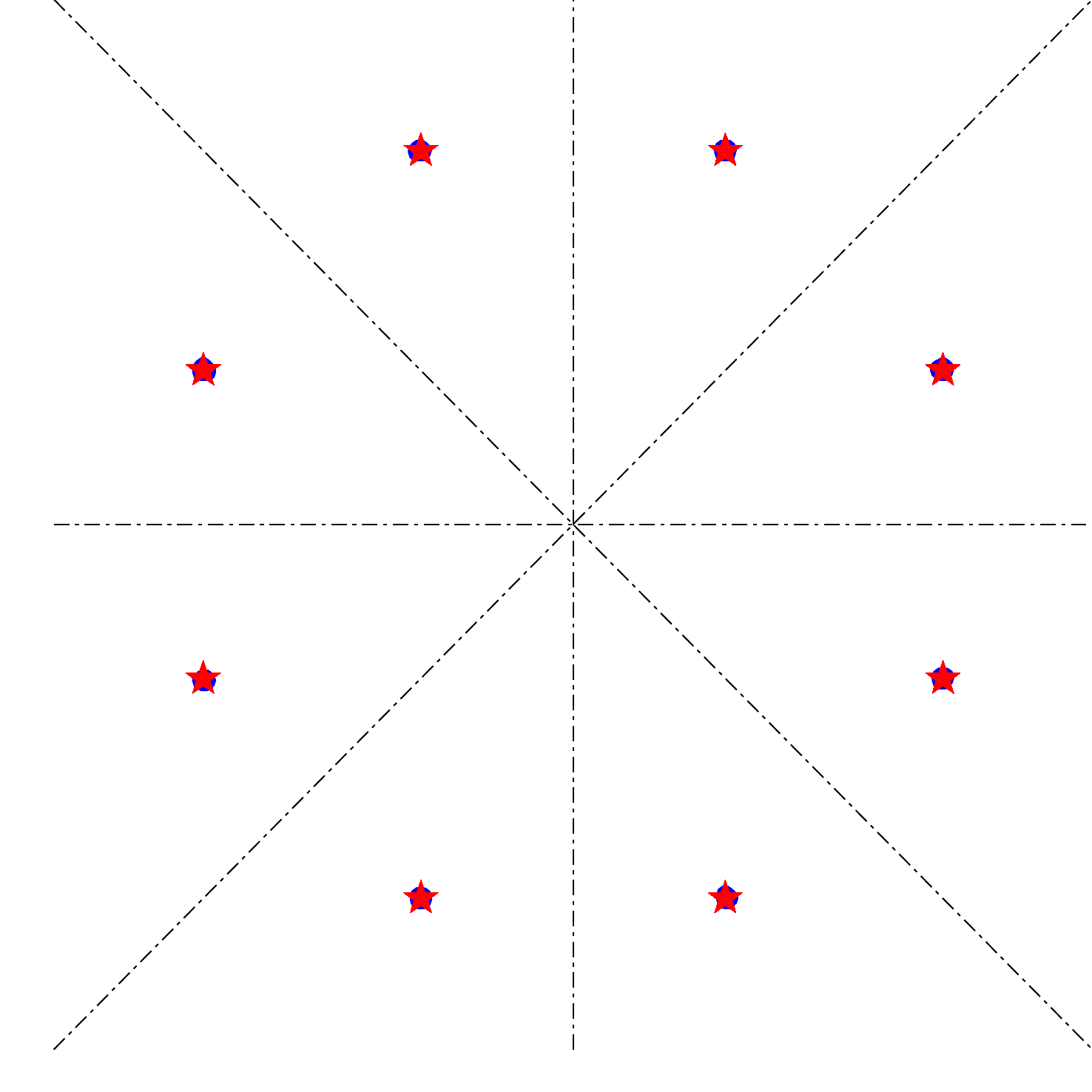}
		\caption{$\theta=90^\circ$,  angle steering}
	\end{subfigure}
	\caption{IQ scatter plots of the basic and angle-steered $\Sigma\Delta$ MRT schemes for $\theta=90^\circ$ and $N=512$. }\label{fig:con_angle}
\end{figure}

Next, we consider the angle-steered $\Sigma\Delta$ MRT scheme in Section \ref{sec:angles}.
The simulation settings are the same above, and we consider $\theta= 90^\circ$.
The result is shown in Fig.~\ref{fig:con_angle}.
We see that
the angle-steered $\Sigma\Delta$ MRT scheme dramatically reduces the symbol shaping errors; visually we see no errors.

\medskip

%
%
\begin{figure}[H]
	\centering	
\begin{subfigure}[b]{0.4\linewidth}
	\includegraphics[width=\textwidth]{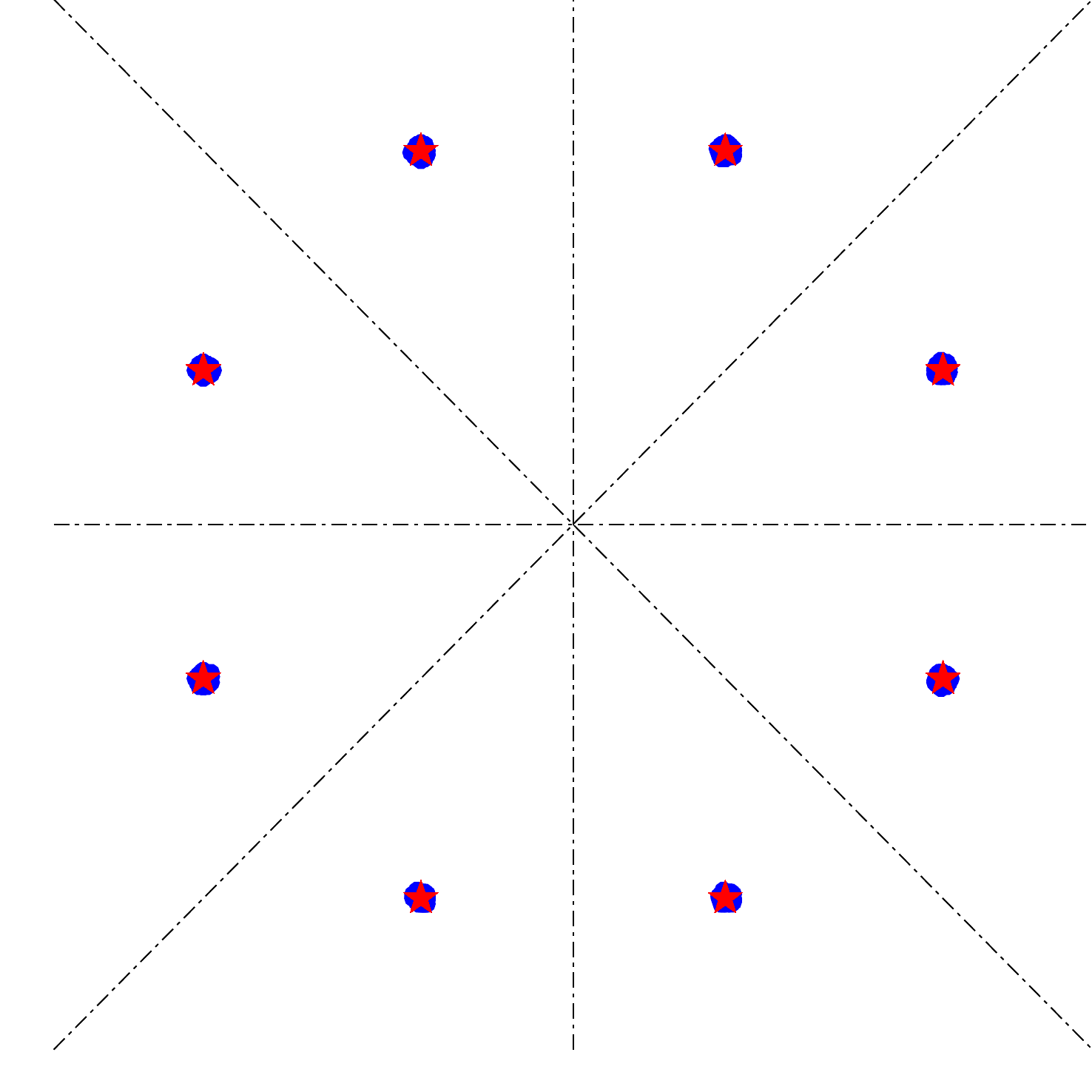}
		\caption*{$\theta=0^\circ$, $N=64$}
	\end{subfigure}
~
\begin{subfigure}[b]{0.4\linewidth}
		\includegraphics[width=\textwidth]{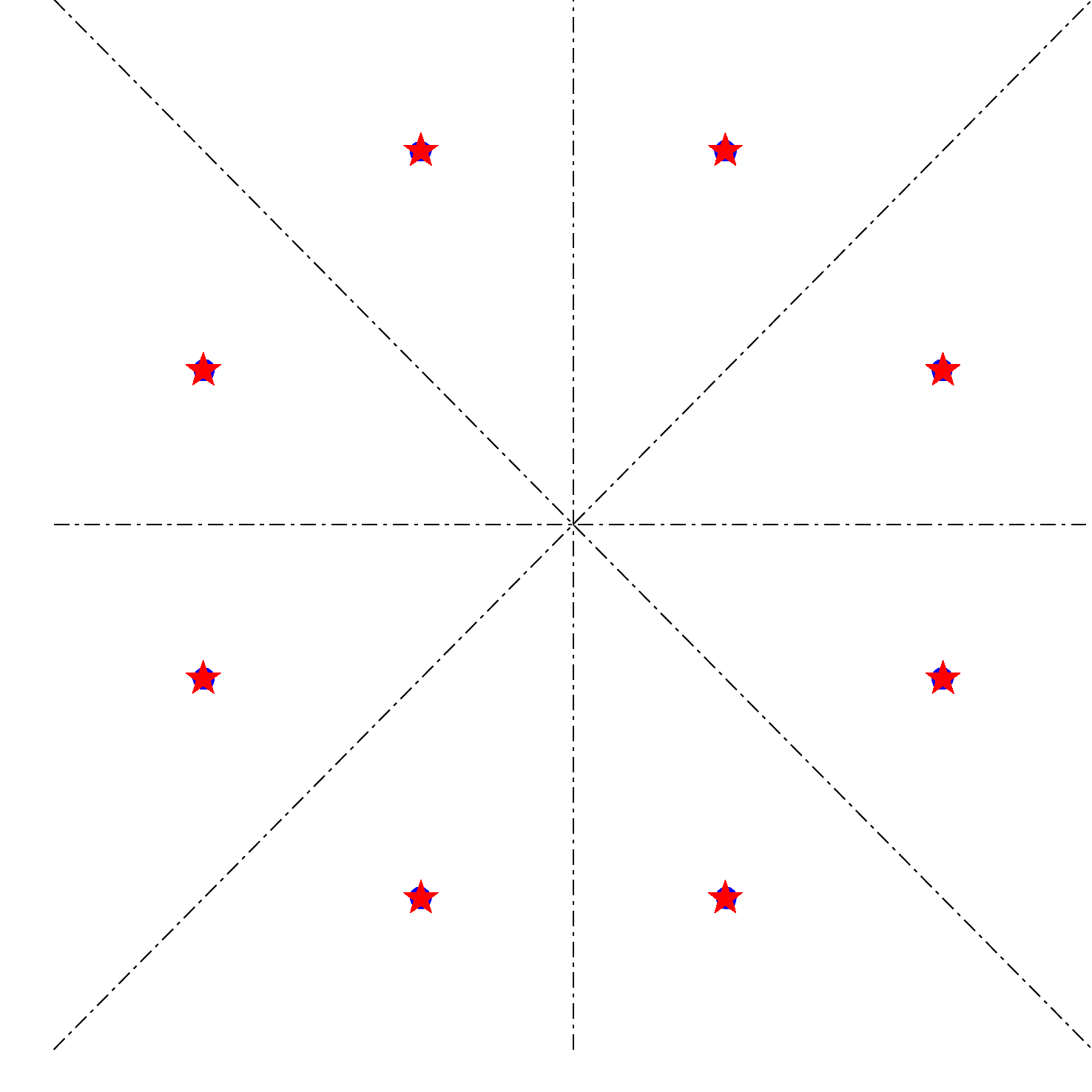}
		\caption*{$\theta=0^\circ$, $N=256$}
	\end{subfigure}
~
\begin{subfigure}[b]{0.4\linewidth}
	\includegraphics[width=\textwidth]{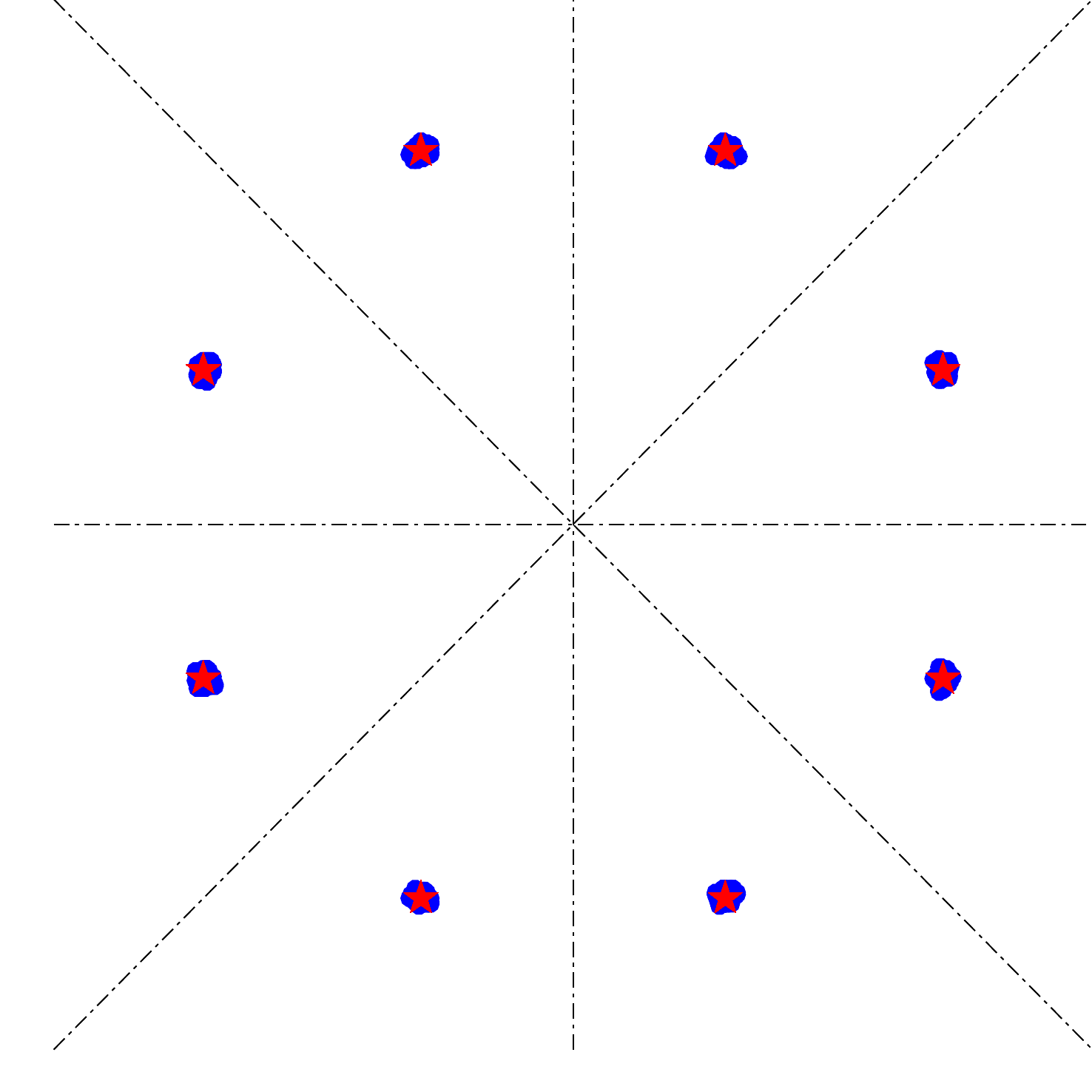}
		\caption*{$\theta=5^\circ$, $N=64$}
	\end{subfigure}
~
\begin{subfigure}[b]{0.4\linewidth}
		\includegraphics[width=\textwidth]{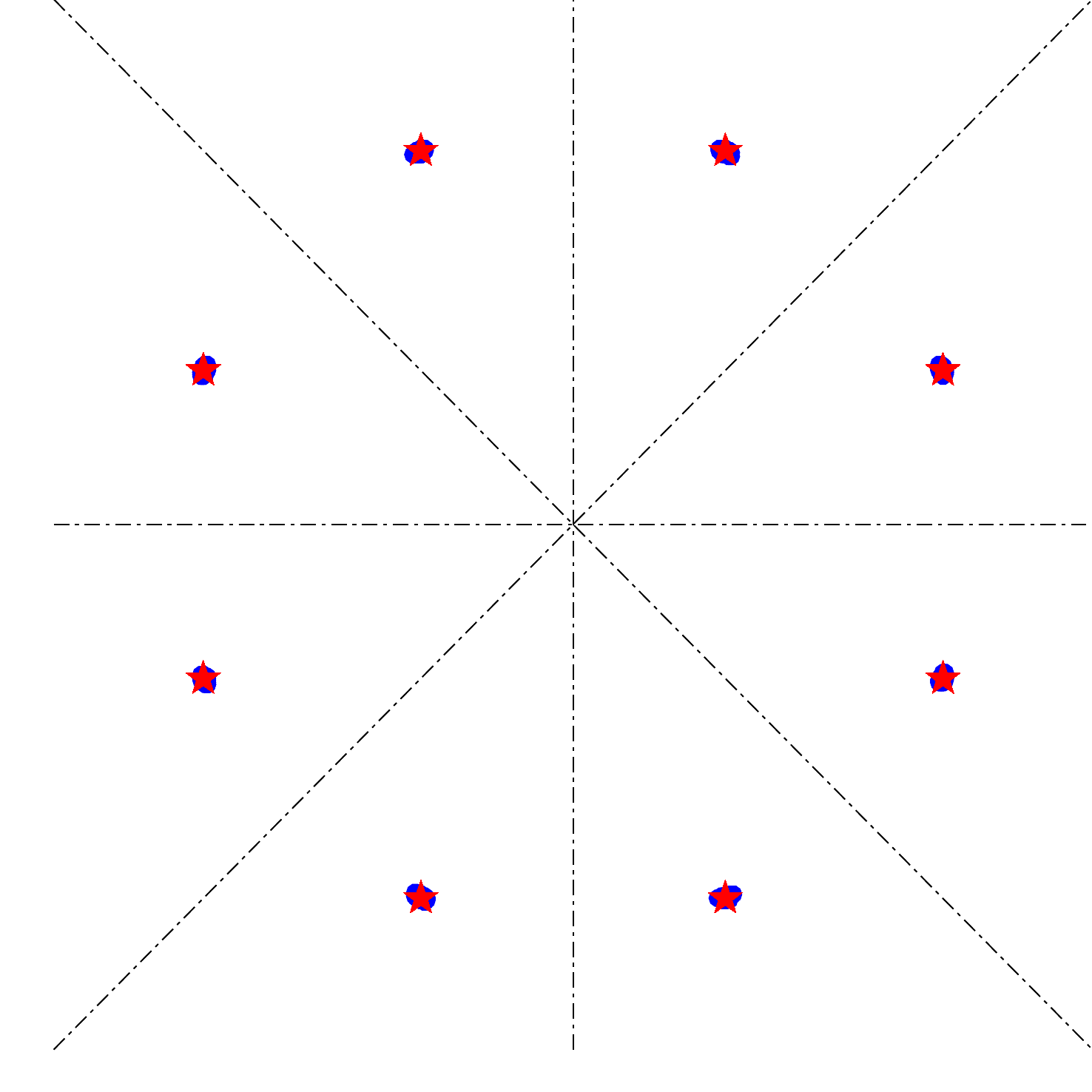}
		\caption*{$\theta=5^\circ$, $N=256$}
	\end{subfigure}
~
\begin{subfigure}[b]{0.4\linewidth}
	\includegraphics[width=\textwidth]{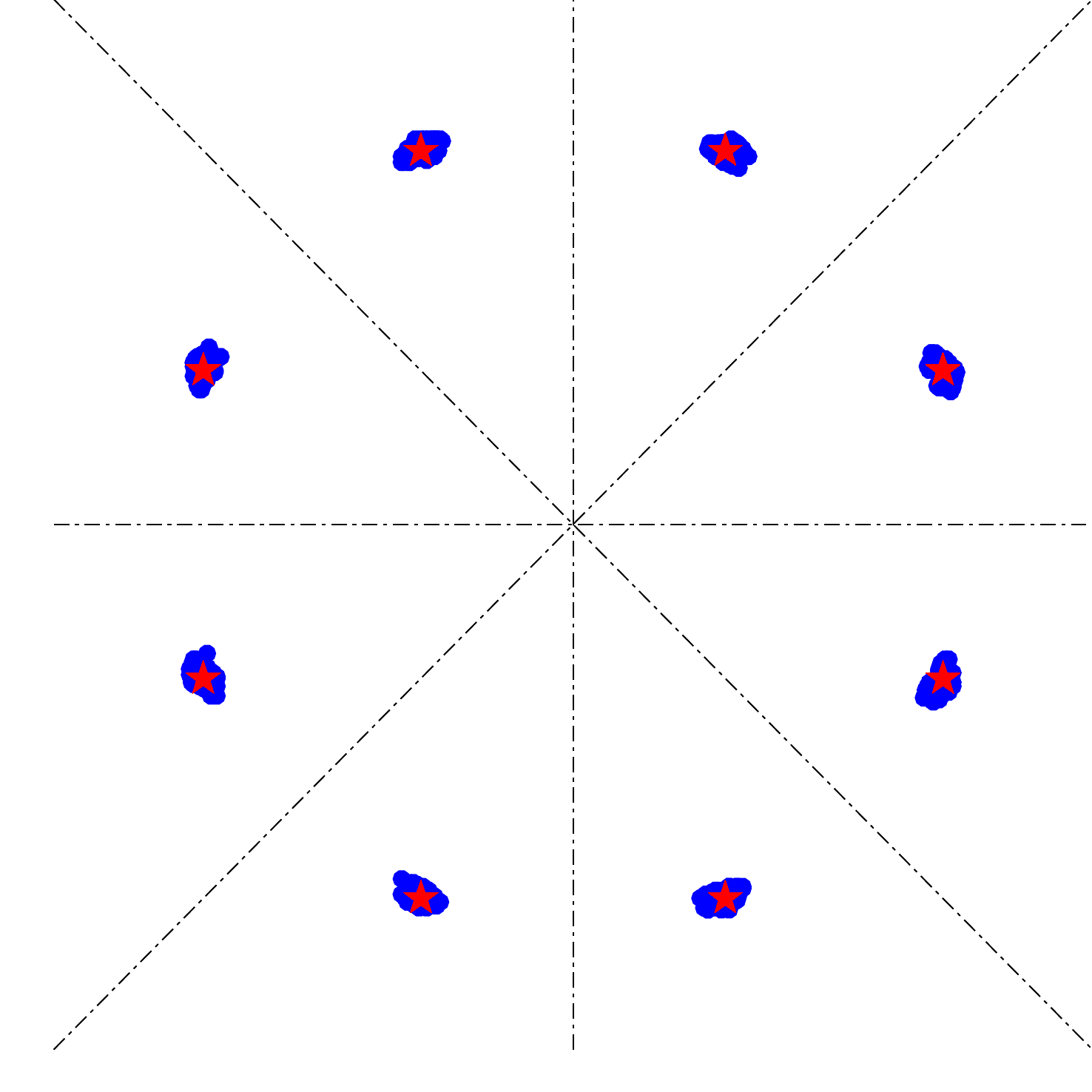}
		\caption*{$\theta=10^\circ$, $N=64$}
	\end{subfigure}
~
\begin{subfigure}[b]{0.4\linewidth}
		\includegraphics[width=\textwidth]{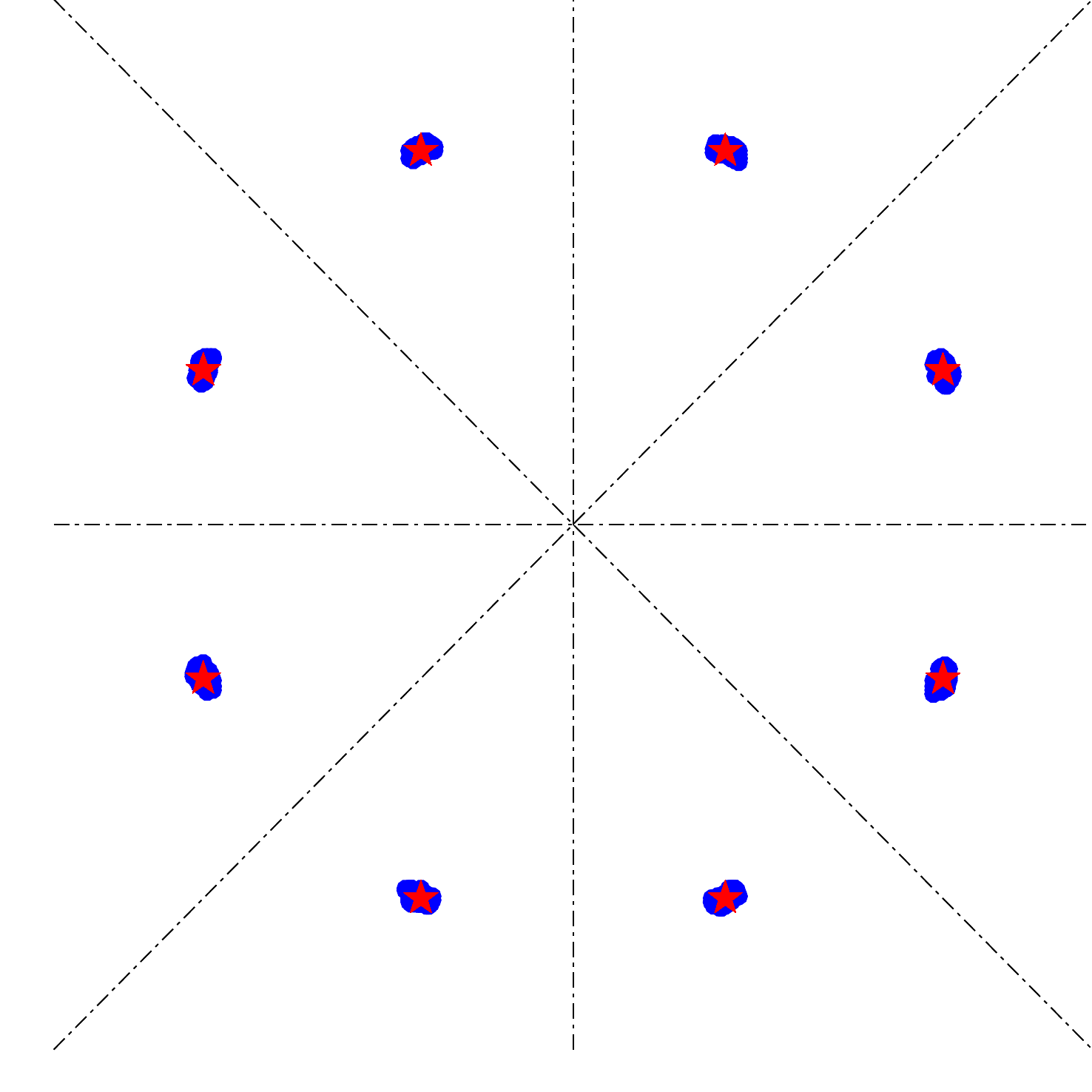}
		\caption*{$\theta=10^\circ$, $N=256$}
	\end{subfigure}
	\caption{IQ scatter plots of the basic sigma-delta MRT scheme for different $\theta$ and $N$; $8$-ary PSK.} \label{fig:PSK_con1}
\end{figure}

\begin{figure}[H]\ContinuedFloat
	\centering
\begin{subfigure}[b]{0.4\linewidth}
	\includegraphics[width=\textwidth]{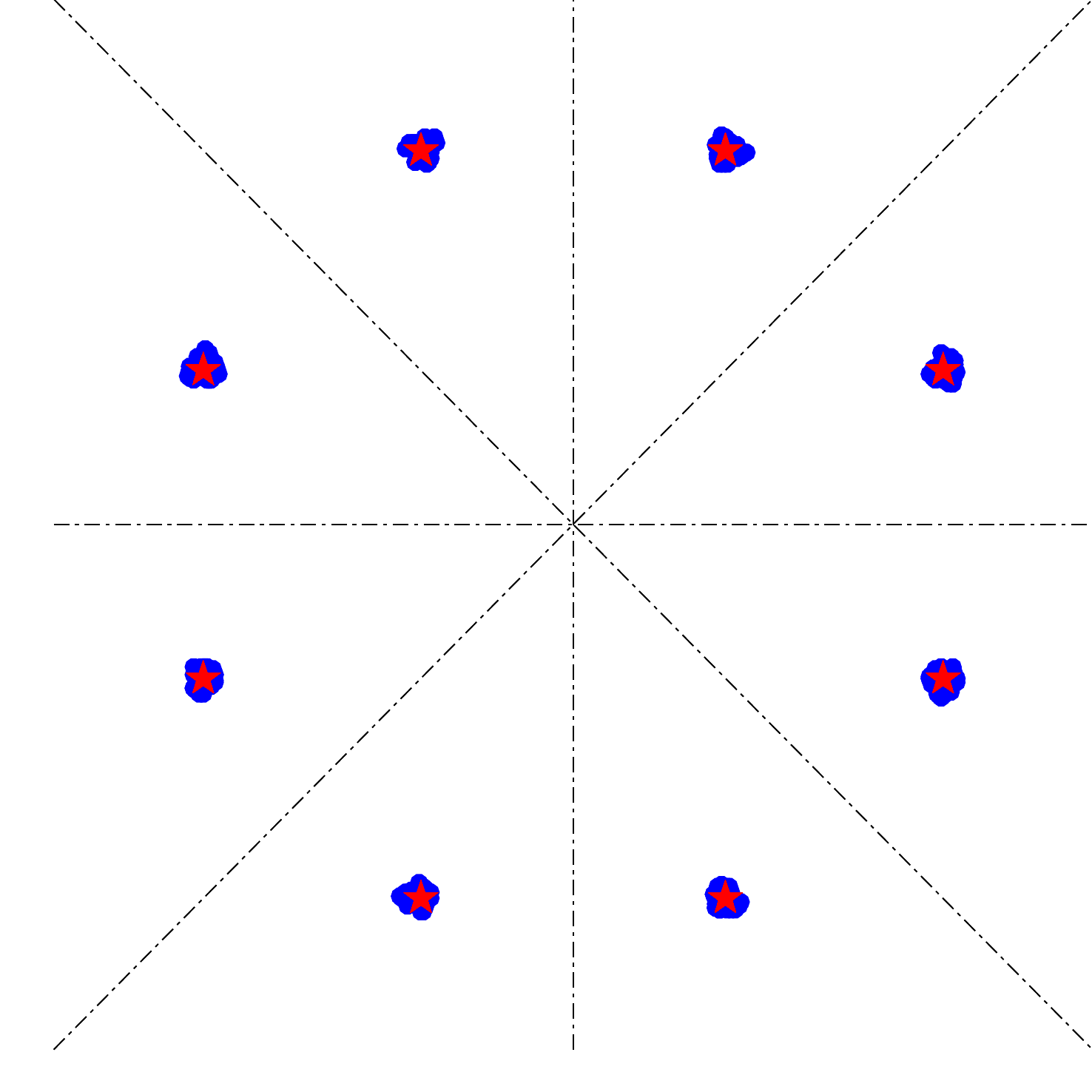}
		\caption*{$\theta=15^\circ$, $N=64$}
	\end{subfigure}
~
\begin{subfigure}[b]{0.4\linewidth}
		\includegraphics[width=\textwidth]{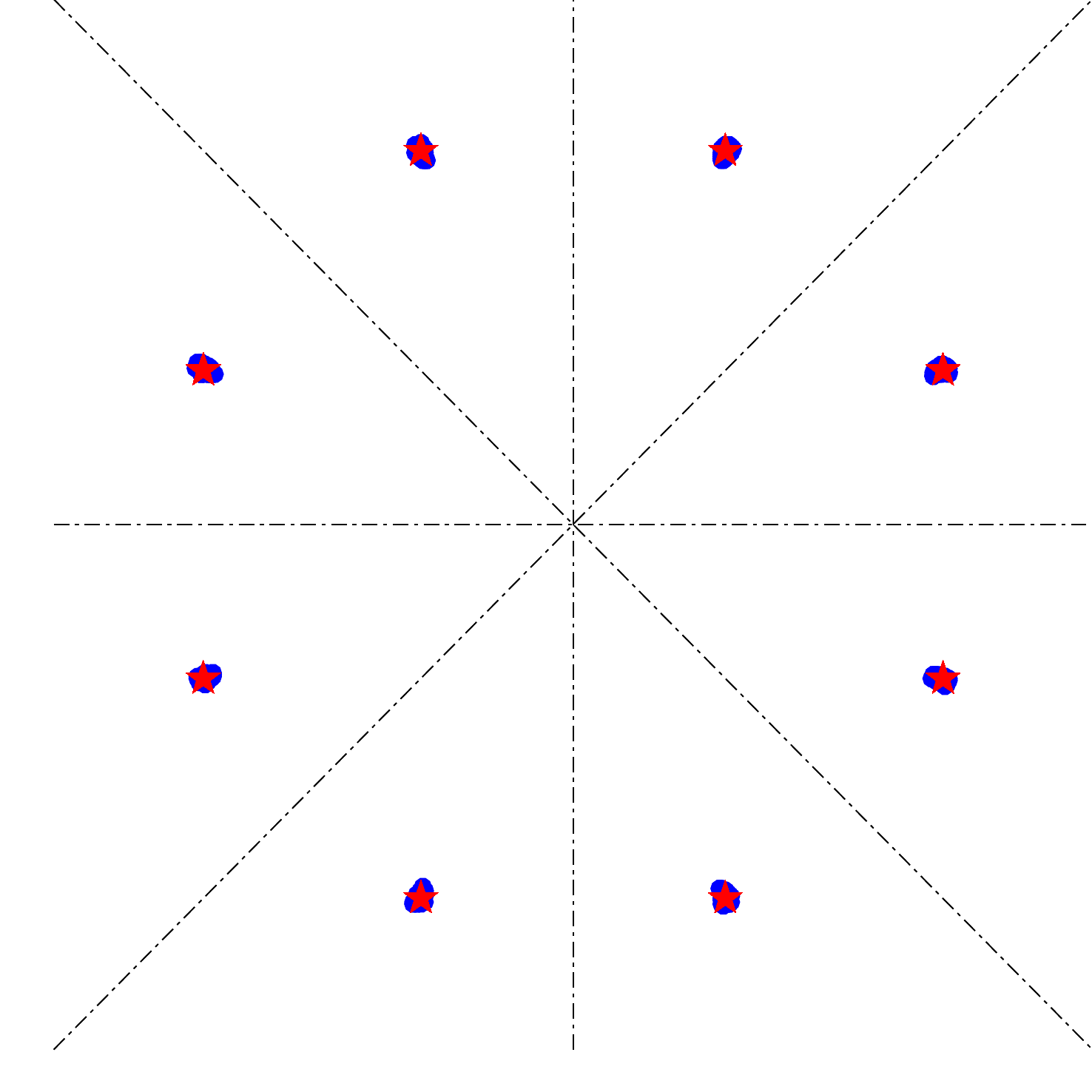}
		\caption*{$\theta=15^\circ$, $N=256$}
	\end{subfigure}
~
\begin{subfigure}[b]{0.4\linewidth}
	\includegraphics[width=\textwidth]{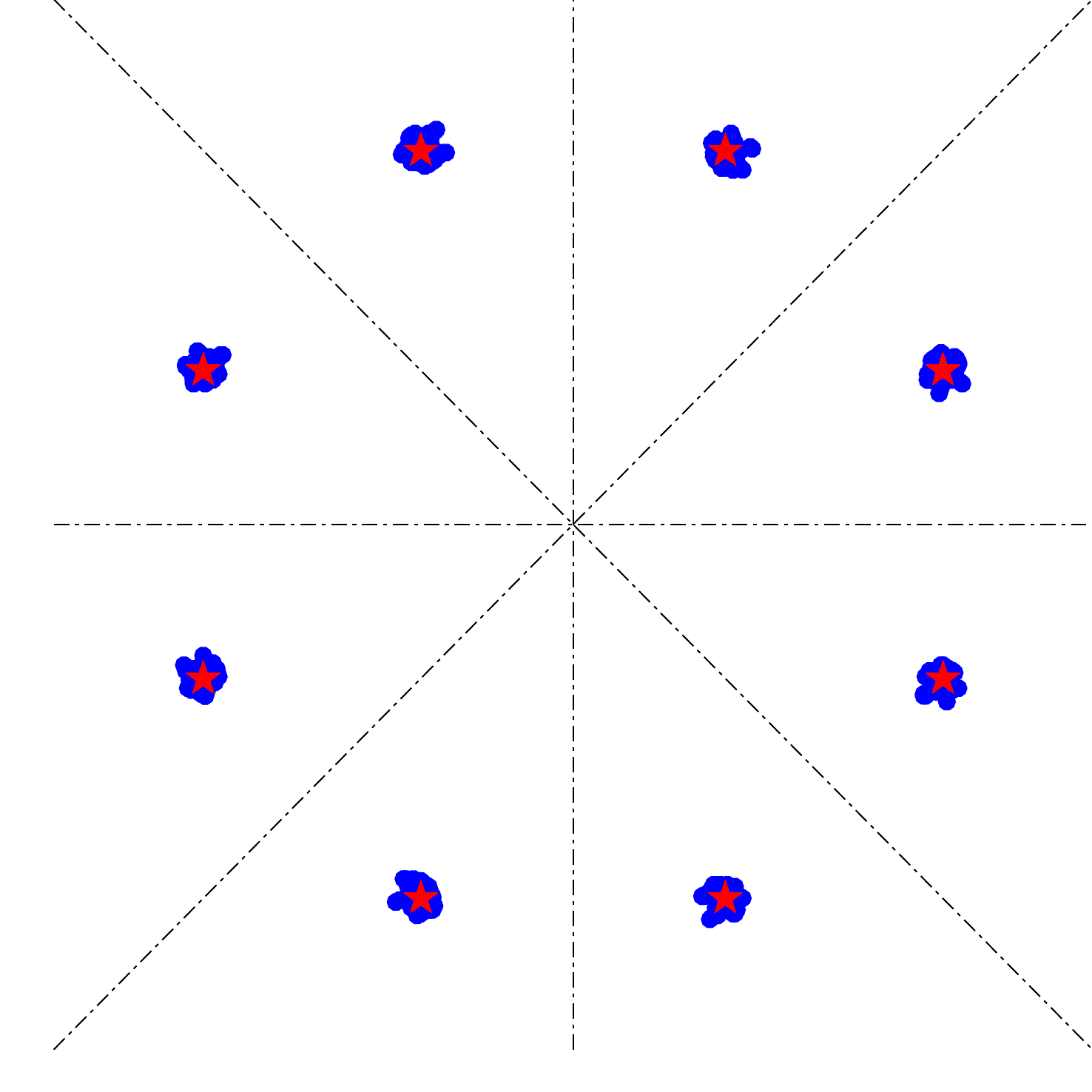}
		\caption*{$\theta=20^\circ$, $N=64$}
	\end{subfigure}
~
\begin{subfigure}[b]{0.4\linewidth}
		\includegraphics[width=\textwidth]{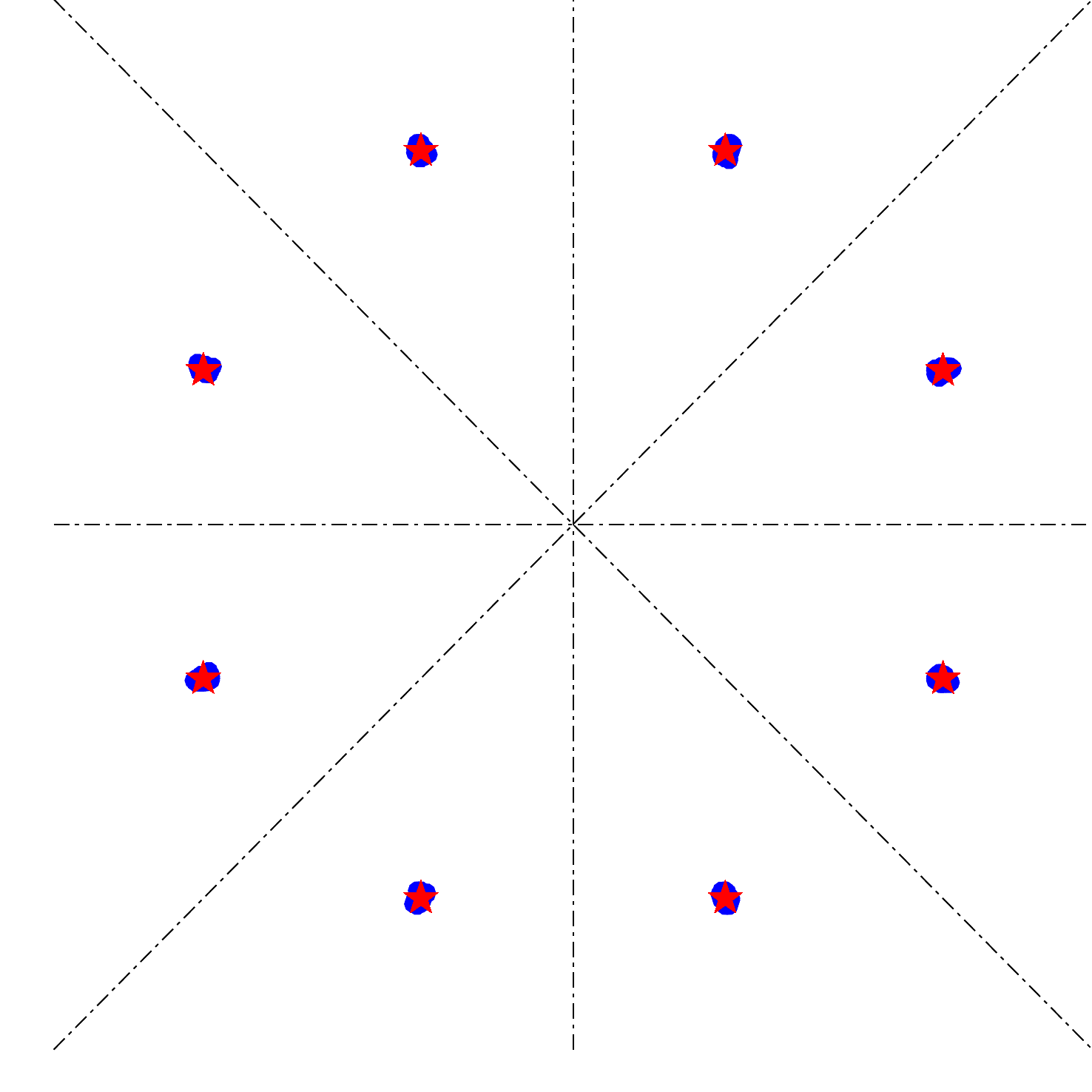}
		\caption*{$\theta=20^\circ$, $N=256$}
	\end{subfigure}
~
\begin{subfigure}[b]{0.4\linewidth}
	\includegraphics[width=\textwidth]{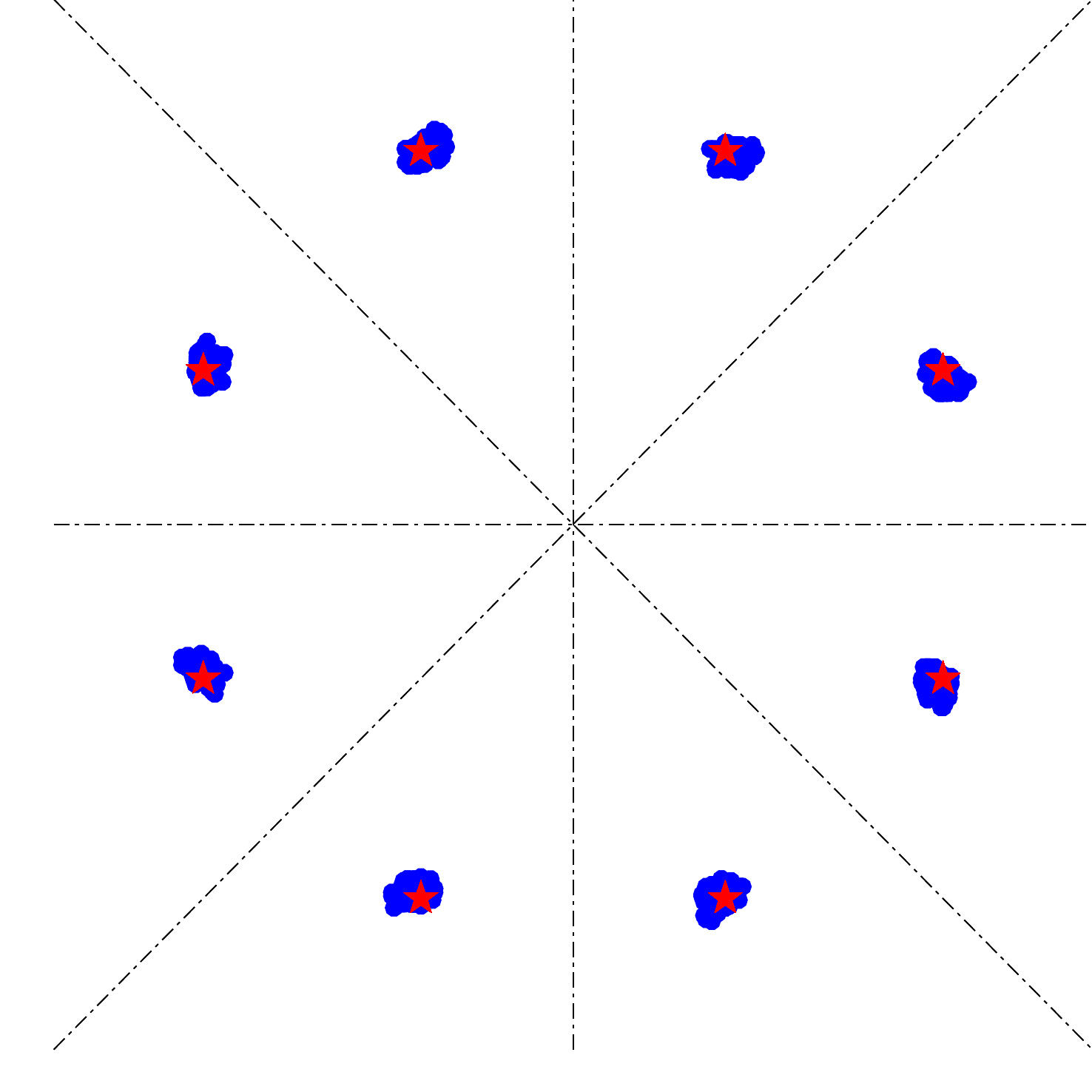}
		\caption*{$\theta=25^\circ$, $N=64$}
	\end{subfigure}
~
\begin{subfigure}[b]{0.4\linewidth}
		\includegraphics[width=\textwidth]{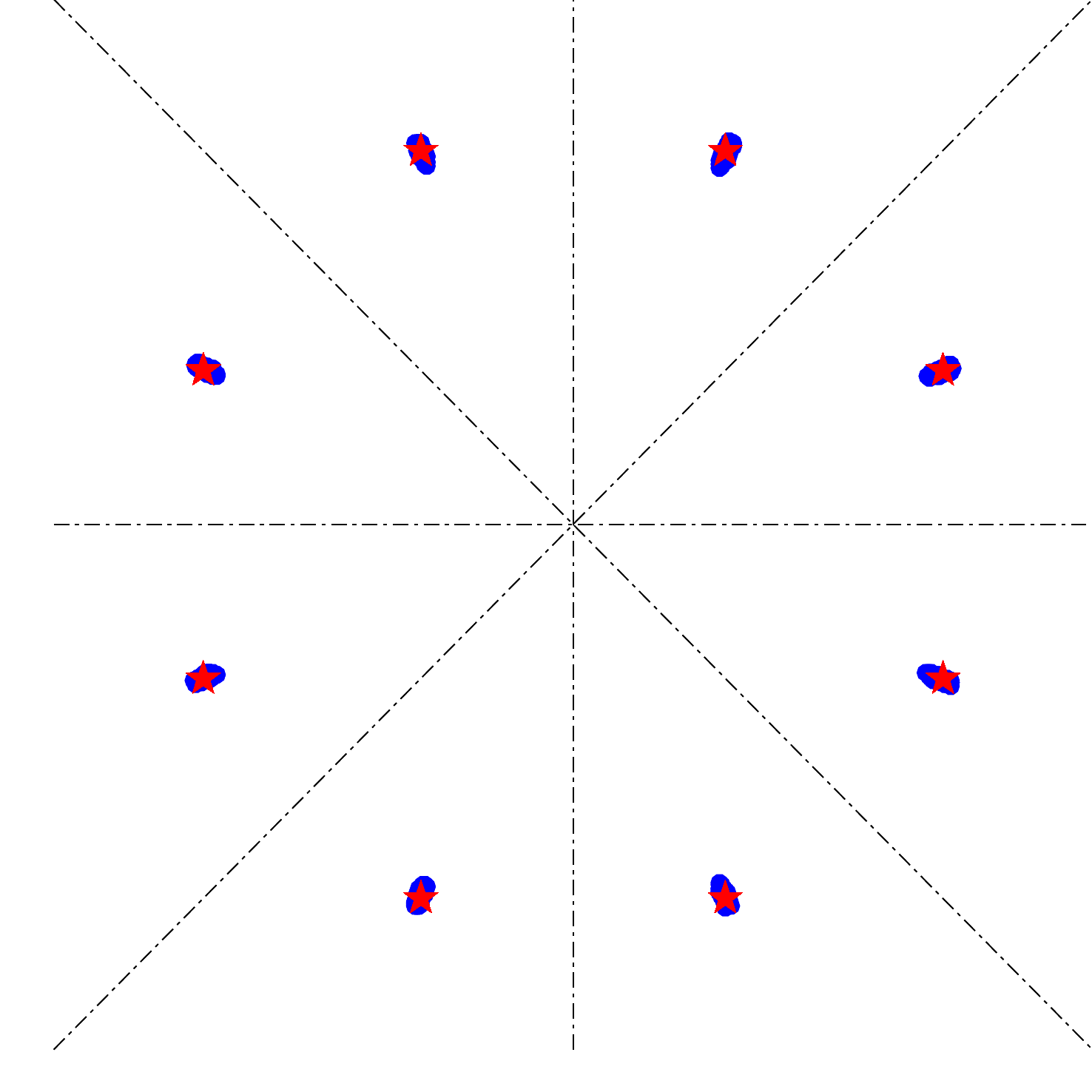}
		\caption*{$\theta=25^\circ$, $N=256$}
	\end{subfigure}
	\caption{IQ scatter plots of the basic sigma-delta MRT scheme for different $\theta$ and $N$; $8$-ary PSK. (cont.) }
\end{figure}

\begin{figure}[H]\ContinuedFloat
	\centering
\begin{subfigure}[b]{0.4\linewidth}
	\includegraphics[width=\textwidth]{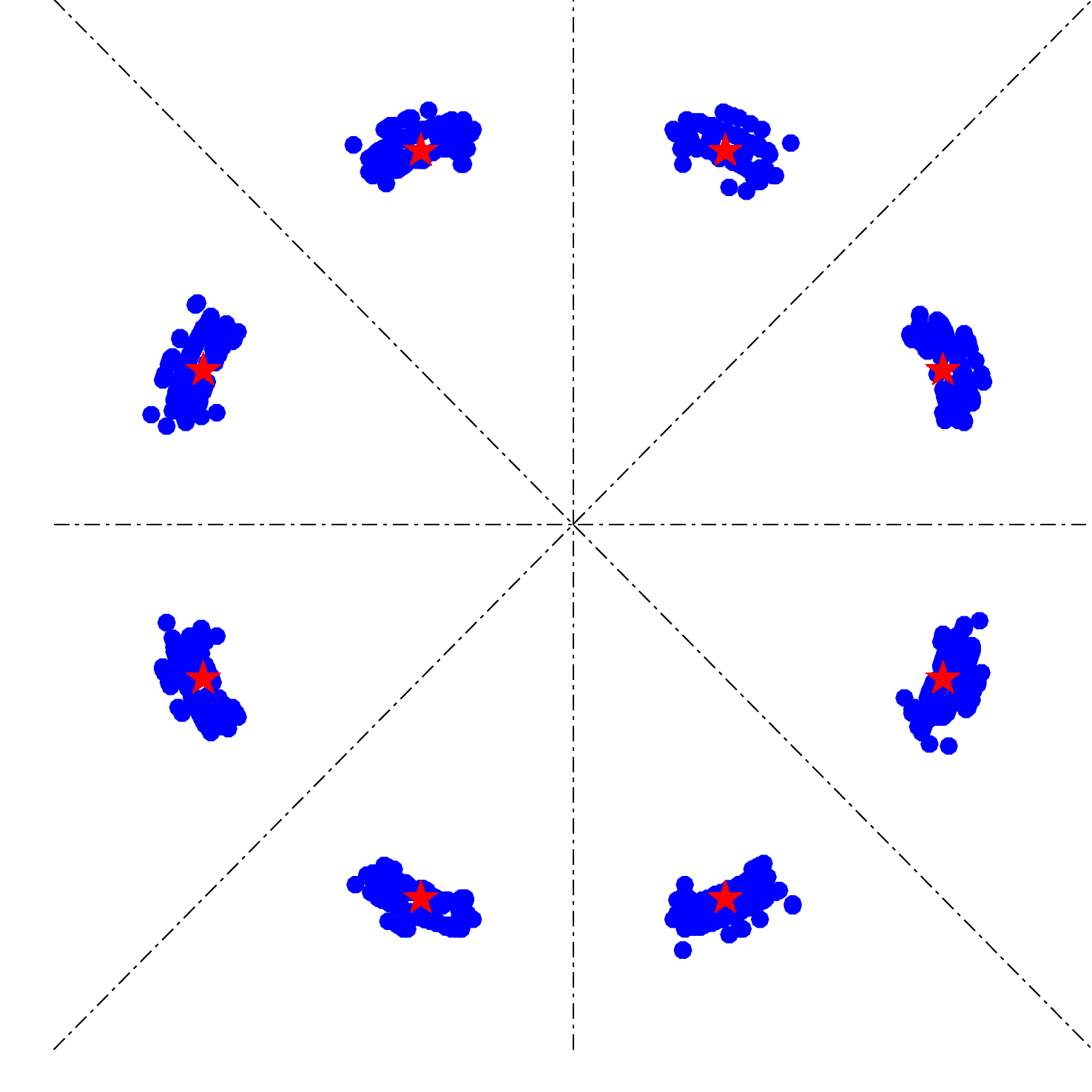}
		\caption*{$\theta=30^\circ$, $N=64$}
	\end{subfigure}
~
\begin{subfigure}[b]{0.4\linewidth}
		\includegraphics[width=\textwidth]{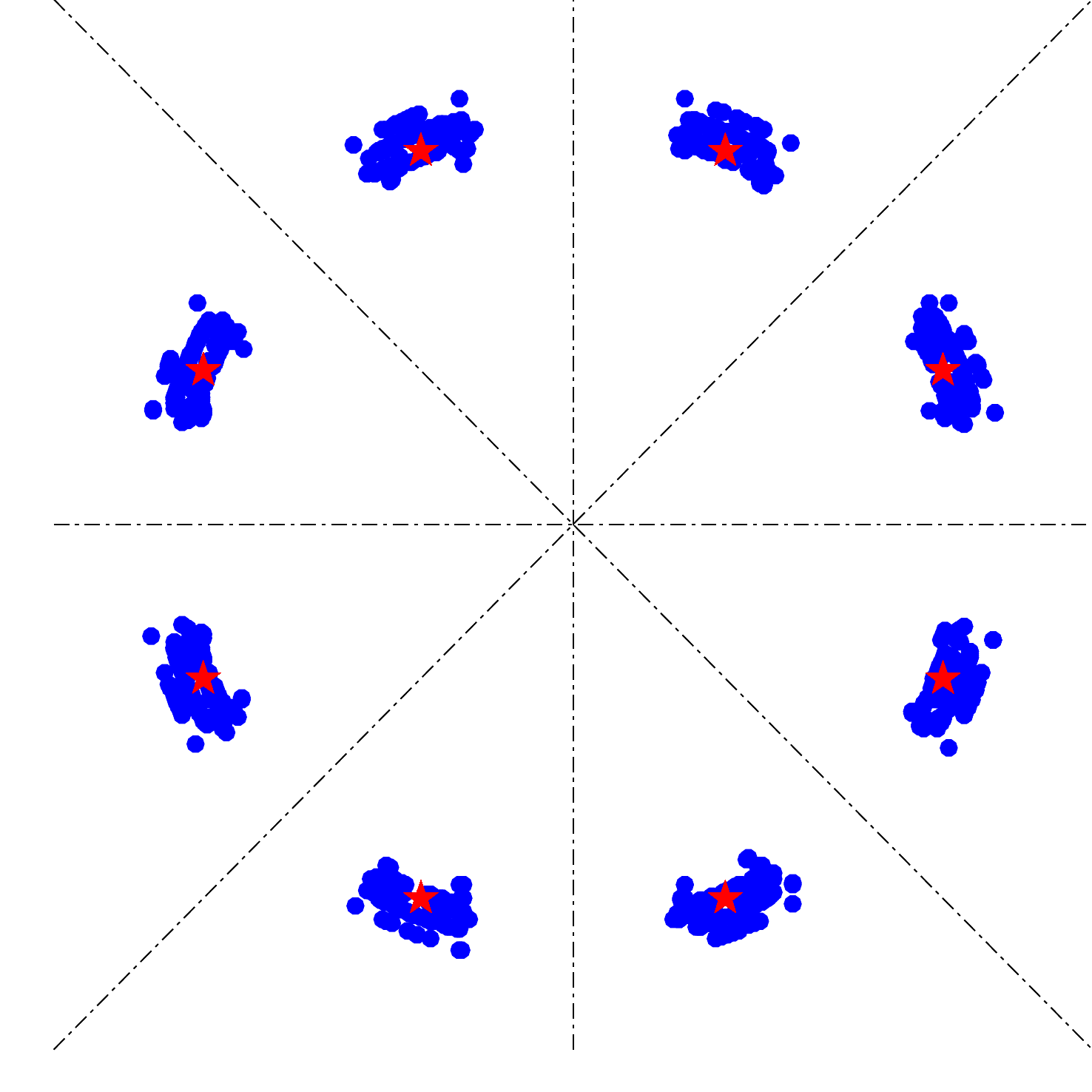}
		\caption*{$\theta=30^\circ$, $N=256$}
	\end{subfigure}
~
\begin{subfigure}[b]{0.4\linewidth}
	\includegraphics[width=\textwidth]{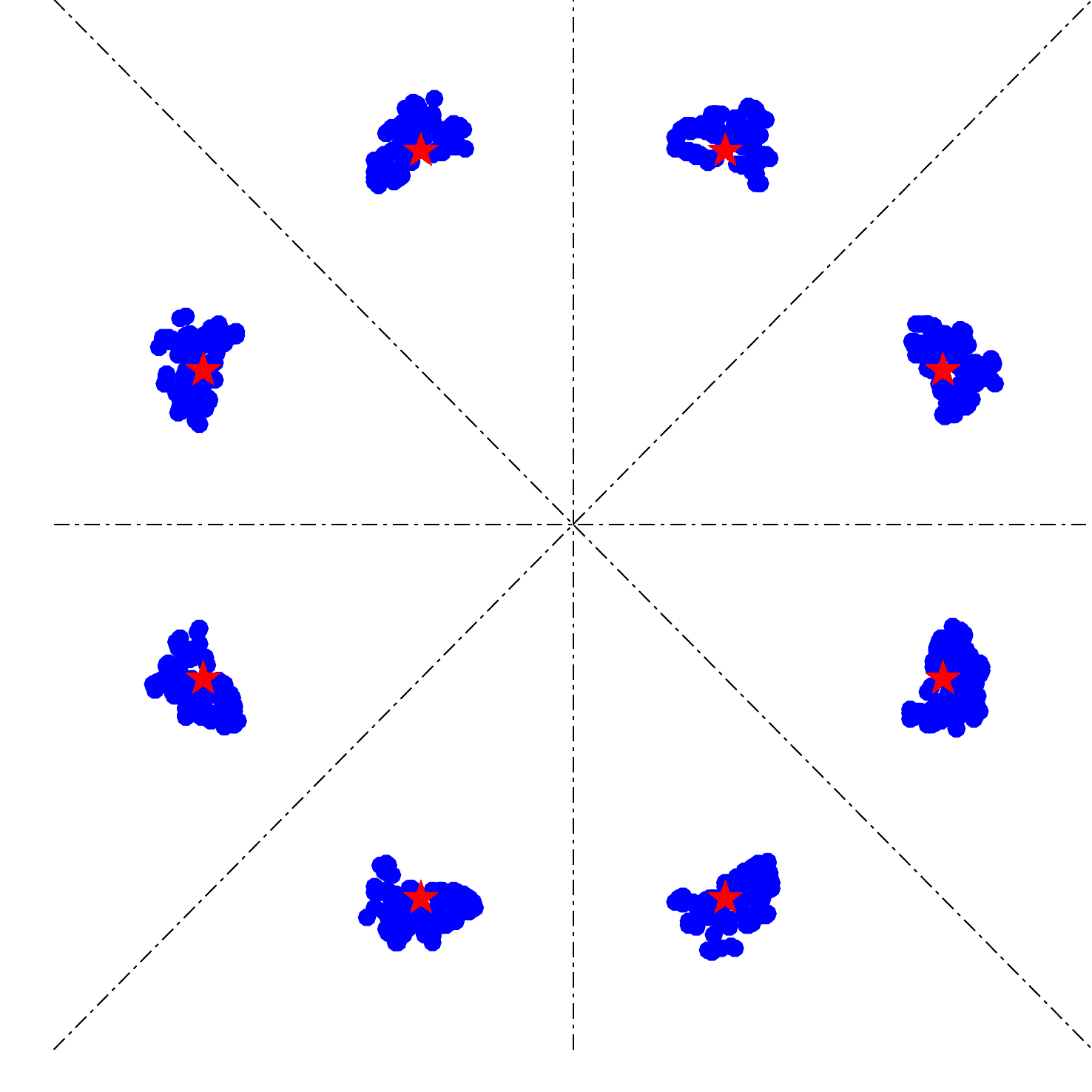}
		\caption*{$\theta=35^\circ$, $N=64$}
	\end{subfigure}
~
\begin{subfigure}[b]{0.4\linewidth}
		\includegraphics[width=\textwidth]{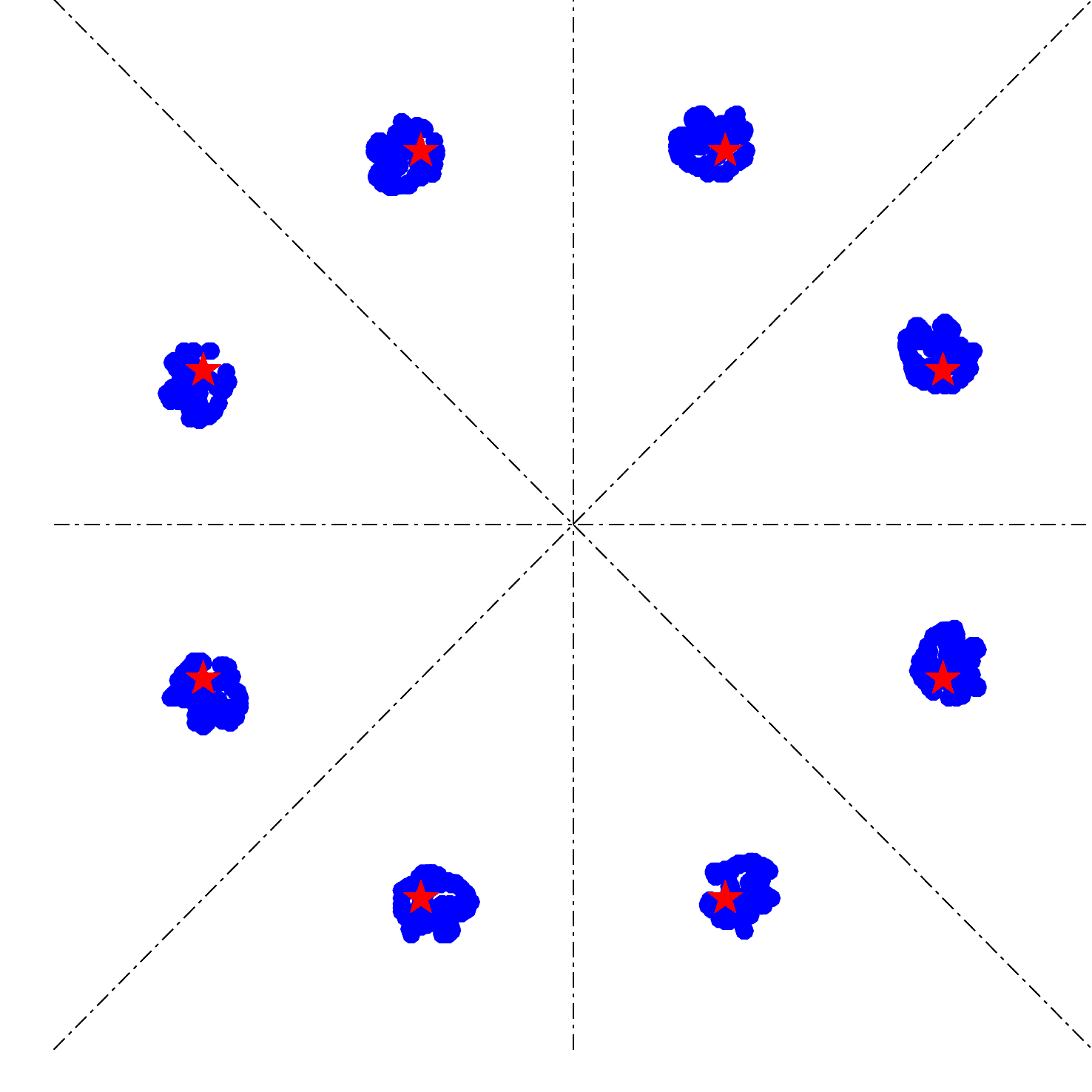}
		\caption*{$\theta=35^\circ$, $N=256$}
	\end{subfigure}
~
\begin{subfigure}[b]{0.4\linewidth}
	\includegraphics[width=\textwidth]{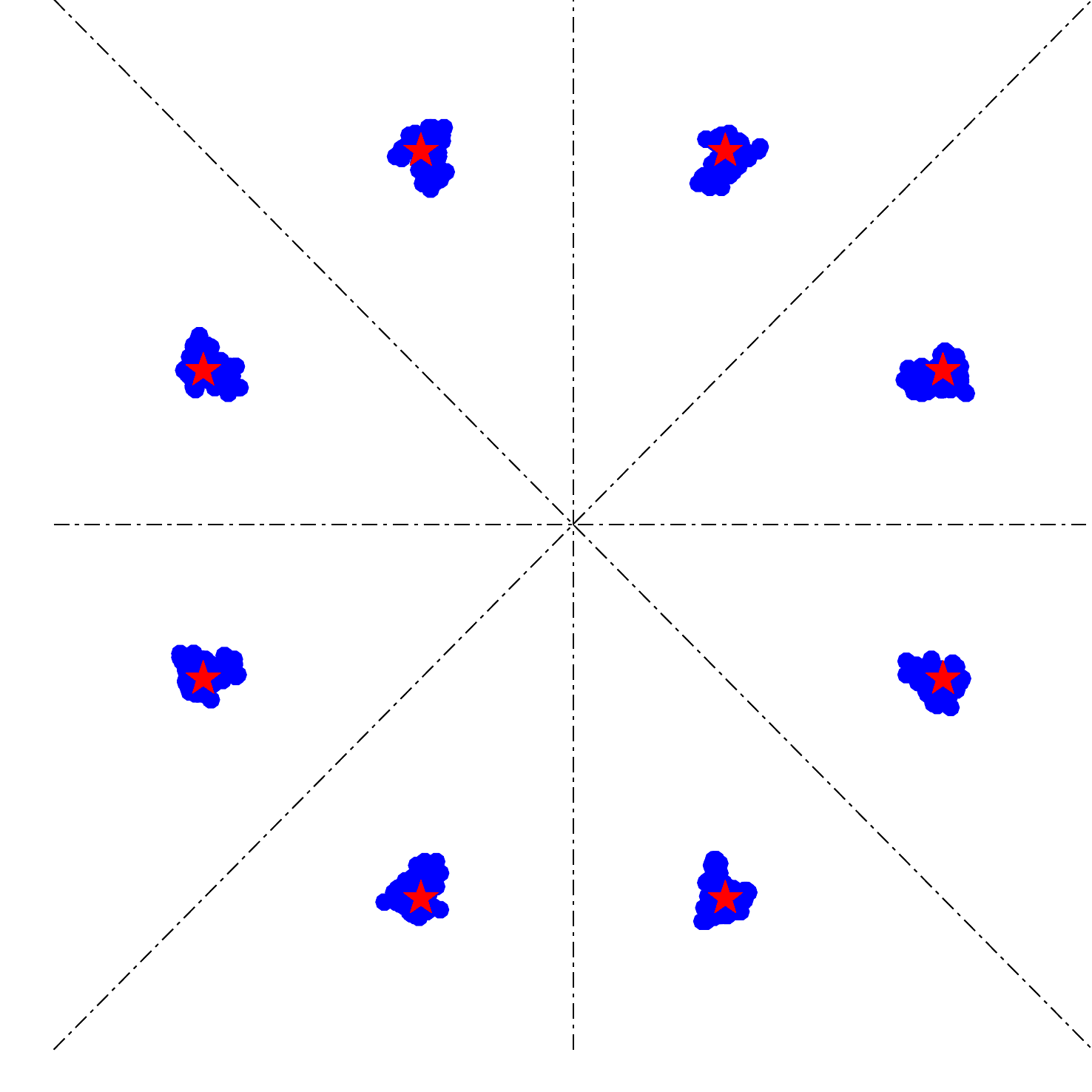}
		\caption*{$\theta=40^\circ$, $N=64$}
	\end{subfigure}
~
\begin{subfigure}[b]{0.4\linewidth}
		\includegraphics[width=\textwidth]{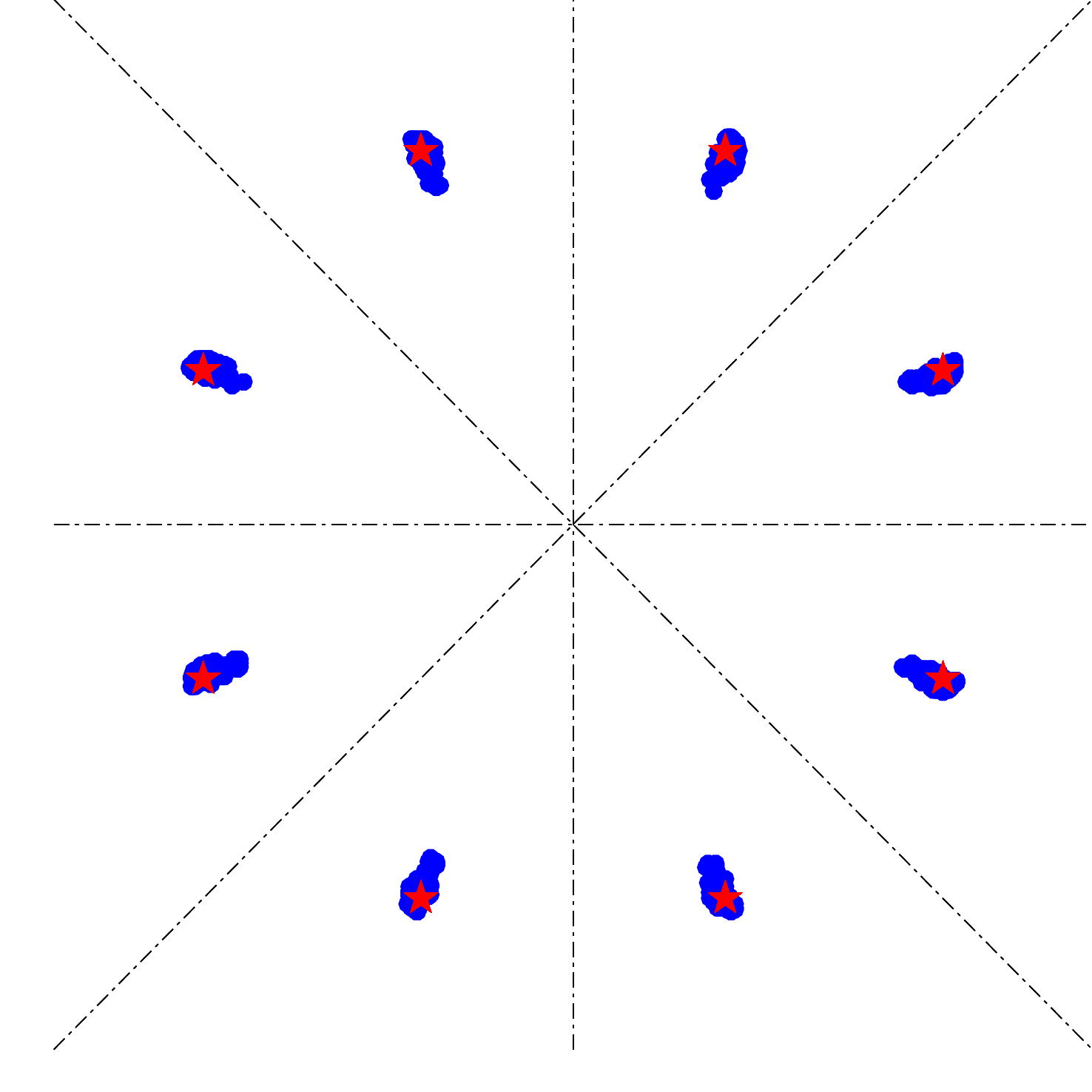}
		\caption*{$\theta=40^\circ$, $N=256$}
	\end{subfigure}
	\caption{IQ scatter plots of the basic sigma-delta MRT scheme for different $\theta$ and $N$; $8$-ary PSK. (cont.)}
\end{figure}

\begin{figure}[H]\ContinuedFloat
	\centering
\begin{subfigure}[b]{0.4\linewidth}
	\includegraphics[width=\textwidth]{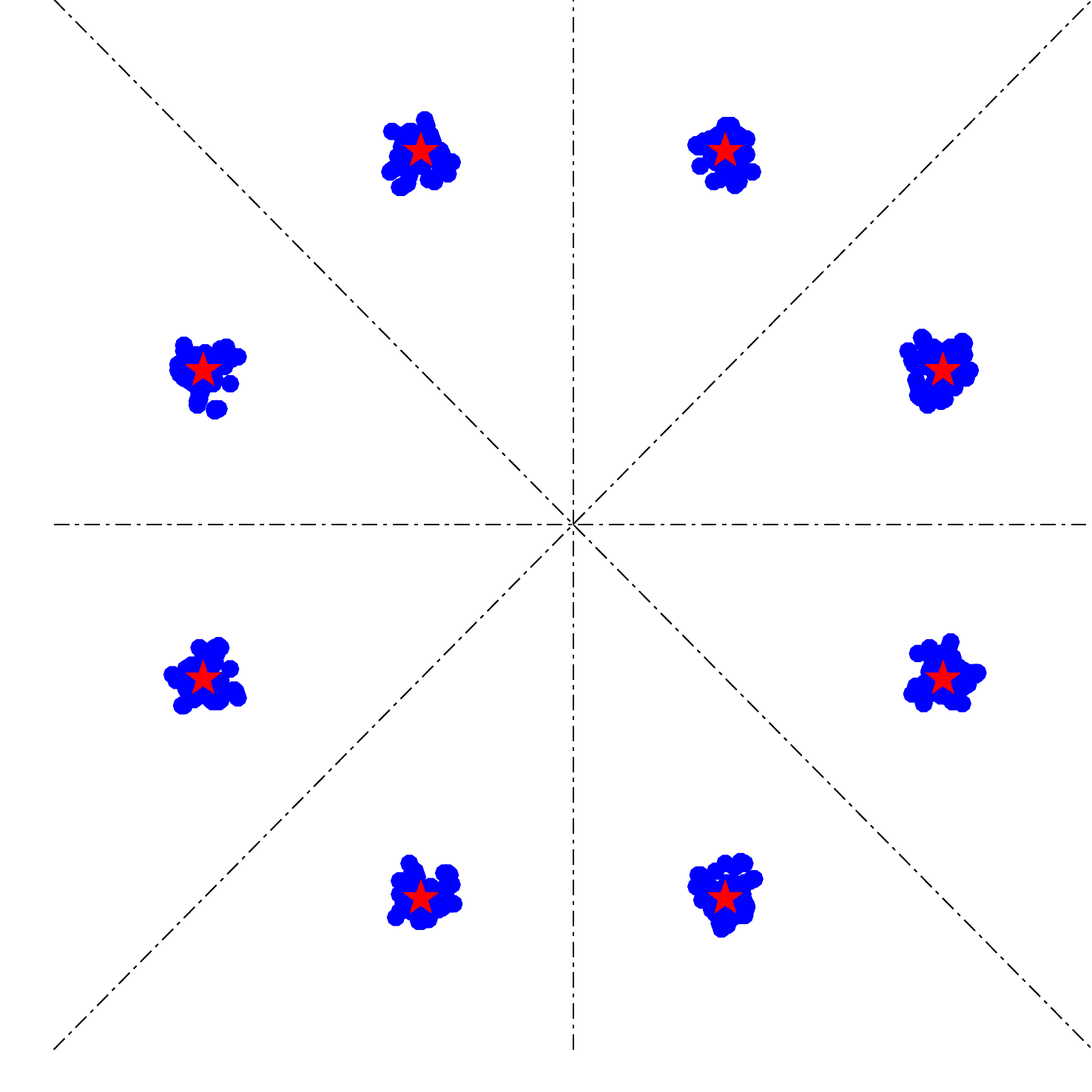}
		\caption*{$\theta=45^\circ$, $N=64$}
	\end{subfigure}
~
\begin{subfigure}[b]{0.4\linewidth}
		\includegraphics[width=\textwidth]{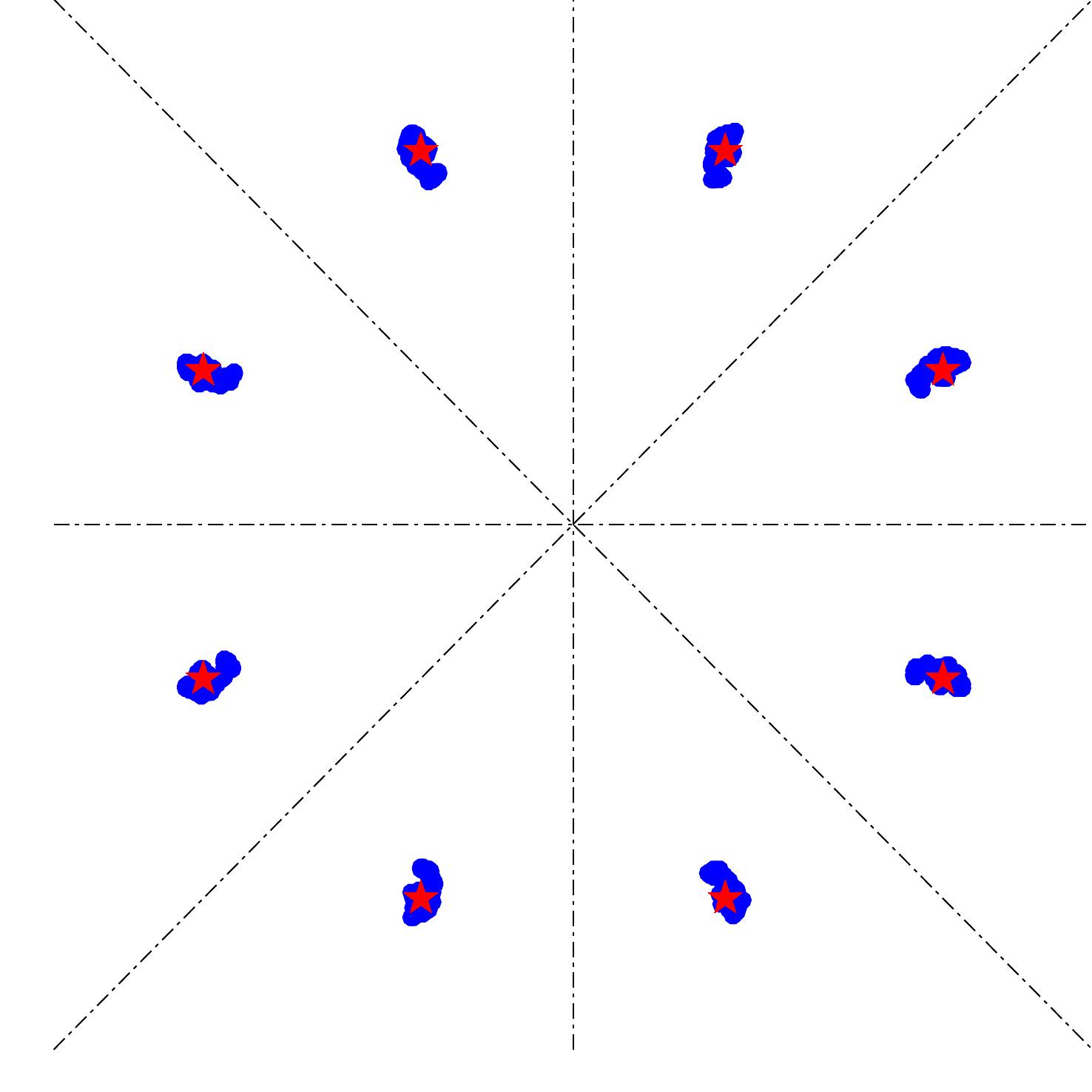}
		\caption*{$\theta=45^\circ$, $N=256$}
	\end{subfigure}
~
\begin{subfigure}[b]{0.4\linewidth}
	\includegraphics[width=\textwidth]{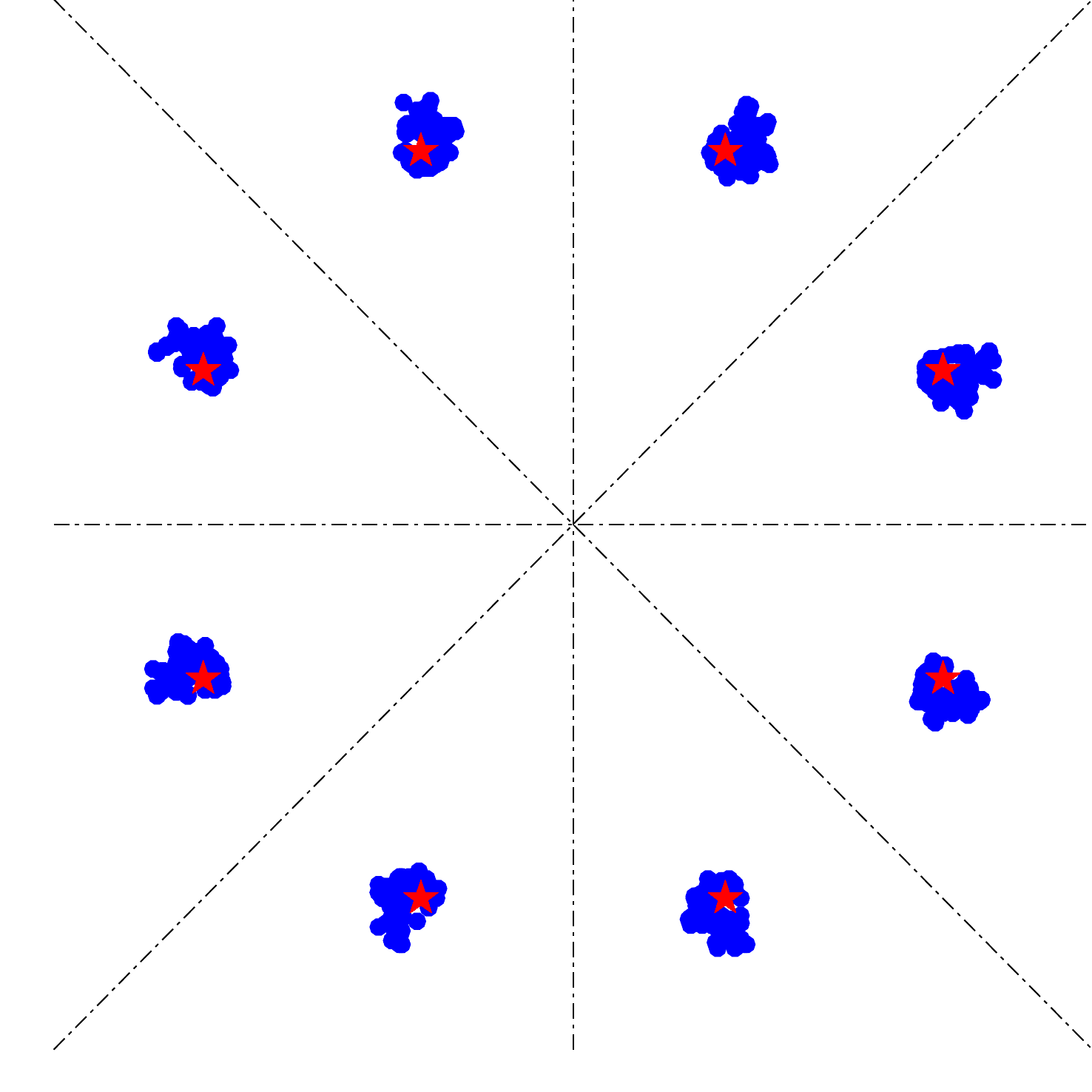}
		\caption*{$\theta=50^\circ$, $N=64$}
	\end{subfigure}
~
\begin{subfigure}[b]{0.4\linewidth}
		\includegraphics[width=\textwidth]{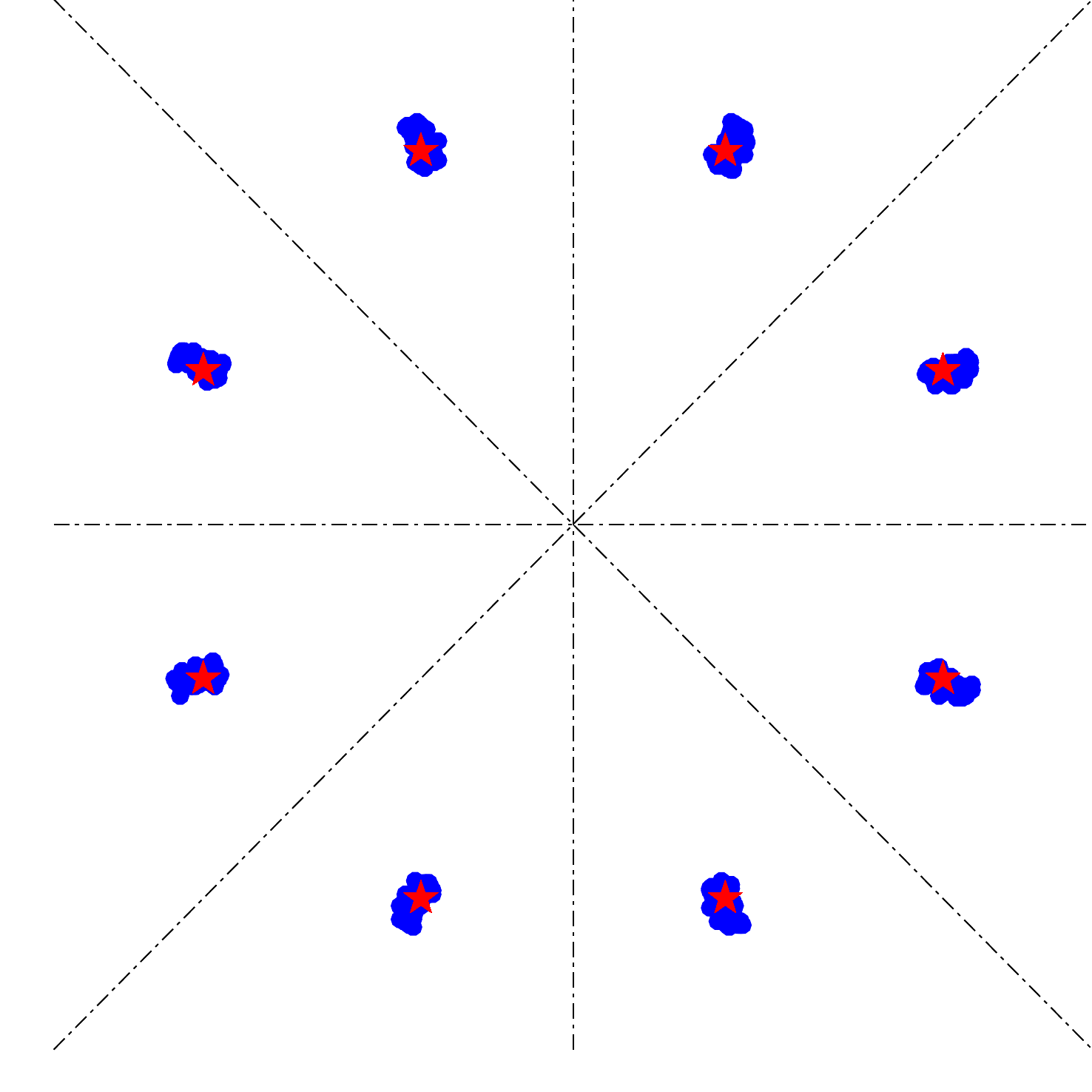}
		\caption*{$\theta=50^\circ$, $N=256$}
	\end{subfigure}
~
\begin{subfigure}[b]{0.4\linewidth}
	\includegraphics[width=\textwidth]{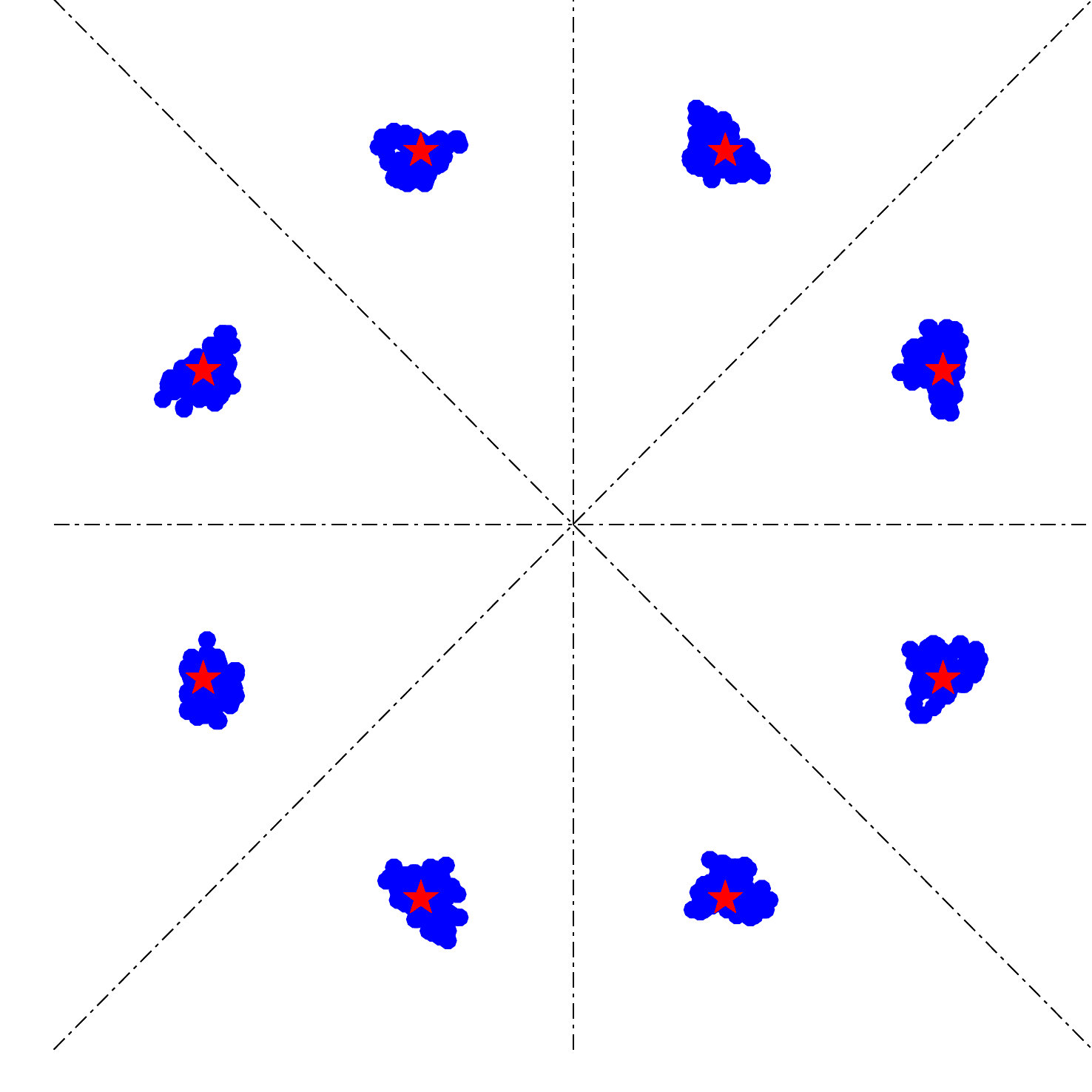}
		\caption*{$\theta=55^\circ$, $N=64$}
	\end{subfigure}
~
\begin{subfigure}[b]{0.4\linewidth}
		\includegraphics[width=\textwidth]{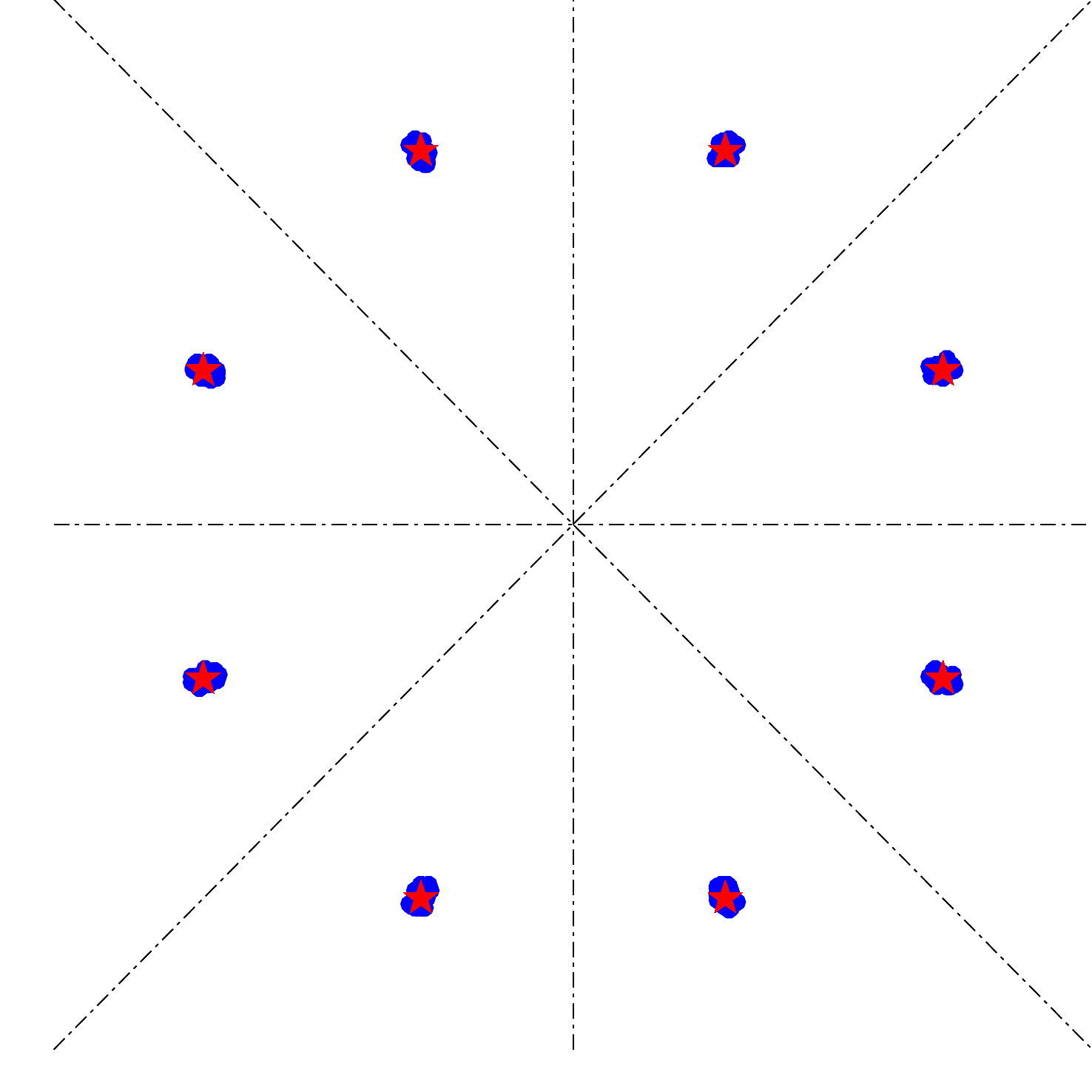}
		\caption*{$\theta=55^\circ$, $N=256$}
	\end{subfigure}
	\caption{IQ scatter plots of the basic sigma-delta MRT scheme for different $\theta$ and $N$; $8$-ary PSK. (cont.)}
\end{figure}

\begin{figure}[H]\ContinuedFloat
	\centering
\begin{subfigure}[b]{0.4\linewidth}
	\includegraphics[width=\textwidth]{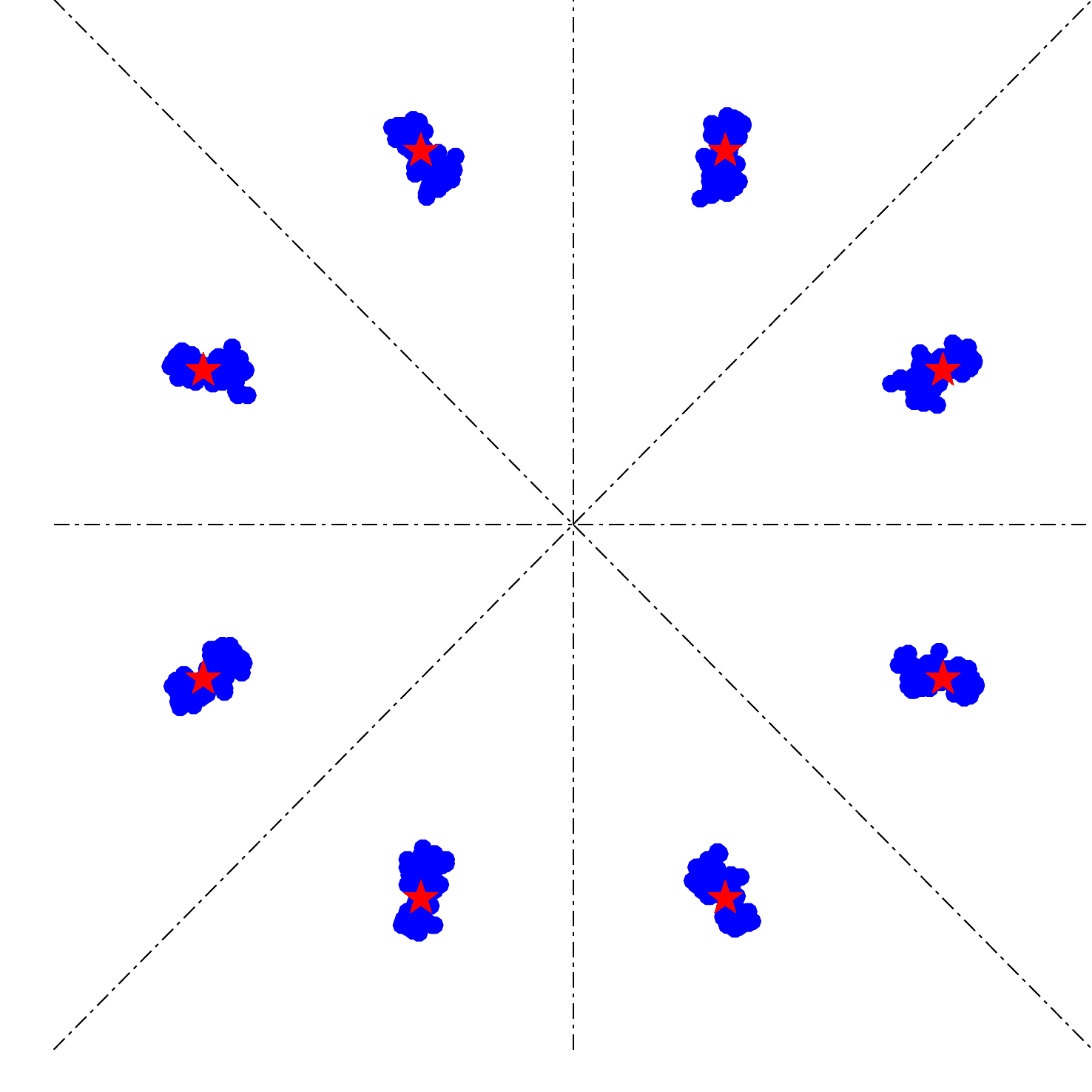}
		\caption*{$\theta=60^\circ$, $N=64$}
	\end{subfigure}
~
\begin{subfigure}[b]{0.4\linewidth}
		\includegraphics[width=\textwidth]{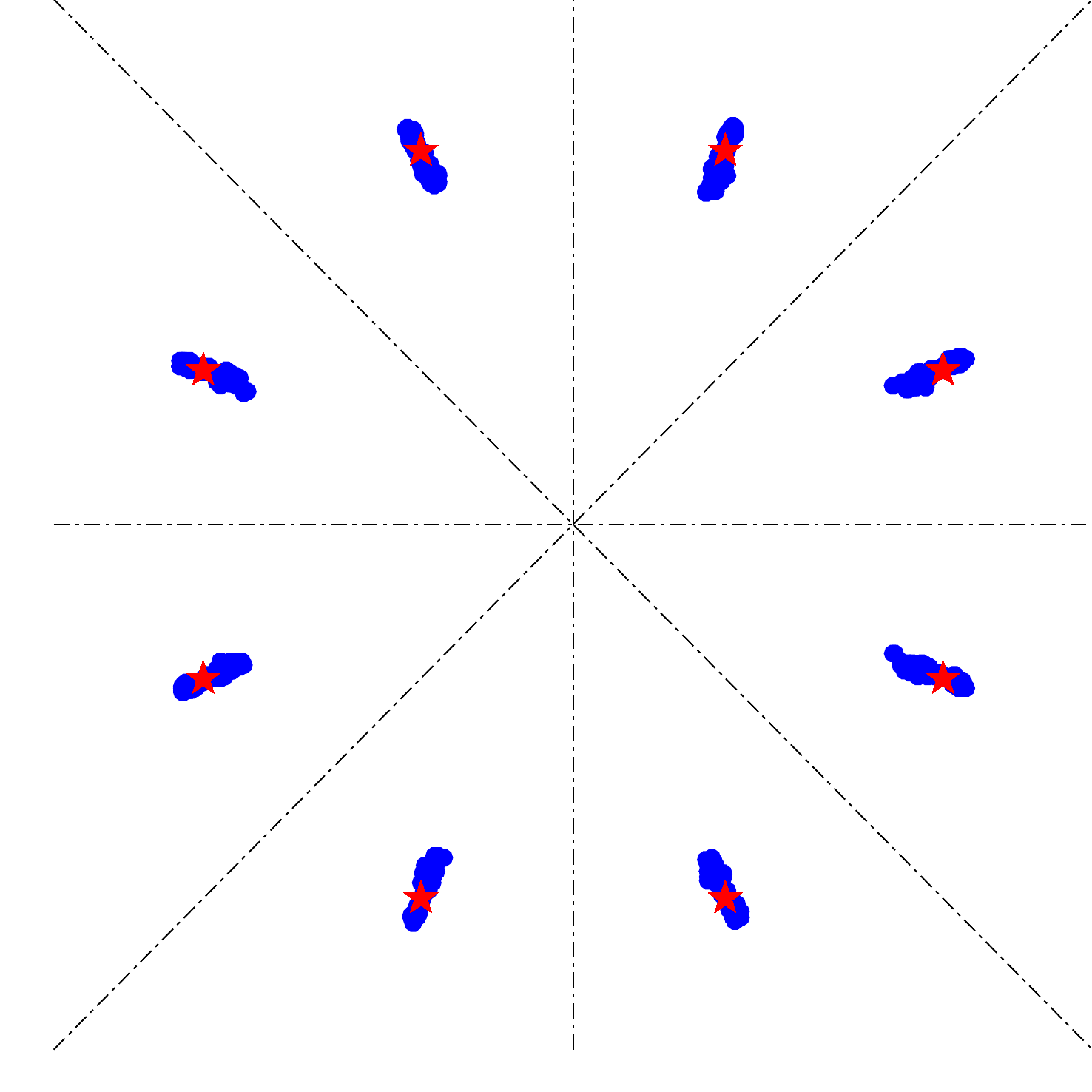}
		\caption*{$\theta=60^\circ$, $N=256$}
	\end{subfigure}
~
\begin{subfigure}[b]{0.4\linewidth}
	\includegraphics[width=\textwidth]{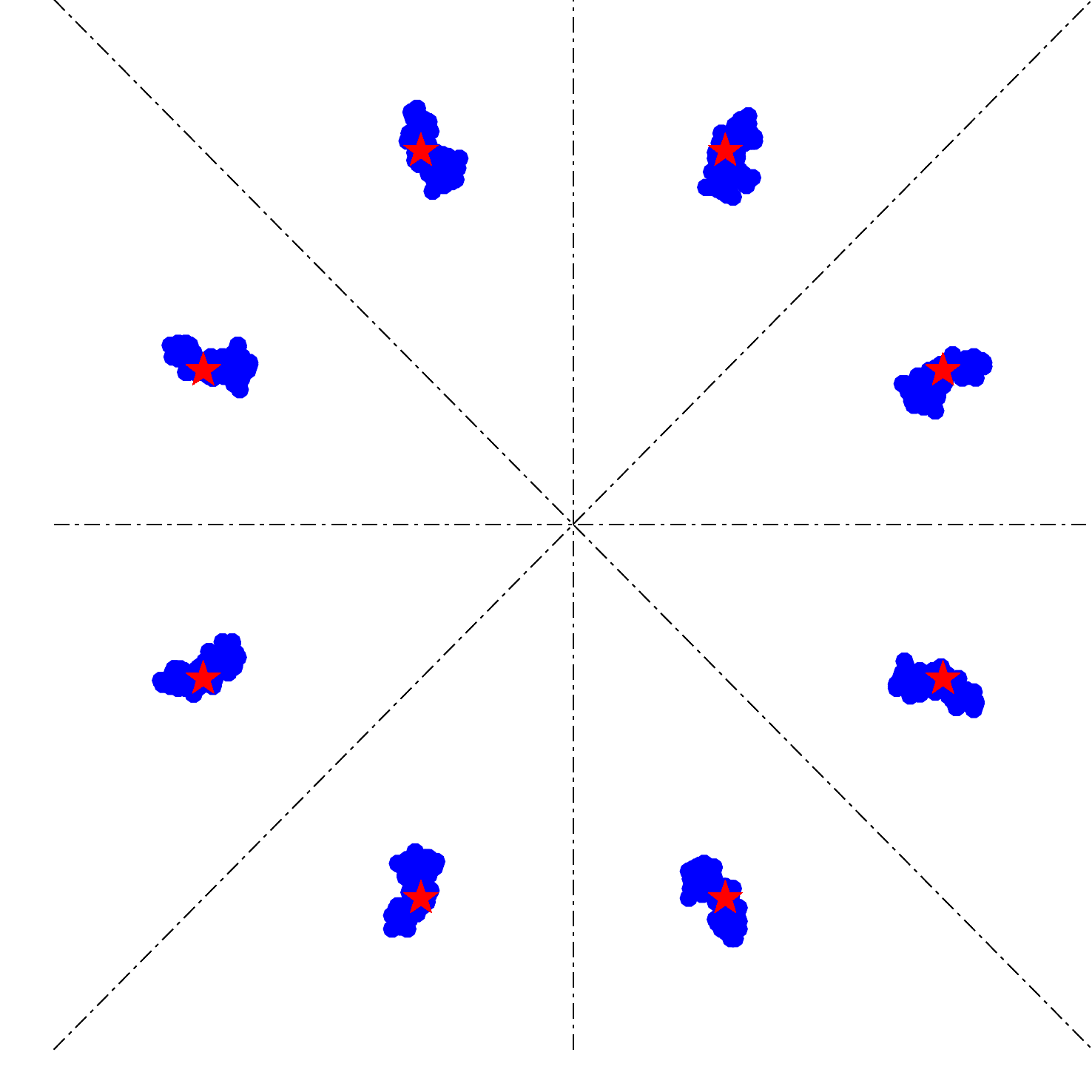}
		\caption*{$\theta=65^\circ$, $N=64$}
	\end{subfigure}
~
\begin{subfigure}[b]{0.4\linewidth}
		\includegraphics[width=\textwidth]{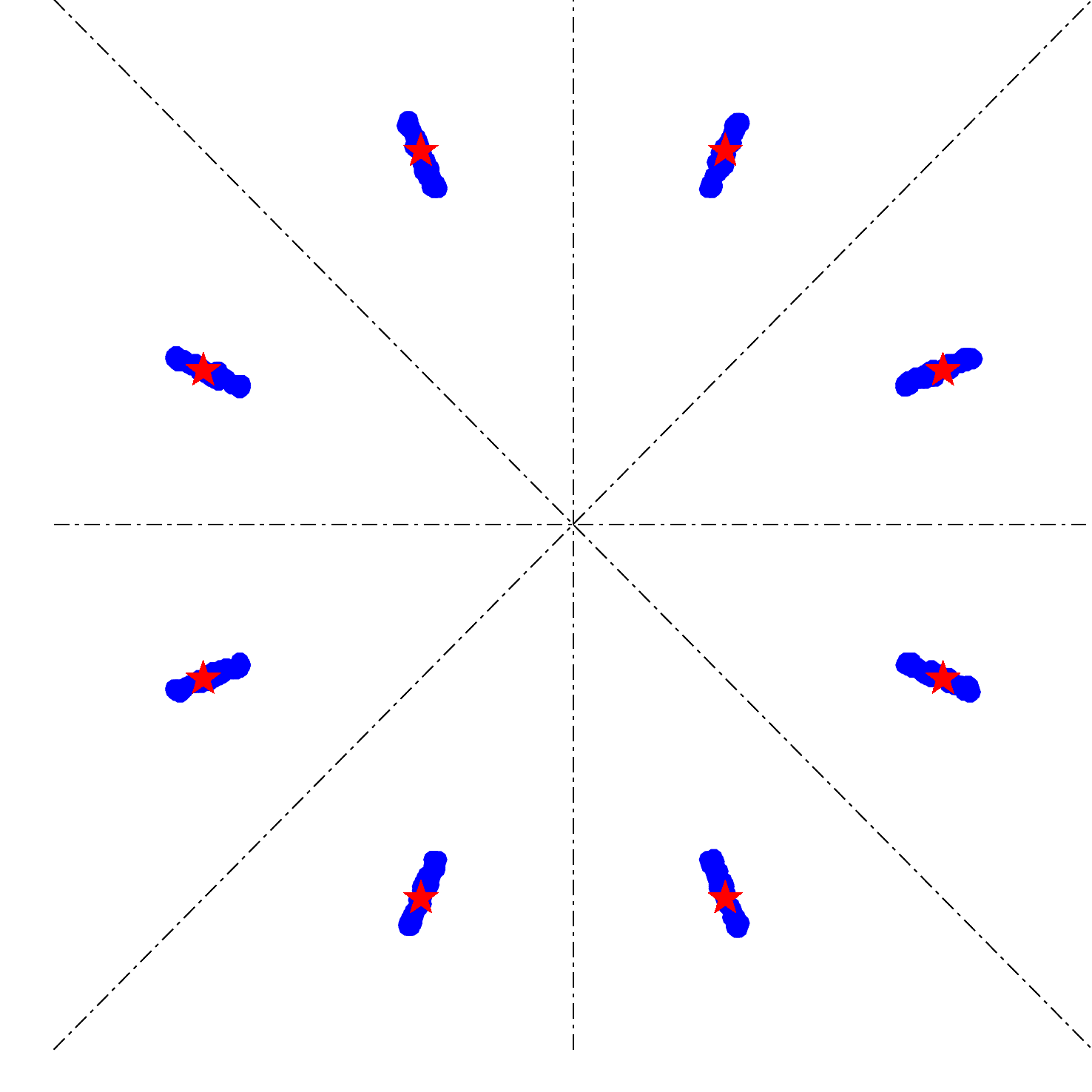}
		\caption*{$\theta=65^\circ$, $N=256$}
	\end{subfigure}
~
\begin{subfigure}[b]{0.4\linewidth}
	\includegraphics[width=\textwidth]{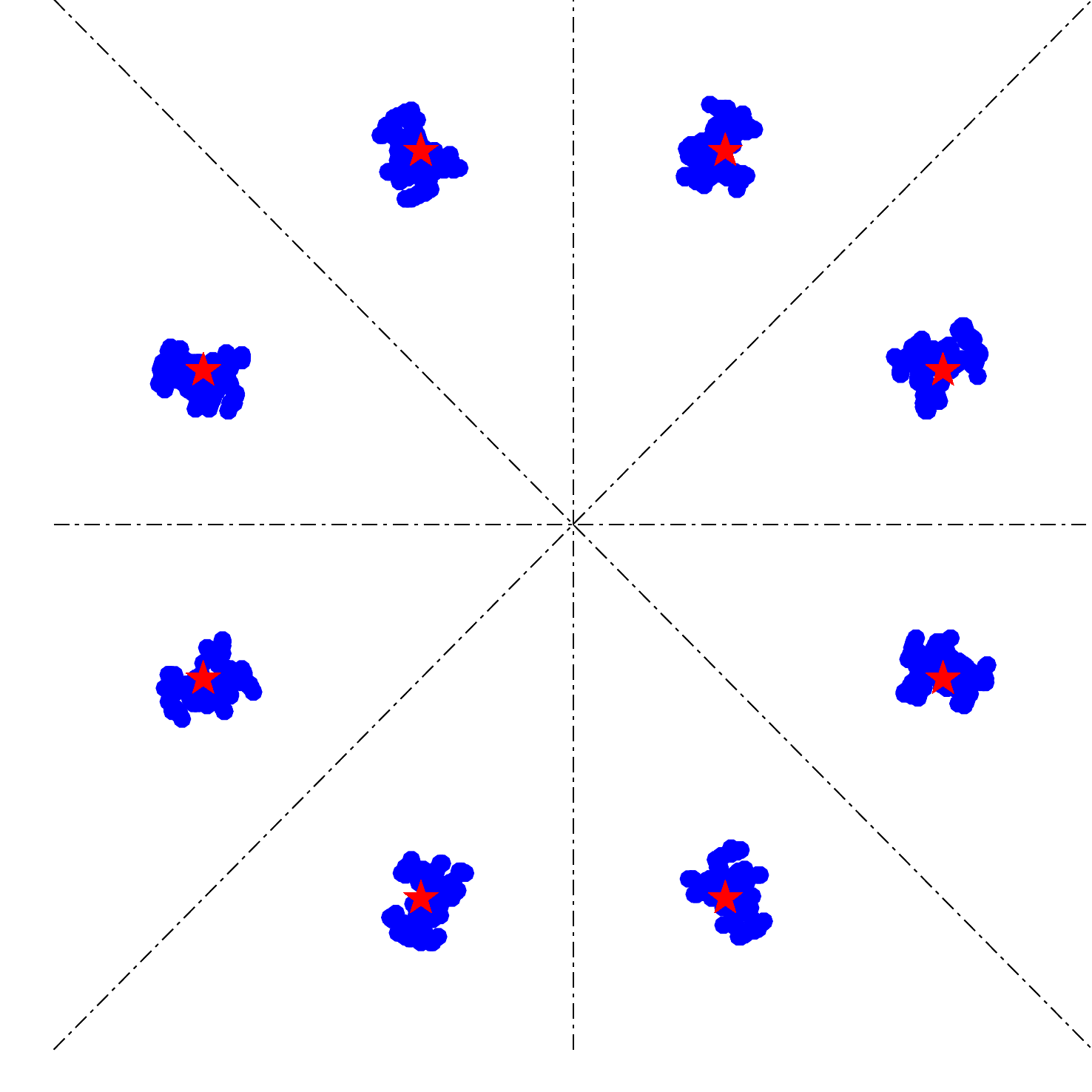}
		\caption*{$\theta=70^\circ$, $N=64$}
	\end{subfigure}
~
\begin{subfigure}[b]{0.4\linewidth}
		\includegraphics[width=\textwidth]{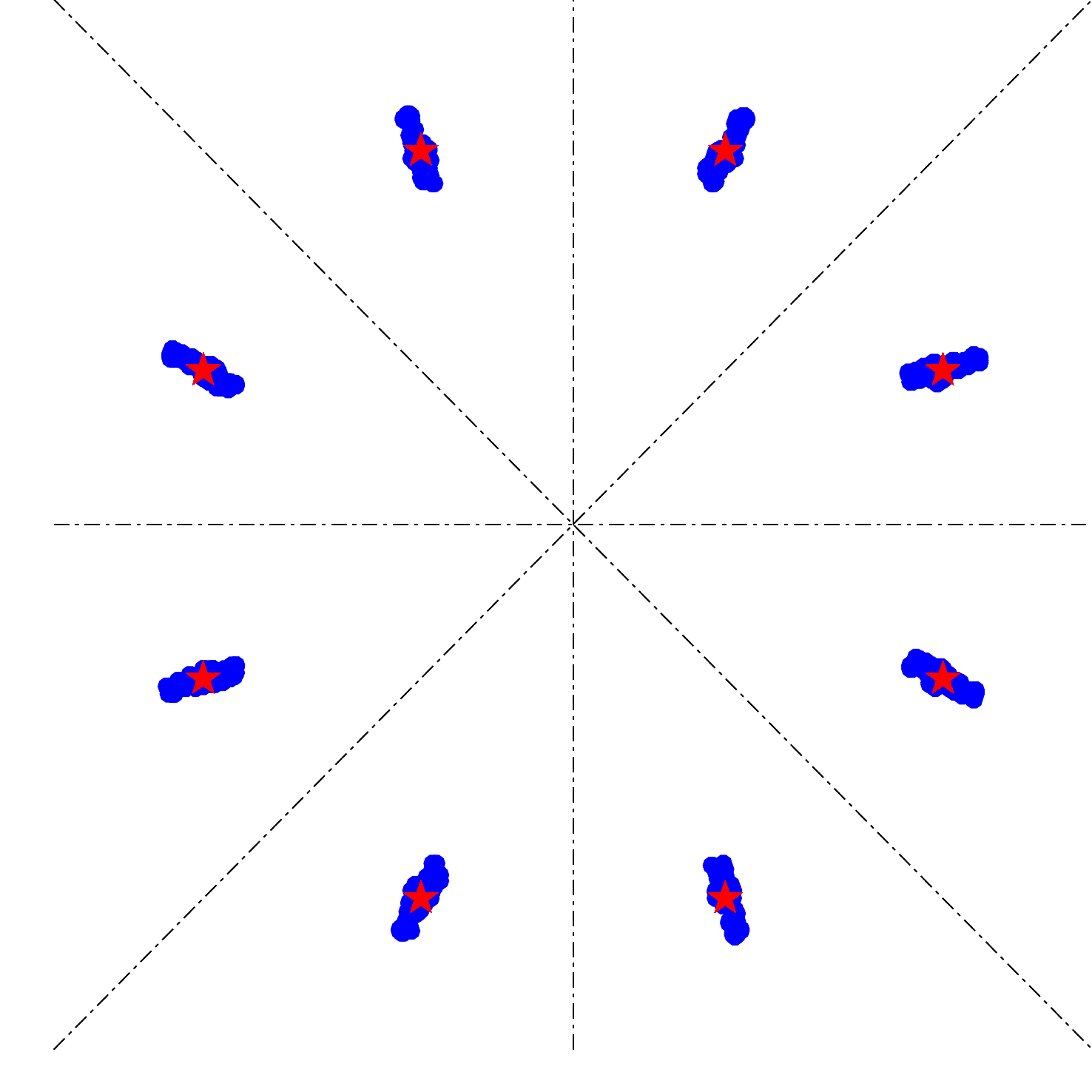}
		\caption*{$\theta=70^\circ$, $N=256$}
	\end{subfigure}
	\caption{IQ scatter plots of the basic sigma-delta MRT scheme for different $\theta$ and $N$; $8$-ary PSK. (cont.)}
\end{figure}

\begin{figure}[H]\ContinuedFloat
\centering
\begin{subfigure}[b]{0.4\linewidth}
	\includegraphics[width=\textwidth]{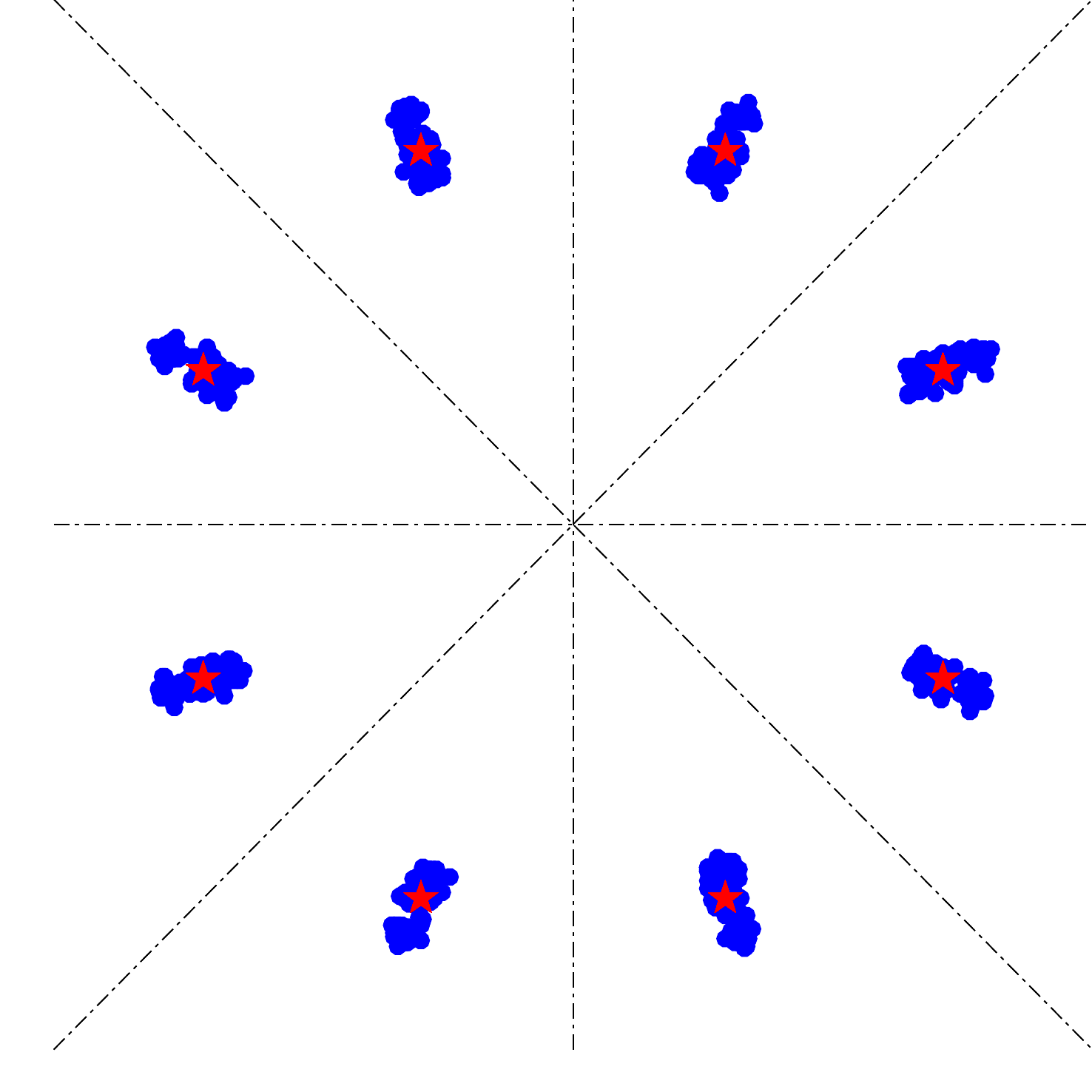}
		\caption*{$\theta=75^\circ$, $N=64$}
	\end{subfigure}
~
\begin{subfigure}[b]{0.4\linewidth}
		\includegraphics[width=\textwidth]{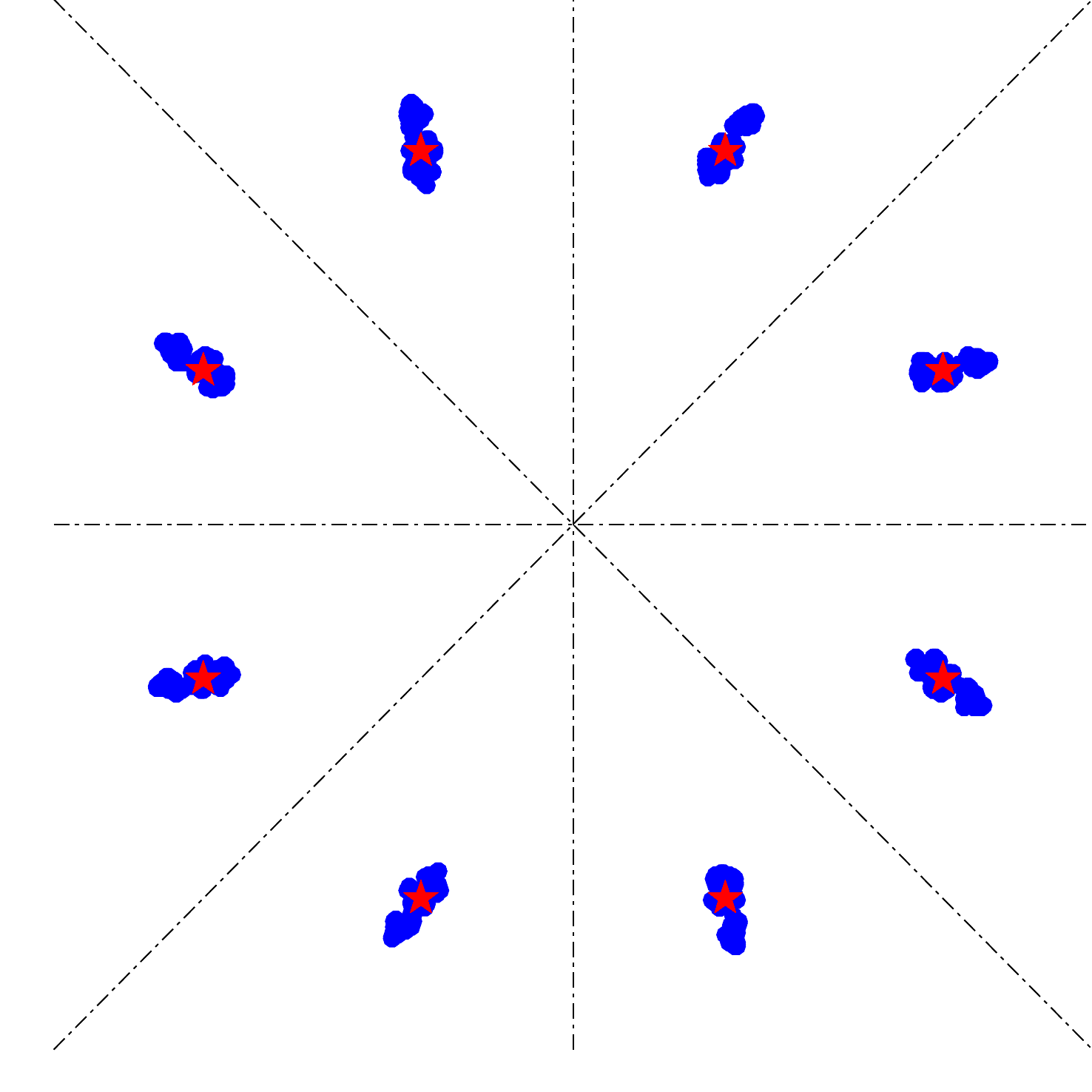}
		\caption*{$\theta=75^\circ$, $N=256$}
	\end{subfigure}
~
\begin{subfigure}[b]{0.4\linewidth}
	\includegraphics[width=\textwidth]{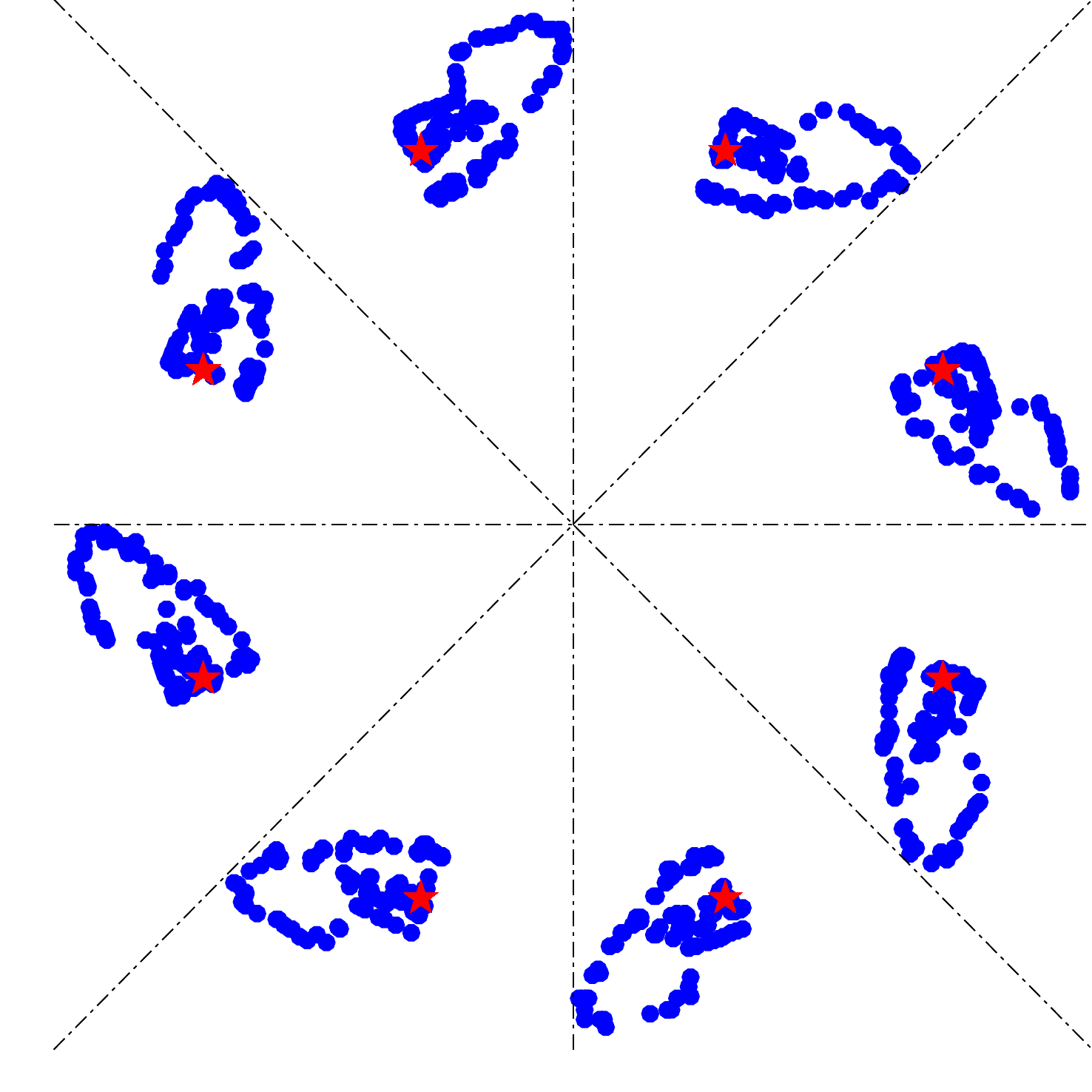}
		\caption*{$\theta=80^\circ$, $N=64$}
	\end{subfigure}
~
\begin{subfigure}[b]{0.4\linewidth}
		\includegraphics[width=\textwidth]{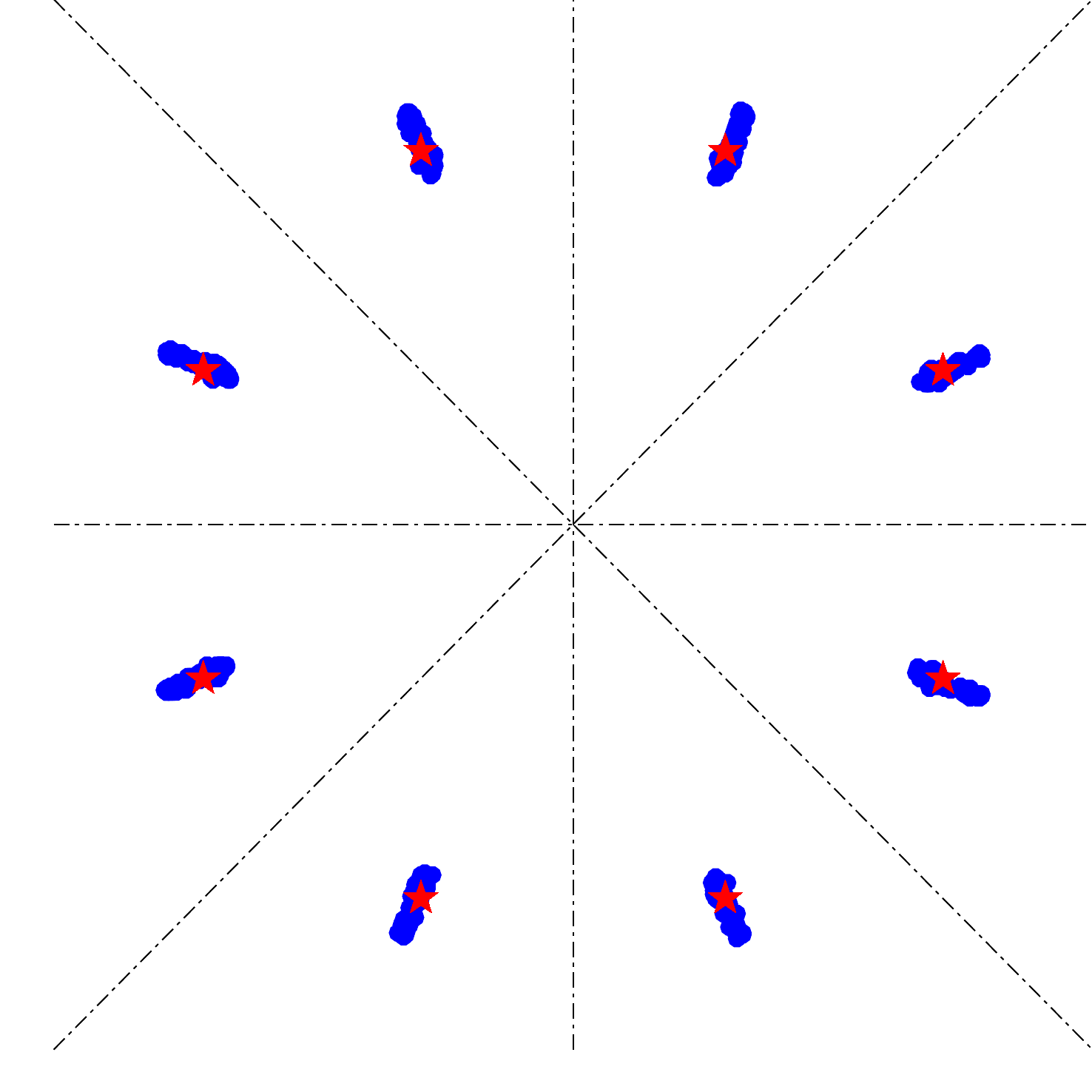}
		\caption*{$\theta=80^\circ$, $N=256$}
	\end{subfigure}
~
\begin{subfigure}[b]{0.4\linewidth}
	\includegraphics[width=\textwidth]{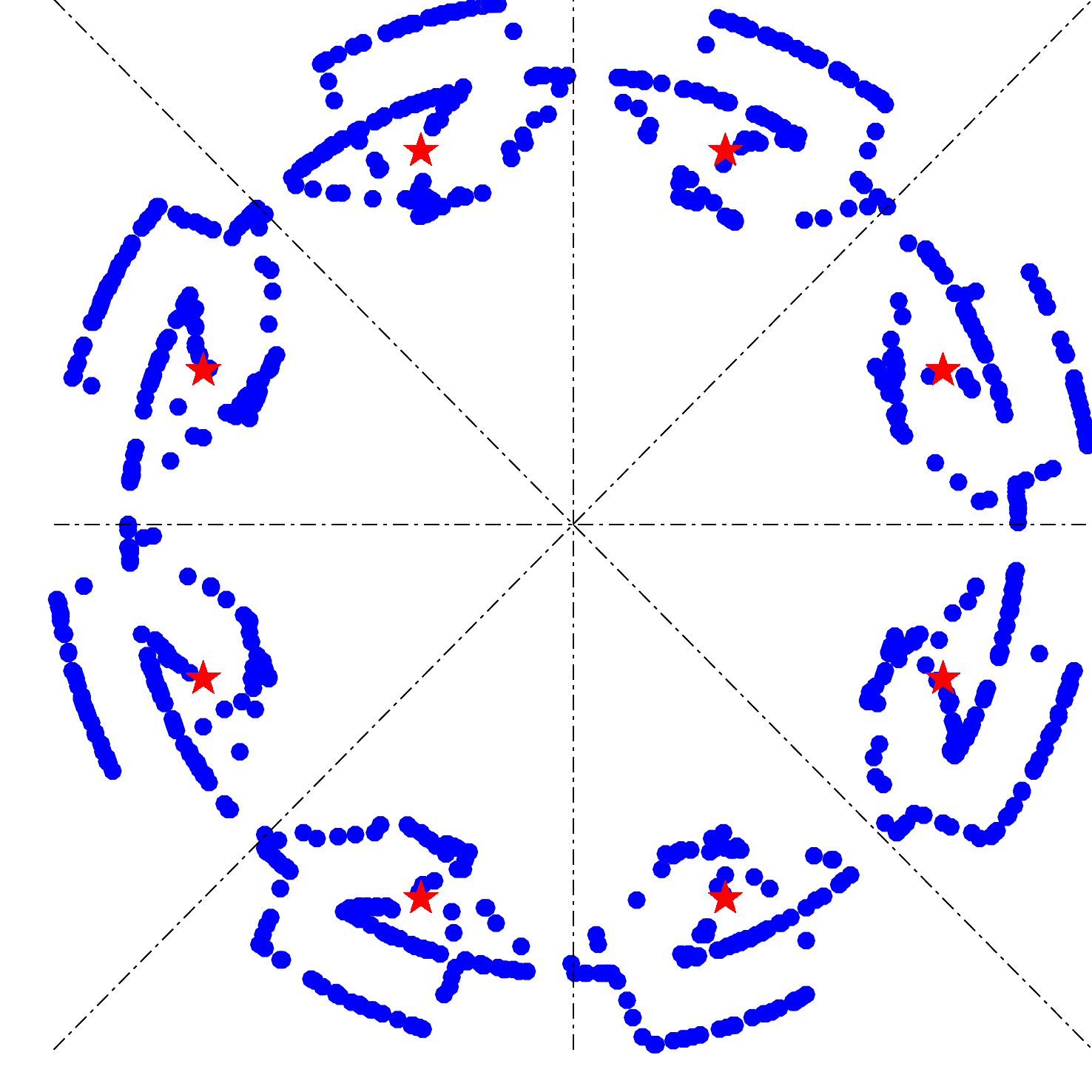}
		\caption*{$\theta=85^\circ$, $N=64$}
	\end{subfigure}
~
\begin{subfigure}[b]{0.4\linewidth}
		\includegraphics[width=\textwidth]{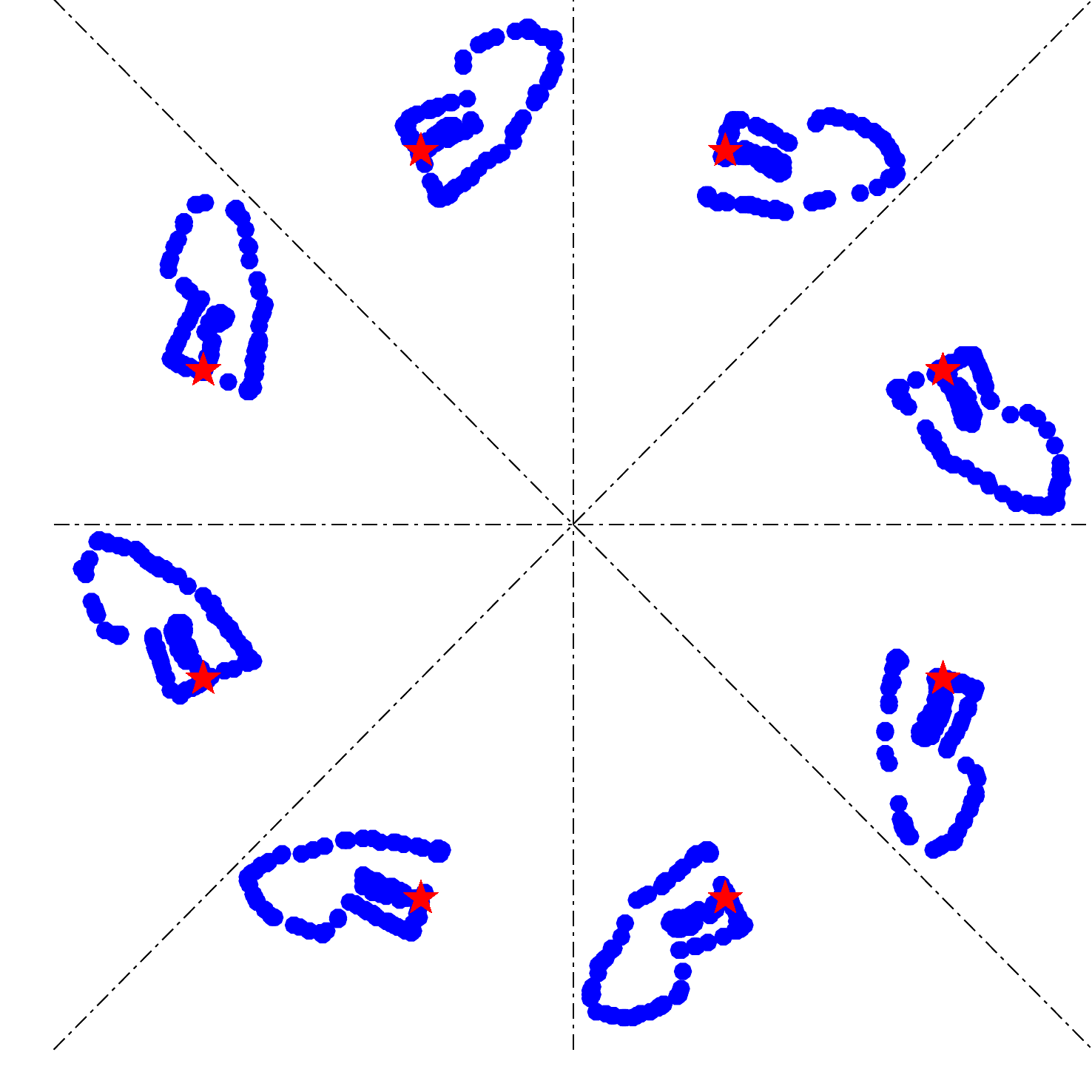}
		\caption*{$\theta=85^\circ$, $N=256$}
	\end{subfigure}
	\caption{IQ scatter plots of the basic sigma-delta MRT scheme for different $\theta$ and $N$; $8$-ary PSK. (cont.)}
\end{figure}

\begin{figure}[H]\ContinuedFloat
	\centering
\begin{subfigure}[b]{0.4\linewidth}
	\includegraphics[width=\textwidth]{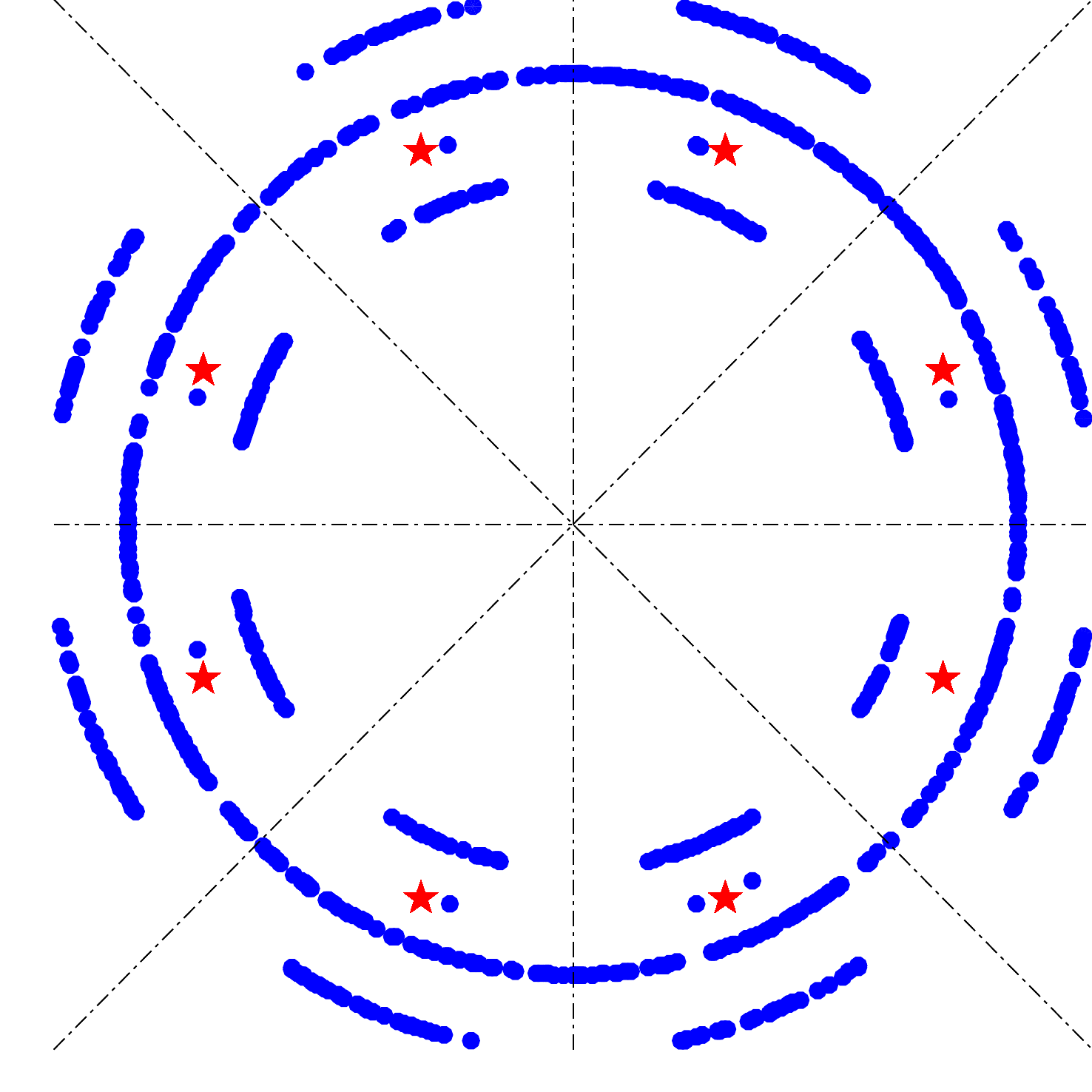}
		\caption*{$\theta=90^\circ$, $N=64$}
	\end{subfigure}
~
\begin{subfigure}[b]{0.4\linewidth}
		\includegraphics[width=\textwidth]{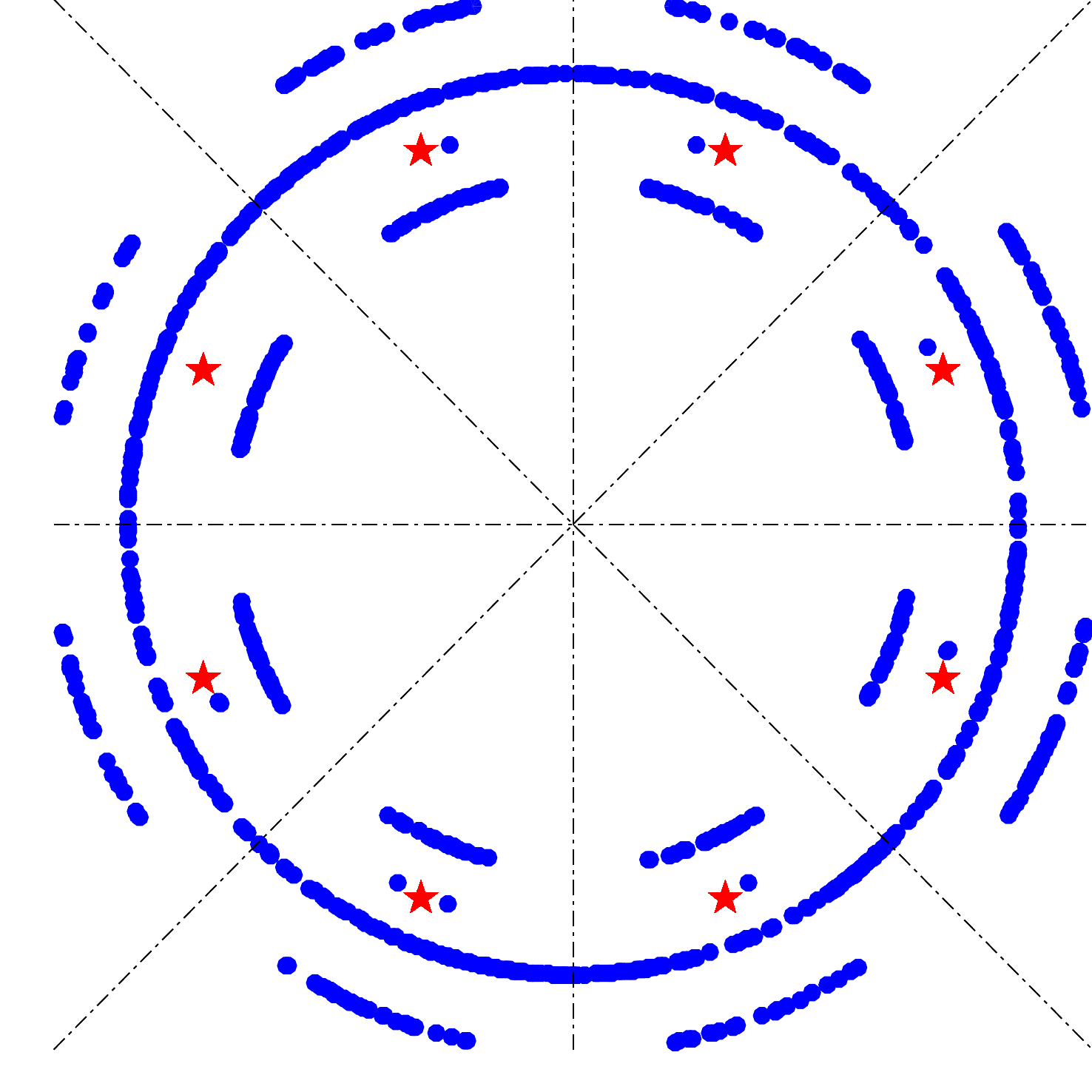}
		\caption*{$\theta=90^\circ$, $N=256$}
	\end{subfigure}
~
	\caption{IQ scatter plots of the basic sigma-delta MRT scheme for different $\theta$ and $N$; $8$-ary PSK. (cont.)}
\end{figure}

\begin{figure}[H]
	\centering	
\begin{subfigure}[b]{0.3\linewidth}
		\includegraphics[width=\textwidth]{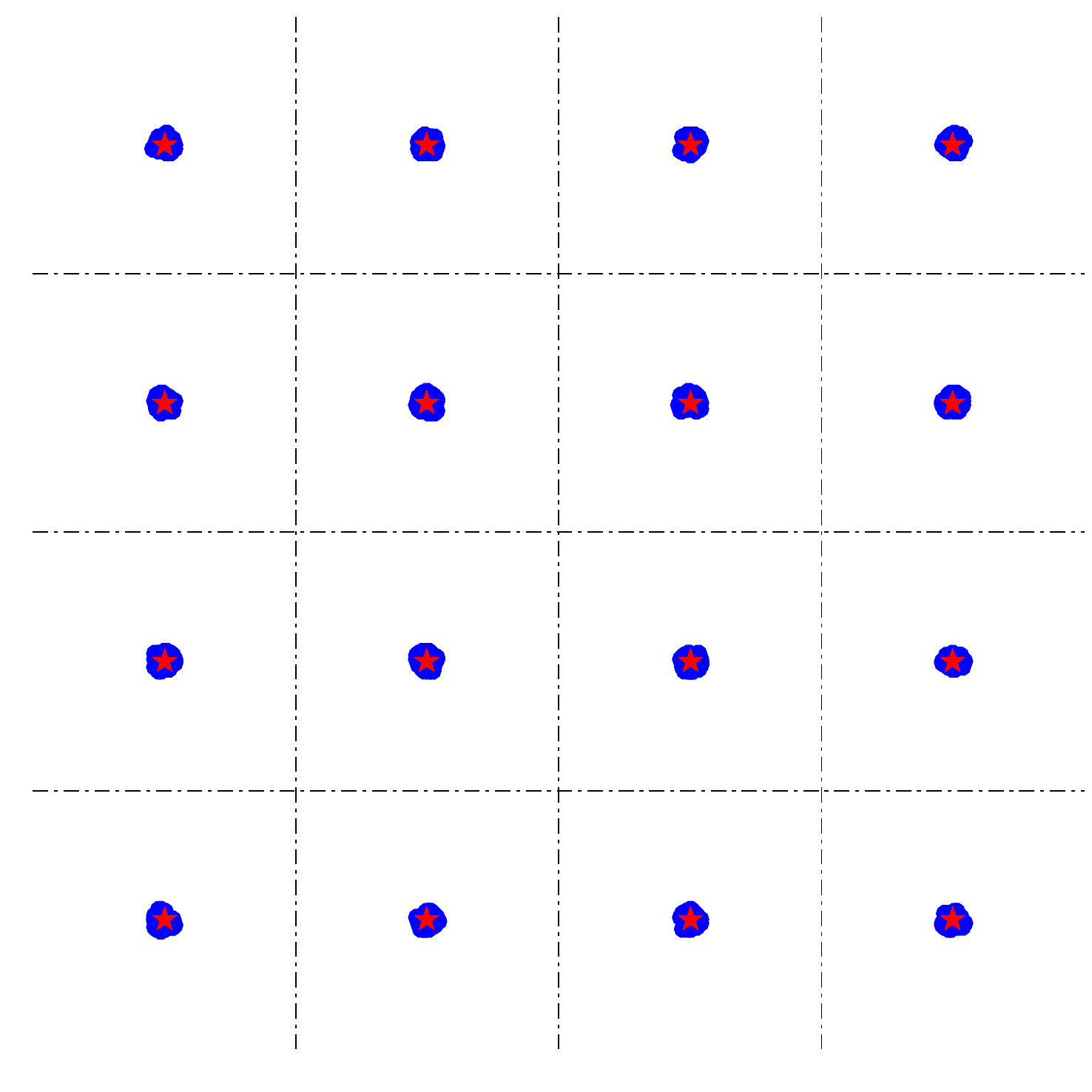}
		\caption*{$\theta=0^\circ$, $N=64$}
	\end{subfigure}
~
\begin{subfigure}[b]{0.3\linewidth}
		\includegraphics[width=\textwidth]{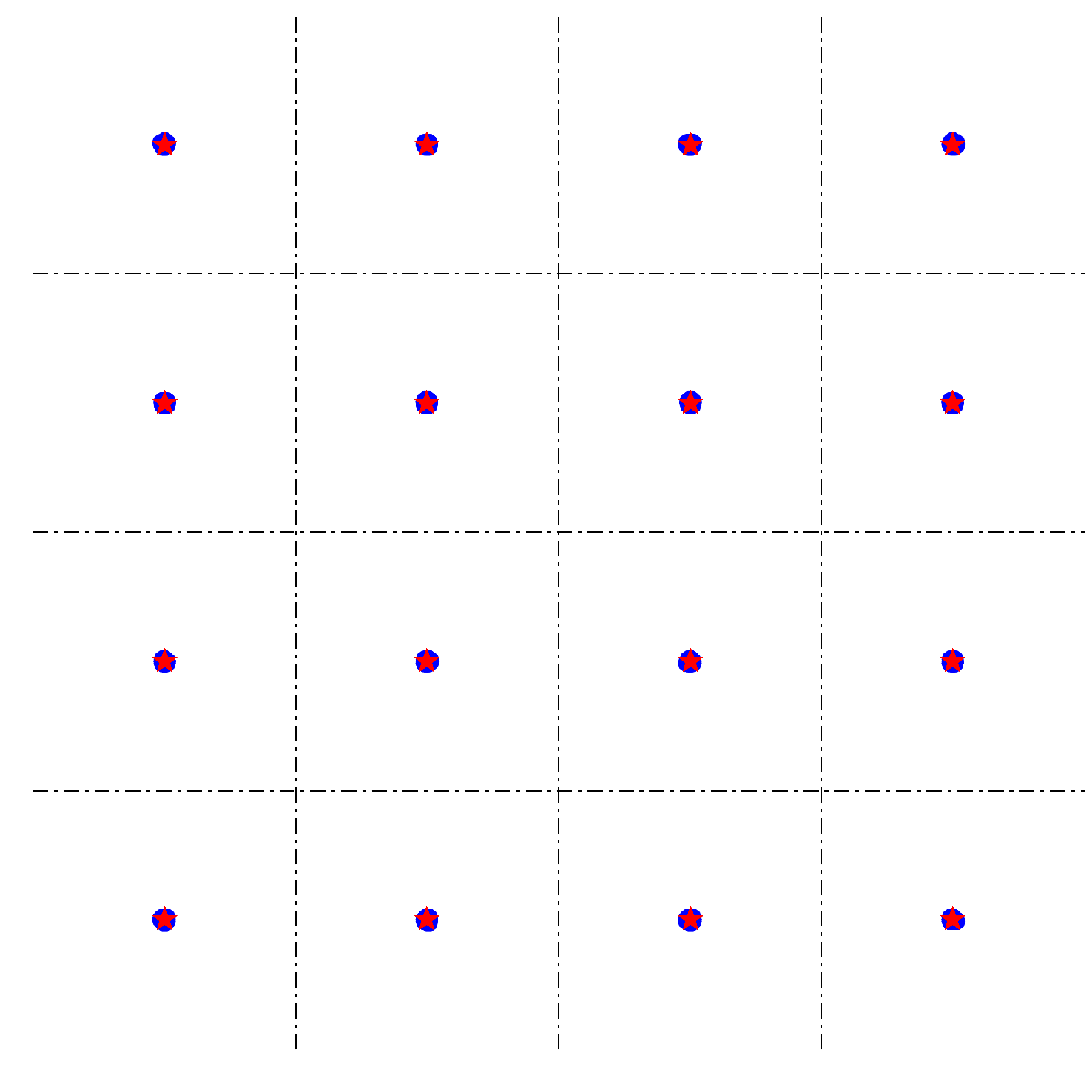}
		\caption*{$\theta=0^\circ$, $N=256$}
	\end{subfigure}
~
\begin{subfigure}[b]{0.3\linewidth}
		\includegraphics[width=\textwidth]{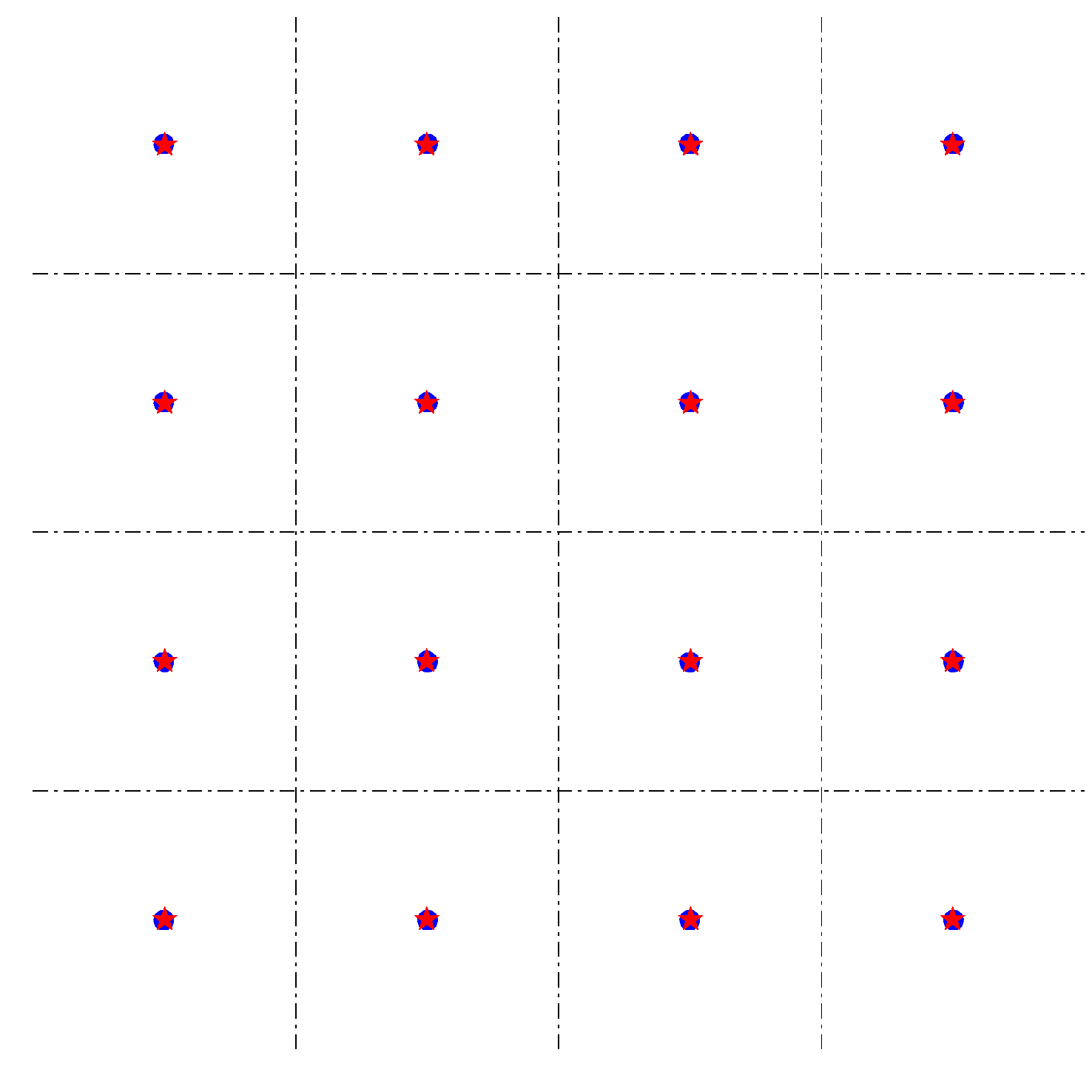}
		\caption*{$\theta=0^\circ$, $N=512$}
	\end{subfigure}
~
\begin{subfigure}[b]{0.3\linewidth}
		\includegraphics[width=\textwidth]{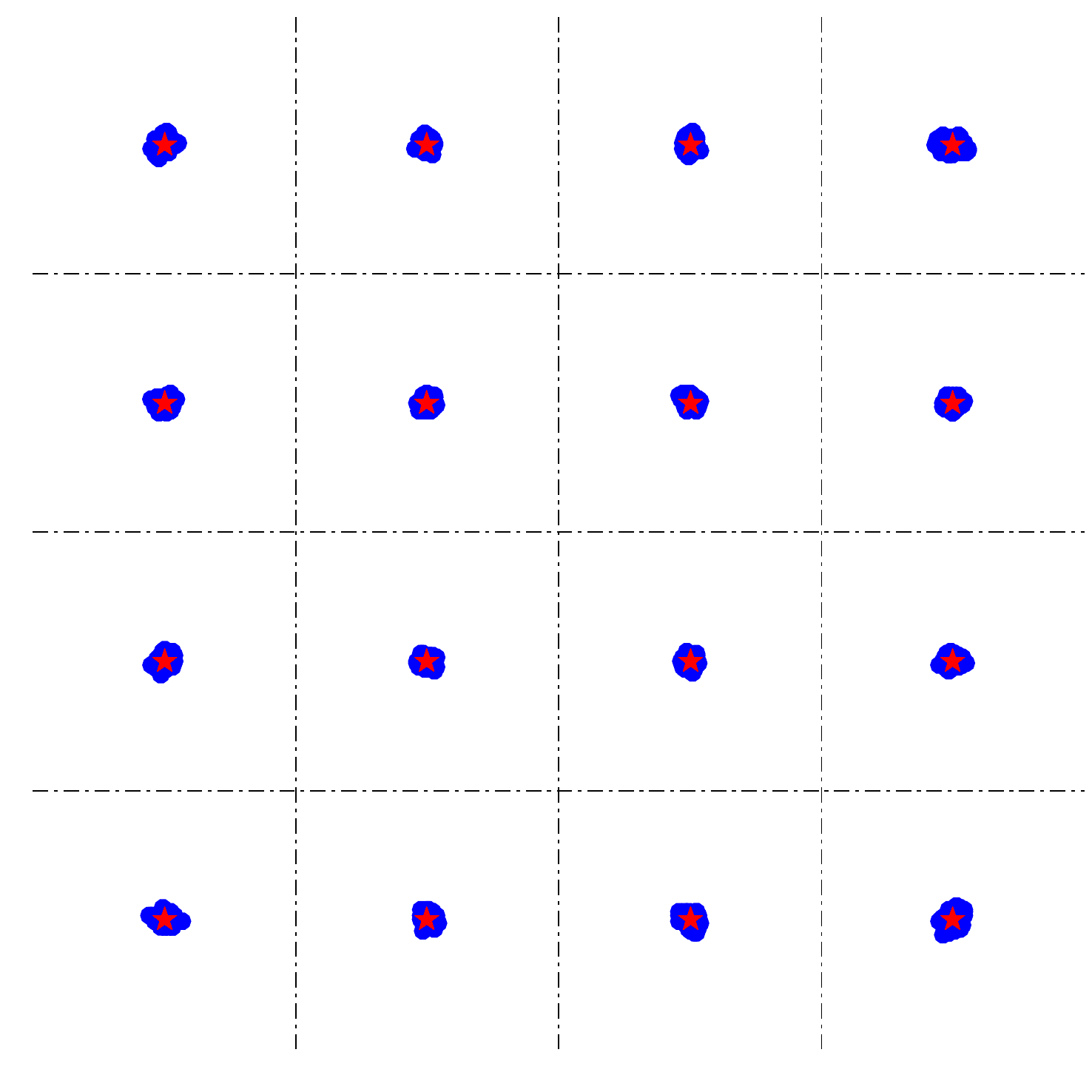}
		\caption*{$\theta=5^\circ$, $N=64$}
	\end{subfigure}
~
\begin{subfigure}[b]{0.3\linewidth}
		\includegraphics[width=\textwidth]{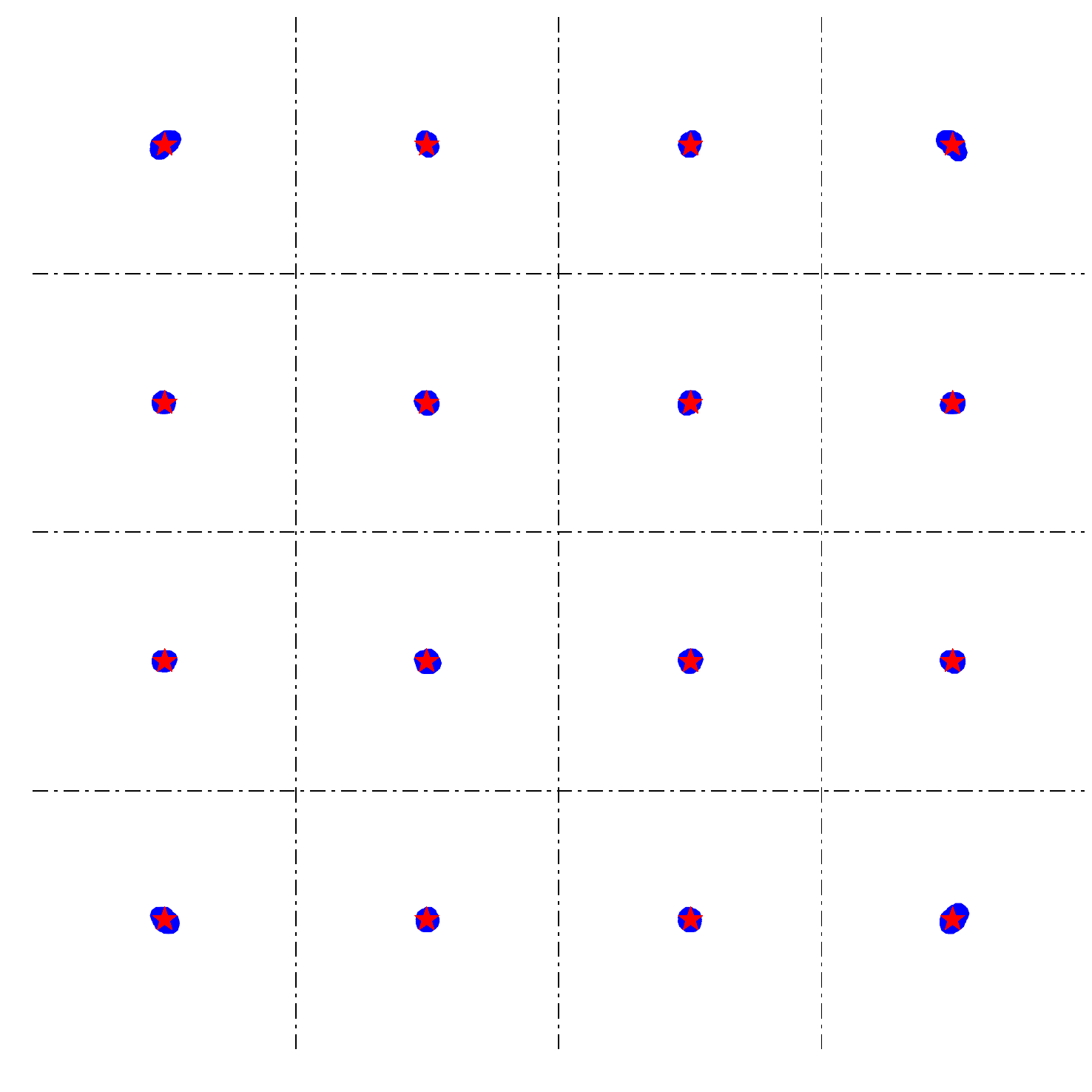}
		\caption*{$\theta=5^\circ$, $N=256$}
	\end{subfigure}
~
\begin{subfigure}[b]{0.3\linewidth}
		\includegraphics[width=\textwidth]{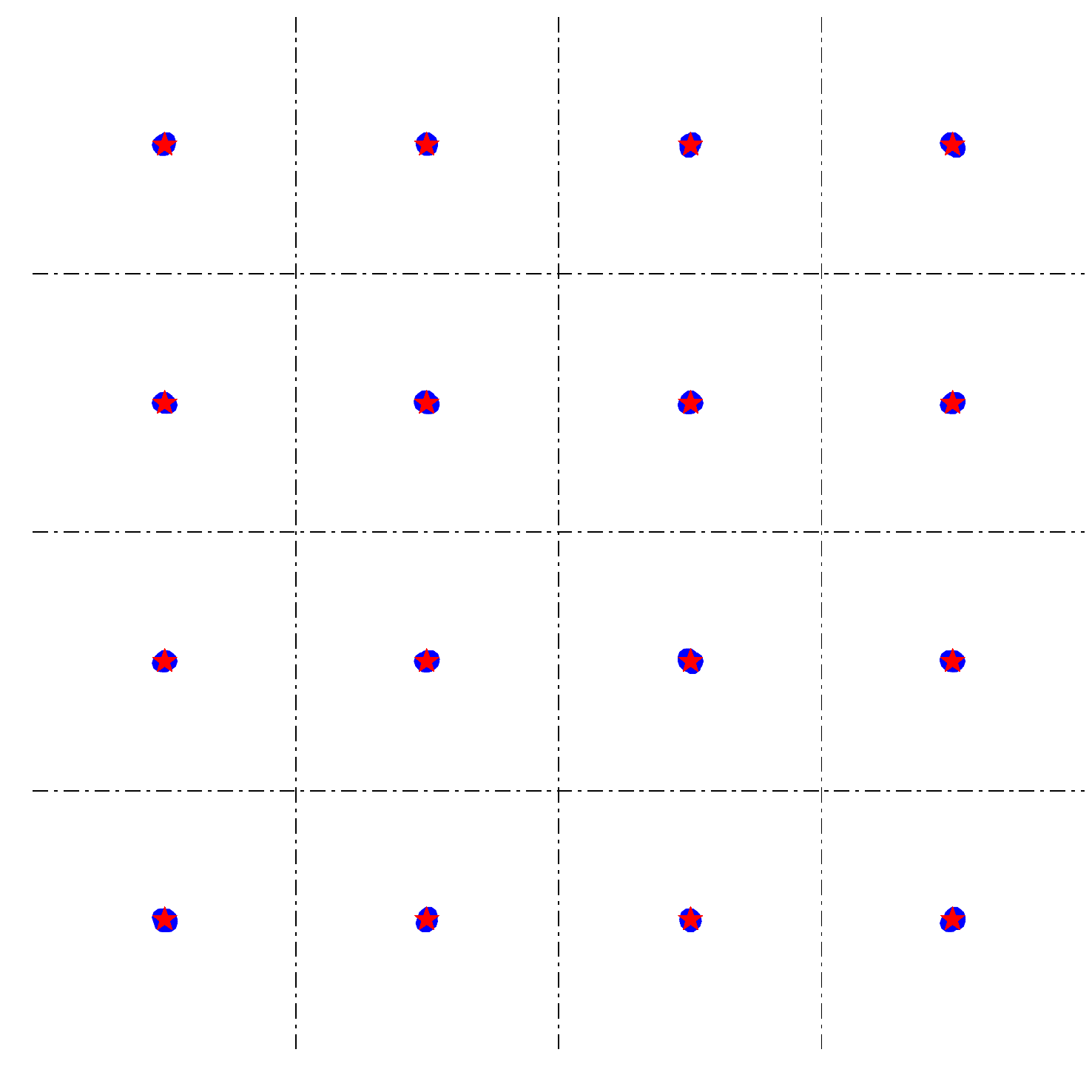}
		\caption*{$\theta=5^\circ$, $N=512$}
	\end{subfigure}
~
\begin{subfigure}[b]{0.3\linewidth}
		\includegraphics[width=\textwidth]{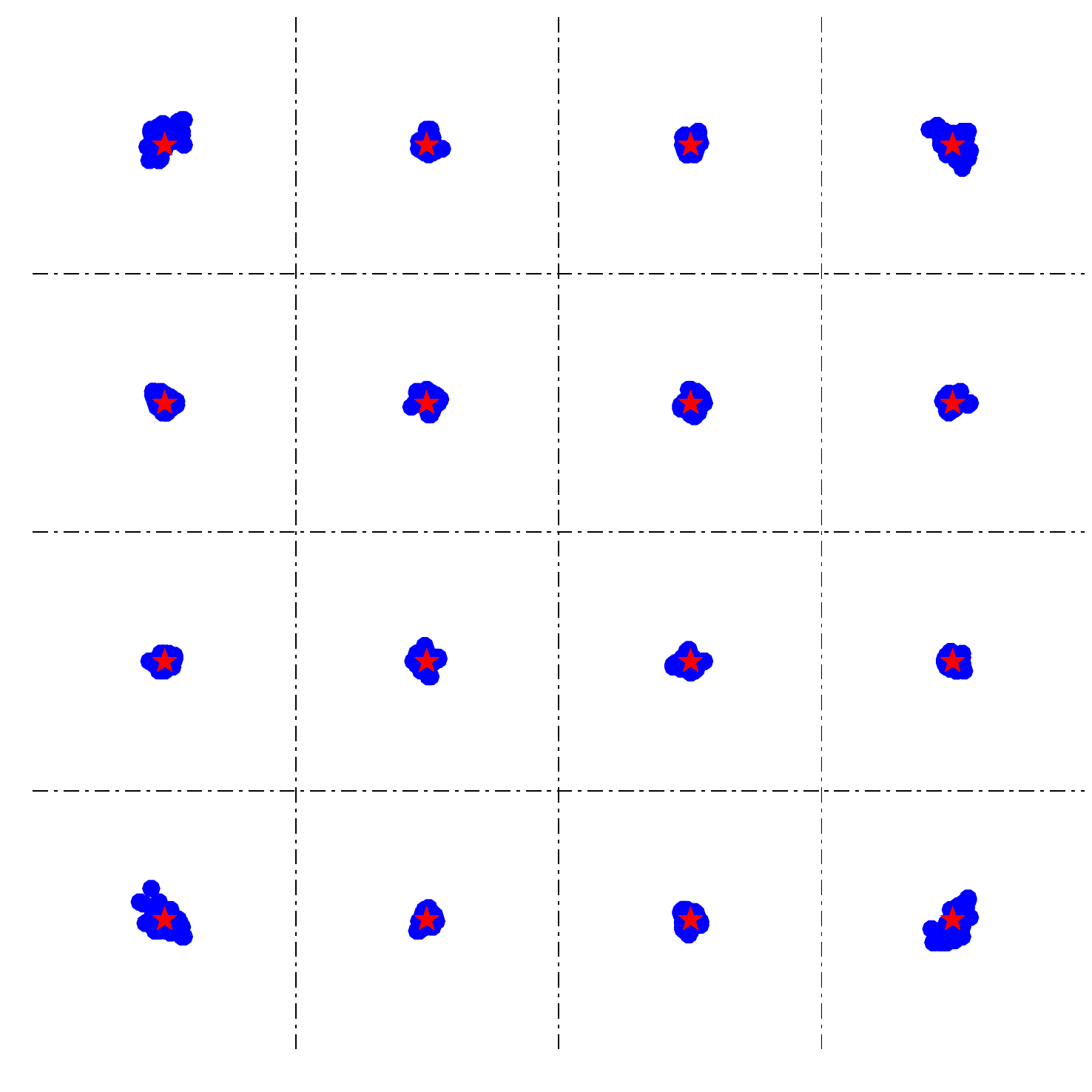}
		\caption*{$\theta=10^\circ$, $N=64$}
	\end{subfigure}
~
\begin{subfigure}[b]{0.3\linewidth}
		\includegraphics[width=\textwidth]{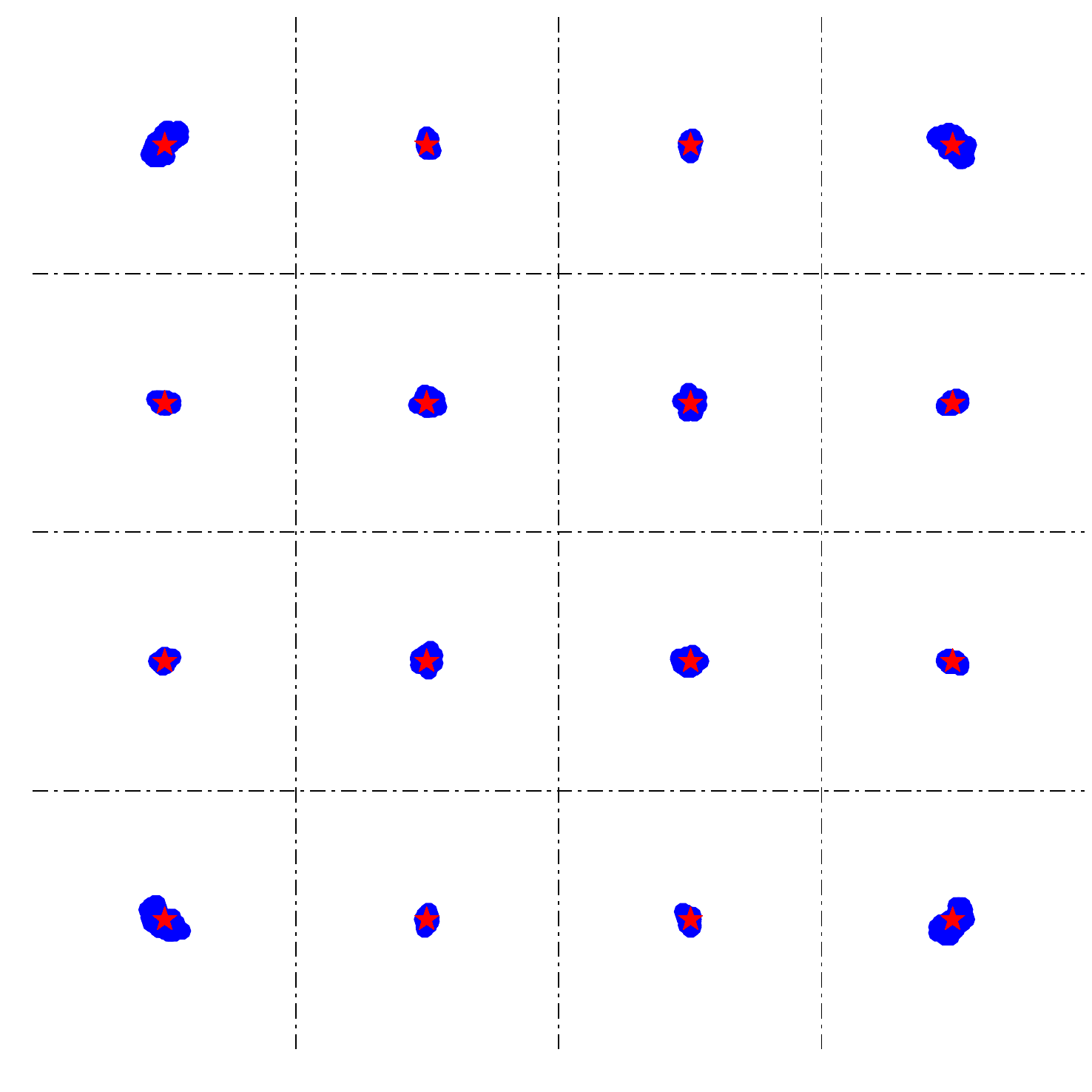}
		\caption*{$\theta=10^\circ$, $N=256$}
	\end{subfigure}
~
\begin{subfigure}[b]{0.3\linewidth}
		\includegraphics[width=\textwidth]{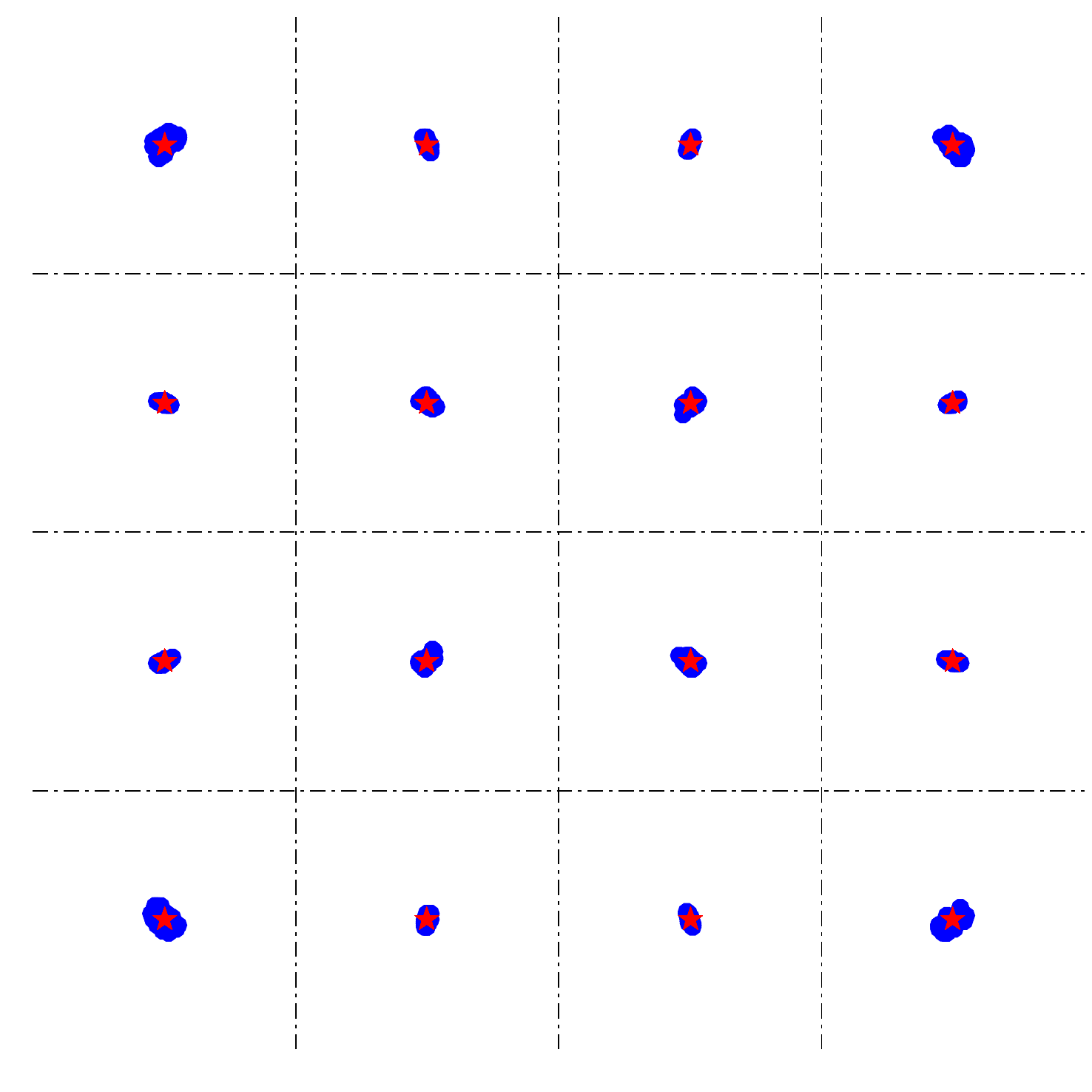}
		\caption*{$\theta=10^\circ$, $N=512$}
	\end{subfigure}
	\caption{IQ scatter plots of the basic sigma-delta MRT scheme for different $\theta$ and $N$; $16$-ary QAM.} \label{fig:QAM_con1}
\end{figure}

\begin{figure}[H]\ContinuedFloat
	\centering	
\begin{subfigure}[b]{0.3\linewidth}
		\includegraphics[width=\textwidth]{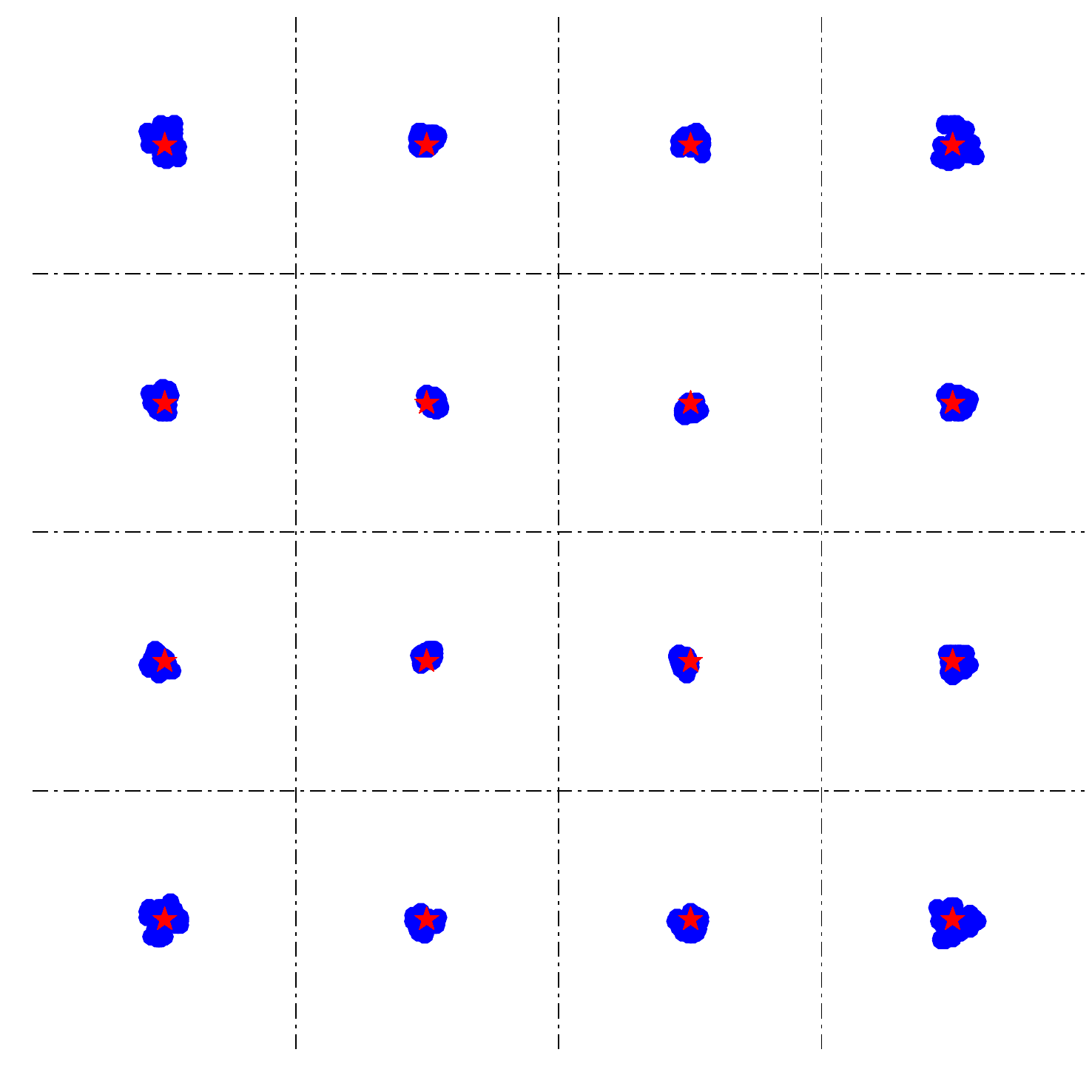}
		\caption*{$\theta=15^\circ$, $N=64$}
	\end{subfigure}
~
\begin{subfigure}[b]{0.3\linewidth}
		\includegraphics[width=\textwidth]{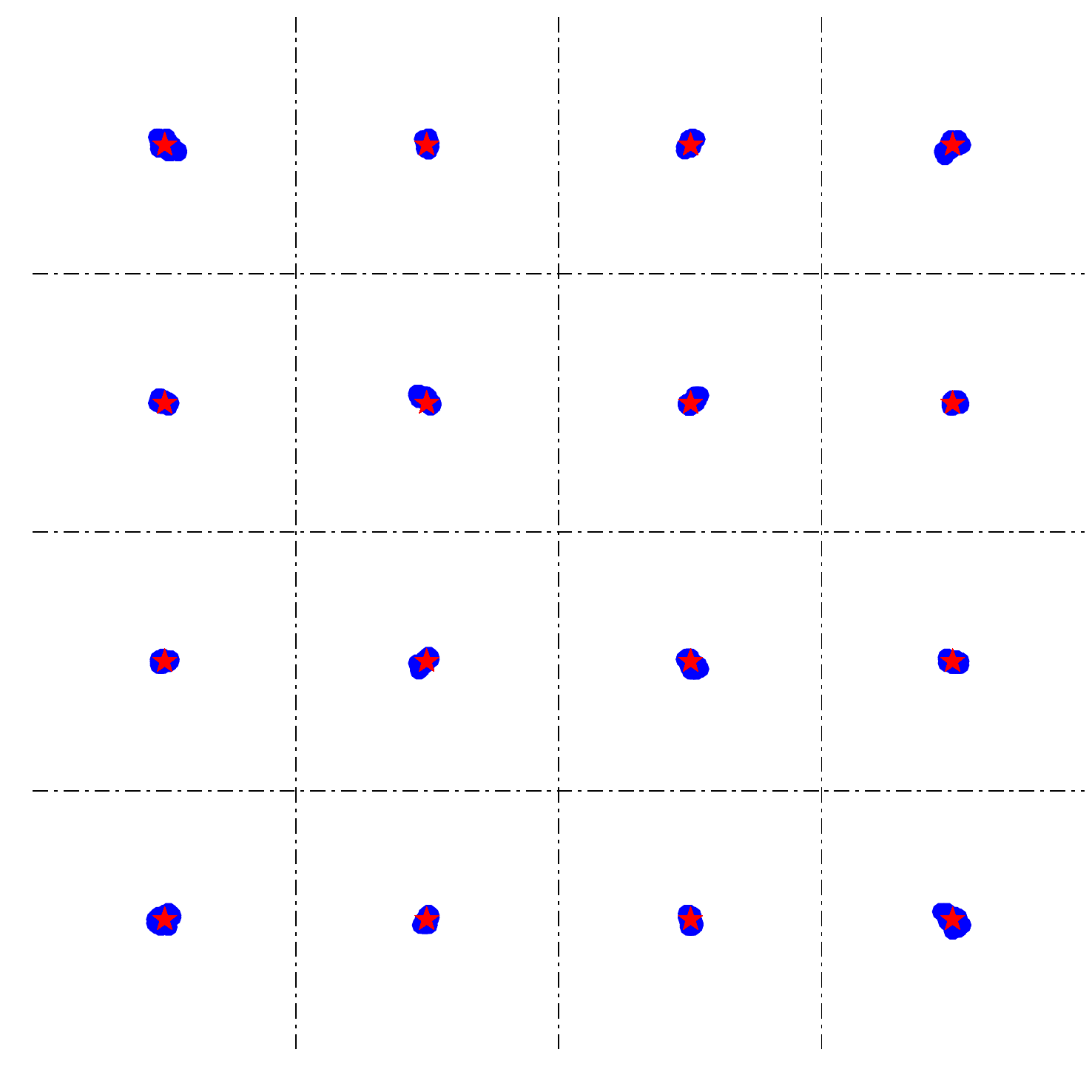}
		\caption*{$\theta=15^\circ$, $N=256$}
	\end{subfigure}
~
\begin{subfigure}[b]{0.3\linewidth}
		\includegraphics[width=\textwidth]{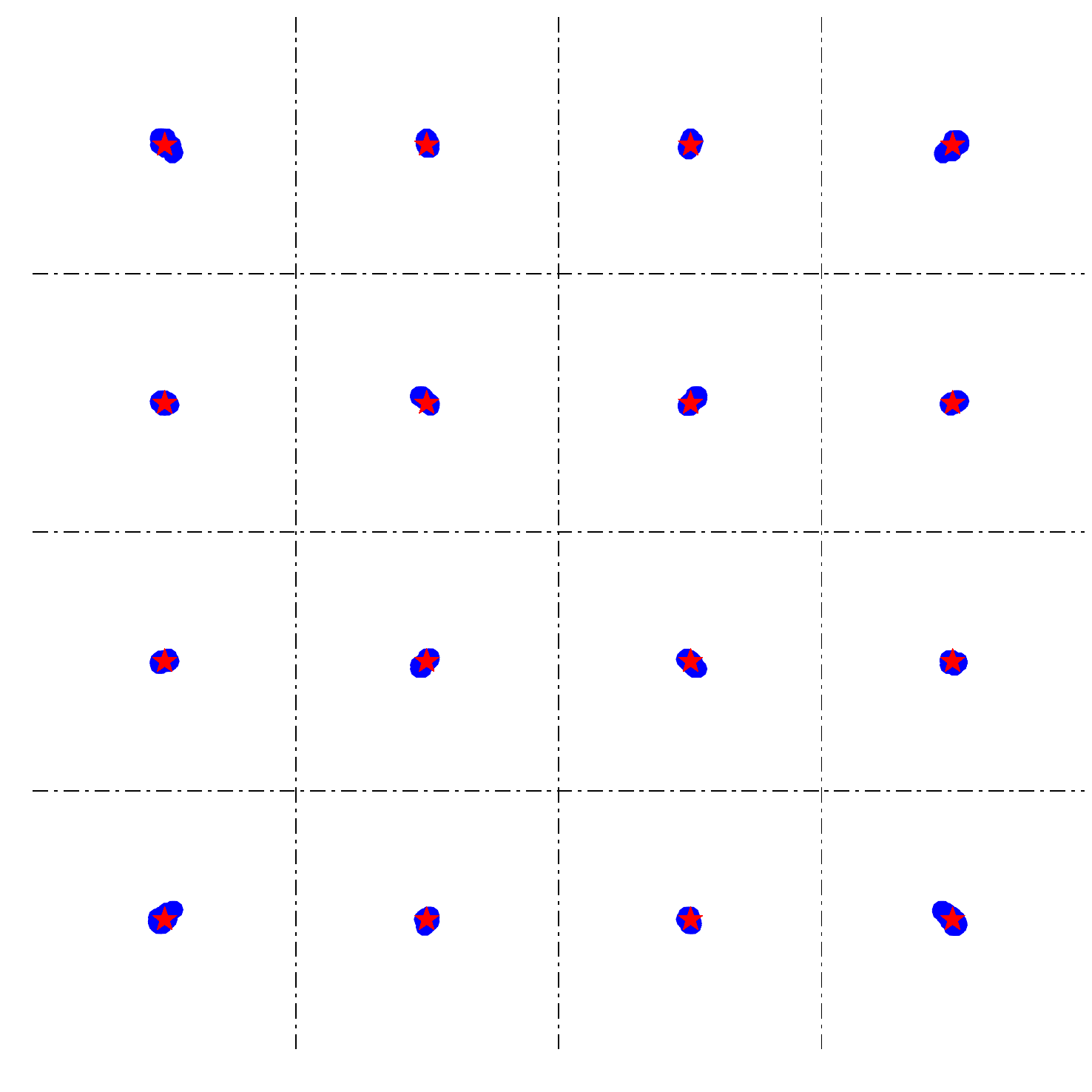}
		\caption*{$\theta=15^\circ$, $N=512$}
	\end{subfigure}
~
\begin{subfigure}[b]{0.3\linewidth}
		\includegraphics[width=\textwidth]{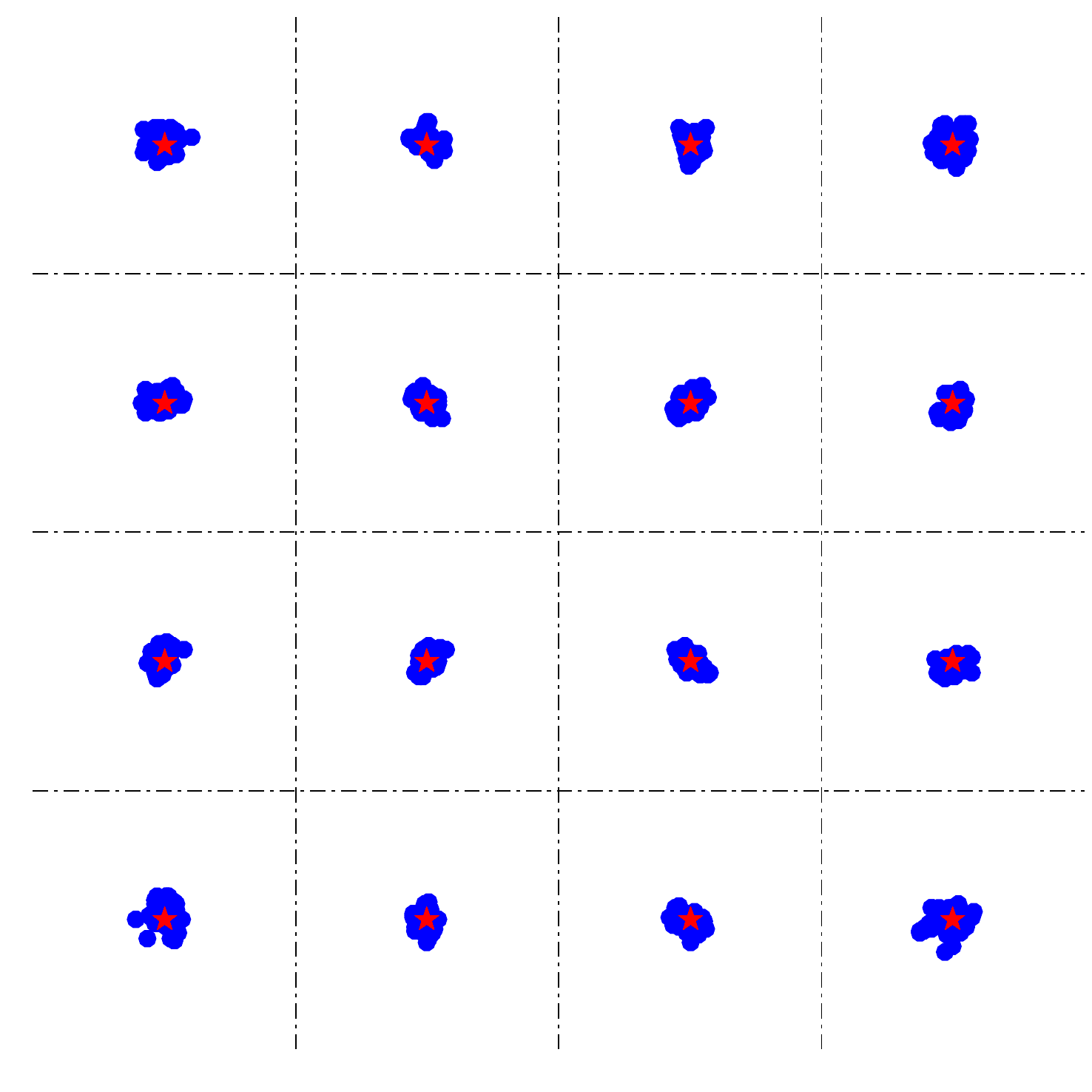}
		\caption*{$\theta=20^\circ$, $N=64$}
	\end{subfigure}
~
\begin{subfigure}[b]{0.3\linewidth}
		\includegraphics[width=\textwidth]{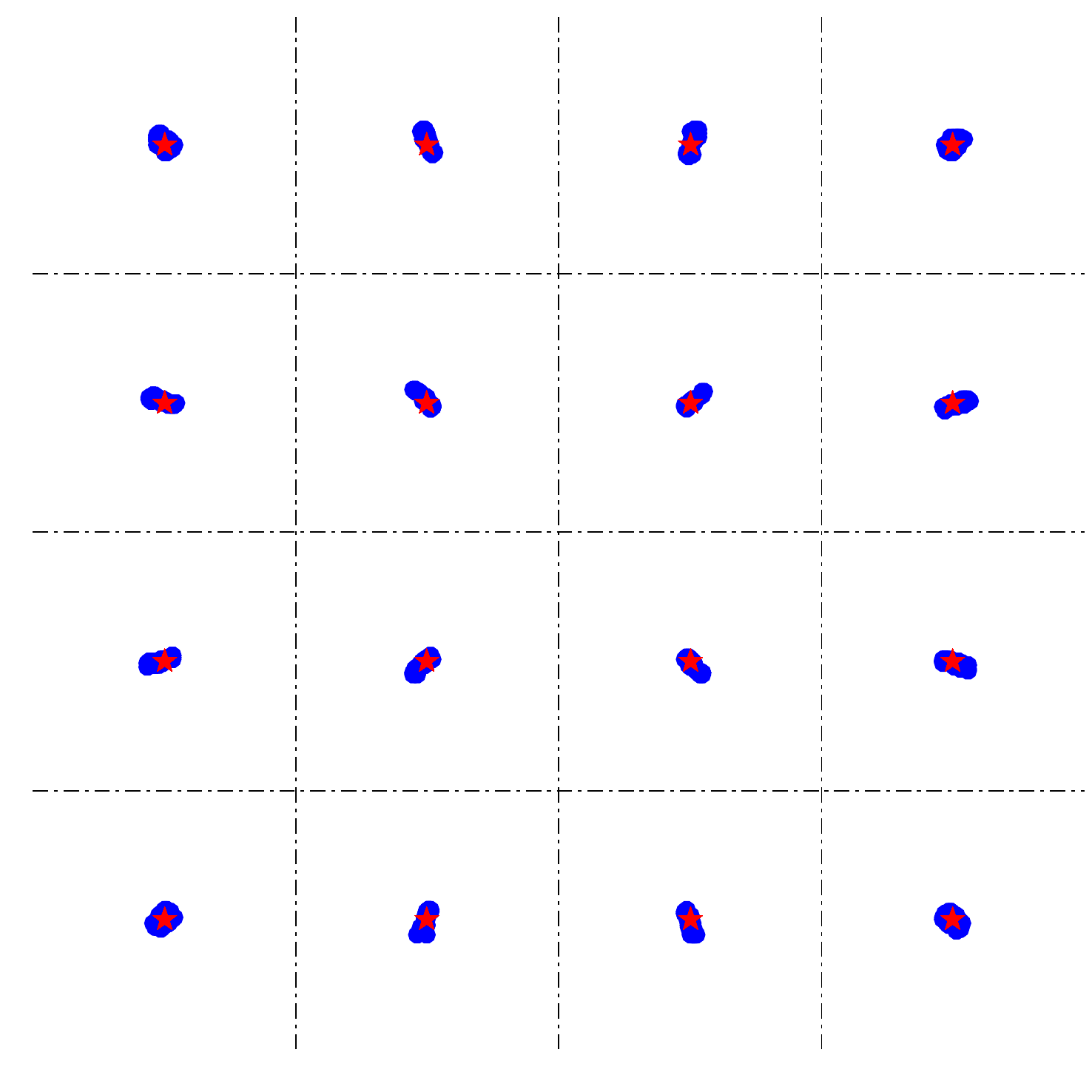}
		\caption*{$\theta=20^\circ$, $N=256$}
	\end{subfigure}
~
\begin{subfigure}[b]{0.3\linewidth}
		\includegraphics[width=\textwidth]{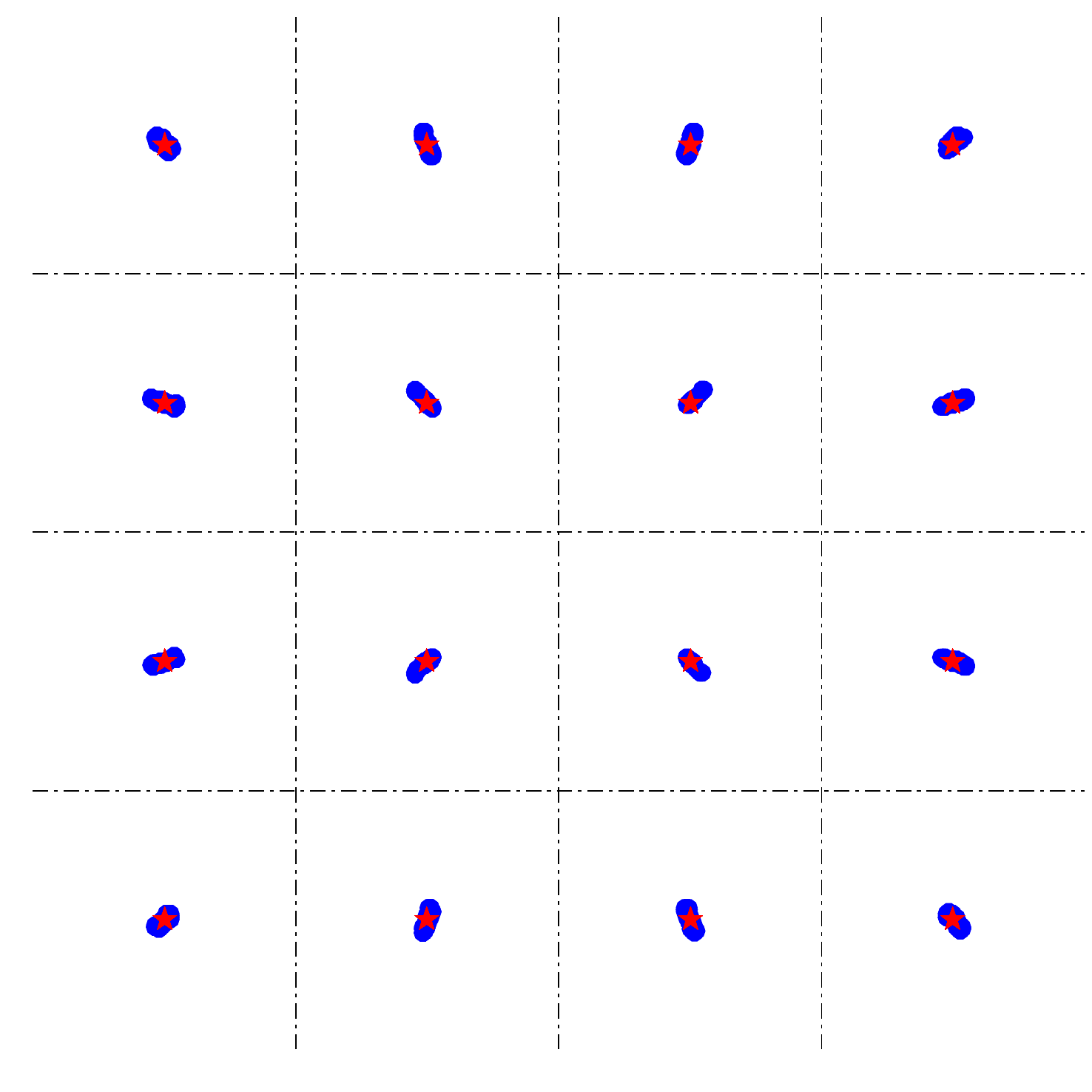}
		\caption*{$\theta=20^\circ$, $N=512$}
	\end{subfigure}
~
\begin{subfigure}[b]{0.3\linewidth}
		\includegraphics[width=\textwidth]{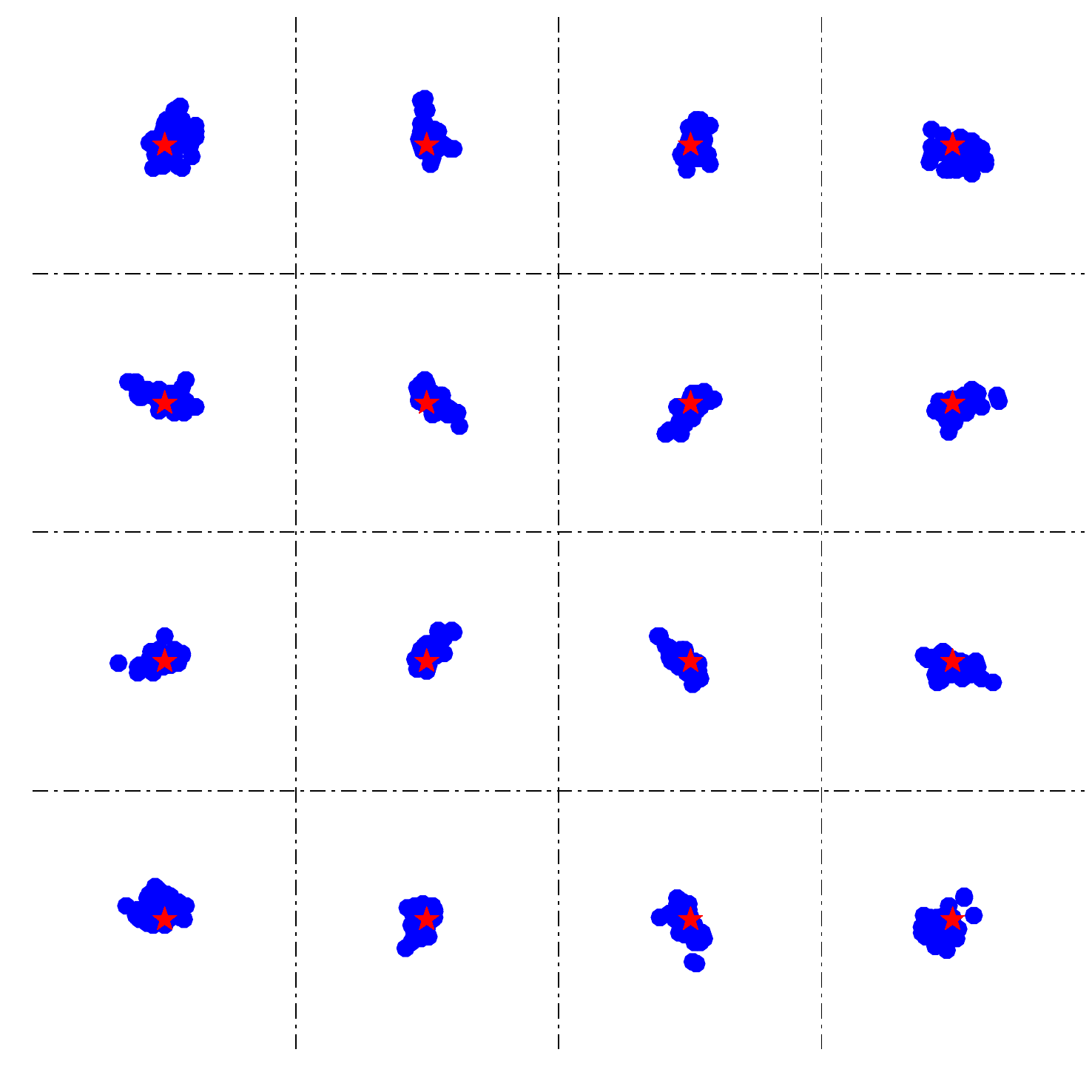}
		\caption*{$\theta=25^\circ$, $N=64$}
	\end{subfigure}
~
\begin{subfigure}[b]{0.3\linewidth}
		\includegraphics[width=\textwidth]{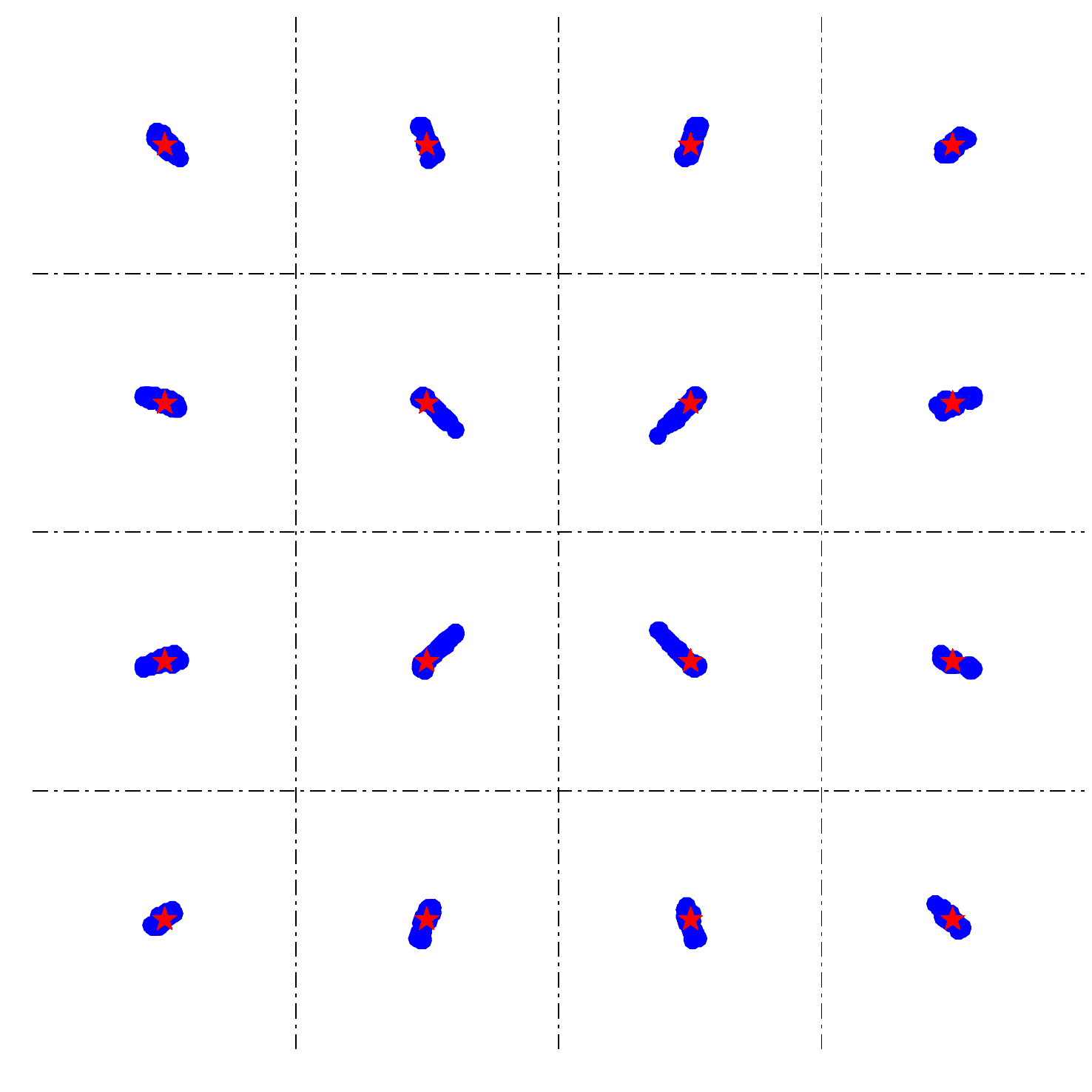}
		\caption*{$\theta=25^\circ$, $N=256$}
	\end{subfigure}
~
\begin{subfigure}[b]{0.3\linewidth}
		\includegraphics[width=\textwidth]{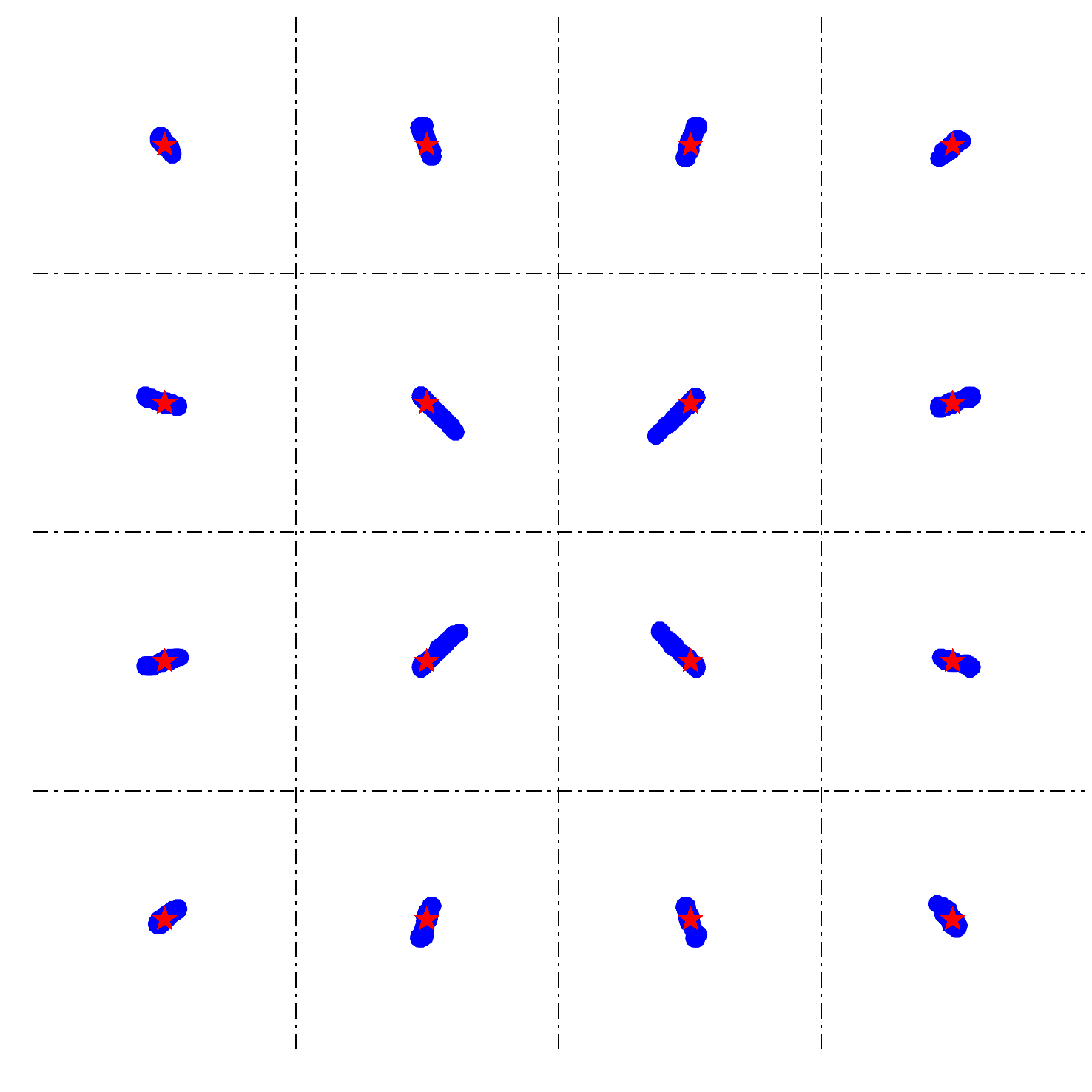}
		\caption*{$\theta=25^\circ$, $N=512$}
	\end{subfigure}
	\caption{IQ scatter plots of the basic sigma-delta MRT scheme for different $\theta$ and $N$; $16$-ary QAM. (cont.)}
\end{figure}

\begin{figure}[H]\ContinuedFloat
	\centering	
\begin{subfigure}[b]{0.3\linewidth}
		\includegraphics[width=\textwidth]{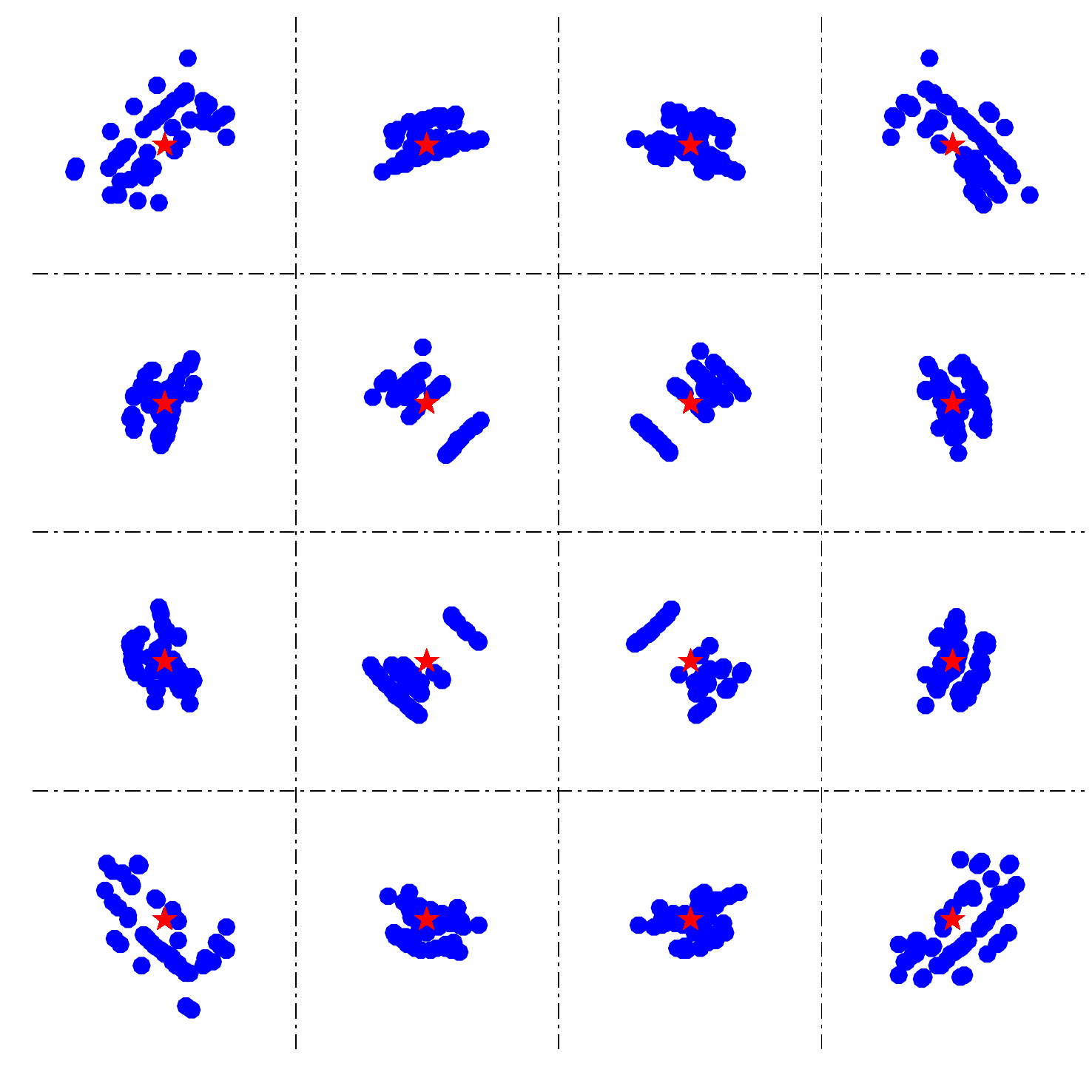}
		\caption*{$\theta=30^\circ$, $N=64$}
	\end{subfigure}
~
\begin{subfigure}[b]{0.3\linewidth}
		\includegraphics[width=\textwidth]{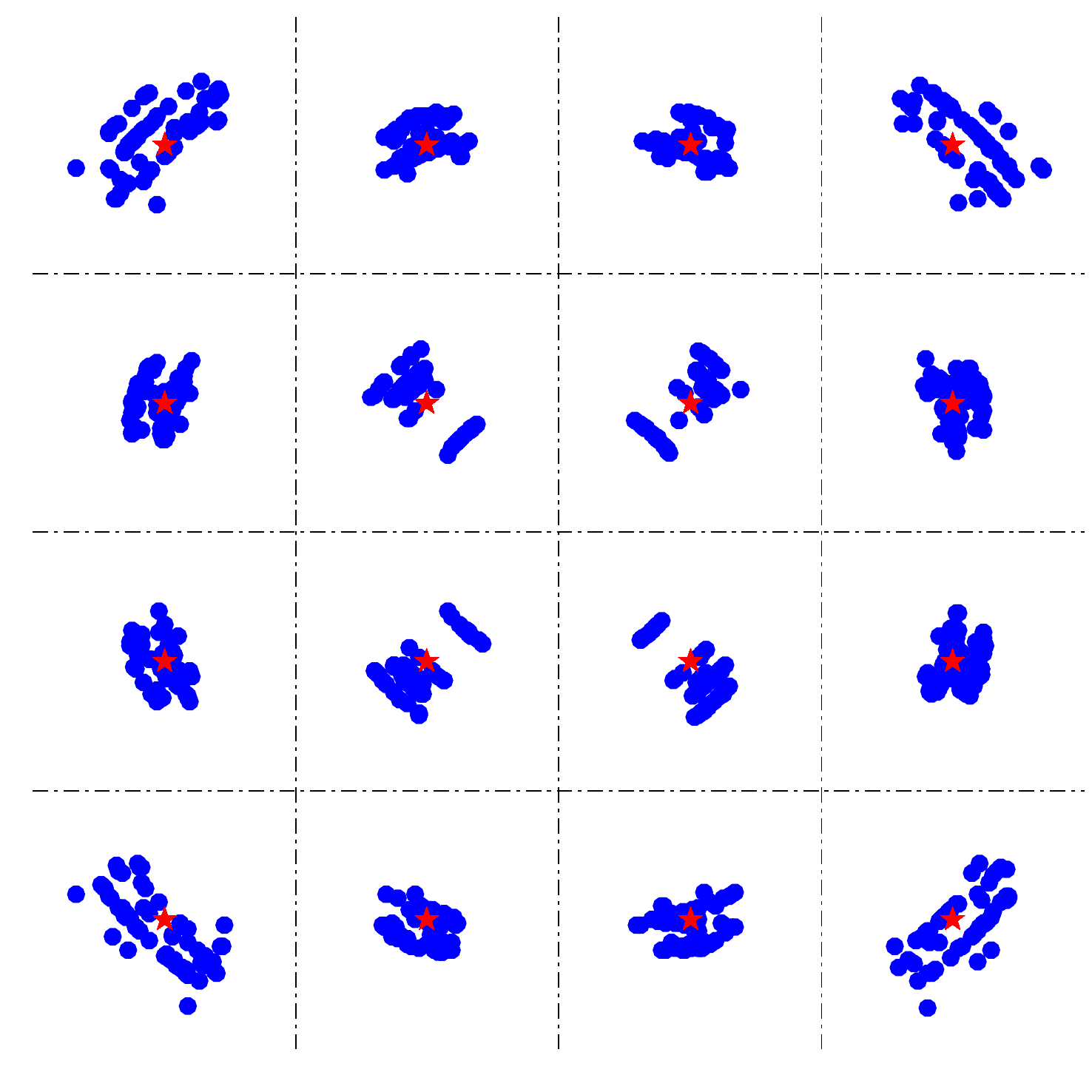}
		\caption*{$\theta=30^\circ$, $N=256$}
	\end{subfigure}
~
\begin{subfigure}[b]{0.3\linewidth}
		\includegraphics[width=\textwidth]{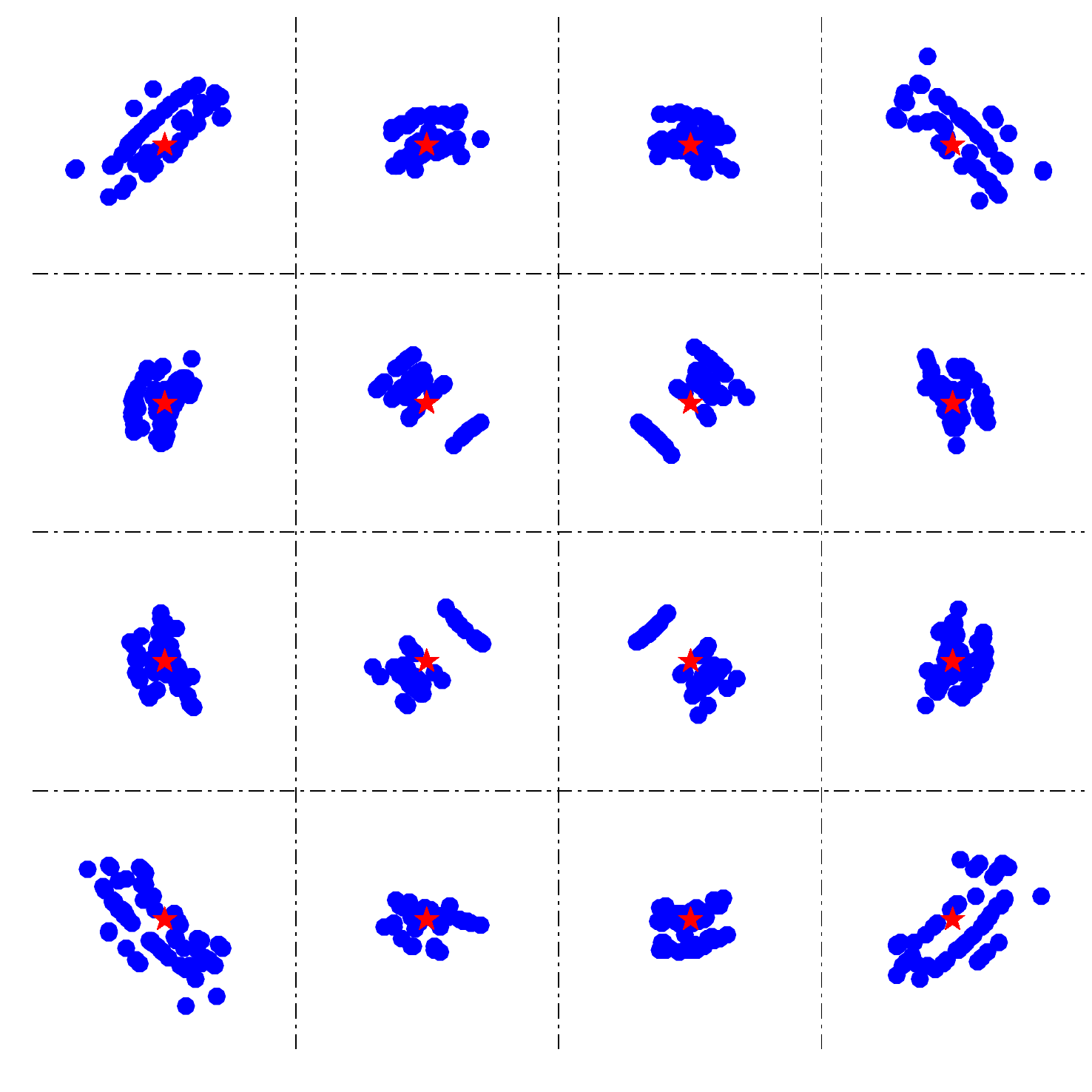}
		\caption*{$\theta=30^\circ$, $N=512$}
	\end{subfigure}
~
\begin{subfigure}[b]{0.3\linewidth}
		\includegraphics[width=\textwidth]{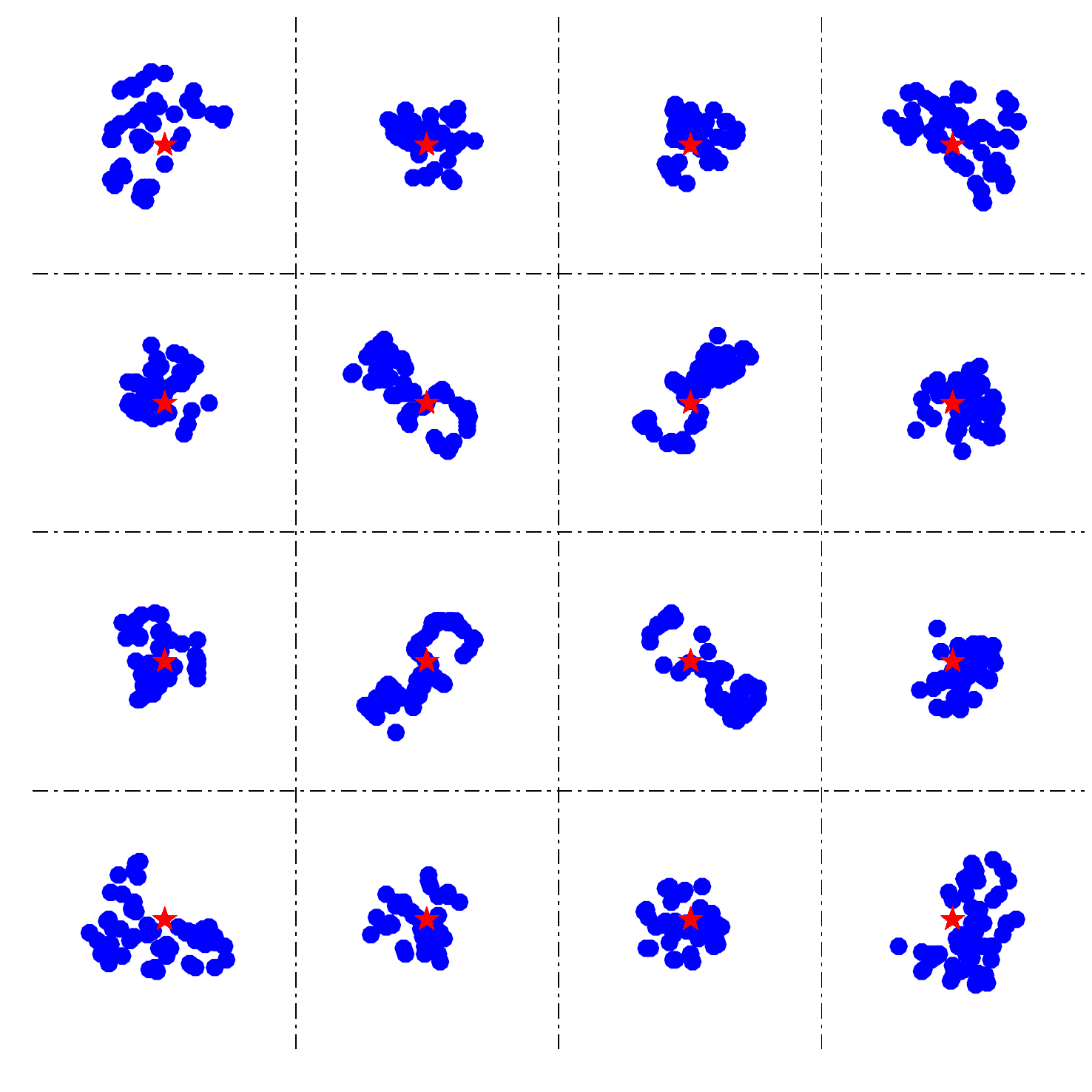}
		\caption*{$\theta=35^\circ$, $N=64$}
	\end{subfigure}
~
\begin{subfigure}[b]{0.3\linewidth}
		\includegraphics[width=\textwidth]{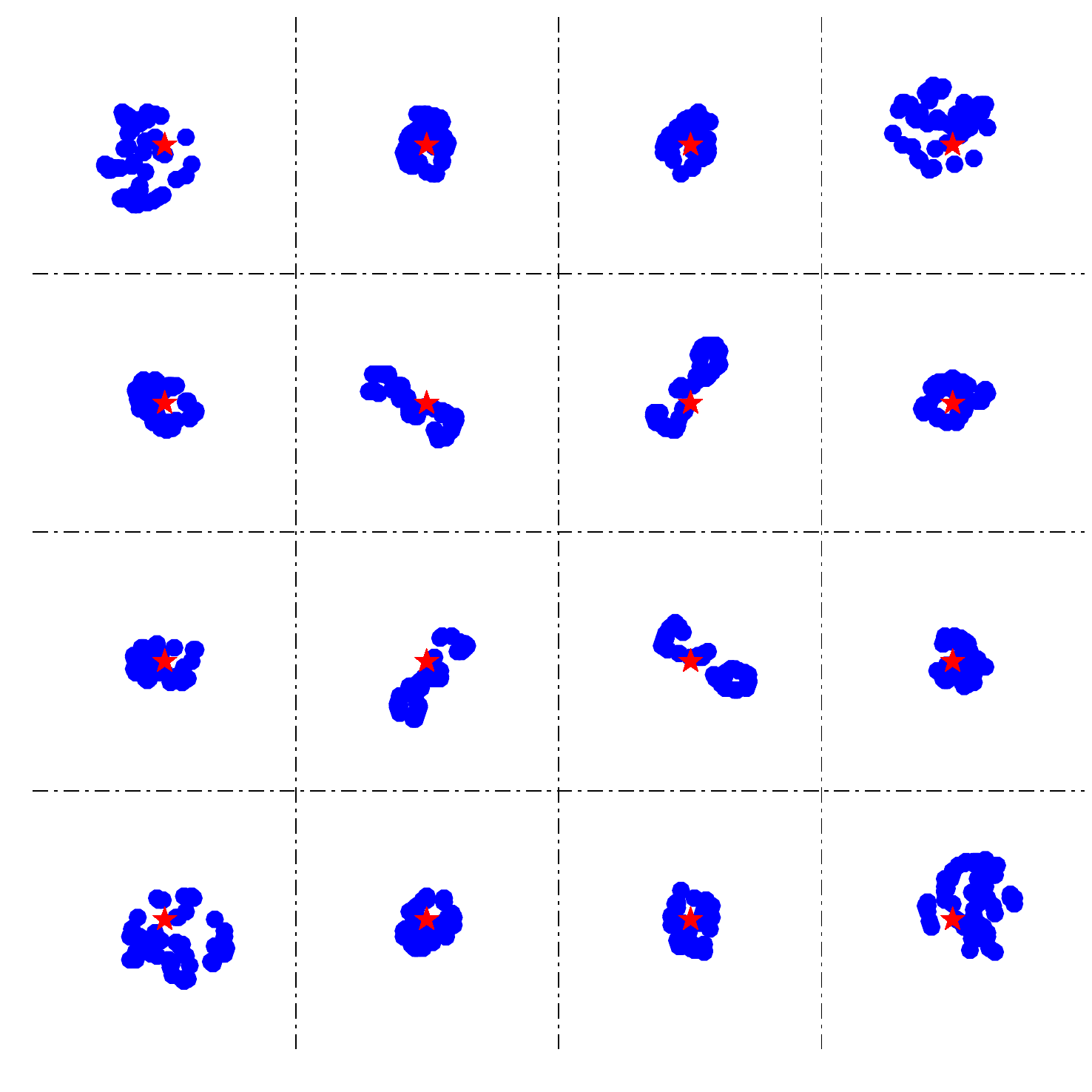}
		\caption*{$\theta=35^\circ$, $N=256$}
	\end{subfigure}
~
\begin{subfigure}[b]{0.3\linewidth}
		\includegraphics[width=\textwidth]{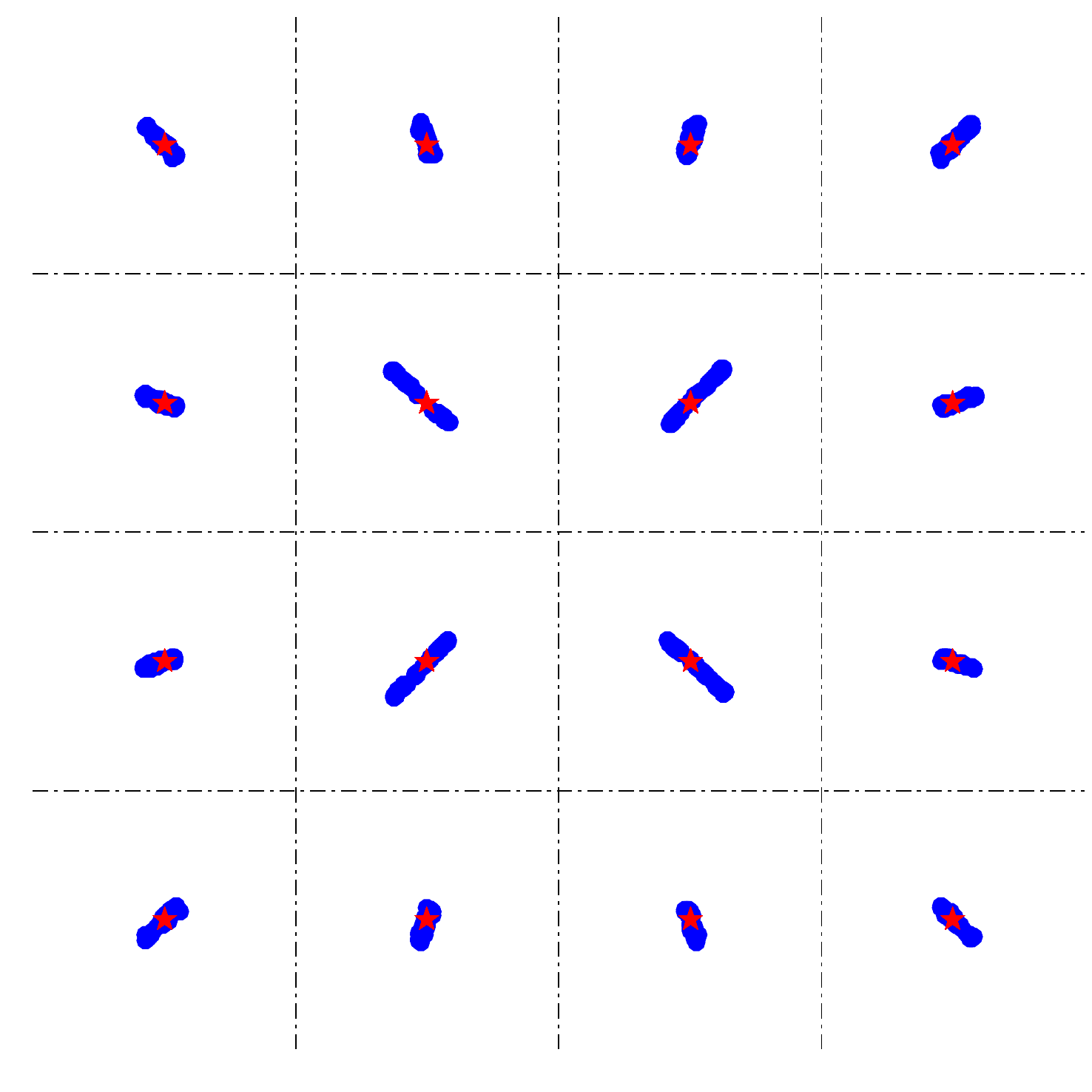}
		\caption*{$\theta=35^\circ$, $N=512$}
	\end{subfigure}
~
\begin{subfigure}[b]{0.3\linewidth}
		\includegraphics[width=\textwidth]{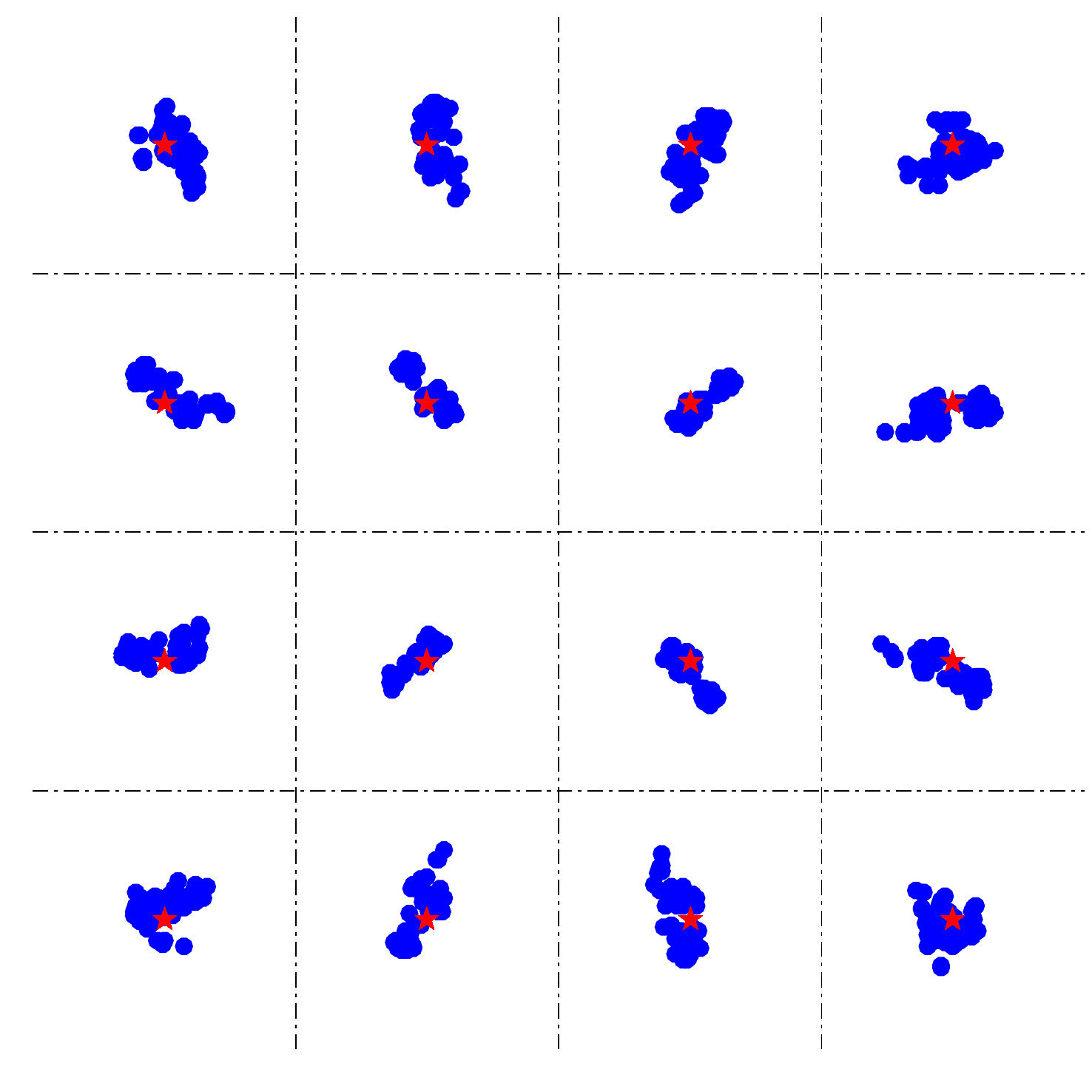}
		\caption*{$\theta=40^\circ$, $N=64$}
	\end{subfigure}
~
\begin{subfigure}[b]{0.3\linewidth}
		\includegraphics[width=\textwidth]{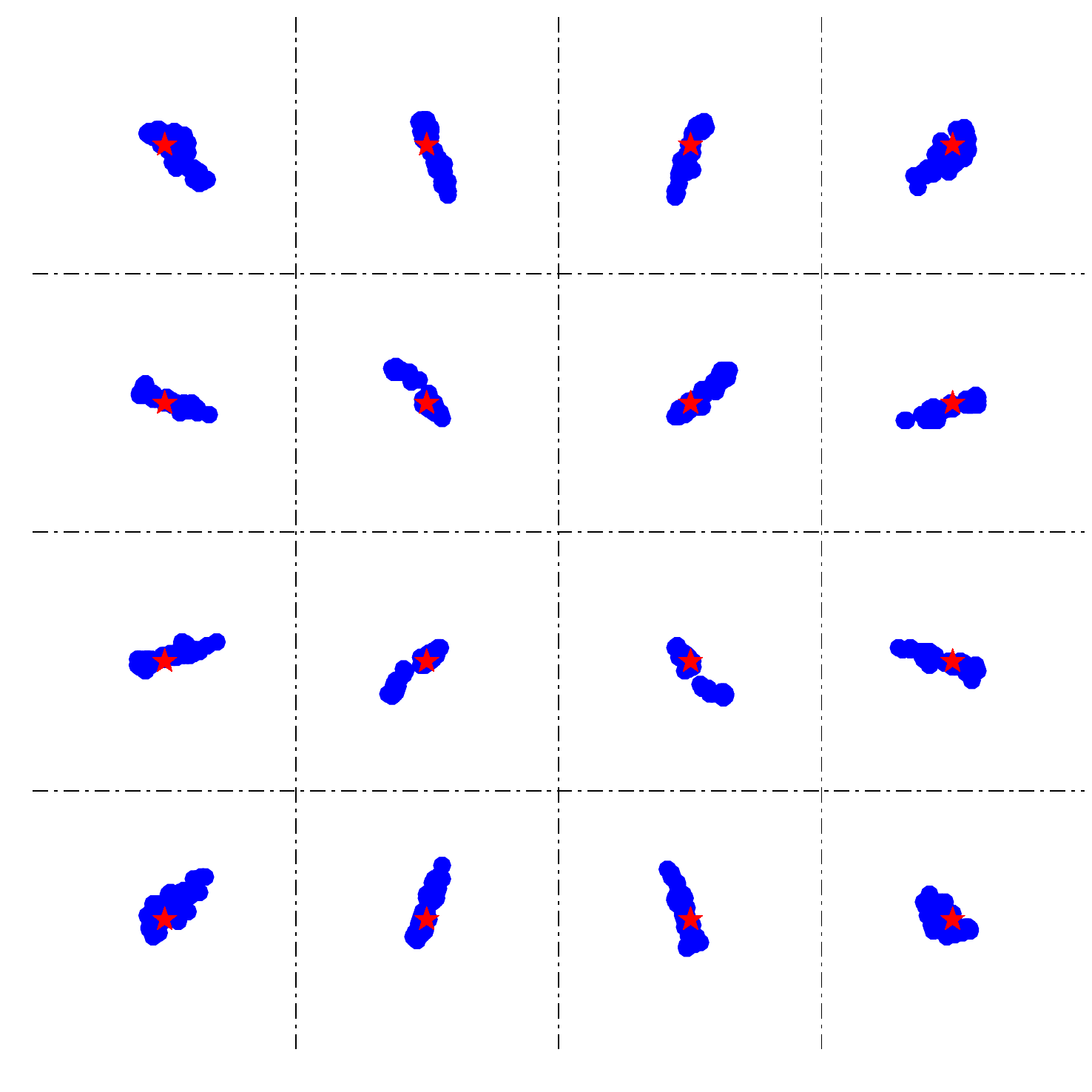}
		\caption*{$\theta=40^\circ$, $N=256$}
	\end{subfigure}
~
\begin{subfigure}[b]{0.3\linewidth}
		\includegraphics[width=\textwidth]{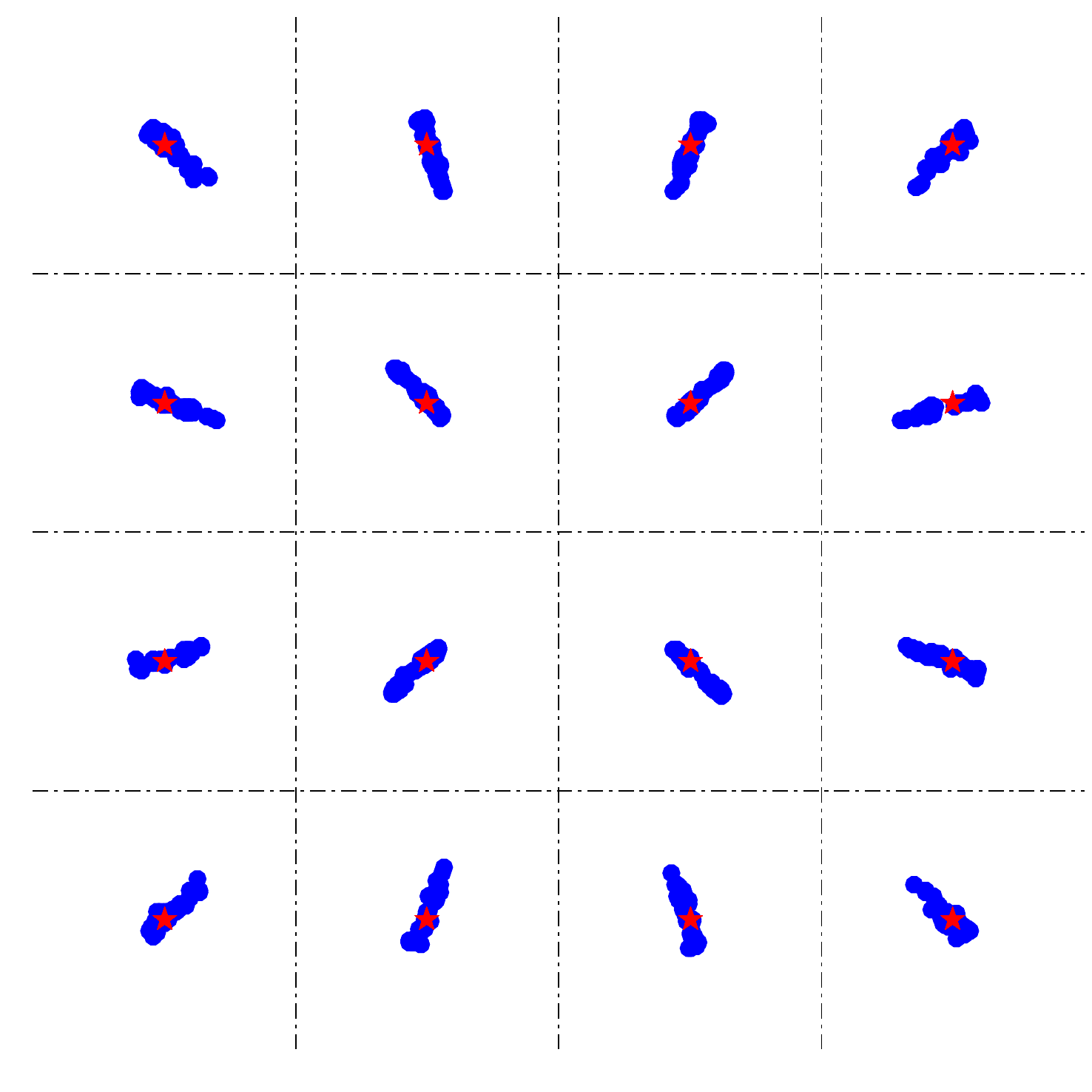}
		\caption*{$\theta=40^\circ$, $N=512$}
	\end{subfigure}
	\caption{IQ scatter plots of the basic sigma-delta MRT scheme for different $\theta$ and $N$; $16$-ary QAM. (cont.) }
\end{figure}

\begin{figure}[H]\ContinuedFloat
	\centering	
\begin{subfigure}[b]{0.3\linewidth}
		\includegraphics[width=\textwidth]{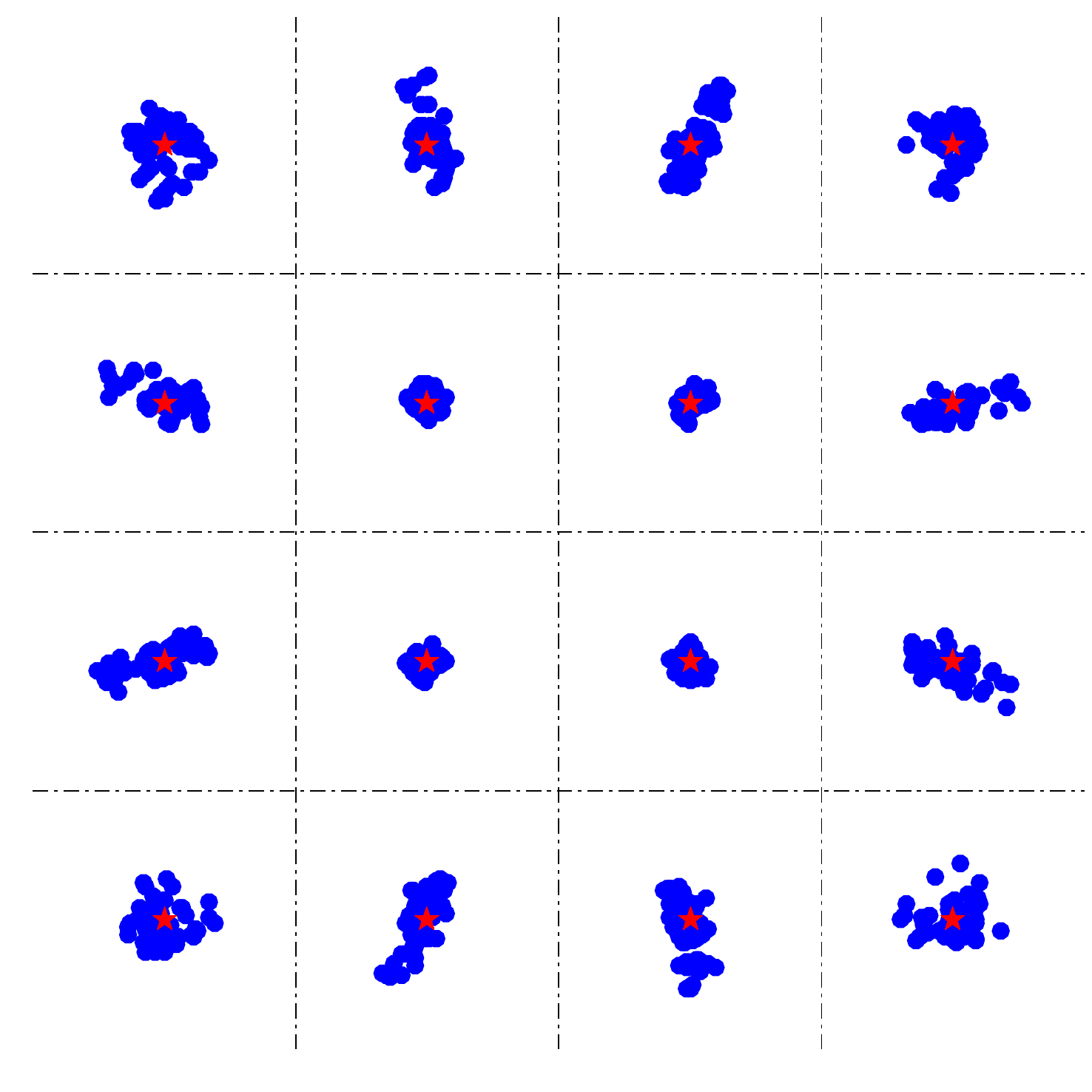}
		\caption*{$\theta=45^\circ$, $N=64$}
	\end{subfigure}
~
\begin{subfigure}[b]{0.3\linewidth}
		\includegraphics[width=\textwidth]{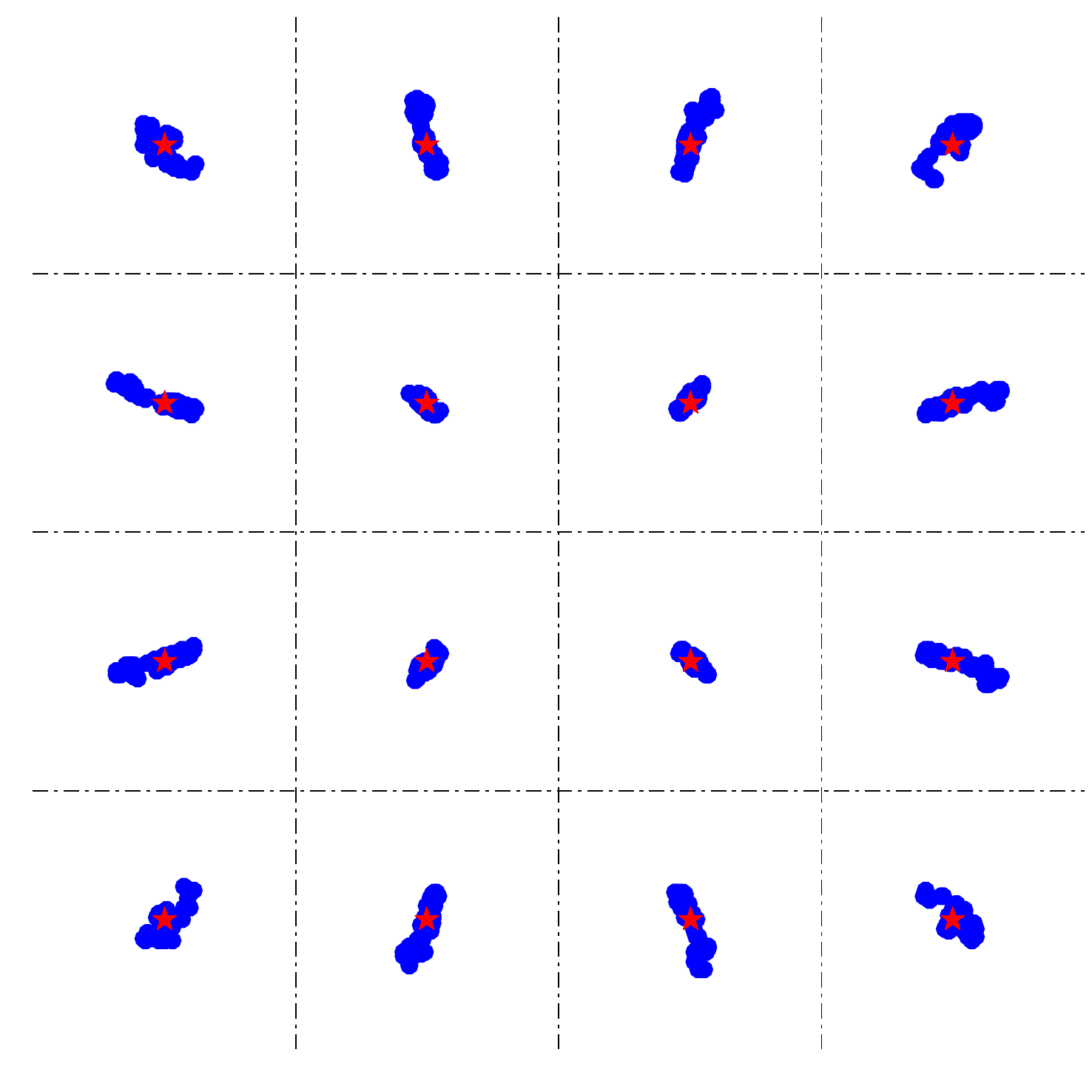}
		\caption*{$\theta=45^\circ$, $N=256$}
	\end{subfigure}
~
\begin{subfigure}[b]{0.3\linewidth}
		\includegraphics[width=\textwidth]{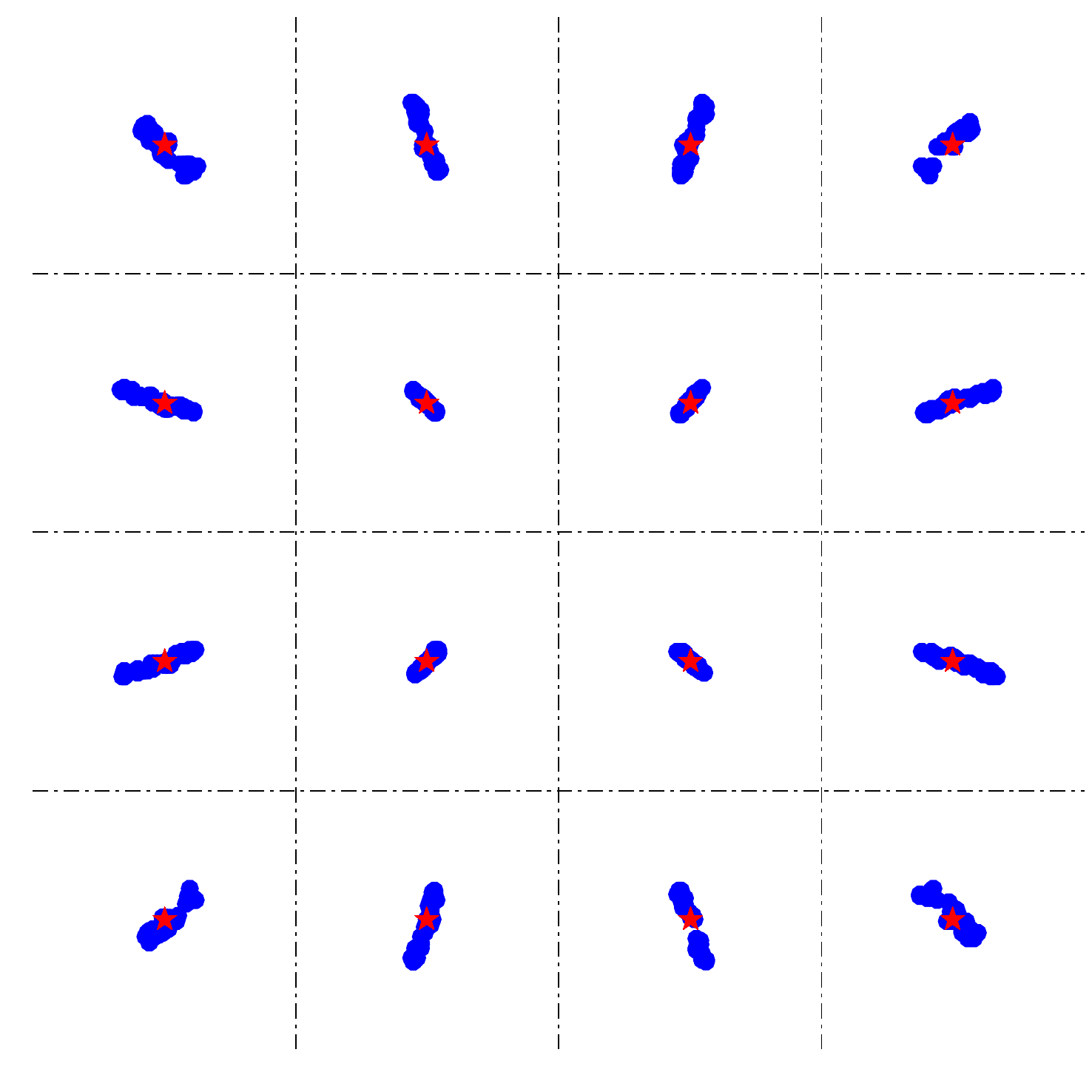}
		\caption*{$\theta=45^\circ$, $N=512$}
	\end{subfigure}
~
\begin{subfigure}[b]{0.3\linewidth}
		\includegraphics[width=\textwidth]{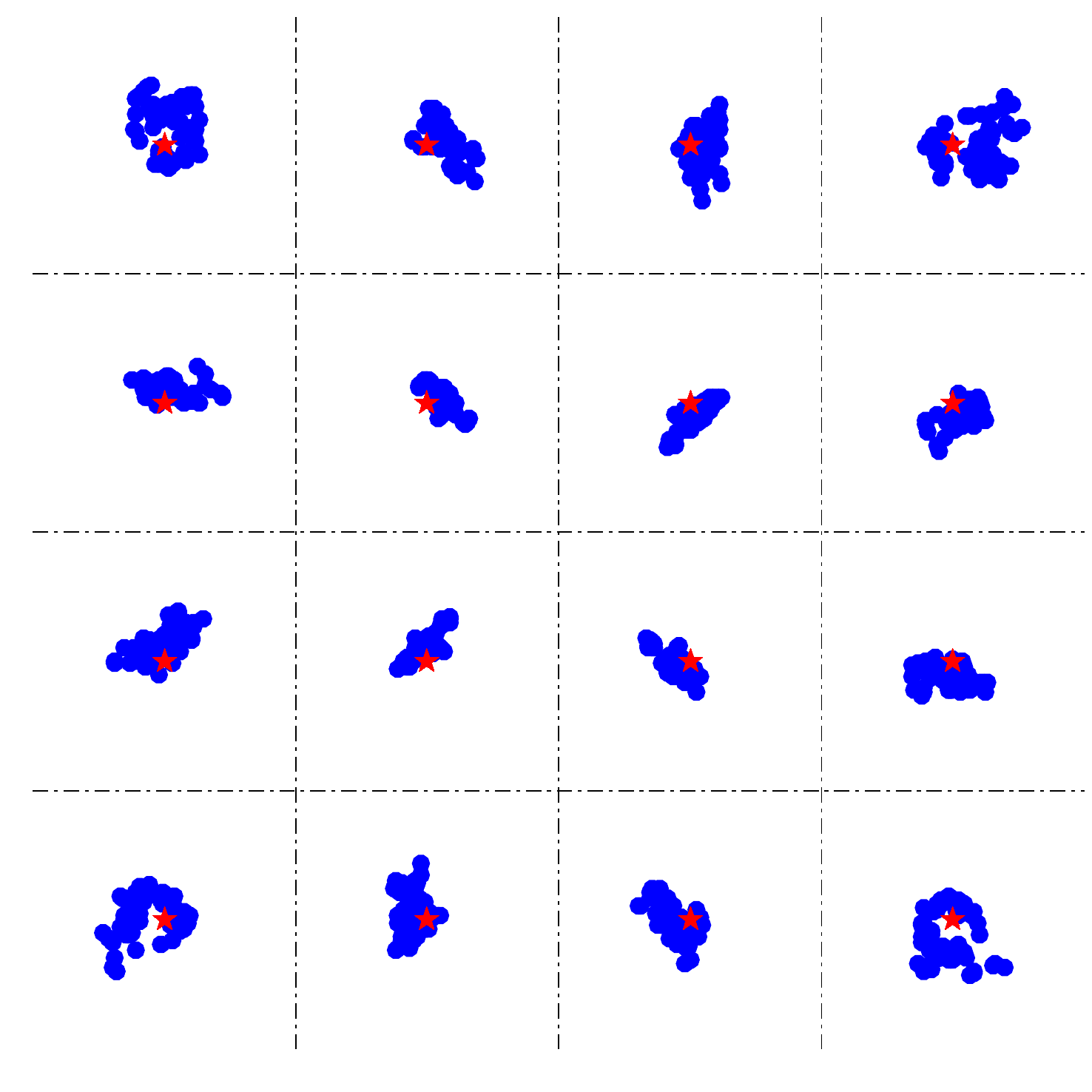}
		\caption*{$\theta=50^\circ$, $N=64$}
	\end{subfigure}
~
\begin{subfigure}[b]{0.3\linewidth}
		\includegraphics[width=\textwidth]{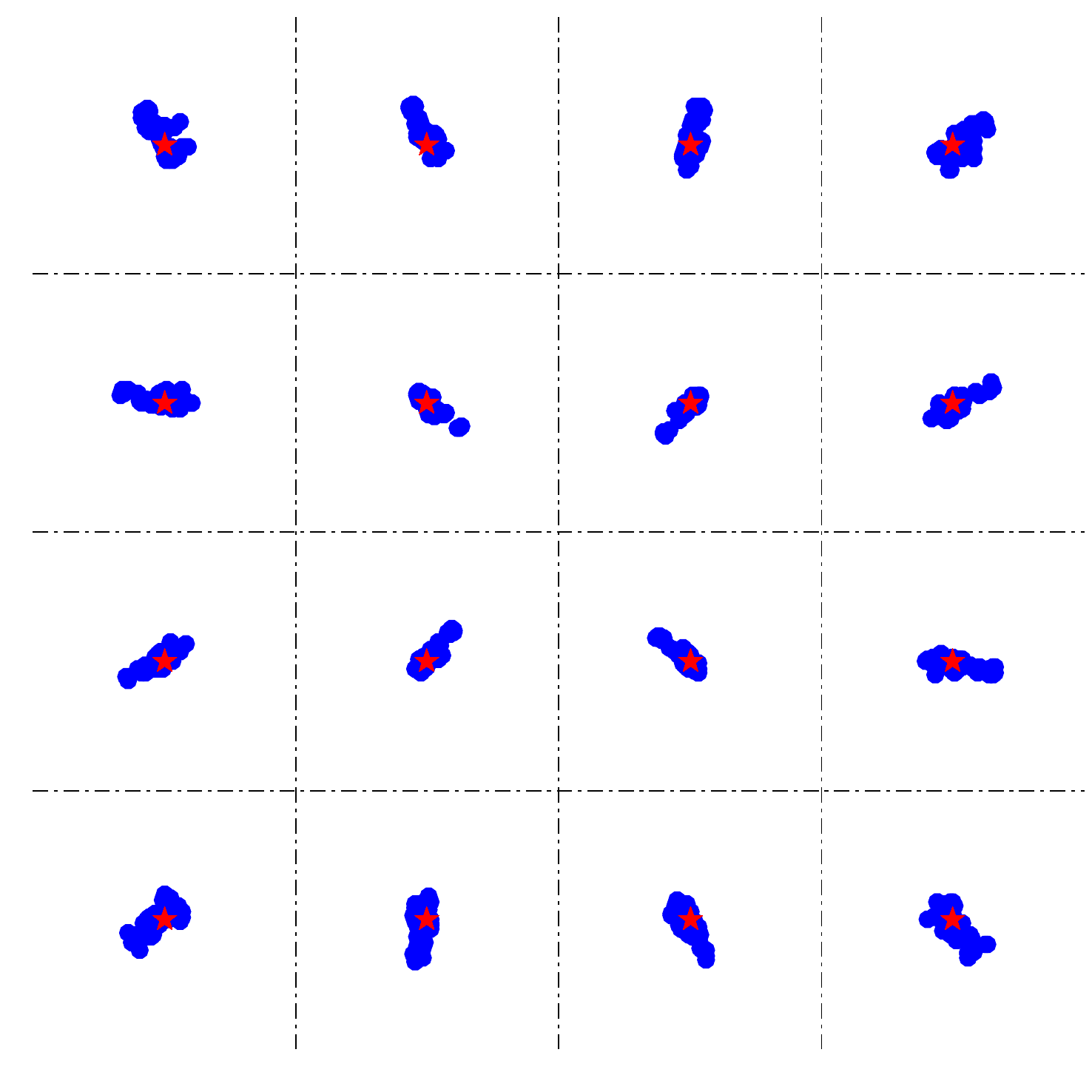}
		\caption*{$\theta=50^\circ$, $N=256$}
	\end{subfigure}
~
\begin{subfigure}[b]{0.3\linewidth}
		\includegraphics[width=\textwidth]{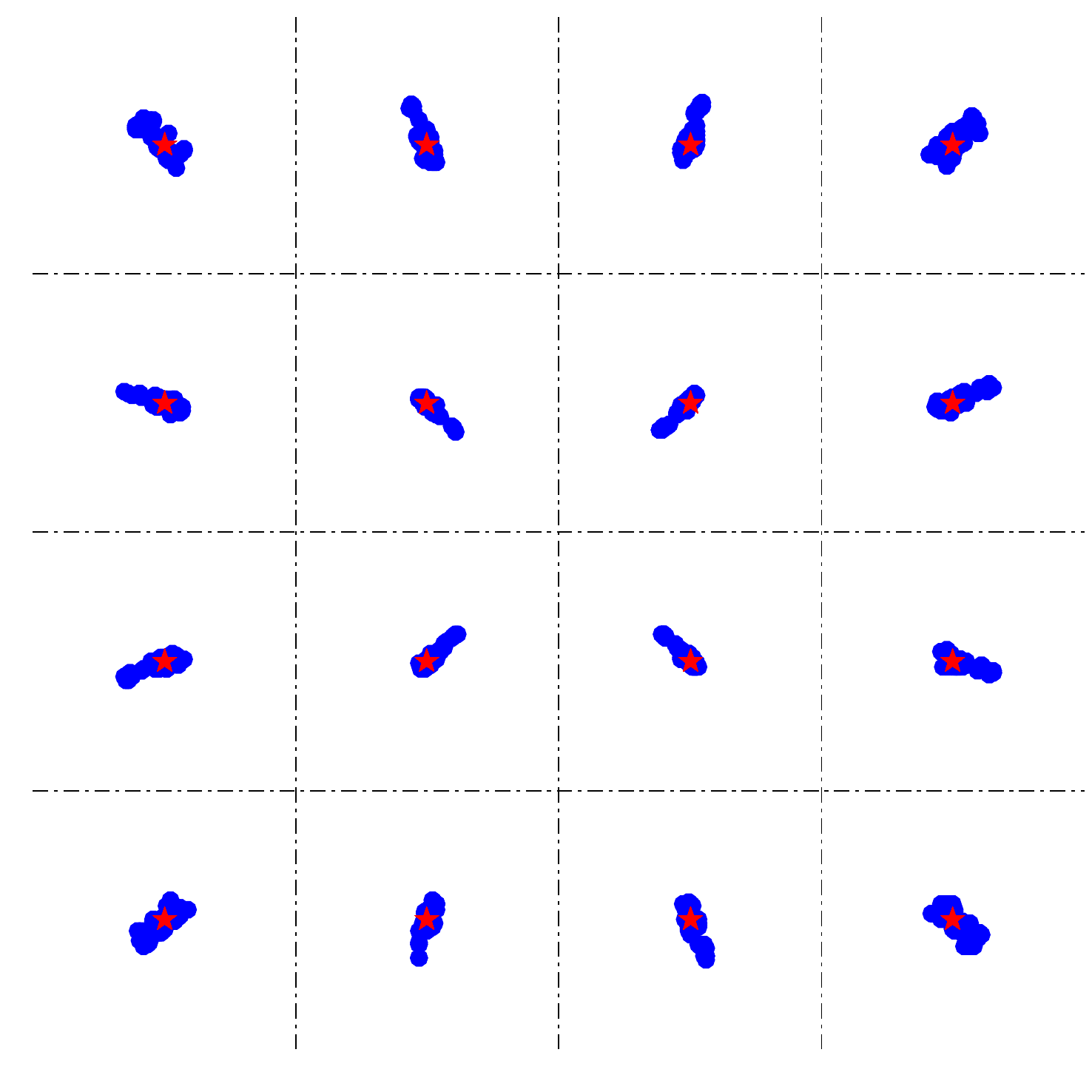}
		\caption*{$\theta=50^\circ$, $N=512$}
	\end{subfigure}
~
\begin{subfigure}[b]{0.3\linewidth}
		\includegraphics[width=\textwidth]{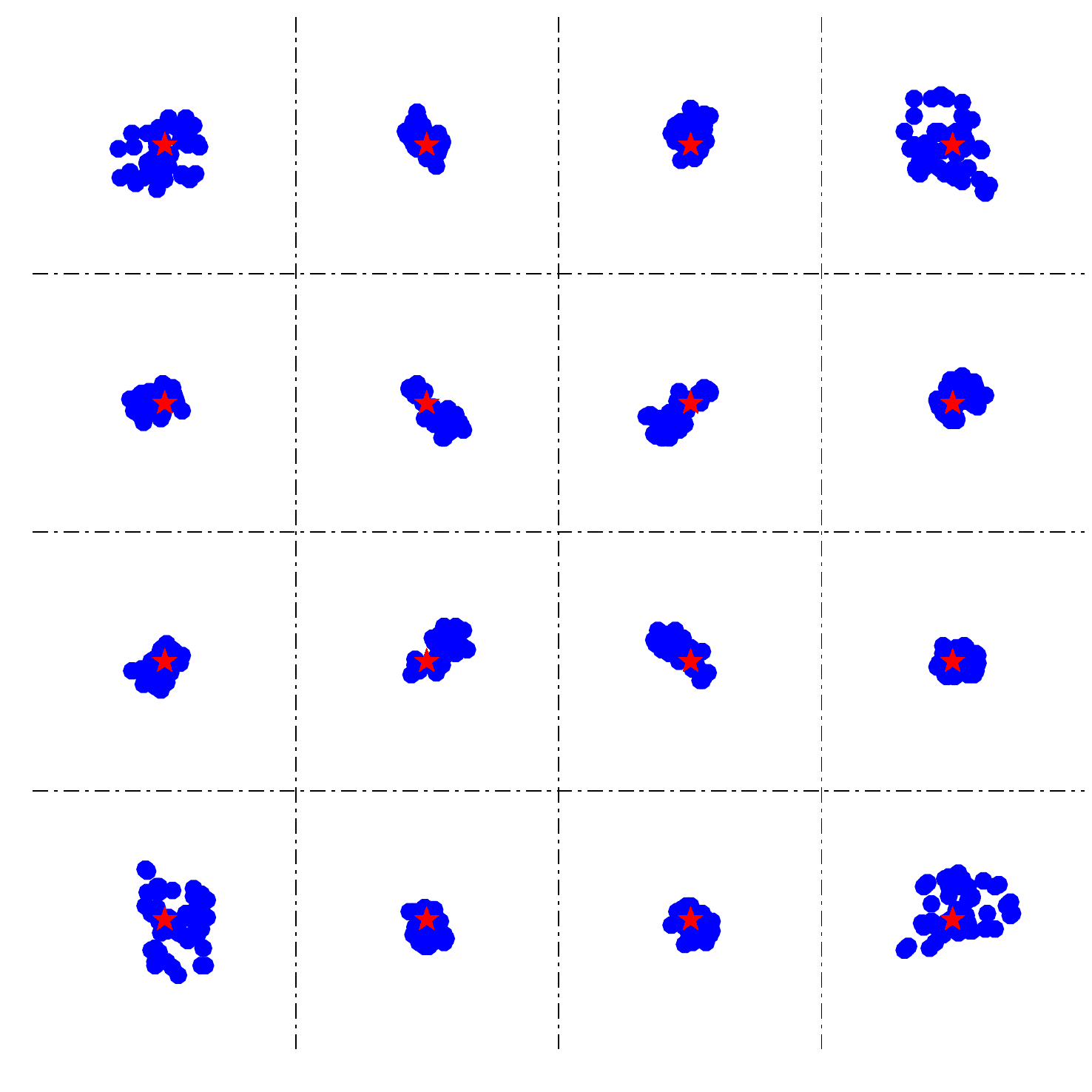}
		\caption*{$\theta=55^\circ$, $N=64$}
	\end{subfigure}
~
\begin{subfigure}[b]{0.3\linewidth}
		\includegraphics[width=\textwidth]{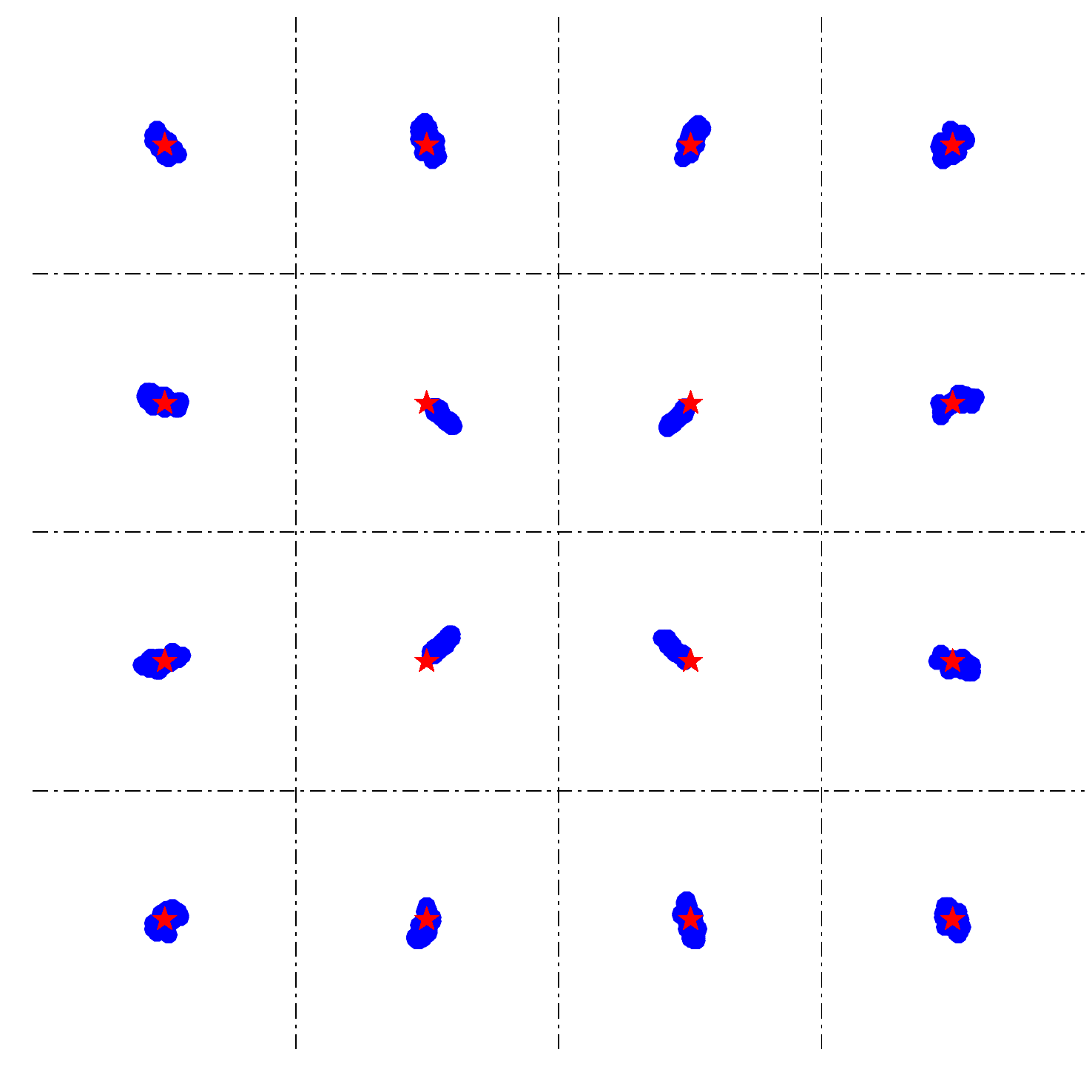}
		\caption*{$\theta=55^\circ$, $N=256$}
	\end{subfigure}
~
\begin{subfigure}[b]{0.3\linewidth}
		\includegraphics[width=\textwidth]{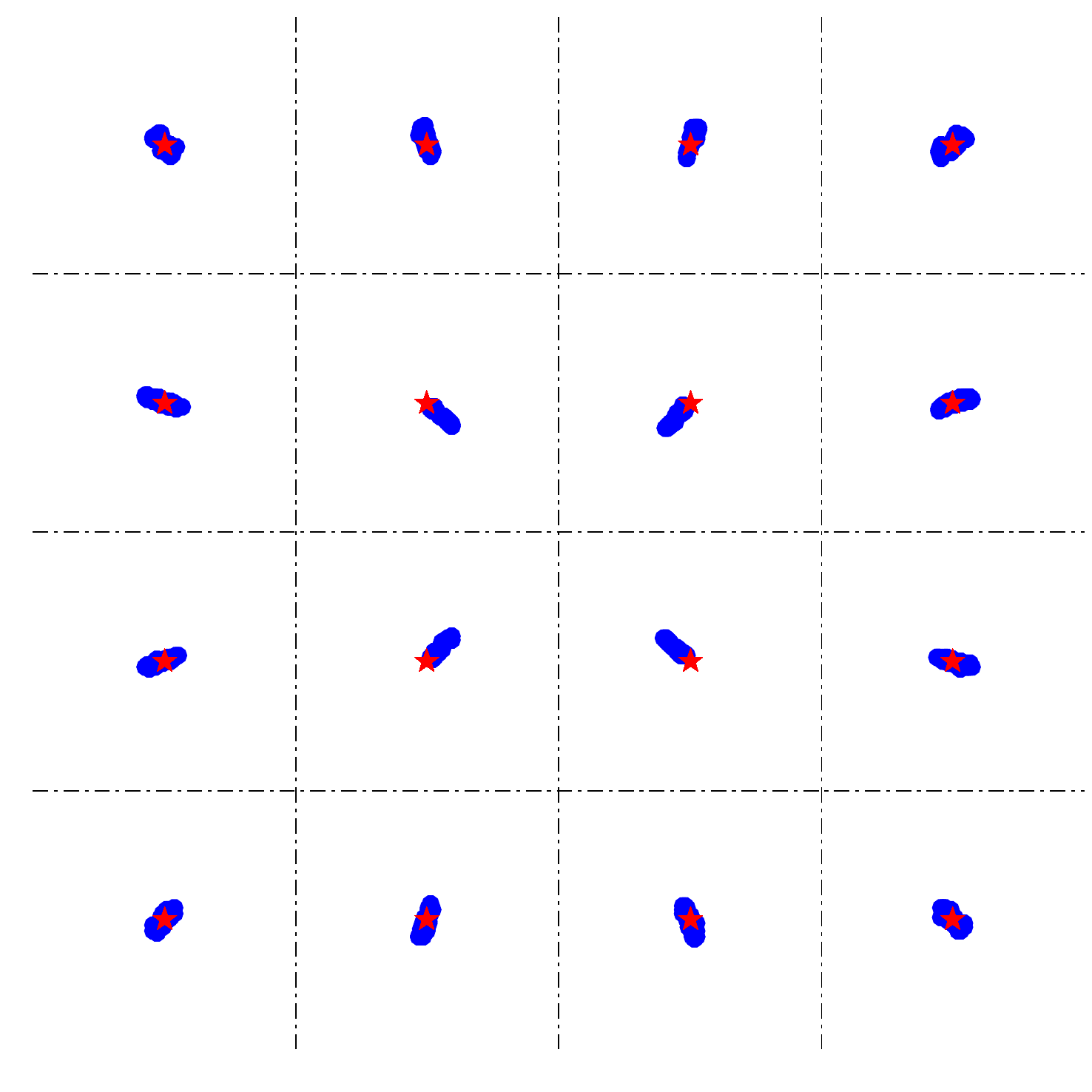}
		\caption*{$\theta=55^\circ$, $N=512$}
	\end{subfigure}
	\caption{IQ scatter plots of the basic sigma-delta MRT scheme for different $\theta$ and $N$; $16$-ary QAM. (cont.)}
\end{figure}

\begin{figure}[H]\ContinuedFloat
	\centering	
\begin{subfigure}[b]{0.3\linewidth}
		\includegraphics[width=\textwidth]{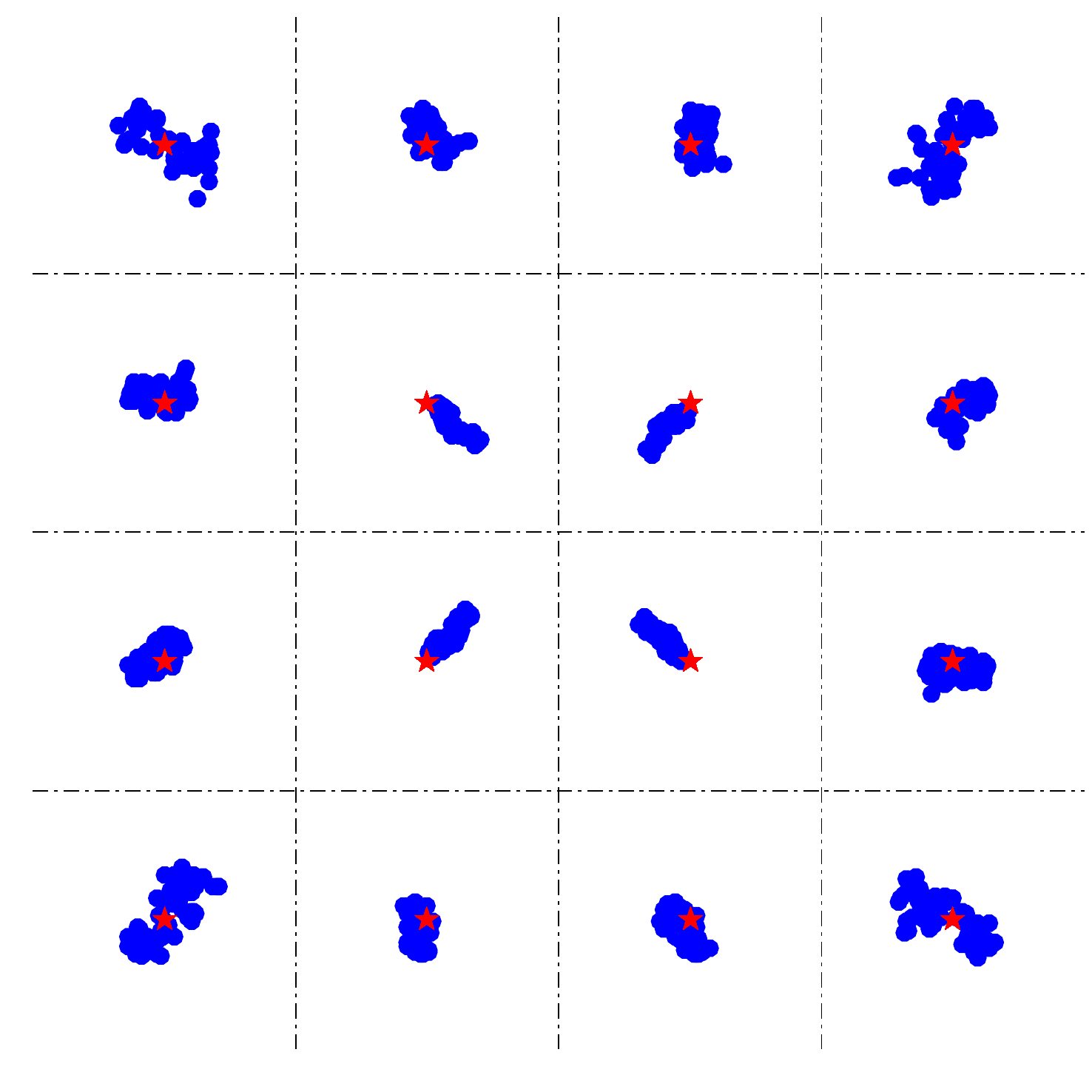}
		\caption*{$\theta=60^\circ$, $N=64$}
	\end{subfigure}
~
\begin{subfigure}[b]{0.3\linewidth}
		\includegraphics[width=\textwidth]{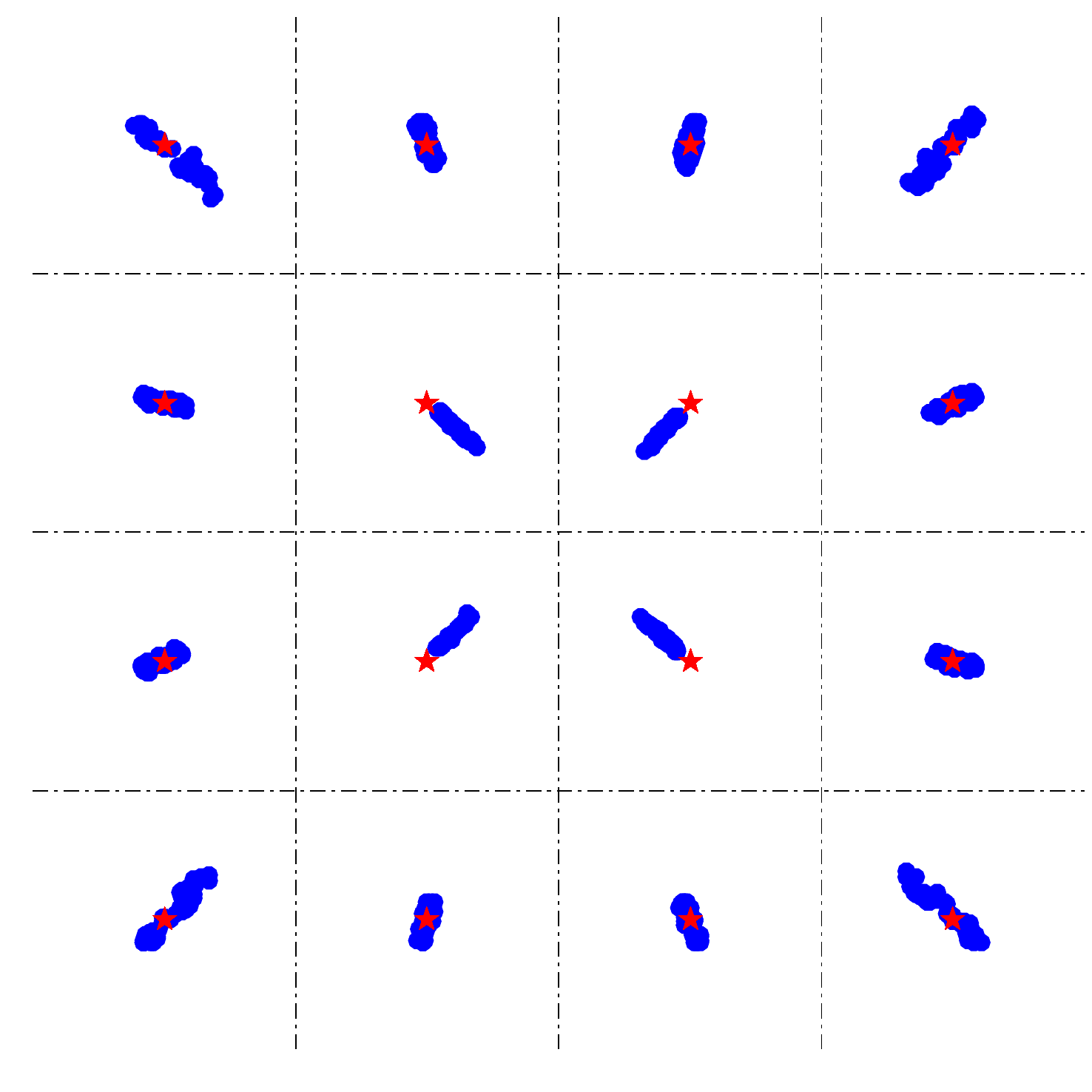}
		\caption*{$\theta=60^\circ$, $N=256$}
	\end{subfigure}
~
\begin{subfigure}[b]{0.3\linewidth}
		\includegraphics[width=\textwidth]{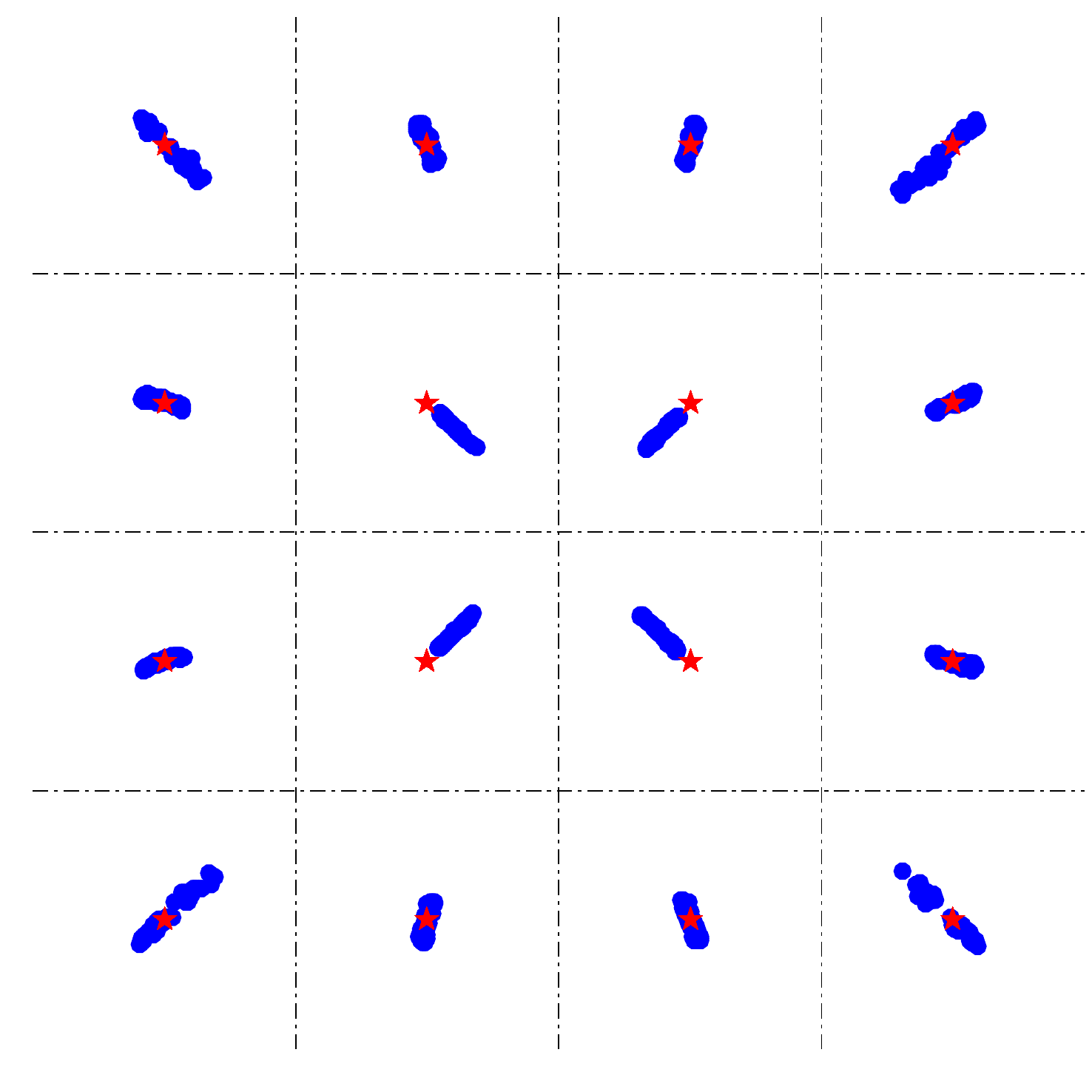}
		\caption*{$\theta=60^\circ$, $N=512$}
	\end{subfigure}
~
\begin{subfigure}[b]{0.3\linewidth}
		\includegraphics[width=\textwidth]{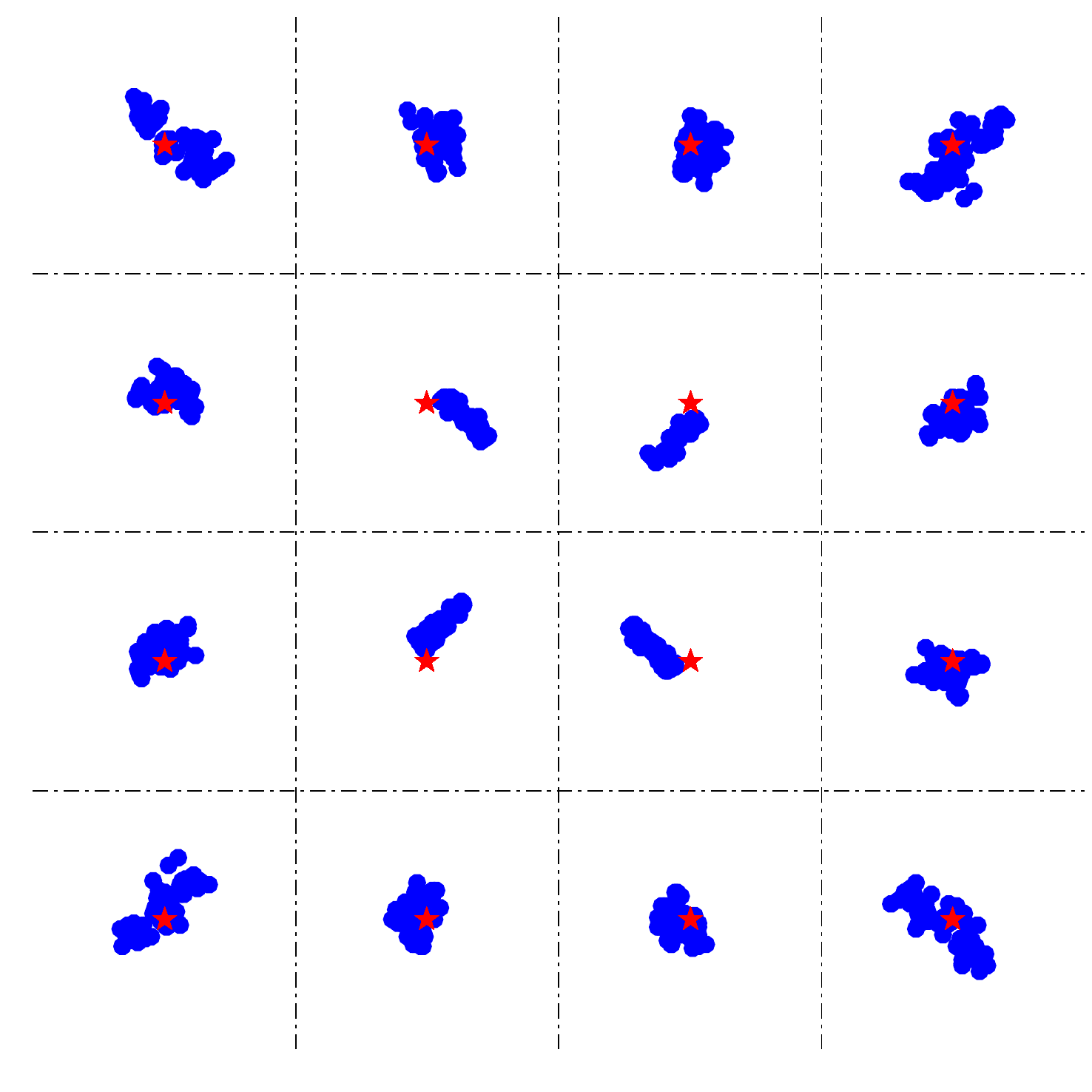}
		\caption*{$\theta=65^\circ$, $N=64$}
	\end{subfigure}
~
\begin{subfigure}[b]{0.3\linewidth}
		\includegraphics[width=\textwidth]{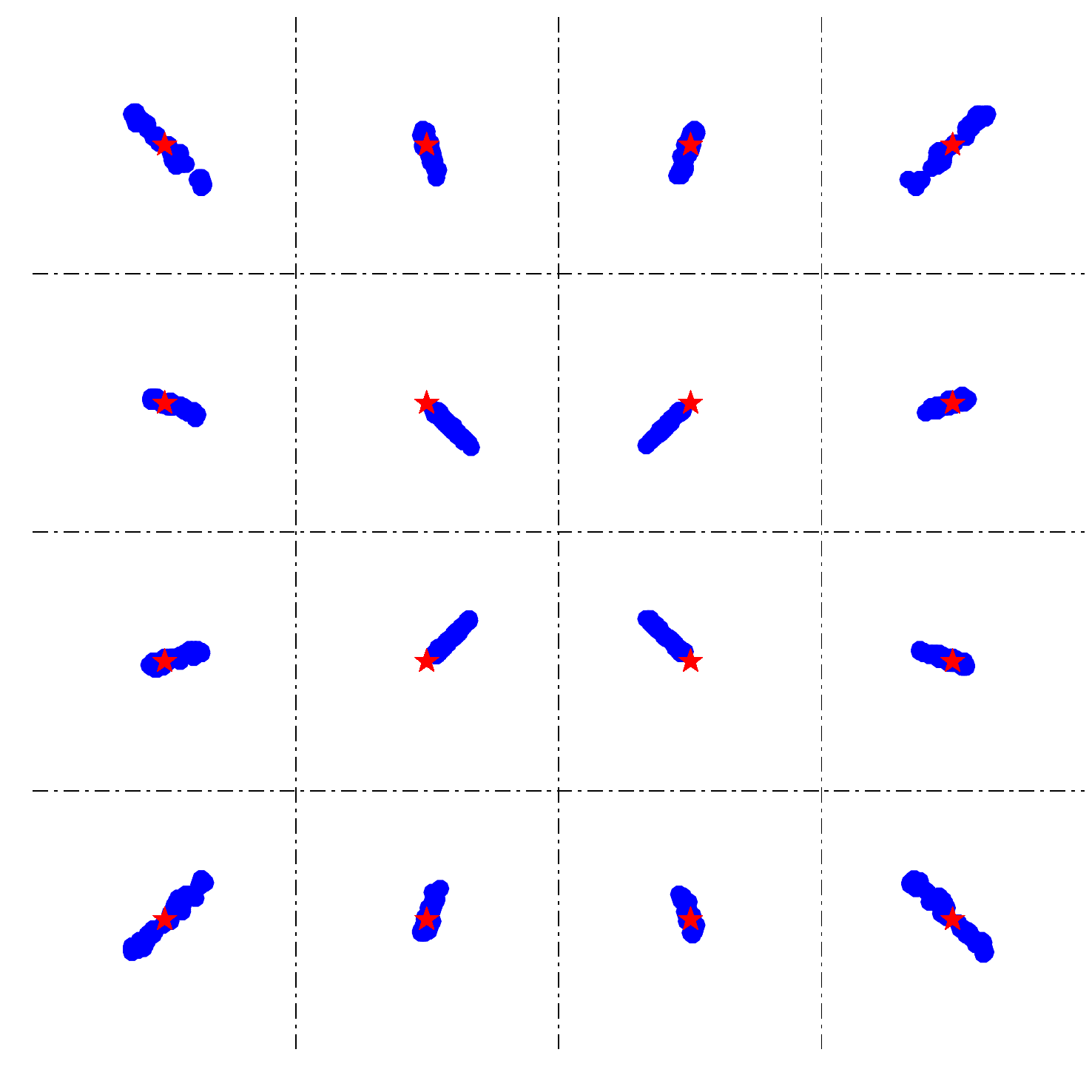}
		\caption*{$\theta=65^\circ$, $N=256$}
	\end{subfigure}
~
\begin{subfigure}[b]{0.3\linewidth}
		\includegraphics[width=\textwidth]{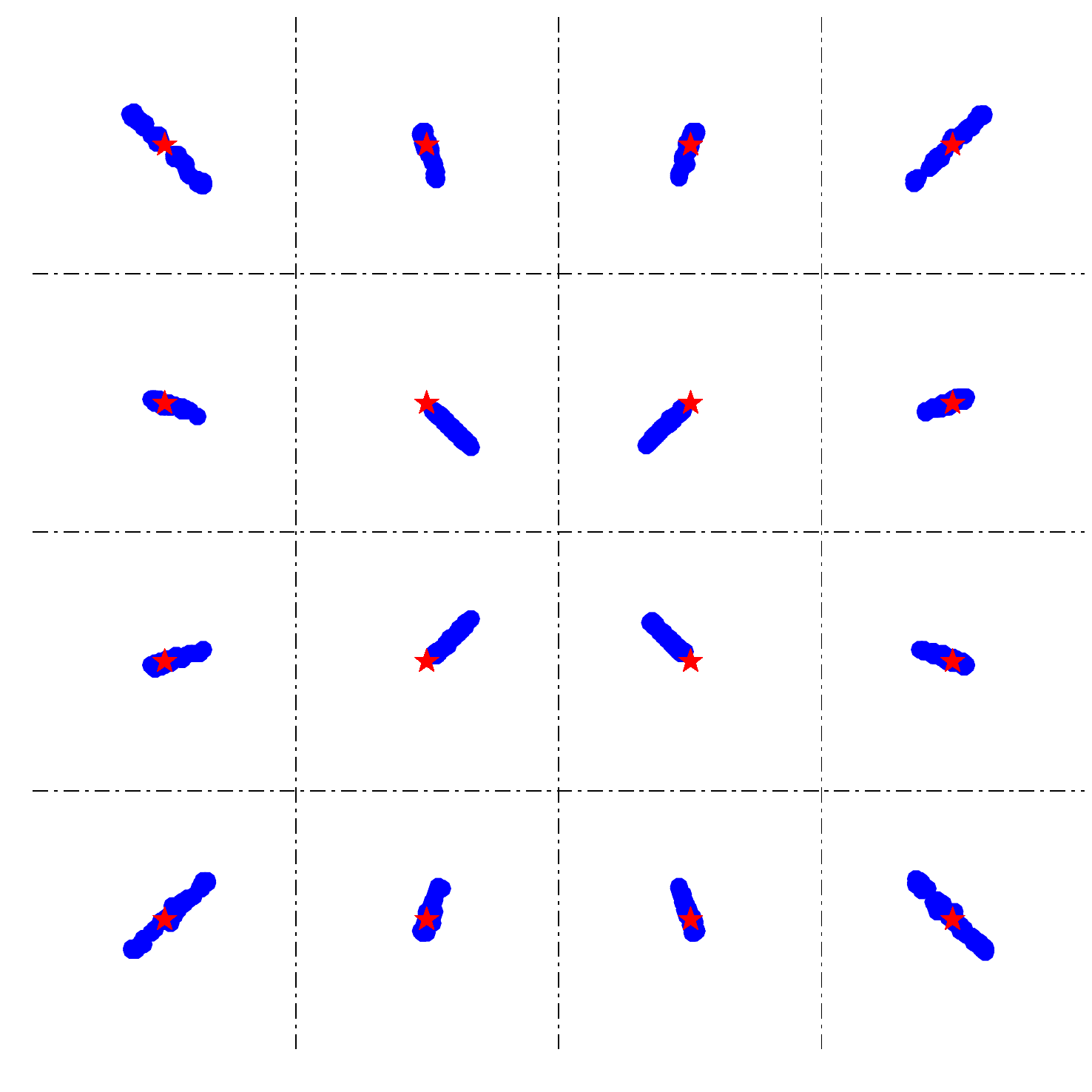}
		\caption*{$\theta=65^\circ$, $N=512$}
	\end{subfigure}
~
\begin{subfigure}[b]{0.3\linewidth}
		\includegraphics[width=\textwidth]{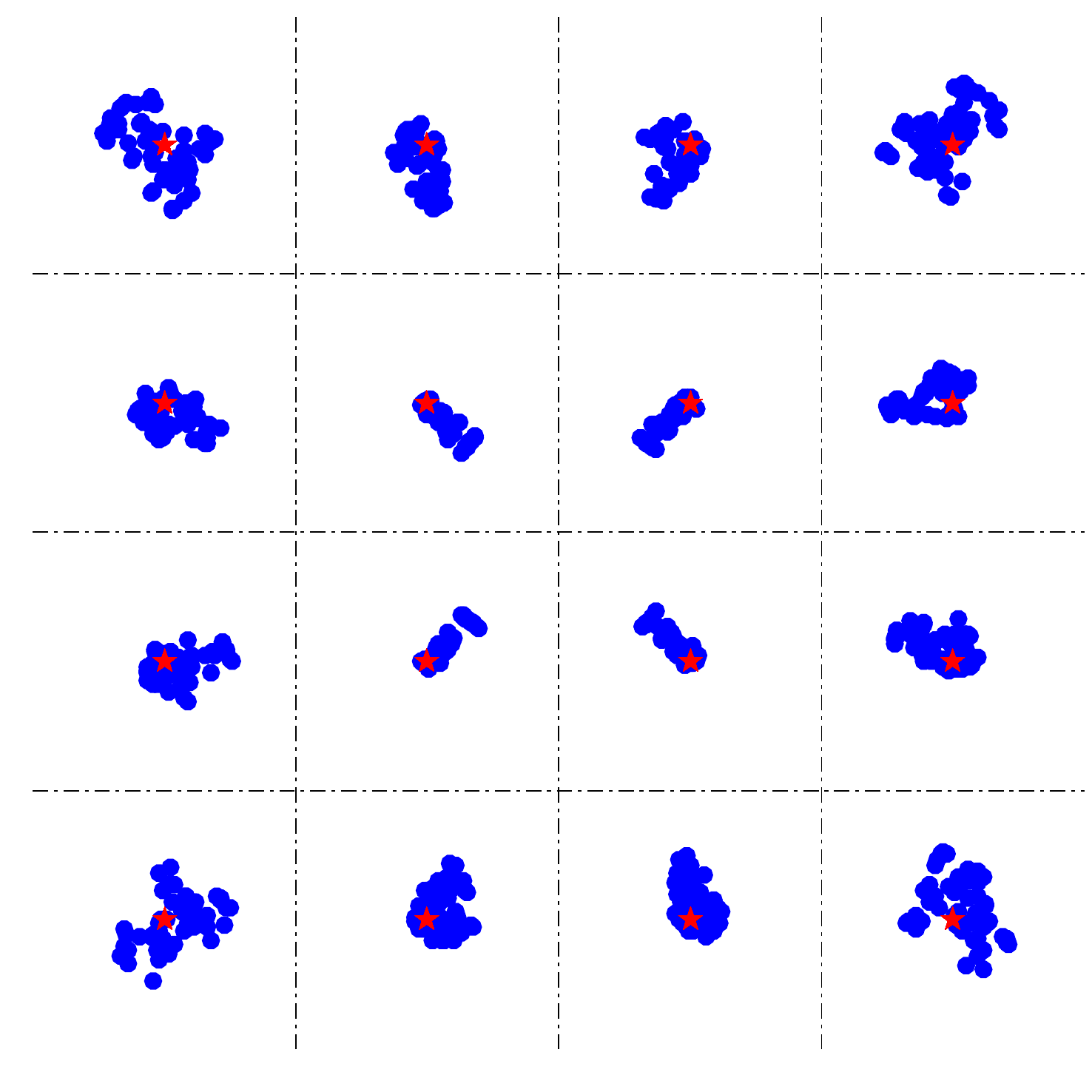}
		\caption*{$\theta=70^\circ$, $N=64$}
	\end{subfigure}
~
\begin{subfigure}[b]{0.3\linewidth}
		\includegraphics[width=\textwidth]{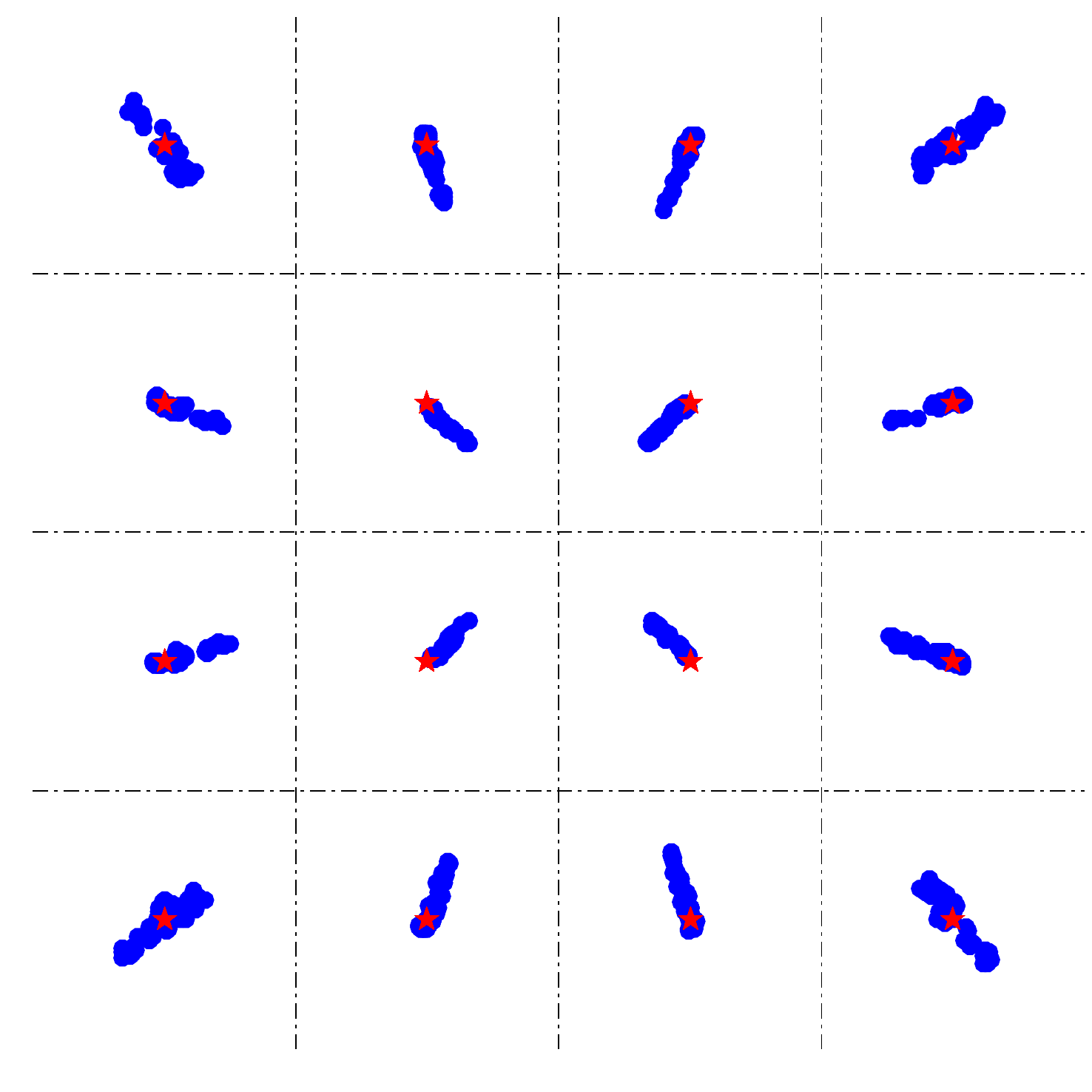}
		\caption*{$\theta=70^\circ$, $N=256$}
	\end{subfigure}
~
\begin{subfigure}[b]{0.3\linewidth}
		\includegraphics[width=\textwidth]{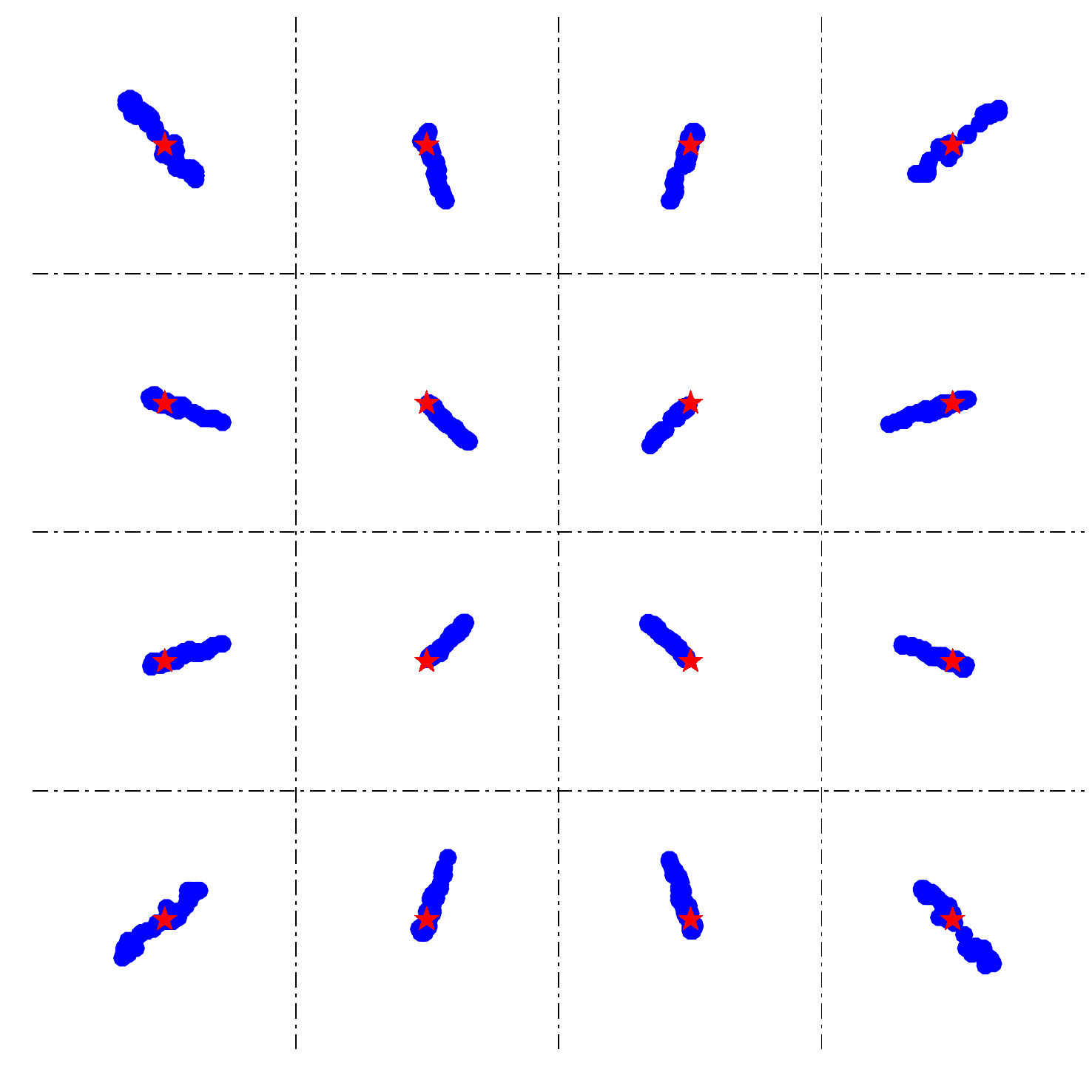}
		\caption*{$\theta=70^\circ$, $N=512$}
	\end{subfigure}
	\caption{IQ scatter plots of the basic sigma-delta MRT scheme for different $\theta$ and $N$; $16$-ary QAM. (cont.) }
\end{figure}

\begin{figure}[H]\ContinuedFloat
	\centering	
\begin{subfigure}[b]{0.3\linewidth}
		\includegraphics[width=\textwidth]{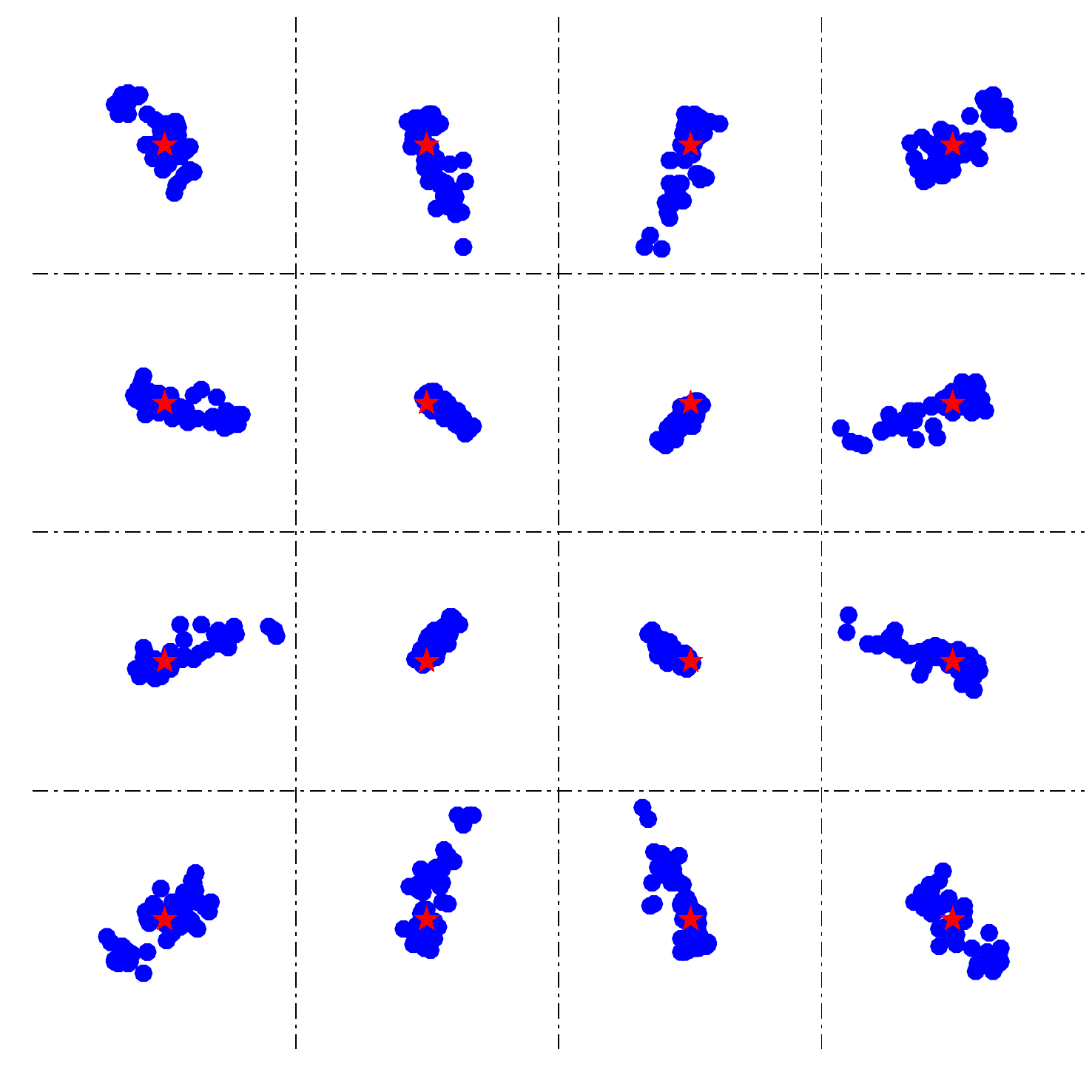}
		\caption*{$\theta=75^\circ$, $N=64$}
	\end{subfigure}
~
\begin{subfigure}[b]{0.3\linewidth}
		\includegraphics[width=\textwidth]{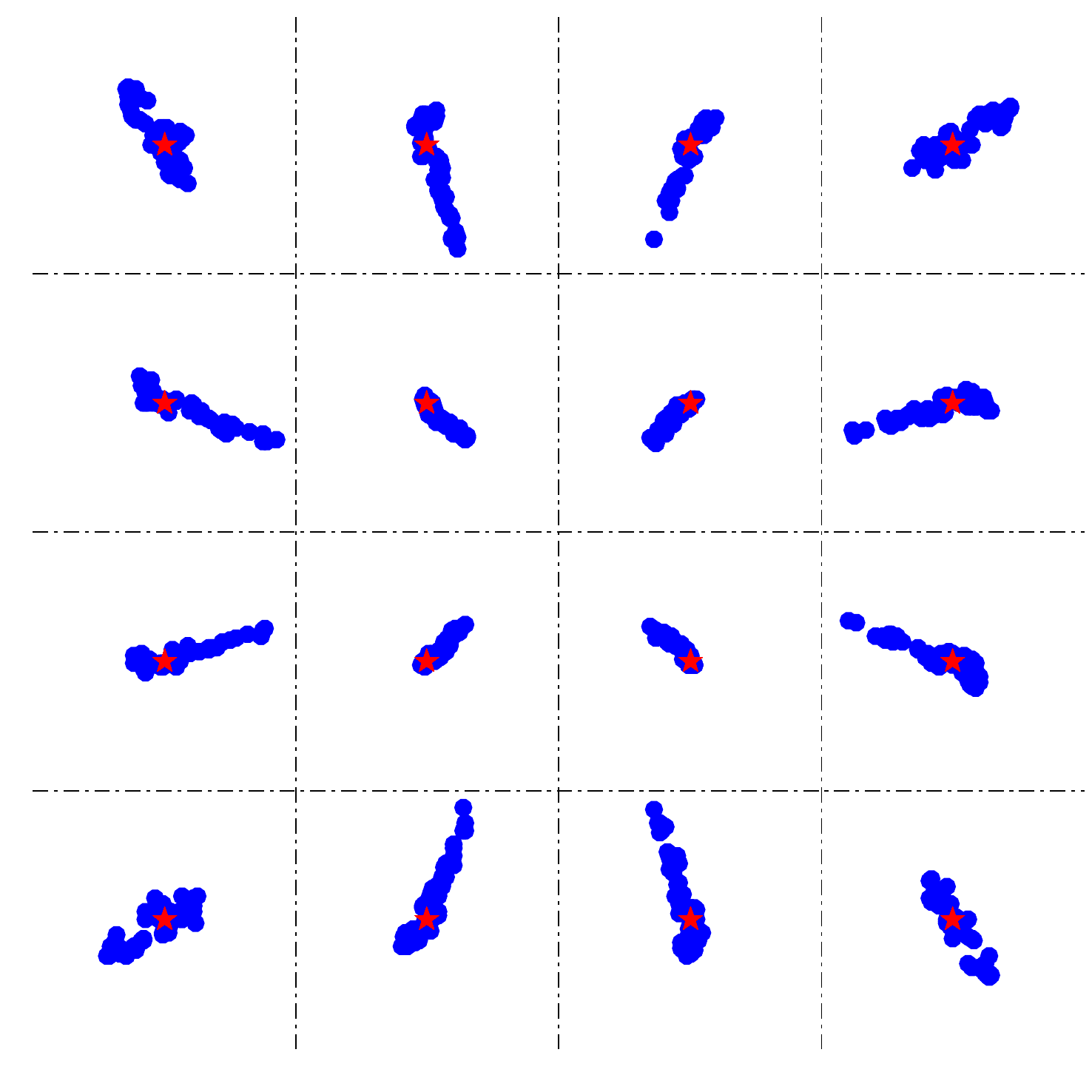}
		\caption*{$\theta=75^\circ$, $N=256$}
	\end{subfigure}
~
\begin{subfigure}[b]{0.3\linewidth}
		\includegraphics[width=\textwidth]{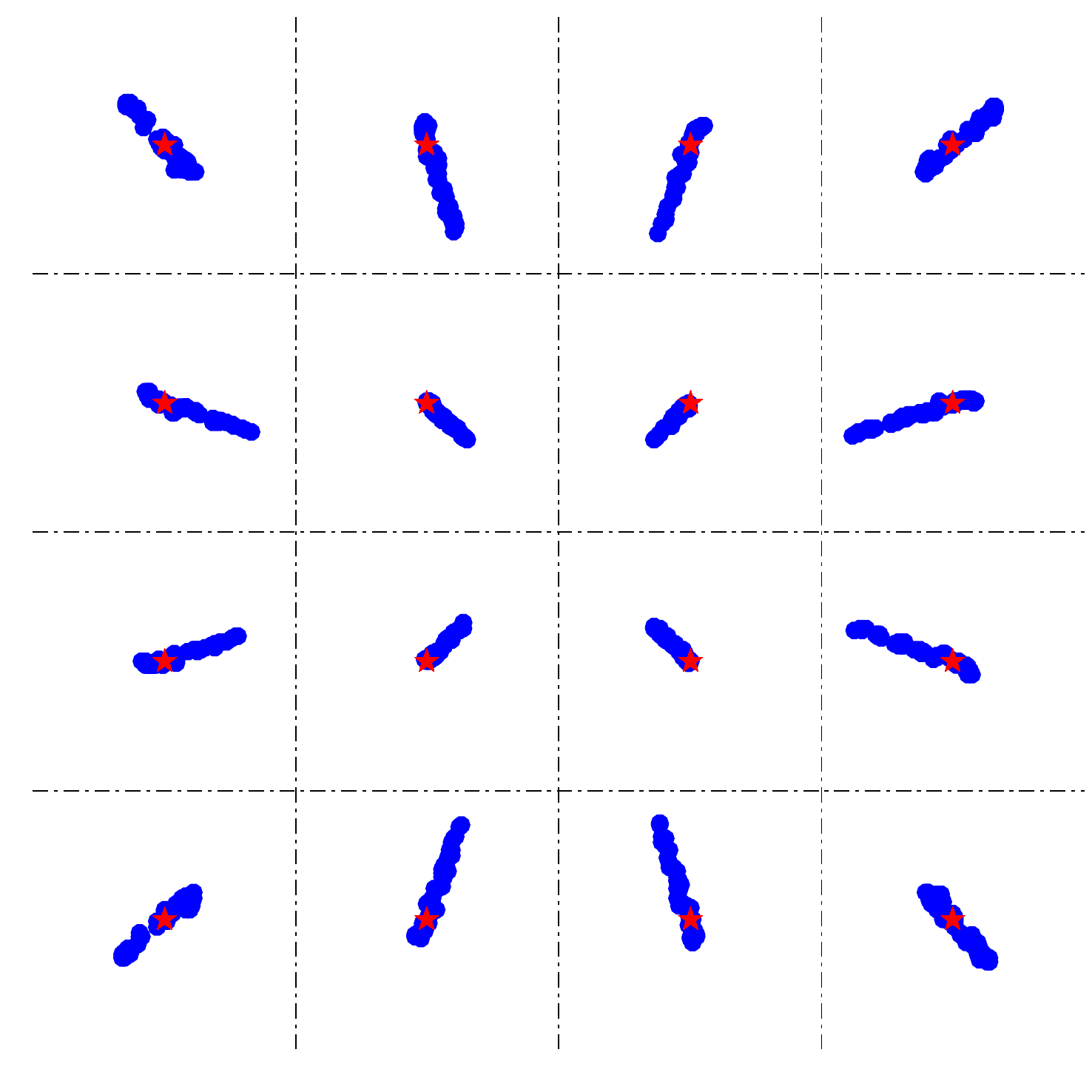}
		\caption*{$\theta=75^\circ$, $N=512$}
	\end{subfigure}
~
\begin{subfigure}[b]{0.3\linewidth}
		\includegraphics[width=\textwidth]{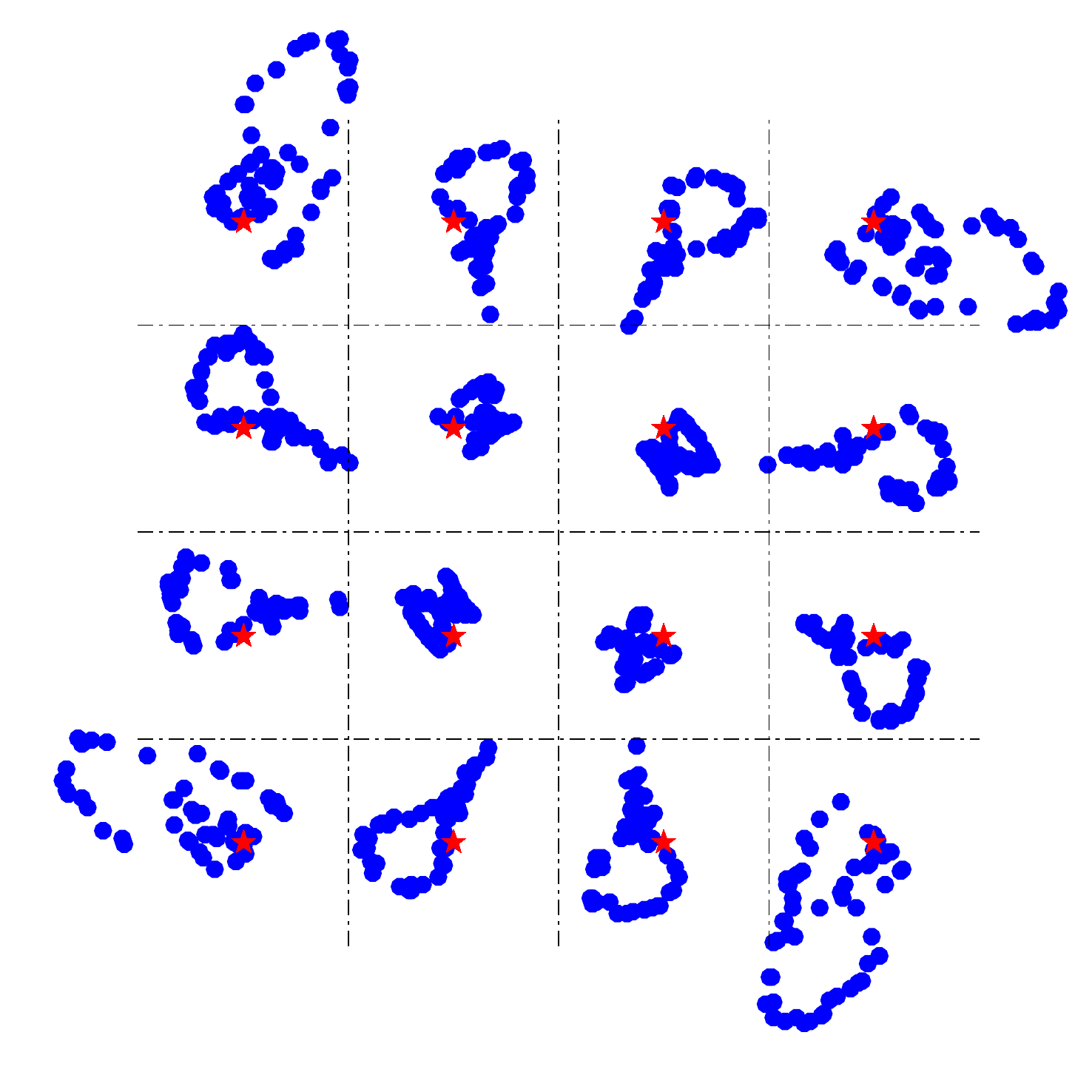}
		\caption*{$\theta=80^\circ$, $N=64$}
	\end{subfigure}
~
\begin{subfigure}[b]{0.3\linewidth}
		\includegraphics[width=\textwidth]{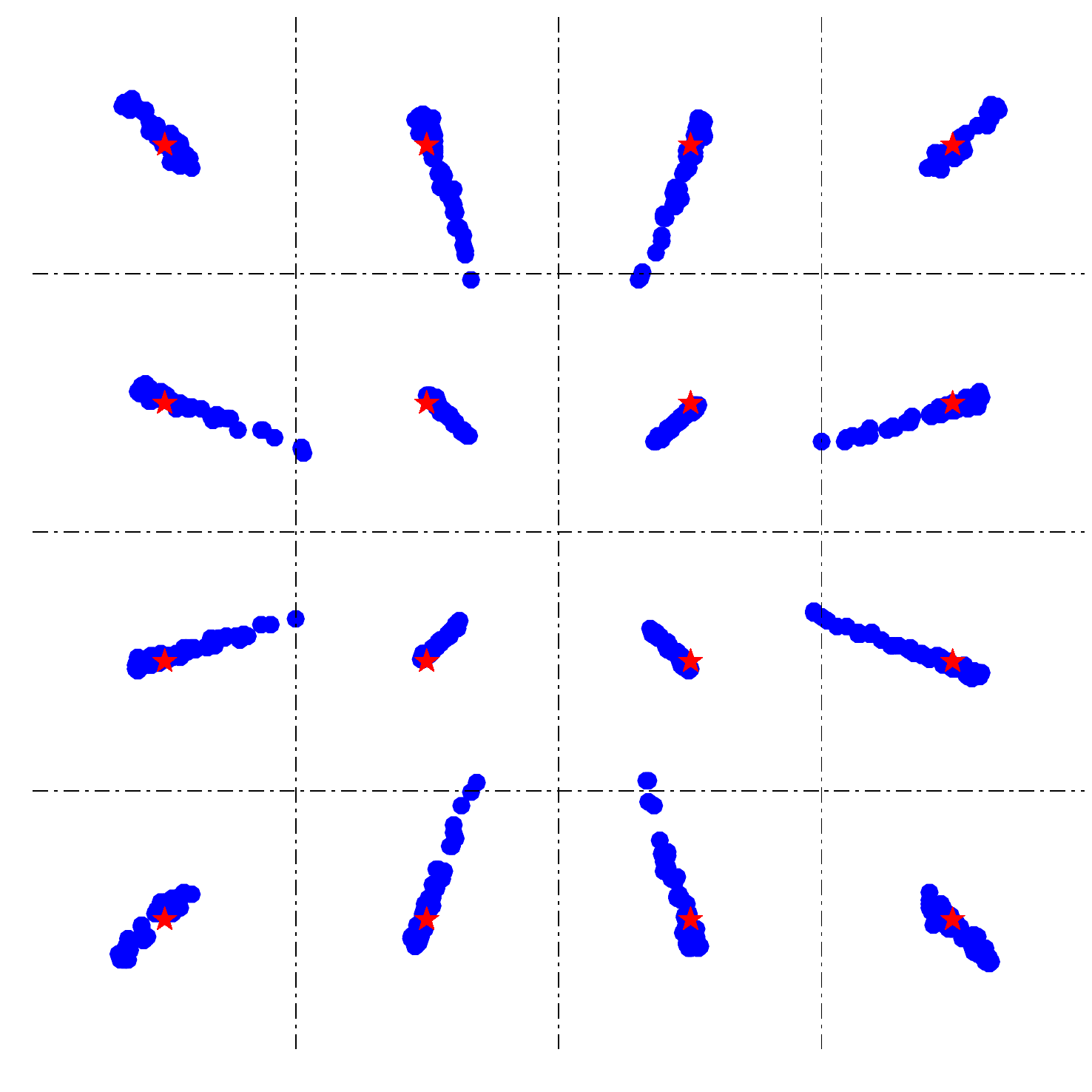}
		\caption*{$\theta=80^\circ$, $N=256$}
	\end{subfigure}
~
\begin{subfigure}[b]{0.3\linewidth}
		\includegraphics[width=\textwidth]{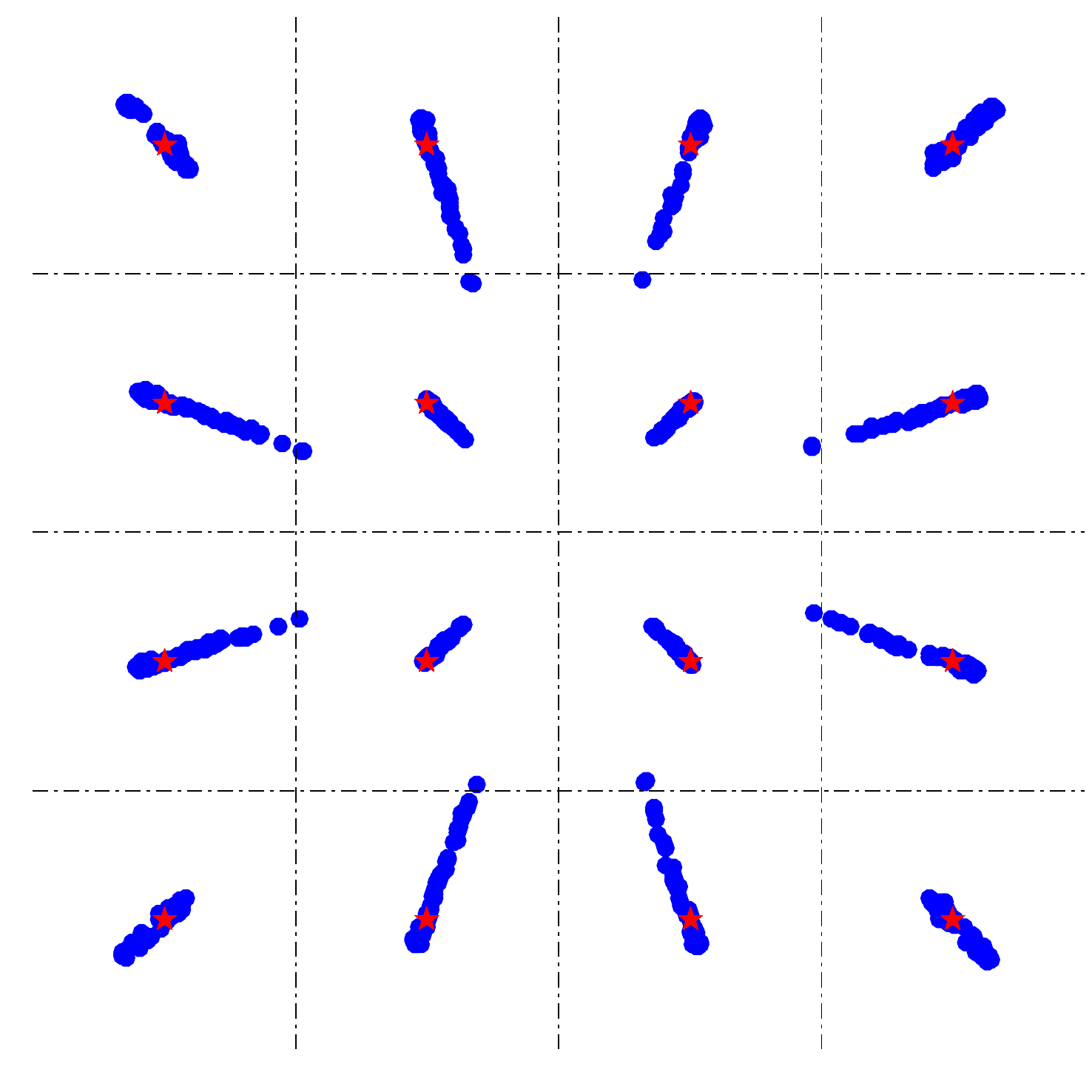}
		\caption*{$\theta=80^\circ$, $N=512$}
	\end{subfigure}
~
\begin{subfigure}[b]{0.3\linewidth}
		\includegraphics[width=\textwidth]{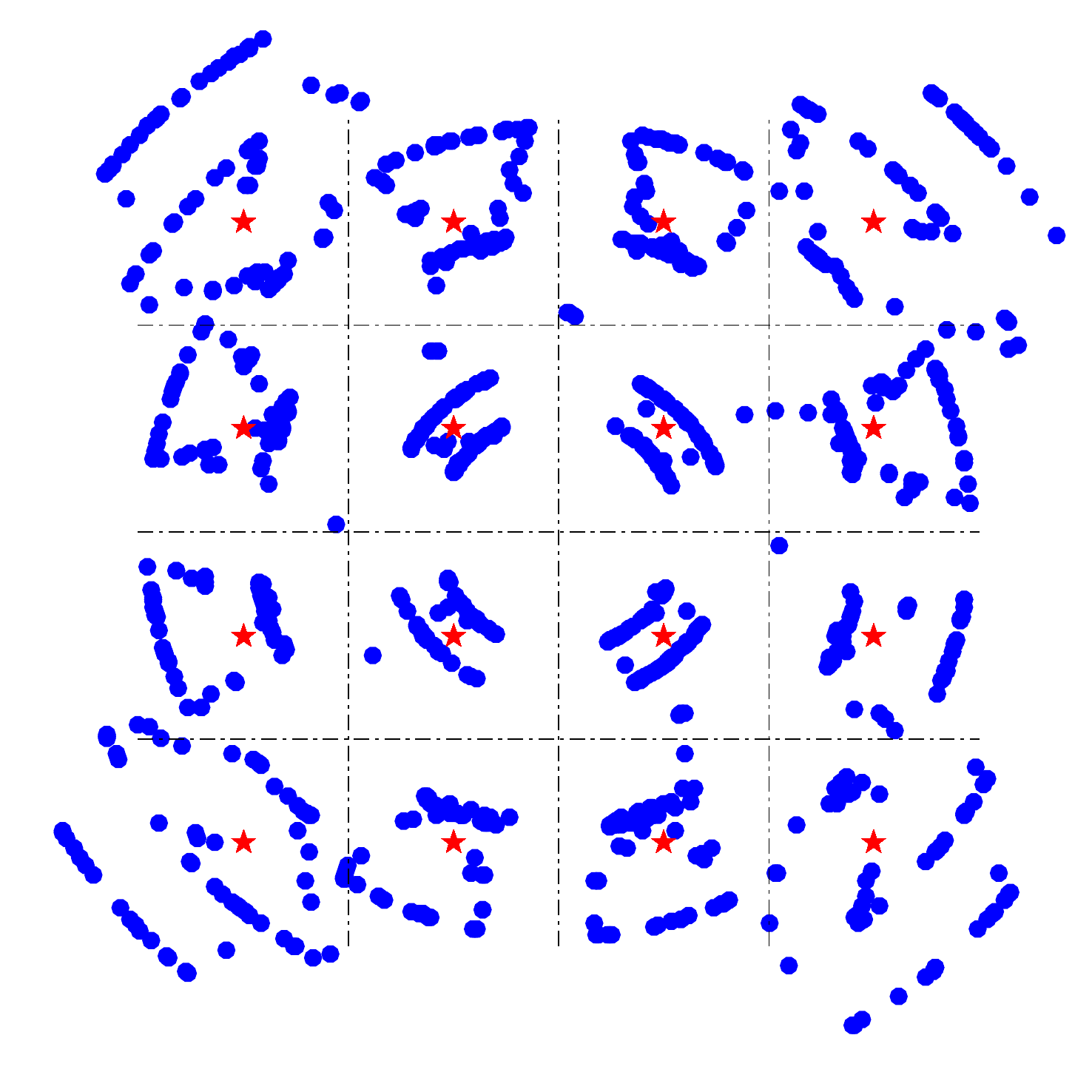}
		\caption*{$\theta=85^\circ$, $N=64$}
	\end{subfigure}
~
\begin{subfigure}[b]{0.3\linewidth}
		\includegraphics[width=\textwidth]{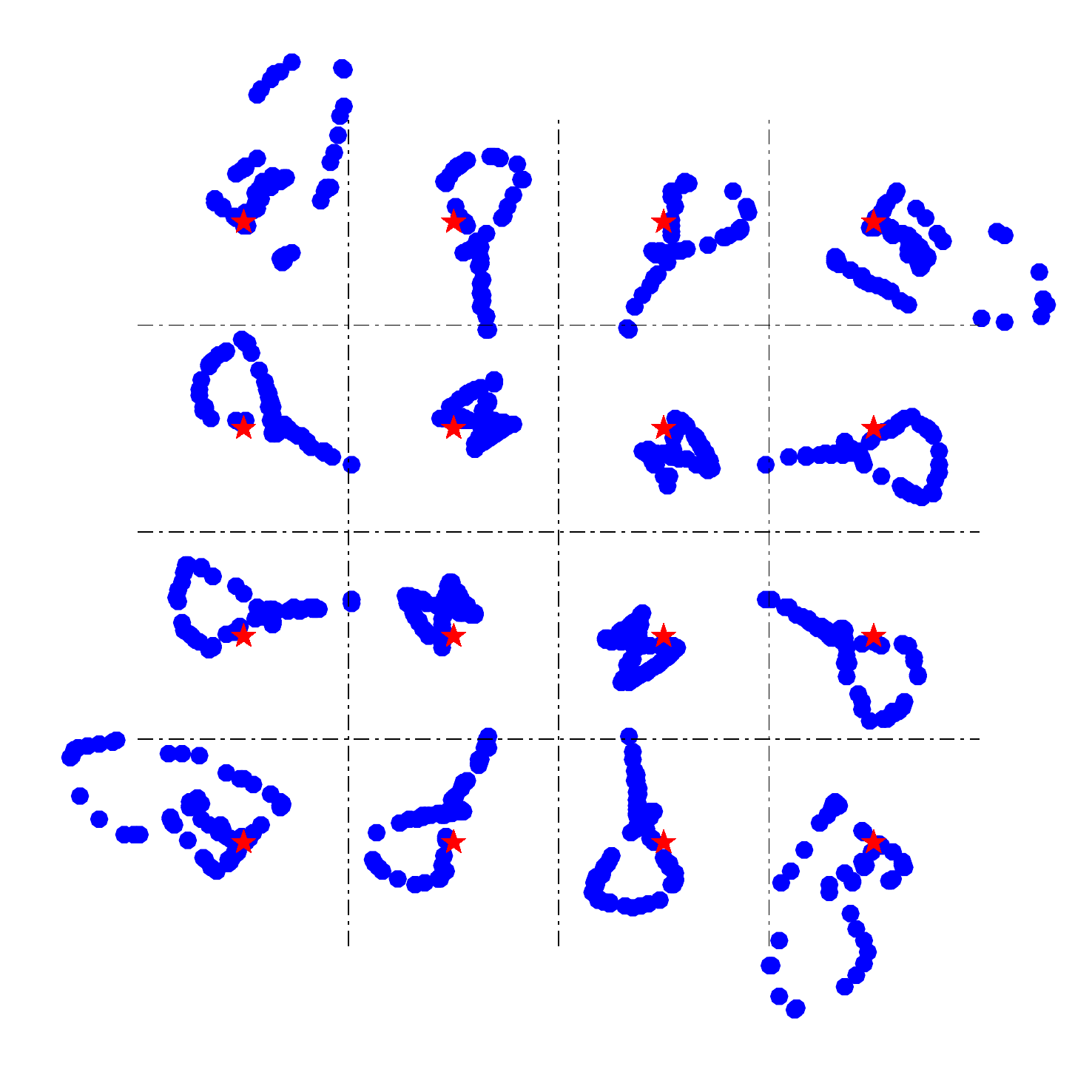}
		\caption*{$\theta=85^\circ$, $N=256$}
	\end{subfigure}
~
\begin{subfigure}[b]{0.3\linewidth}
		\includegraphics[width=\textwidth]{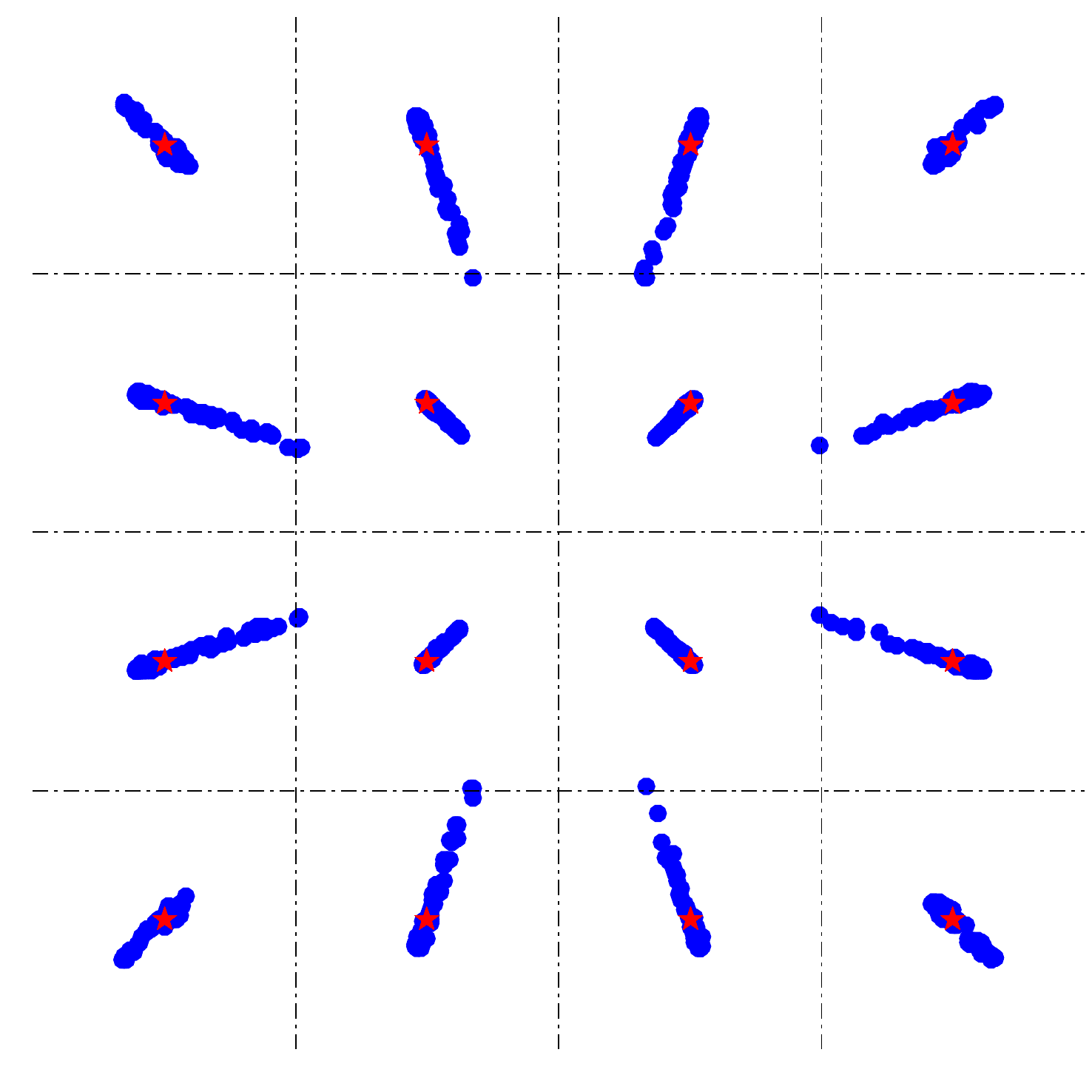}
		\caption*{$\theta=85^\circ$, $N=512$}
	\end{subfigure}
%
%
\begin{subfigure}[b]{0.3\linewidth}
		\includegraphics[width=\textwidth]{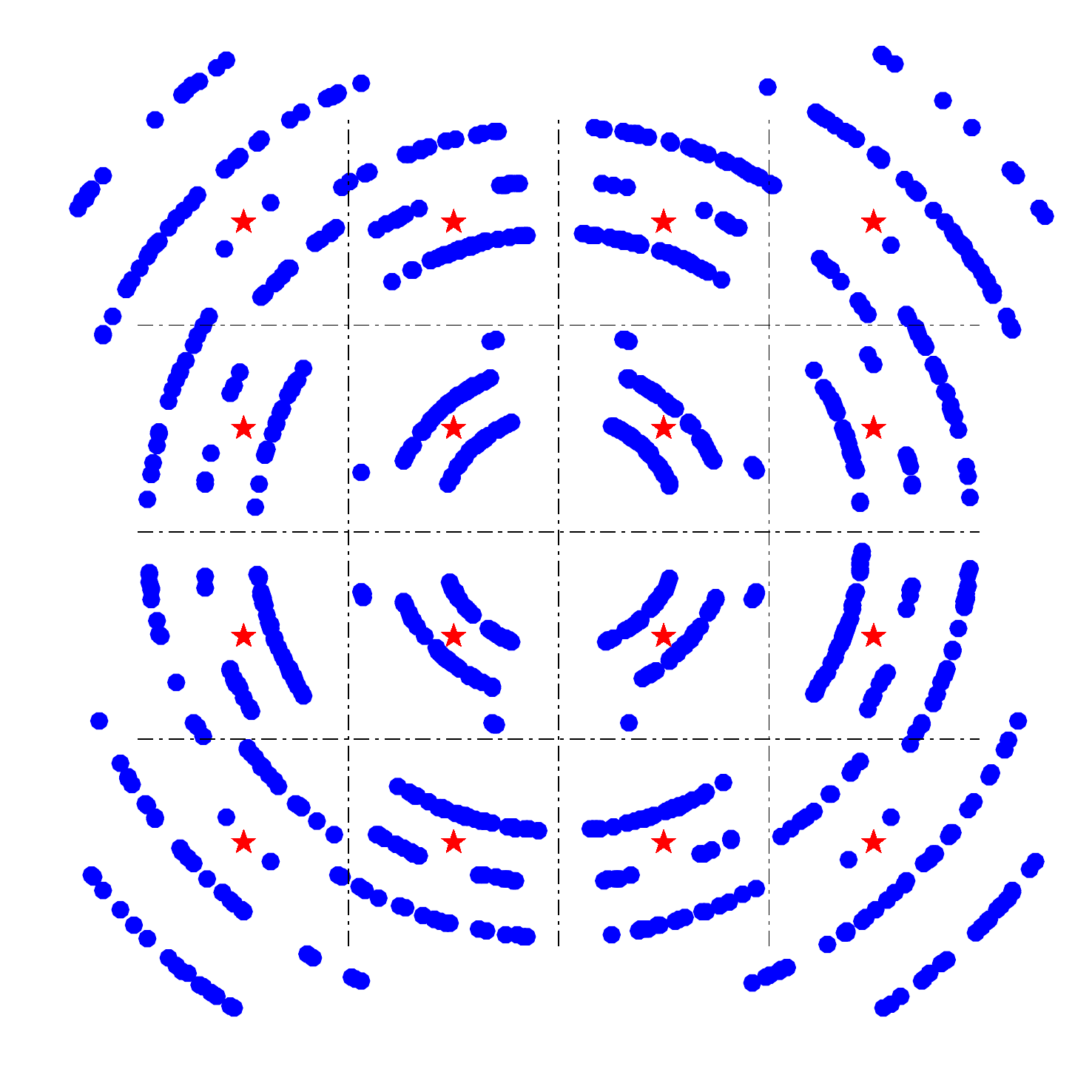}
		\caption*{$\theta=90^\circ$, $N=64$}
	\end{subfigure}
~
\begin{subfigure}[b]{0.3\linewidth}
		\includegraphics[width=\textwidth]{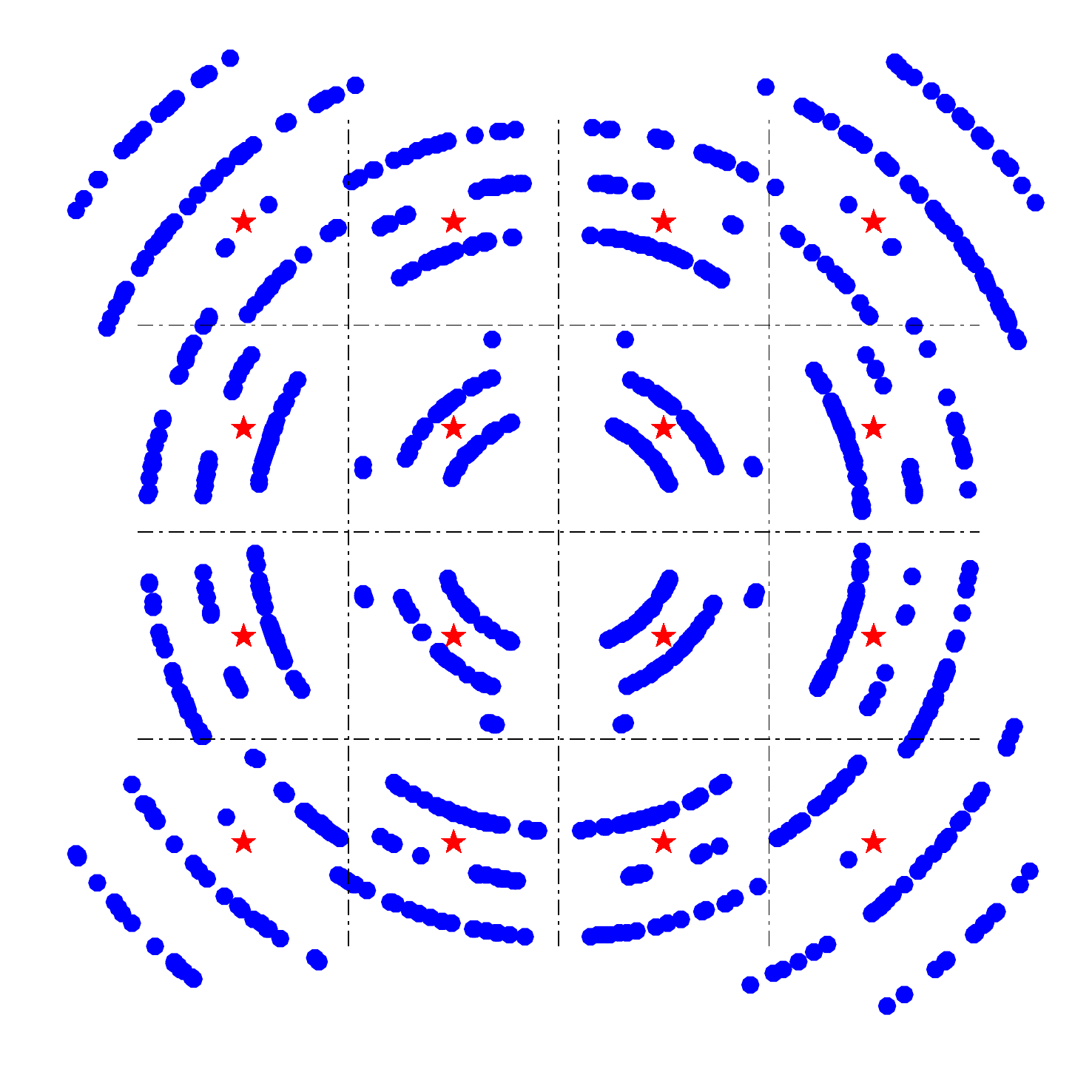}
		\caption*{$\theta=90^\circ$, $N=256$}
	\end{subfigure}
~
\begin{subfigure}[b]{0.3\linewidth}
		\includegraphics[width=\textwidth]{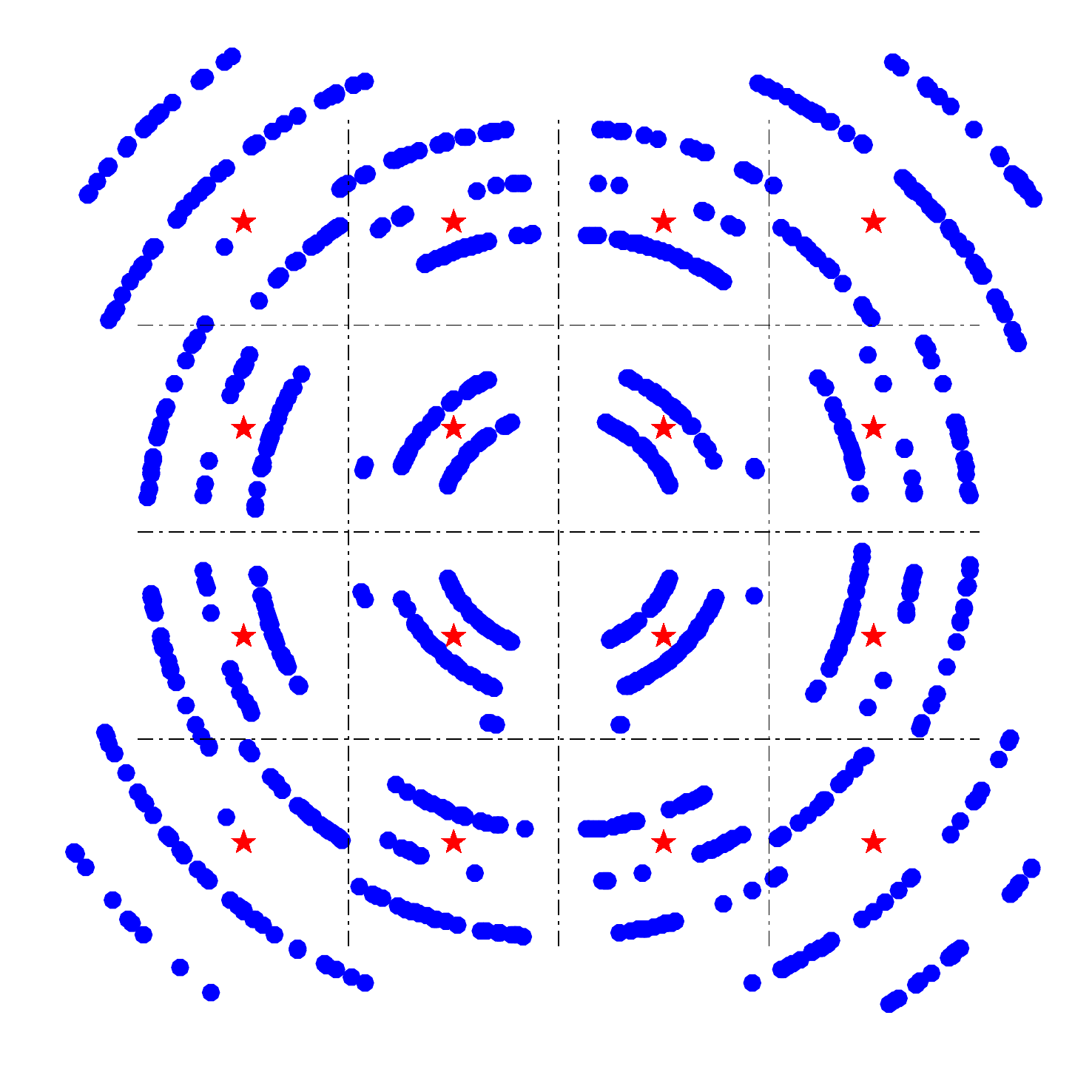}
		\caption*{$\theta=90^\circ$, $N=512$}
	\end{subfigure}
	\caption{IQ scatter plots of the basic sigma-delta MRT scheme for different $\theta$ and $N$; $16$-ary QAM. (cont.)}
\end{figure}

%
%

\section{Proof of \eqref{eq:lam_R}} \label{appen:1}

The proof of \eqref{eq:lam_R} in Proposition \ref{prop:eff_snr_zf} is as follows.
Since $\bR$ is Hermitian,
we can write
\[
\lambda_{\rm min}(\bR) = \min_{ \| \bx \|_2 = 1} \bx^H \bR \bx.
\]
Let $\phi_i = 2 \pi d \sin(\theta_i)/\lambda$, and let $z_i = e^{\jj \phi}$.
Since $\ba_i = [~ 1, z_i^{-1}, \ldots, z_i^{-(N-1)} ~]^T$,
we have
\begin{align*}
r_{ij} = \frac{1}{N} \ba_i^T \ba_j^* = \frac{1}{N} \sum_{n=0}^{N-1} ( z_i^{-1} z_j )^{n}
= \frac{1 - (z_i^{-1} z_j)^N}{N(1- (z_i^{-1} z_j))}.
\end{align*}
Thus, the elements of $\bR$ satisfy $r_{ii}=1$ and
\[
|r_{ij}| = \left| D_N\left( \frac{\phi_1 - \phi_2}{2}  \right) \right| \leq \rho.
\]
Since $r_{ii}= 1$, we have
\[
\lambda_{\rm min}(\bR) \leq \be_1^H \bR \be_1 = 1,
\]
where $\be_1 = [~ 1, 0,\ldots,0~ ]^T$.
Also, it holds that
\begin{align*}
\bx^H \bR \bx & \geq \sum_{i=1}^K |x_i|^2 - \sum_{i \neq j} |r_{ij}| |x_i||x_j| \\
& \geq \sum_{i=1}^K |x_i|^2 - \rho \sum_{i \neq j} |x_i||x_j| \\
& = | \bx |^T ( (1+\rho)\bI - \rho \bone \bone^T ) | \bx | \\
& \geq ( 1 + (1-K) \rho ) \| \bx \|_2^2,
\end{align*}
where we denote $| \bx | = [~ |x_1|, \ldots, |x_n| ~]^T$ in the third equation;
the last equation is due to $\lambda_{\min}( (1+\rho)\bI - \rho \bone \bone^T ) = 1 + \rho - \rho \| \bone \|_2^2 = 1 +(1-K)\rho$.
It therefore follows that $\lambda_{\rm min}(\bR) \geq 1 + (1-K)\rho$.
The proof of \eqref{eq:lam_R} is thus complete.

\section{Derivation of \eqref{eq:sigma_w_mp}}
\label{app:noise_va_mpath}

Following the single-user development in Section~\ref{sec:sig_del_mod_basic},
the noise term $w_i$ in the model \eqref{eq:model_mu} of the multiuser case is given by
\begin{align*}
w_i & = \sqrt{\frac{P}{2N}} \bh^T(\bq - \bq^-) + v_i;
\end{align*}
recall $\bq^- =  [~ 0, q_1, \ldots, q_{N-1} ~]^T$.
When the channel $\bh_i$ takes the multipath form in \eqref{eq:ch_mpath},
$w_i$ can be expressed as
\begin{align*}
w_i & = \sqrt{\frac{P}{2N}}
	\Bigg[
		\sum_{n=0}^{N-2} \left( \sum_{l=1}^{L_i} \alpha_{il} (z_{il}^{-n} - z_{il}^{-n-1})  \right) q_{n+1} \\
	& \quad 	
		+  \left( \sum_{l=1}^{L_i} \alpha_{il} z_{il}^{-N-1} \right) q_N
	\Bigg] + v_i,
\end{align*}
where $z_{il} =  e^{\jj \frac{2\pi d}{\lambda} \sin(\theta_{il})}$.
Under the uniform i.i.d. assumption with $\bq$, we have $\Exp[ w_i ] = 0$ and
\begin{align*}
\sigma_{w,i}^2 & = \frac{P \sigma_q^2}{2N} \Bigg[
	\sum_{n=0}^{N-2} \left| \sum_{l=1}^{L_i} \alpha_{il} (z_{il}^{-n} - z_{il}^{-n-1}) \right|^2  \\
	& \quad
	+ \left| \sum_{l=1}^{L_i} \alpha_{il} z_{il}^{-N-1} \right|^2
	\Bigg] + \sigma_v^2,
\end{align*}
where $\sigma_q^2 = \Exp[ |q_n|^2 ] = 2/3$.
By approximating the above expression as
\begin{align*}
\sigma_{w,i}^2 & \approx \frac{P \sigma_q^2}{2N} \Bigg[
\sum_{n=0}^{N-1} \left| \sum_{l=1}^{L_i} \alpha_{il} (z_{il}^{-n} - z_{il}^{-n-1}) \right|^2 \Bigg]  + \sigma_v^2,
\end{align*}
which is reasonable for large $N$, we arrive at the noise variance formula in \eqref{eq:sigma_w_mp}.

\bibliographystyle{IEEEtran}
\bibliography{refs}

\begin{thebibliography}{10}
\providecommand{\url}[1]{#1}
\csname url@samestyle\endcsname
\providecommand{\newblock}{\relax}
\providecommand{\bibinfo}[2]{#2}
\providecommand{\BIBentrySTDinterwordspacing}{\spaceskip=0pt\relax}
\providecommand{\BIBentryALTinterwordstretchfactor}{4}
\providecommand{\BIBentryALTinterwordspacing}{\spaceskip=\fontdimen2\font plus
\BIBentryALTinterwordstretchfactor\fontdimen3\font minus
  \fontdimen4\font\relax}
\providecommand{\BIBforeignlanguage}[2]{{%
\expandafter\ifx\csname l@#1\endcsname\relax
\typeout{** WARNING: IEEEtran.bst: No hyphenation pattern has been}%
\typeout{** loaded for the language `#1'. Using the pattern for}%
\typeout{** the default language instead.}%
\else
\language=\csname l@#1\endcsname
\fi
#2}}
\providecommand{\BIBdecl}{\relax}
\BIBdecl

\bibitem{FanJWZ5}
L.~Fan, S.~Jin, C.~Wen, and H.~Zhang, ``Uplink achievable rate for massive
  {MIMO} systems with low-resolution {ADC},'' \emph{IEEE Commun. Lett.},
  vol.~19, no.~12, pp. 2186--2189, Dec 2015.

\bibitem{juncil2015near}
J.~Choi, J.~Mo, and R.~W. Heath, ``Near maximum-likelihood detector and channel
  estimator for uplink multiuser massive {MIMO} systems with one-bit {ADCs},''
  \emph{IEEE Trans. Commun.}, vol.~64, no.~5, pp. 2005--2018, May 2016.

\bibitem{MollenCLH17}
C.~Moll\'{e}n, J.~Choi, E.~G. Larsson, and R.~W. Heath, ``Uplink performance of
  wideband massive {MIMO} with one-bit {ADCs},'' \emph{IEEE Trans. Wireless
  Commun.}, vol.~16, no.~1, pp. 87--100, Jan 2017.

\bibitem{LiTSMSL17}
Y.~Li, C.~Tao, G.~Seco-Granados, A.~Mezghani, A.~L. Swindlehurst, and L.~Liu,
  ``Channel estimation and performance analysis of one-bit massive {MIMO}
  systems,'' \emph{IEEE Trans.\ Signal Process.}, vol.~65, no.~15, pp.
  4075--4089, Aug 2017.

\bibitem{JacobssonDCGS17}
S.~Jacobsson, G.~Durisi, M.~Coldrey, U.~Gustavsson, and C.~Studer, ``Throughput
  analysis of massive {MIMO} uplink with low-resolution {ADCs},'' \emph{IEEE
  Trans. Wireless Commun.}, vol.~16, no.~6, pp. 4038--4051, June 2017.

\bibitem{ShaoLL18}
Z.~Shao, R.~C. de~Lamare, and L.~T.~N. Landau, ``Iterative detection and
  decoding for large-scale multiple-antenna systems with 1-bit {ADCs},''
  \emph{IEEE Wireless Commun. Lett.}, vol.~7, no.~3, pp. 476--479, June 2018.

\bibitem{JeonLHH18}
Y.~Jeon, N.~Lee, S.~Hong, and R.~W. Heath, ``One-bit sphere decoding for uplink
  massive {MIMO} systems with one-bit {ADCs},'' \emph{IEEE Trans.\ Wireless
  Commun.}, vol.~17, no.~7, pp. 4509--4521, July 2018.

\bibitem{Nossek2009Transmit}
A.~Mezghani, R.~Ghiat, and J.~A. Nossek, ``Transmit processing with low
  resolution {D/A}-converters,'' in \emph{Proc. 16th IEEE Int. Conf. Electron.,
  Circuits, Syst.}, Dec 2009, pp. 683--686.

\bibitem{SaxenaFS17}
A.~K. Saxena, I.~Fijalkow, and A.~Swindlehurst, ``Analysis of one-bit quantized
  precoding for the multiuser massive {MIMO} downlink,'' \emph{IEEE Trans.\
  Signal Process.}, vol.~65, no.~17, pp. 4624--4634, Sept 2017.

\bibitem{LiTSML17}
Y.~Li, C.~Tao, A.~L. Swindlehurst, A.~Mezghani, and L.~Liu, ``Downlink
  achievable rate analysis in massive {MIMO} systems with one-bit {DACs},''
  \emph{IEEE Commun. Lett.}, vol.~21, no.~7, pp. 1669--1672, July 2017.

\bibitem{CastanedaJDCGS17}
O.~Castaneda, S.~Jacobsson, G.~Durisi, M.~Coldrey, T.~Goldstein, and C.~Studer,
  ``{1-bit massive MU-MIMO precoding in VLSI},'' \emph{IEEE J. Emerg. Sel.
  Topics Circuits Syst.}, vol.~7, no.~4, pp. 508--522, Dec 2017.

\bibitem{Jacobsson2017}
S.~Jacobsson, G.~Durisi, M.~Coldrey, T.~Goldstein, and C.~Studer, ``{Quantized
  precoding for massive MU-MIMO},'' \emph{IEEE Trans. Commun.}, vol.~65,
  no.~11, pp. 4670--4684, Nov 2017.

\bibitem{Swindlehurst2017}
A.~Swindlehurst, A.~Saxena, A.~Mezghani, and I.~Fijalkow, ``{Minimum
  probability-of-error perturbation precoding for the one-bit massive {MIMO}
  downlink},'' in \emph{Proc. IEEE Int. Conf. Acous., Speech, Signal Process.
  (ICASSP)}, Mar. 2017, pp. 6483--6487.

\bibitem{LandauL17}
L.~T.~N. Landau and R.~C. de~Lamare, ``Branch-and-bound precoding for multiuser
  {MIMO} systems with 1-bit quantization,'' \emph{IEEE Wireless Commun. Lett.},
  vol.~6, no.~6, pp. 770--773, Dec 2017.

\bibitem{JeddaMSN18}
H.~Jedda, A.~Mezghani, A.~L. Swindlehurst, and J.~A. Nossek, ``{Quantized
  constant envelope precoding with PSK and QAM signaling},'' \emph{IEEE Trans.
  Wireless Commun.}, vol.~17, no.~12, pp. 8022--8034, Dec 2018.

\bibitem{li2018massive}
A.~Li, C.~Masouros, F.~Liu, and A.~L. Swindlehurst, ``{Massive MIMO 1-bit DAC
  transmission: A low-complexity symbol scaling approach},'' \emph{IEEE Trans.
  Wireless Commun.}, vol.~17, no.~11, pp. 7559--7575, 2018.

\bibitem{Sohrabi2018}
F.~Sohrabi, Y.-F. Liu, and W.~Yu, ``One-bit precoding and constellation range
  design for massive {MIMO} with {QAM} signaling,'' \emph{IEEE J. Sel. Topics
  Signal Process.}, vol.~12, no.~3, pp. 557--570, 2018.

\bibitem{shao2018framework}
M.~Shao, Q.~Li, W.-K. Ma, and A.~M.-C. So, ``{A framework for one-bit and
  constant-envelope precoding over multiuser massive {MISO} channels},''
  \emph{{\rm to appear in} IEEE Trans. Signal Process.}, 2019.

\bibitem{aziz1996overview}
P.~M. Aziz, H.~V. Sorensen, and J.~Van Der~Spiegel, ``An overview of
  sigma-delta converters: How a 1-bit {ADC} achieves more than 16-bit
  resolution,'' \emph{IEEE Signal Process. Mag.}, vol.~13, no.~1, pp. 61--84,
  1996.

\bibitem{Corey_Sig}
R.~M. Corey and A.~C. Singer, ``Spatial sigma-delta signal acquisition for
  wideband beamforming arrays,'' in \emph{Proc. Int. ITG Workshop Smart
  Antennas (WSA)}, March 2016.

\bibitem{baracspatial}
D.~Barac and E.~Lindqvist, ``Spatial sigma-delta modulation in a massive {MIMO}
  cellular system,'' Master's thesis, Department of Computer Science and
  Engineering, Chalmers University of Technology, 2016.

\bibitem{nikoofard}
A.~Nikoofard, J.~Liang, M.~Twieg, S.~Handagala, A.~Madanayake, L.~Belostotski,
  and S.~Mandal, ``Low-complexity {N-port ADCs} using {2-D} sigma-delta
  noise-shaping for {N-element} array receivers,'' in \emph{Proc. Int. Midwest
  Symposium Circuits Syst. (MWSCAS)}, 2017, pp. 301--304.

\bibitem{madan2017}
A.~Madanayake, N.~Akram, S.~Mandal, J.~Liang, and L.~Belostotski, ``Improving
  {ADC} figure-of-merit in wideband antenna array receivers using
  multidimensional space-time delta-sigma multiport circuits,'' in \emph{Proc.
  Int. Workshop Multidimensional (nD) Syst. (nDS)}, Sept 2017.

\bibitem{Venk2011}
V.~Venkateswaran and A.~van~der Veen, ``Multichannel {$\Sigma \Delta$ ADCs}
  with integrated feedback beamformers to cancel interfering communication
  signals,'' \emph{IEEE Trans.\ Signal Process.}, vol.~59, no.~5, pp.
  2211--2222, May 2011.

\bibitem{palguna2016}
D.~S. Palguna, D.~J. Love, T.~A. Thomas, and A.~Ghosh, ``Millimeter wave
  receiver design using low precision quantization and parallel {$\Delta
  \Sigma$} architecture,'' \emph{IEEE Trans.\ Wireless Commun.}, vol.~15,
  no.~10, pp. 6556--6569, Oct 2016.

\bibitem{GaltonJ95}
I.~Galton and H.~T. Jensen, ``Delta-sigma modulator based {A/D} conversion
  without oversampling,'' \emph{IEEE Trans. Circuits Syst. II, Analog Digital
  Signal Process.}, vol.~42, no.~12, pp. 773--784, Dec 1995.

\bibitem{scholnik2004spatio}
D.~P. Scholnik, J.~O. Coleman, D.~Bowling, and M.~Neel, ``Spatio-temporal
  delta-sigma modulation for shared wideband transmit arrays,'' in \emph{Proc.
  IEEE Radar Conf.}, 2004, pp. 85--90.

\bibitem{kriegerDense}
J.~D. Krieger, C.~P. Yeang, and G.~W. Wornell, ``Dense delta-sigma phased
  arrays,'' \emph{{IEEE} Trans. Antennas Propag.}, vol.~61, no.~4, pp.
  1825--1837, April 2013.

\bibitem{Jacobsson2016Nonlinear}
S.~{Jacobsson}, G.~{Durisi}, M.~{Coldrey}, T.~{Goldstein}, and C.~{Studer},
  ``{Nonlinear 1-bit precoding for massive MU-MIMO with higher-order
  modulation},'' in \emph{Proc. Asilomar Conf. Signals, Syst. Comp.}, Nov 2016,
  pp. 763--767.

\bibitem{Jedda2017}
H.~Jedda, A.~Mezghani, J.~A. Nossek, and A.~L. Swindlehurst, ``Massive {MIMO}
  downlink 1-bit precoding with linear programming for {PSK} signaling,'' in
  \emph{2017 IEEE 18th Int. Workshop Signal Process. Advances Wireless Commun.
  (SPAWC)}, July 2017, pp. 1--5.

\bibitem{Swindlehurst2018}
A.~Swindlehurst, H.~Jedda, and I.~Fijalkow, ``Reduced dimension minimum {BER
  PSK} precoding for constrained transmit signals in massive {MIMO},'' in
  \emph{Proc. IEEE Int. Conf. Acous., Speech, Signal Process. (ICASSP)}, 2018,
  pp. 3584--3588.

\bibitem{gray1990quantization}
R.~M. Gray, ``Quantization noise spectra,'' \emph{IEEE Trans. Inf. Theory},
  vol.~36, no.~6, pp. 1220--1244, 1990.

\bibitem{norsworthy1992effective}
S.~R. Norsworthy, ``Effective dithering of sigma-delta modulators,'' in
  \emph{Proc. 1992 IEEE Int. Symposium Circuits Syst. (ISCAS)}, vol.~3, 1992,
  pp. 1304--1307.

\bibitem{proakis2001digital}
J.~Proakis, \emph{Digital Communications}, ser. Electrical engineering
  series.\hskip 1em plus 0.5em minus 0.4em\relax McGraw-Hill, 2001.

\bibitem{liu2018symbol}
Y.~Liu and W.-K. Ma, ``Symbol-level precoding is symbol-perturbed {ZF} when
  energy efficiency is sought,'' in \emph{Proc. IEEE Int. Conf. Acous., Speech,
  Signal Process. (ICASSP)}, 2018, pp. 3869--3873.

\bibitem{wiesel2008zero}
A.~Wiesel, Y.~C. Eldar, and S.~Shamai, ``Zero-forcing precoding and generalized
  inverses,'' \emph{IEEE Trans. Signal Process.}, vol.~56, no.~9, pp.
  4409--4418, 2008.

\bibitem{shao2018globsip}
M.~Shao, Q.~Li, Y.~Liu, and W.-K. Ma, ``Multiuser one-bit massive {MIMO}
  precoding under {MPSK} signaling,'' in \emph{Proc. IEEE Global Conf. Signal
  and Inf. Process. (GlobalSIP)}, Nov. 2018, pp. 833--837.

\bibitem{nesterov1983method}
Y.~Nesterov, ``A method for unconstrained convex minimization problem with the
  rate of convergence $\mathcal{O}(1/k^2)$,'' in \emph{Doklady AN USSR}, vol.
  269, 1983, pp. 543--547.

\bibitem{beck2009fast}
A.~Beck and M.~Teboulle, ``A fast iterative shrinkage-thresholding algorithm
  for linear inverse problems,'' \emph{SIAM J. Imaging Sci.}, vol.~2, no.~1,
  pp. 183--202, 2009.

\bibitem{beck2017first}
A.~Beck, \emph{First-Order Methods in Optimization}.\hskip 1em plus 0.5em minus
  0.4em\relax SIAM, 2017, vol.~25.

\bibitem{sion1958general}
M.~Sion, ``On general minimax theorems,'' \emph{Pacific J. Math.}, vol.~8,
  no.~1, pp. 171--176, 1958.

\bibitem{condat2016fast}
L.~Condat, ``Fast projection onto the simplex and the $l_1$ ball,''
  \emph{Mathematical Programming}, vol. 158, no. 1-2, pp. 575--585, 2016.

\bibitem{bertsekas2003convex}
D.~P. Bertsekas, A.~Nedi, and A.~E. Ozdaglar, \emph{Convex Analysis and
  Optimization}.\hskip 1em plus 0.5em minus 0.4em\relax Athena Scientific,
  2003.

\end{thebibliography}

\end{document}